\documentclass[paper=a4,twoside,openany,titlepage,chapterprefix=true,pagesize,12pt,bibtotoc,pointlessnumbers,normalheadings,table]{book}
\usepackage[a4paper,left=2cm,right=2cm,top=3cm]{geometry}
\usepackage[utf8]{inputenc}
\usepackage[english]{babel}

\usepackage{setspace}

\usepackage[T1]{fontenc}
\usepackage{palatino}

\usepackage{euscript}

\usepackage{microtype,indentfirst}
\usepackage[titletoc,title]{appendix}

\usepackage[font=small]{caption}

\usepackage{physics,amsmath,amssymb,mathtools,cancel,slashbox,bm, tensor,subcaption,tcolorbox,graphicx,xhfill,float,slashed,dsfont,enumerate,upgreek}
\numberwithin{equation}{chapter}

\usepackage[hidelinks,colorlinks=true,linkcolor=blue,citecolor=teal]{hyperref}
\usepackage{pdfpages,epstopdf,epigraph,bookmark}

\usepackage{tabularx,colortbl,multicol,multirow,colortbl}

\usepackage{tikz}
\usetikzlibrary{angles,quotes,arrows.meta}
\usetikzlibrary{decorations.pathmorphing, decorations.markings, calc, shapes.geometric}

\graphicspath{ {Images/} }

\usepackage{titlesec}
\titleformat{\chapter}[display]
{\centering\bfseries\Large}
{\noindent\xrfill[0.7ex]{2pt} \ \filcenter\MakeUppercase{\chaptertitlename}  \ {\huge \thechapter} \;\xrfill[0.7ex]{2pt}}
{-2mm}
{\vspace{2mm} \huge\MakeUppercase}
[\vspace{2mm}\hrule height 2pt]

\titleformat{\section}[hang]
{\bfseries}
{}
{0mm}
{\vspace{0mm} \Large \thesection. }
[\vspace{1mm}\hrule height 1pt]

\usepackage{fancyhdr,ragged2e}
\usepackage{blindtext,pifont}
\usepackage{xcolor}

\definecolor{ufabc}{RGB}{0,90,60}

\usepackage[nameinlink]{cleveref}

\usepackage[square,numbers,sort&compress]{natbib}
\usepackage{imakeidx}
\makeindex[intoc]

\newenvironment{pagina}{}{}

\newcommand{\red}{\color{red}}

\def\mean#1{\left< #1 \right>}

\newcommand{\ce}{\text{CE}}
\newcommand{\dce}{\text{DCE}}
\newcommand{\cc}{\text{CC}}
\newcommand{\dcc}{\text{DCC}}
\newcommand{\ltau}{\text{\large $\tau$}}

\newcommand{\smallT}{{\scalebox{.5}{\textsc{$T$}}}}

 \usepackage{array}

\newcommand{\abbr}[1]{\makebox[2.6cm][l]{\textbf{#1}\leaders\hbox to 0.7em{\hss.\hss}\hfill\kern0.3em}}
 \newcommand{\symb}[1]{\makebox[2.2cm][l]{$#1$\leaders\hbox to 0.7em{\hss.\hss}\hfill\kern0.3em}}
 \newcolumntype{S}{@{}l@{\hspace{4pt}}>{\raggedright\arraybackslash}p{\dimexpr\linewidth-2.2cm-4pt\relax}@{}}

\Crefname{subsection}{Subsection}{Subsections}

\newcommand{\cev}[1]{\reflectbox{\ensuremath{\vec{\reflectbox{\ensuremath{#1}}}}}}

\def\shrinkage{2.1mu}
\def\vecsign{\mathchar"017E}
\def\dvecsign{\smash{\stackon[-2.2pt]{\mkern-\shrinkage\vecsign}{\rotatebox{180}{$\mkern-\shrinkage\vecsign$}}}}
\def\dvec#1{\def\useanchorwidth{T}\stackon[-4.2pt]{#1}{\,\dvecsign}}
\usepackage{stackengine}
\stackMath

\title{Configurational Information and Hadron Spectroscopy in Holographic QCD}
\author{P.H.O. da Silva}

\begin{document}
	
	\frontmatter
	
\begin{titlepage}
		\begin{center}
			{\large
				
				\includegraphics[width=0.15\linewidth]{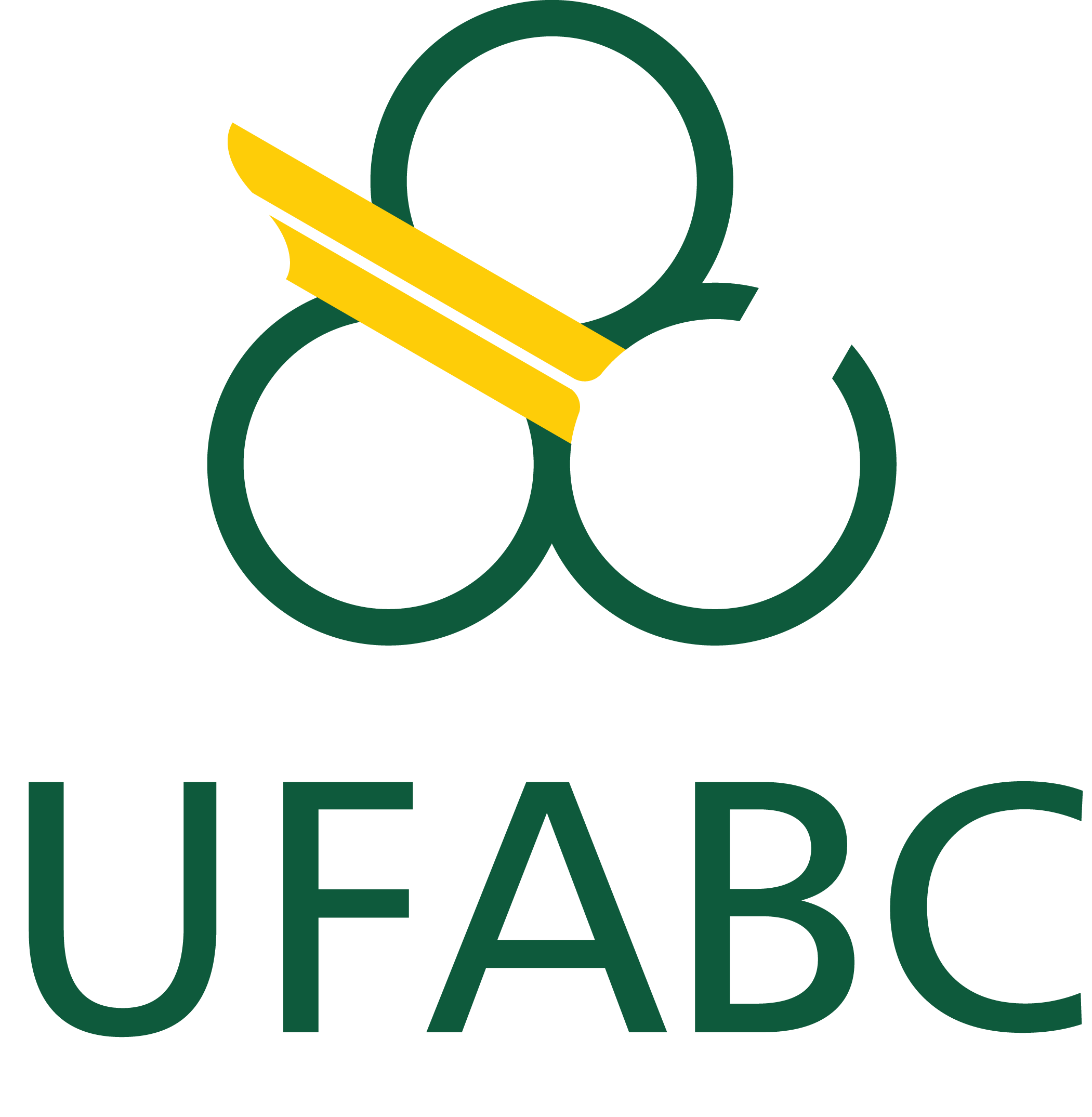}
				
					UNIVERSIDADE FEDERAL DO ABC \\ Programa de Pós-Graduação em Física
							
				\vspace{3cm}
				
				Pedro Henrique Oliveira da Silva} \\[3cm]
			
			\vspace{5mm}
			{\Huge \bf Configurational Information and Hadron Spectroscopy in Holographic QCD} \\
			\vspace{5mm}
			\vspace*{\fill}

			{\large Santo André - SP
				
				2026}
		\end{center}
	\end{titlepage}

\begin{pagina}
	\begin{center}
		{\large Pedro Henrique Oliveira da Silva}\\[6cm]
		
		{\Huge \bf Configurational Information and Hadron Spectroscopy in Holographic QCD}\\[4cm]

		\hspace{.37\textwidth}
		\begin{minipage}{.45\textwidth}
			{Tese apresentada ao Programa de Pós-Graduação em Física da Universidade Federal do ABC
				(UFABC), como requisito parcial à obtenção do título de Doutor em Física.
				\\
				
				Orientador: Prof. Dr. Roldão da Rocha Junior \\}

		\end{minipage}
		
		\vfill
		{\large Santo André - SP
			
			2026}
		
		\setcounter{page}{1}
		
		\thispagestyle{empty}
	\end{center}
\end{pagina}

\begin{pagina}
	
	\includepdf[pages=-]{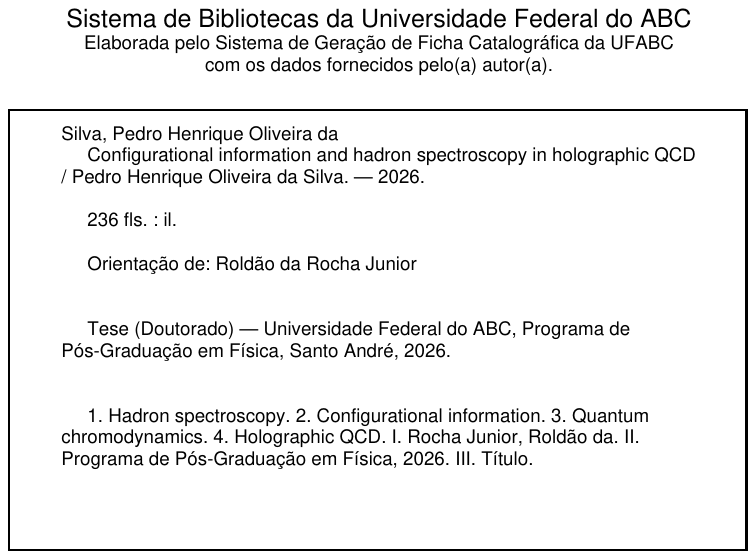}
	
\end{pagina}

\begin{pagina}
	
	\includepdf[pages=-]{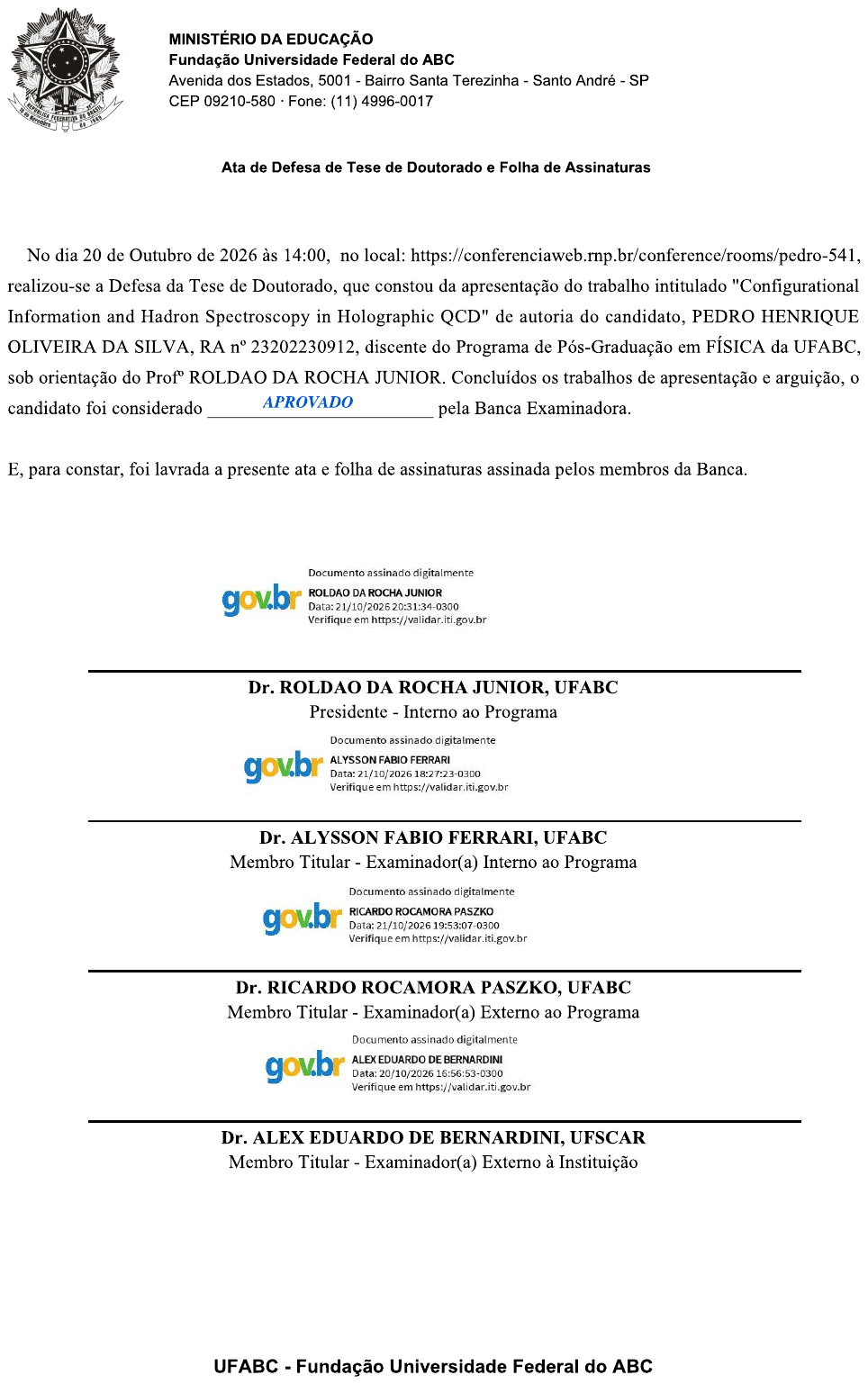}
	
\end{pagina}

\begin{pagina}
	\pagebreak
	
	\thispagestyle{empty}
	
	\mbox{}
	\vfill	
	\begin{center}
		Este exemplar foi revisado e alterado em relação à versão original, de acordo com as observações levantadas pela banca examinadora no dia da defesa, sob responsabilidade única do autor e com a anuência do orientador.
	\end{center}

		\vfill
\end{pagina}

\begin{pagina}
	\pagebreak
	
	\thispagestyle{empty}
	
	\mbox{}
	\vfill	
	\begin{center}
		This study was financed in part by the Coordenação de Aperfeiçoamento de Pessoal de Nível Superior – Brasil (CAPES) – Finance Code 001.
	\end{center}
	
	\vfill
\end{pagina}

\begin{pagina}
	\pagebreak	
	
	\thispagestyle{empty}
	
	\mbox{}
	\vfill
	
	\begin{minipage}{0.4\textwidth}
		\hspace{2mm}
	\end{minipage}
	\begin{minipage}{0.45\textwidth}
		\it Aos meus pais, que essas poucas palavras sejam suficientes para expressar toda minha eterna gratidão.
	\end{minipage}
	
\end{pagina}

\begin{pagina}
	\begin{center}
		\chapter*{Acknowledgements}
	\end{center}
	
		First and foremost, I would like to thank my family for their unconditional support throughout all these years, without which I could never have even dreamed of reaching this point. Even without understanding exactly my work, you never doubted me or stopped supporting my dreams. Few things in life compare to the peace of mind of knowing there is a home to always return to, regardless of the circumstances. This assurance was, without a doubt, one of the greatest foundations of my journey.
	
		To my partner, I am deeply grateful for your love and understanding. Thank you for accepting the countless occasions when the time we could have shared had to be dedicated to this work, and for turning that sacrifice into encouragement.
	
		To my advisor, I am grateful for all the lessons taught and the partnership we have built over these years. I am especially thankful for your patience and willingness to answer even the simplest questions with clarity. Your attention to detail, your care with every stage of the work, and your respect for my pace demonstrated a commitment that went far beyond academic advising. I cannot imagine the completion of this work, nor my development as a researcher, without your guidance.
		
		To the UFABC professors who have been part of my trajectory since my undergraduate studies, thank you for contributing, each in your own way, to my academic and professional development. I would also like to thank UFABC for welcoming me and providing an environment of learning, growth, and opportunity.
		
		To my friends, thank you for making this journey lighter and for providing moments of peace and joy amidst the exhaustion and uncertainties. Your friendship was fundamental in allowing me to walk this path without losing sight of the importance of taking care of my physical and mental health.
		
		As someone who was able to remain at the university thanks to student assistance policies, I would like to express my special gratitude for this support throughout a significant part of my education. I also thank UFABC and CAPES for the financial support granted throughout this work.
		
		Finally, I extend my gratitude to everyone who, directly or indirectly, contributed to making this journey possible. To all of you, thank you very much.

	{\red

	}
	
	\thispagestyle{plain}
	\pagebreak
\end{pagina}

\begin{pagina}
	\thispagestyle{plain}
	
	\epigraph{There is a theory which states that if ever anyone discovers exactly what the Universe is for and why it is here, it will instantly disappear and be replaced by something even more bizarre and inexplicable. There is another theory which states that this has already happened.}{Douglas Adams}
	
\end{pagina}

\begin{pagina}
	
	\begin{center}
		\chapter*{Resumo}
	\end{center}

	\begin{otherlanguage}{portuguese}
		Baseadas nos conceitos de entropia e complexidade, as Medidas de Informação Configuracional (CIMs, do inglês \textit{Configurational Information Measures}) são ferramentas capazes de extrair informações sobre sistemas físicos a partir de sua densidade de energia espacialmente localizada, determinando propriedades como limites de estabilidade e transições de fase.
		Nesta tese, as CIMs são empregadas para investigar a espectroscopia de hádrons. Utilizando modelos fenomenológicos de AdS/QCD em conjunto com as CIMs, as ressonâncias catalogadas de mésons e bárions são interpoladas em estruturas análogas às trajetórias de Regge, de onde se derivam expressões analíticas que permitem estimar a massa de estados excitados de ordem superior. 
		Tais estimativas são comparadas com dados experimentais em busca de possíveis correspondências que podem elucidar as propriedades de partículas detectadas, mas ainda não identificadas.
		A partir desse método híbrido de AdS/QCD baseado em CIMs, este trabalho analisa mésons leves em um modelo dinâmico, kaons vetoriais estranhos em um modelo que leva em consideração a massa dos quarks constituintes e bárions de sabor leve em um regime de temperatura finita. 
	
	\vspace{5mm}
	\textbf{Palavras-chave:} Medidas de Informação Configuracional, Entropia Configuracional, Complexidade Configuracional, Espectroscopia de Hádrons, Trajetórias de Regge, AdS/QCD.
	
	\end{otherlanguage}

	\thispagestyle{plain}	
\end{pagina}

\begin{pagina}
	
	\begin{center}
		\chapter*{Abstract}
	\end{center}
	
	Based on the concepts of entropy and complexity, Configurational Information Measures (CIMs) are tools capable of extracting information regarding physical systems from their spatially localized energy density, determining properties such as stability limits and phase transitions.
	In this thesis, CIMs are employed to investigate hadron spectroscopy. By utilizing phenomenological AdS/QCD models in conjunction with CIMs, cataloged meson and baryon resonances are interpolated into structures analogous to Regge trajectories, from which analytical expressions are derived to estimate the masses of higher-order excited states. These estimates are compared with experimental data to search for possible correspondences that may elucidate the properties of particles that have been experimentally detected but remain unidentified.
	Using this CIM-based AdS/QCD hybrid method, this work analyzes light mesons within a dynamical model, strange vector kaons in a framework that accounts for constituent quark masses, and light-flavor baryons at finite temperature.
	
	\vspace{5mm}
	\textbf{Keywords:} Configurational Information Measures, Configurational Entropy, Configurational Complexity, Hadron Spectroscopy, Regge Trajectories, AdS/QCD.

	\thispagestyle{plain}	
	
\end{pagina}

\begin{pagina}
	\thispagestyle{plain}	
	
	{
		\chapter{List of Published Papers}
		\hypersetup{linkcolor=black}
		
		\begin{enumerate}
			\item R. da Rocha and P.H.O. Silva.\\					
				Configurational entropy and shape complexity of strange vector kaons in AdS/QCD. \\
				\textit{Eur. Phys. J. Plus} \textbf{138}, pp. 729, 2023. 
			\item R. da Rocha and P.H.O. Silva.\\					
				Deformed AdS/QCD, mesonic mass spectra, and DCE: Still a margin for heavier resonances. \\
				\textit{Phys. Rev. D} \textbf{110}, pp. 126019, 2024. 
			\item R. da Rocha and P.H.O. Silva.\\					
				Higher-spin light-flavor baryonic spectroscopy in AdS/QCD at finite temperature. \\
				\textit{Phys. Lett. B} \textbf{870}, pp. 139917, 2025.
			\item H. Almeida, R. da Rocha, P.H.O. Silva, B. Toniato.\\
				Holographic information measures for spin-$3/2$ $\Delta$ baryons in AdS/QCD.\\ 
				\textit{Nucl. Phys. B} \textbf{1072}, pp. 117458, 2026.
		\end{enumerate}	
		
	}
	
\end{pagina}

\begin{pagina}
	\thispagestyle{plain}	
	
	\clearpage
	\phantomsection
	\addcontentsline{toc}{chapter}{List of Figures}
	\hypersetup{linkcolor=black}
	\listoffigures
		
\end{pagina}

\begin{pagina}
	
	\clearpage
	\phantomsection
	\addcontentsline{toc}{chapter}{List of Tables}
	\hypersetup{linkcolor=black}
	\listoftables

\end{pagina}

\begin{pagina}
	
\chapter{List of Abbreviations}
	
		\begingroup
		\small\renewcommand{\arraystretch}{1.2}\setlength{\tabcolsep}{4pt}\noindent
		\begin{minipage}{0.485\textwidth}
			\begin{tabular}{@{}ll@{}}
				\abbr{AdS}     & Anti-de~Sitter \\
				\abbr{AdS/CFT} & \begin{tabular}[t]{@{}l@{}}Anti-de~Sitter/Conformal\\ field theory\end{tabular} \\
				\abbr{AdS/QCD} & \begin{tabular}[t]{@{}l@{}}Anti-de~Sitter/Quantum\\ chromodynamics\end{tabular} \\
				\abbr{BH}      & Black hole \\
				\abbr{CC}      & Configurational complexity \\
				\abbr{CE}      & Configurational entropy \\
				\abbr{CFT}     & Conformal field theory \\
				\abbr{CIM}     & \begin{tabular}[t]{@{}l@{}}Configurational information\\ measure\end{tabular} \\
				\abbr{DCC}     & \begin{tabular}[t]{@{}l@{}}Differential configurational\\ complexity\end{tabular} \\
				\abbr{DCE}     & \begin{tabular}[t]{@{}l@{}}Differential configurational\\ entropy\end{tabular} \\
				\abbr{DSW}     & Dynamical soft-wall (model) \\
				\abbr{GKPW}    & \begin{tabular}[t]{@{}l@{}}Gubser-Klebanov-Polyakov-\\ Witten\end{tabular}
			\end{tabular}
		\end{minipage}\hfill
		\begin{minipage}{0.485\textwidth}
			\begin{tabular}{@{}ll@{}}
				\abbr{HW}      & Hard-wall (model) \\
				\abbr{IR}      & Infrared \\
				\abbr{LHC}     & Large Hadron Collider \\
				\abbr{PDG}     & Particle Data Group \\
				\abbr{QCD}     & Quantum chromodynamics \\
				\abbr{QED}     & Quantum electrodynamics \\
				\abbr{QFT}     & Quantum field theory \\
				\abbr{QGP}     & Quark--gluon plasma \\
				\abbr{RE}      & Relative error \\
				\abbr{RMSD}    & Root mean square deviation \\
				\abbr{SAdS}    & Schwarzschild--AdS \\
				\abbr{SCT}     & \begin{tabular}[t]{@{}l@{}}Special conformal\\ transformation\end{tabular} \\
				\abbr{SW}      & Soft-wall (model) \\
				\abbr{SYM}     & Super Yang--Mills \\
				\abbr{ThAdS}   & Thermal AdS \\
				\abbr{UV}      & Ultraviolet \\
				\abbr{VEV}     & Vacuum expectation value
			\end{tabular}
		\end{minipage}
		\endgroup

\end{pagina}

\begin{pagina}
	\newpage
	\renewcommand\contentsname{Table of Contents}
	\pdfbookmark[chapter]{\contentsname}{toc}
	\hypersetup{linkcolor=black}
	\tableofcontents
	
	\pagebreak
\end{pagina}

\mainmatter

    \chapter{Introduction}

	\section{Background}
	
	The second half of the 20th century was the scene of an unprecedented and accelerated expansion in the development of physics. Driven by experimental discoveries and the creation of new mathematical tools, many models and theories emerged with the aim of not only describing phenomena but also bringing some order to the structure of nature. As part of this search for order, describing particles as excitations of fields allowed for their dynamics to be described through a \textit{quantum field theory} (QFT). One of the consequences of this approach is that it becomes possible to simplify the variety of forces in nature into just four fundamental interactions --- electromagnetic, weak, strong, and gravitational --- which are responsible for determining how the fields interact with each other.
	
	The electromagnetic interaction, which had already been well understood in the classical regime since the previous century, was quickly characterized by the first successful QFT, \textit{quantum electrodynamics }(QED). Subsequently, the principles of the weak interaction were also unraveled.
	Meanwhile, the strong interaction presented some challenges that made the task of investigating it considerably more laborious. As the name suggests, the interaction is strongly coupled, which prevents the use of the main perturbative methods employed in the analysis of other QFTs. Moreover, its most fundamental particles, quarks and gluons, possess the property of confinement, which prevents them from being detected individually. Therefore, from a certain energy scale, the realm governed by the strong interaction is impregnable.
	
	The quest for a theory for the strong interaction led to the creation of string theory, which managed to explain some phenomena and properties of the known particles using its one-dimensional objects, the strings, as a way to model the flux tubes that confine quarks inside hadrons. However, it also presented debilitating problems, such as the need for extra dimensions and the appearance of a never-detected spin-2 particle. Eventually, \textit{quantum chromodynamics} (QCD), the most complete and robust QFT for the strong interaction ever developed, put an end to this quest, although some of the same challenges from before still haunt the theory to this day.
	
	One of the triumphs of QCD was its seamless integration with the other two interactions. Electromagnetism and weak interaction had not only been explained but also unified in the electroweak theory. Later, QCD was also unified with the other two. The essence of this unification lies in detecting the symmetries of the interactions and identifying them as different aspects of the same type of field theory, a gauge theory.
	This was an indication that the four interactions should be unified at some point in the history of the universe, so there should be some unique and self-contained theory capable of describing all interactions. The best we currently have in this regard is the standard model, but it still has some unresolved issues, among them the absence of a QFT that explains gravitational interaction, a task currently performed by general relativity, a classical field theory.
	
	In this context, string theory returned even more powerful. Throughout the 1980s and 1990s, new understandings (even called revolutions, given their impact) about strings expanded their applicability and elevated it as a candidate for a theory of everything, capable of consistently unifying the four interactions through a quantum theory of gravity. What was previously considered an aberration, the spin-2 particle, could, in this framework, be interpreted as the graviton, the elementary particle responsible for transmitting gravity.
	
	Regardless of its eventual success or failure in this endeavor, string theory has already produced very relevant derivatives. One of them is the AdS/CFT correspondence proposed by Juan Maldacena in 1997 \cite{Maldacena:1997re}. Originally, Maldacena found that a string theory in the 10-dimensional spacetime of AdS$_5\times S^5$ (anti-de Sitter space in 5 dimensions with a 5-dimensional sphere) was equivalent to a theory that is both gauge and conformal existing on the boundary of that same space, which in that case is the familiar 4-dimensional Minkowski spacetime.
	
	At first, this result suggests a gauge/string duality, but within appropriate limits, string theory behaves as a theory of gravity. Hence, what emerges is a gauge/gravity duality. The specifics of each side of the correspondence, as well as the conditions necessary for its validity, will be explored in more detail throughout this thesis. In general terms, the correspondence establishes that a gravitational theory can be equivalent to a gauge theory defined on the boundary of its spacetime. The key to this duality lies in the fact that its construction allows one theory to be strongly coupled while the other remains weakly coupled. This property enables calculations to be performed on the weakly coupled side using well-established methods, with the results then translated into their equivalents on the strongly coupled side, where standard analytical techniques would otherwise be inapplicable \cite{Gubser:1998bc,Witten:1998qj,Aharony:1999ti}.
	
	The AdS/CFT correspondence appears ideally suited for investigating QCD in its strongly coupled regime, as long as a corresponding higher-dimensional theory can be identified. However, constructing such a theory remains a highly nontrivial task. Different approaches can be employed to build an AdS/QCD model (also known as holographic QCD), each with its own advantages and limitations.
	Despite these challenges, AdS/QCD models have proven capable of reproducing and estimating many properties confirmed by experiments and phenomenology, as will be discussed extensively throughout this thesis.
	
	This work presents applications of holographic QCD in combination with configurational information measures, originally developed by Marcelo Gleiser and Nikitas Stamatopoulos \cite{Gleiser:2011di}. Initially designed to study objects with spatially-localized energy, such as solitons, these tools have found broader applications in measuring the informational content of physical systems using a formalism based on information theory.
	
	Particle catalogs are filled with detected states whose properties have not yet been fully identified, as only some characteristics have been determined. With this in mind, we use configurational information measures to extract the properties of mesons and baryons from experimental data, creating a robust foundation. These quantities are then extrapolated to estimate the properties of new, higher excited states, which may correspond to some of the as-yet-unidentified states. Therefore, we aim to construct a method to identify states based on well-established experimental data, thereby enhancing the predictive power of the approach.
	
	\pagebreak
	
	\section{Notations and conventions}
	
	Throughout this thesis, we will follow some notations and conventions summarized in this section, but which can also be reintroduced within a more appropriate context in the text.
	
	We will use Planck's natural units, $c=G=\hbar=1$, unless a specific context establishes otherwise. These exceptions occur in order to highlight a particular property or principle and will be explicitly noted. In some cases, other constants will also be assumed to be unity, especially to avoid repetition of symbols and to make equations cleaner.
	
	The metric signature convention used will be mostly plus, so the 4-dimensional Minkowski spacetime will have signature $(-+++)$ for the 3+1 dimensions, with its metric given by $\eta^{\mu\nu} = \text{diag}(-1,+1,+1,+1)$. In a general case of $d+1$ dimensions, the metric will have signature $(-++\dots+)$ and will be denoted by $g^{MN}\propto \text{diag}(-1,+1,+1,\dots,+1)$.
	In this way, the negative sign is carried exclusively by the temporal coordinate, so that the additional dimensions share the same signature as the remaining spatial ones. The particular details of the metrics will be addressed in their respective contexts.
	
	Latin capital letters ($A,B,\dots$) are reserved to label coordinates in a spacetime with a number of dimensions larger than four.	
	Greek letters ($\mu,\nu,\dots$) label all coordinates in a 4-dimensional spacetime, thus ranging from 0 to 3, encompassing the temporal coordinate and the spatial coordinates.
	Latin letters ($i,j,\dots$) are reserved to label spatial coordinates, regardless of the number of dimensions in spacetime.
	
	Building on these index conventions, we denote vectors in spacetime by lightface letters (e.g., $x$). Their components are expressed using the appropriate index for the given dimensionality, such as $x^\mu$ in 4D or $x^M$ in higher dimensions. In contrast, purely spatial vectors are distinguished by boldface type (e.g., $\boldsymbol{x}$). When referring to their components, the boldface is dropped in favor of a Latin index, such that the components of $\boldsymbol{x}$ are denoted by $x^i$.
	
	\pagebreak
	
	\section{Overview}
	
	This thesis is organized as follows:
	
	\begin{itemize}
		\item[\ding{43}] \Cref{chapter:fundamentals} provides an introductory review of key concepts necessary for understanding this work.
		It begins with a review of the holographic principle, followed by explanations of the duality's two sides: anti-de Sitter space and conformal field theories, along with their respective symmetries. The chapter also presents a concise overview of string theory and super Yang-Mills theory, culminating in the development of the AdS/CFT correspondence. In the second part, the chapter introduces some basic elements of quantum chromodynamics, such as chiral symmetry and hadron structure.
		
		\item[\ding{43}] \Cref{chapter:cims} introduces the main tools used in this thesis. The first section is dedicated to an exposition about AdS/QCD models, with a focus on the soft-wall model. Then, after a brief introduction to some concepts from information theory, the configurational information measures are detailed and how to integrate them with AdS/QCD models is presented. The chapter then presents applications of these approaches to different types of mesons at zero temperature.
		
		\item[\ding{43}] In \Cref{chapter:thermalqcd}, the discussion is expanded to a finite-temperature scenario. First, the introduction of temperature in quantum field theory is presented and then extended to AdS spacetime, including a glimpse of phase transitions. Next, the treatment of hadrons within thermal AdS/QCD models is discussed. Finally, the configurational entropy is applied to baryons, more specifically to nucleons with different spin. In addition to the standard spectroscopic analysis, the investigation is further extended to employ configurational entropy as a tool to obtain insights into the phase transition temperature.
		
		\item[\ding{43}] \Cref{chapter:conclusion} summarizes the main results obtained for the three scenarios investigated. It highlights the possible correspondences between the estimated masses of higher excited states and unidentified states in the PDG, as well as the role of configurational entropy and complexity as indicators of hadronic stability.

	\end{itemize}
	
\chapter{AdS/QCD Fundamentals}
\label{chapter:fundamentals}

	The AdS/QCD correspondence refers to models that utilize the gauge/gravity duality to study properties of quantum chromodynamics, more specifically to analyze its non-perturbative regime, where other methods are often ineffective. This concise definition, however, is built upon a vast framework of knowledge, without which it is impossible to fully comprehend the applications presented in this thesis.
	For this reason, this chapter aims to introduce the key elements that constitute the AdS/QCD correspondence, emphasizing the aspects most crucial for a proper understanding of this work.

	The starting point will be the proposition of the holographic principle, which offers a first approach to encoding information between spaces of different dimensions. \Cref{section:holographic} provides a brief historical overview of the holographic principle, tracing its origins in the investigation of information in black holes.
	
	Introducing the two sides of the gauge/gravity correspondence in more detail, \Cref{section:adsspace} will address the concept of higher-dimensional spaces, focusing specifically on anti-de Sitter space, while \Cref{section:cft} provides a concise review of symmetries and conformal field theories.
	Next, \Cref{section:adscftcorrespondence} will cover in more detail the formulation of the AdS/CFT correspondence, including a very brief overview of both $\mathcal{N}=4$ super Yang-Mills theory and string theory, which are essential to the original conjecture. The section concludes by presenting the core of the correspondence and its different limits.
		
	Finally, \Cref{section:qcd} presents a review of quantum chromodynamics, with subsections focusing on topics pertinent to this work, such as chiral symmetry breaking and a brief general overview of hadrons and hadron spectroscopy.

	\section{The holographic principle}
		\label{section:holographic}\index{holographic principle}
	Einstein's \textit{general relativity}\index{general relativity} (GR) establishes gravity as the result of the dynamics of matter and energy on a curved spacetime. This relation between energy distribution and the geometry of spacetime is mathematically described by the field equations
	\begin{flalign}
		R_{\mu\nu} - \frac{1}{2}Rg_{\mu\nu}+\Lambda g_{\mu\nu} = 8\pi T_{\mu\nu},
		\label{einstein}
	\end{flalign}
	where the Ricci curvature tensor $R_{\mu\nu}$ and the Ricci scalar $R=g^{\mu\nu}R_{\mu\nu}$ represents the geometric side, while the energy-momentum tensor $T_{\mu\nu}$ represents the density and flux of energy and momentum. The parameter $\Lambda$ is the cosmological constant. 
	
	The first exact solution of (\ref{einstein}) (other than the trivial flat space with $R_{\mu\nu}=0$) was found by Schwarzschild assuming $\Lambda=0$. It represents a homogeneous, uncharged, static and spherically symmetric body and is given by \cite{Schwarzschild:1916ae}
	\begin{flalign}
		ds^2 = -\left(1-\frac{2M}{r}\right)dt^2 + \left(1-\frac{2M}{r}\right)^{-1}dr^2 + r^2d\Omega^2,
		\label{schwarzschild}
	\end{flalign}
	where $M$ is the mass of the object and $d\Omega^2=d\theta^2+\sin^2\theta d\phi^2$ is the solid angle element. 
	Years later, Eq. (\ref{schwarzschild}) would be recognized as the simplest black hole \index{black holes} solution, since it describes a singularity at $r=0$ encompassed by an event horizon at $r=r_S=2M$ (the Schwarzschild radius) with area $A=4\pi r_S^2 = 16\pi M^2$. A simple but important result seen here is that the horizon area of a Schwarzschild black hole grows with the square of its mass.
	
	The simplicity of the Schwarzschild solution lies in the fact that only the mass of the black hole is taken into account and fully characterizes the object.		
	Over time, new solutions were developed and new parameters began to characterize black holes, such as electric charge (Reissner-Nordström solution \cite{Reissner:1916cle,Weyl:1917gp,Nordstrom:1918,Jeffery:1921}) and its angular momentum, i.e., its rotation (Kerr solution \cite{Kerr:1963ud}), or a combination of these two (Kerr-Newman solution \cite{Newman:1965my}).
	
	But a curious result with an equally curious name, the \textit{no-hair theorem}\index{black holes!no-hair theorem}, states that only mass, charge and angular momentum characterize a black hole. This theorem means that regardless of what the black hole absorbed (e.g., a star or a gas cloud) or what were its initial formation conditions, its final state will be characterized only by these three properties.		
	
	The no-hair theorem raises an important issue regarding the fate of information and entropy that falls into a black hole. If an object absorbed by a black hole takes all its information with it, then its entropy appears to disappear from the perspective of an observer outside the event horizon. This appears to violate the second law of thermodynamics, as it would imply that the universe is losing entropy.
	
	The solution to this problem may lie in another theorem, the \textit{area theorem}\index{black holes!area theorem}, proposed by Stephen Hawking \cite{Hawking:1971tu}. This theorem states that, since nothing comes out from the black hole (at least not classically), its horizon area never decreases with time:		
	\begin{equation}
		\frac{dA}{dt}\ge 0.
		\label{areatheorem}
	\end{equation}

	The implications of the theorem extend further, showing that if two black holes merge, the area of the resulting horizon will exceed the sum of the areas of the original ones. Similarly, when an object falls into a black hole, its mass increases, which directly causes the horizon area to expand. This leads to the conclusion that the loss of entropy when the object crosses the event horizon can be compensated by the increase in the black hole’s horizon area.

	In Ref. \cite{Bekenstein:1972tm}, Bekenstein proposed a relation between the black hole entropy $S_\text{BH}$ and the horizon area for a Schwarzschild black hole in the form
	\begin{equation}
		S_\text{BH} = \frac{\eta k_B}{\ell_P^2} A,
		\label{blackholeentropy}
	\end{equation}
	where we made a brief exception in the notation to explicit the Boltzmann constant $k_B$ and the Planck length $\ell_P=(\hbar G_N/c^3)^{1/2}$, and $\eta$ is a constant number of order unity, later determined to be $\eta=1/4$ \cite{Hawking:1974rv}.
	
	From thermodynamics, it is known that entropy is an extensive property, meaning it depends on the amount of material in the system, making it additive: the total entropy of two independent systems is equal to the sum of their individual entropies. For this reason, entropy typically depends on the system's volume. However, in expression (\ref{blackholeentropy}), entropy instead depends on a quantity with one less dimension --- specifically, the horizon area of the black hole.
	Therefore, this result suggests that all the information contained within a black hole, as well as everything that has fallen into it, can be entirely encoded on its boundary, the surface defined by the event horizon.
	
	Based on this result, Gerard 't Hooft \cite{tHooft:1993dmi} proposed a generalization of what Bekenstein obtained. He suggested that the physics of any quantum theory with gravity can be completely encoded and described in the lower-dimensional boundary of the same spacetime in which this theory is defined. This idea is known as the \textit{holographic principle} and was later formalized by Leonard Susskind \cite{Susskind:1994vu}, who applied it to string theory. The name "holographic" comes from how a hologram works by encoding all the information in a three-dimensional system into a two-dimensional projection.
	
	The AdS/CFT correspondence is currently the most successful application of the holographic principle. It establishes that quantum gravitational theories in anti-de Sitter (AdS) space are equivalent to conformal field theories (CFT) defined on the boundary of that space. By mapping theories such as quantum chromodynamics to a higher-dimensional AdS space, the holographic principle enables the use of established techniques in this scenario, simplifying the original problems and providing new insights into the underlying physics.
	
	The details of these processes and the extension of this principle to the AdS/QCD correspondence will be discussed in the following sections with appropriate detail.
	
	\section{The anti-de Sitter space}
	\label{section:adsspace}
	
	In \Cref{section:holographic}, it was presented that, through the holographic principle, it is possible to establish a correspondence between a theory defined within the bulk of a spacetime and a theory described on the boundary of that space, such that both theories share corresponding information despite existing in different dimensions.	
	To construct this equivalence, it is first necessary to identify a spacetime whose boundary is itself also a spacetime, with both capable of sustaining a theory.
	
	Initially, it is convenient to work with more familiar elements. An example is a sphere, specifically a 2-sphere (denoted as $S^2$), described by
	\begin{flalign}
		x^2+y^2+z^2=L^2,
	\end{flalign}
	with $L$ being the radius. The sphere is embedded on a 3-dimensional Euclidean space,
	\begin{flalign}
		ds^2 = dx^2 + dy^2 + dz^2 = L^2(d\theta^2+\sin^2\theta d\phi^2).
	\end{flalign}

	The sphere is not merely a simple geometric object but a curved $(d+1)$-dimensional space embedded within a flat $(d+2)$-dimensional space. It has a constant positive curvature and is invariant under the action of the group SO$(3)$, which represents all rotations in $\mathbb{R}^3$. This invariance ensures that any point on the sphere, when subjected to a rotation, will map to another point on the sphere. Consequently, the sphere lacks any preferred direction or specific location, characterizing it as both homogeneous and isotropic.
	
	A space that is simultaneously homogeneous and isotropic is referred to as \textit{maximally symmetric} \cite{Weinberg:1972kfs}. Likewise, there exist maximally symmetric spacetimes\index{maximally symmetric spacetimes}, such as Minkowski spacetime, which serves as a solution to Einstein's equations (\ref{einstein}) in vacuum with constant curvature $R=0$ and a vanishing cosmological constant $\Lambda=0$.
	Two additional maximally symmetric spacetimes of significant interest are the \textit{de Sitter} and the \textit{anti-de Sitter} spaces.
	
	The de Sitter spacetime \index{de Sitter spacetime} dS has a constant positive curvature, so it is defined by a manifold in $(d+1)$-dimensions \cite{Nastase:2015wjb}
	\begin{flalign}
		-X^2_0 + \sum_{i=1}^{d}X^2_i + X^2_{d+1} = L^2,
	\end{flalign} 
	with $L$ being the radius of curvature, embedded in a $(d+2)$-dimensions space,
	\begin{flalign}
		ds^2 = -dX_0^2+\sum_{i=1}^{d}dX^2_i+ dX^2_{d+1},
	\end{flalign}
	where the minus sign in the first term comes from the signature $(-+\dots+)$. One can see that this is a generalization of a $(d+1)$-sphere embedded in a $(d+2)$-dimensional Minkowski spacetime. For this reason, the de Sitter spacetime in $(d+1)$-dimensions is invariant under transformations of the SO$(1,d+1)$ group. 
	
	The details of the dS space will not be covered in this work, but it is worth noting a fundamentally important result: it is a vacuum solution of Einstein's equations with a positive cosmological constant, making it important in cosmological models of the accelerated expansion of the universe and the study of the universe's late-time evolution in the presence of dark energy.\\
	
	The	anti-de Sitter spacetime \index{anti-de Sitter spacetime} AdS in $(d+1)$-dimensions is represented by a hyperboloid of radius $L$ \cite{Nastase:2015wjb}
	\begin{flalign}
		-X^2_0 + \sum_{i=1}^{d}X^2_i - X^2_{d+1} = -L^2
		\label{ads}
	\end{flalign}
	embedded in a $(d+2)$-dimensions space
	\begin{flalign}
		ds^2 = -dX_0^2+\sum_{i=1}^{d-1}dX^2_i- dX^2_{d+1}.
		\label{adsmetric}
	\end{flalign}

	It  has a constant negative curvature and is a vacuum solution of Einstein's equations with a negative cosmological constant. A possible solution to (\ref{ads}) corresponds to the set of so-called \textit{global coordinates}\index{global coordinates} ($\tau,\rho,\Omega_i$), with \cite{Aharony:1999ti}
	\begin{flalign}
		X_0&=L\sec\rho \cos\tau, \nonumber\\
		X_i &= L\tan\rho \; \Omega_i,\nonumber\\
		X_{d+1}&= L\sec\rho\sin\tau,
	\end{flalign}
	where $0\le\rho<\pi/2$, $-\pi<\tau\le\pi$ and $-1\le\Omega_i\le1$, with $i=1,\dots,d$ and $\sum_i^d\Omega_i^2=1$. The line element then becomes
	\begin{flalign}
		ds^2 = \frac{L^2}{\cos^2\rho}\left(-d\tau^2+d\rho^2+\sin^2\rho\; d\Omega^2\right).
	\end{flalign}
	
	These coordinates provide a complete coverage of AdS spacetime. However, the $\tau$ coordinate presents a notable issue: it has a compact range, which can lead to the generation of closed time-like curves, thereby disrupting causality. To fix this problem, the coordinate is unwrapped considering $-\infty<\tau<\infty$. This results in what is known as the \textit{universal cover} of AdS\index{global coordinates!universal cover of AdS}, which is a spacetime structure that is fully covered by coordinates and maintains causality. As usual in the literature, whenever we mention the AdS spacetime in this work, we mean this covering space.
	
	There are at least three coordinate systems (other than the global one) that can be used to solve Eq. (\ref{ads}), each with its own advantages and specific applications. \textit{Static coordinates} ($\tilde{\tau},\tilde{r},\Omega_i$), for example, are recommended for dealing with black holes, since, when defined from global coordinates as $\tilde{r}=L\sinh\rho$ and $\tilde{\tau}=L\tau$, the metric is written as
	\begin{flalign}
		ds^2 = -\left(1+\frac{\tilde{r}^2}{L^2}\right)d\tilde{\tau}^2 + \left(1+\frac{\tilde{r}^2}{L^2}\right)^{-1}d\tilde{r}^2 + \tilde{r}^2d\Omega^2_{d-1},
		\label{staticcoordinates}
	\end{flalign}
	which resembles the Schwarzschild solution (\ref{schwarzschild}).
	
	For the purposes of this work, the best system is the \textit{Poincaré coordinates}\index{Poincaré coordinates} $(t,x^i,z)$, which are commonly used in the context of AdS/CFT \cite{Natsuume:2014sfa}.
	With these coordinates we find a solution to (\ref{ads}) in the form of
	\begin{flalign}
		X_0 &=\frac{1}{2z}\left(z^2+L^2+(x^i)^2-t^2\right),  \nonumber\\
		X_i &=\frac{Lx^i}{z},\nonumber\\
		X_d &=-\frac{1}{2z}\left(z^2-L^2+(x^i)^2-t^2\right),\nonumber\\
		X_{d+1}&=\frac{Lt}{z},	
	\end{flalign}
	with $i=1,2,\dots,d-1$ and $z\ge0$. In these coordinates the line element takes on a much cleaner form:
	\begin{flalign}
		ds^2 = \frac{L^2}{z^2}\left(-dt^2 + (dx^i)^2 + dz^2\right).
		\label{eq:ads_poincare}
	\end{flalign}
	One can see that, given a fixed value of $z$, the resulting line element is the same as in the Minkowski spacetime. In fact, one can also notice that this is a Minkowski space up to a conformal factor, so we say that anti-de Sitter space is \textit{conformally flat}. 
	Therefore, the AdS space can be seen as a stack of Minkowski spacetimes of one lower dimension, each one existing in a different value of $z$. 
	
	To express the geometry more formally using tensor notation, we identify the 5-dimensional coordinates $(t,x^1,x^2,x^3,z)$ as $X^M = (x^0, x^1, x^2, x^3, x^5)$, where the index with Latin capital letters $M$ run over 0, 1, 2, 3, 5.\footnote{The "4" is left out on purpose, so that the fifth dimension/coordinate is more evident.}
	So, the line element $ds^2$ in Eq. (\ref{eq:ads_poincare}) is written as
	\begin{equation}
		ds^2 = \frac{L^2}{z^2}\left(-dt^2 + (dx^1)^2 + (dx^2)^2 + (dx^3)^2 + dz^2\right) = g_{MN}dX^M dX^N,
	\end{equation}
	where $g_{MN}$ is the metric tensor of the AdS spacetime.
	
	Based on what was observed in the previous paragraph, it is possible to define a 5-dimensional Minkowski metric $\eta_{MN}=\text{diag}(-1,+1,+1,+1,+1)$ and then express the AdS metric from it with the inclusion of a conformal factor,
	\begin{equation}
		g_{MN} = \frac{L^2}{z^2} \eta_{MN}, \quad \text{and} \quad g^{MN} = \frac{z^2}{L^2} \eta^{MN}.
	\end{equation}
	While writing the metric this way is convenient, the Minkowski metric $\eta_{MN}$ (and its inverse $\eta^{MN}$) must be used with caution. 
	Specifically, it cannot be used to raise or lower indices for tensors within the curved AdS spacetime. Index manipulation (raising and lowering) must be performed using the proper AdS metric $g_{MN}$ and its inverse $g^{MN}$.
	
	The determinant of the metric is
	\begin{flalign}
		g=\det(g_{MN}) = -\frac{L^{10}}{z^{10}.}
	\end{flalign}
	
	In some cases, it is convenient to treat the $z$ coordinate separately. In this scenario, Greek letters are used for the indices of the 4D Minkowski spacetime, such that $\mu = 0, 1, 2, 3$. Thus, the line element is written as
	\begin{flalign}
		ds^2 = \frac{L^2}{z^2}(\eta_{\mu\nu}dx^\mu d^\nu + dz^2).
		\label{poincaremetric}
	\end{flalign}
	
	The $z$ coordinate imposes two limits on spacetime in the Poincaré coordinates. Firstly, there is a singularity at $z=0$. Secondly, as we approach $z=\infty$, spacetime concentrates at a single point. Therefore, the Poincaré coordinates only cover a portion of the AdS spacetime, known as the \textit{Poincaré patch}\index{Poincaré coordinates!Poincaré patch}, with the extreme regions at $z=0$ and $z=\infty$ acting as its boundaries \cite{Nastase:2015wjb,Bayona:2005nq}.
	
	\begin{figure}[H]
		\centering
			\begin{minipage}{0.5\textwidth}
				\begin{tikzpicture}[>=Stealth, scale=1]
						
						\fill[ufabc, opacity=0.3] plot[domain=0:360, samples=100] ({2*cos(\x)}, {1.75 - 1.75*cos(\x) + 0.6*sin(\x)});
						
						\fill[ufabc, opacity=0.3] plot[domain=0:360, samples=100] ({2*cos(\x)}, {-1.75 + 1.75*cos(\x) + 0.6*sin(\x)});

						\draw[thick, ->] (0,-4.5) -- (0,4.5) node[right] {\Large $\tau$}; 
												
						\draw[very thick] (-2,-3.5) -- (-2,3.5); 
						\draw[very thick] (2,-3.5) -- (2,3.5);  
						
						\draw[very thick] (0,3.5) ellipse (2 and 0.6);
						\draw[very thick] (0,-3.5) ellipse (2 and 0.6);
						
						\draw[dashed, thick] (2,0) arc (0:180:2 and 0.6);
						\draw[dashed, thick] (-2,0) arc (180:360:2 and 0.6);
						\draw[dashed, thick] (-2,0) -- (2,0);
						
						\node[left] at (-2.1, 3.5) {$\tau = \pi$};
						\node[left] at (-2.1, -3.5) {$\tau = -\pi$};
						\node[left] at (-2.1, 0) {$\tau = 0$};
						
						\draw[thick, ufabc] plot[domain=0:180, samples=60] ({2*cos(\x)}, {1.75 - 1.75*cos(\x) + 0.6*sin(\x)})	
							node at (1.3,2) {\footnotesize \begin{tabular}{c}
									future \\
									horizon
							\end{tabular}};
						\draw[thick, ufabc] plot[domain=180:360, samples=60] ({2*cos(\x)}, {1.75 - 1.75*cos(\x) + 0.6*sin(\x)});
						
						\draw[very thick, ufabc] plot[domain=0:180, samples=60] ({2*cos(\x)}, {-1.75 + 1.75*cos(\x) + 0.6*sin(\x)})
							node at (1.3,-2) {\footnotesize \begin{tabular}{c}
									past \\
									horizon
							\end{tabular}};
						\draw[very thick, ufabc] plot[domain=180:360, samples=60] ({2*cos(\x)}, {-1.75 + 1.75*cos(\x) + 0.6*sin(\x)});
				
				\end{tikzpicture}
		\end{minipage}\hspace{2mm}
		\begin{minipage}{0.3\textwidth}
			\begin{tikzpicture}[>=Stealth, scale=1]
				
				\draw[very thick] (-2,-3.5) -- (-2,3.5);	
				\draw[very thick] (2,-3.5) -- (2,3.5);		
				\draw[very thick] (-2,3.5) -- (2,3.5);		
				\draw[very thick] (-2,-3.5) -- (2,-3.5);	
				
				\fill[ufabc!30] (-2,-3.5) -- (-2,3.5) -- (2,0) -- cycle;
				
				\draw[ufabc,very thick] (-2,3.5) -- (2,0) node[midway, above, sloped] {$z=+\infty,\; t=+\infty$};
				\draw[ufabc,very thick] (-2,-3.5) -- (2,0) node[midway, below, sloped] {$z=+\infty,\; t=-\infty$};
				
				\node[purple,rotate=90] at (-2.4,0) {\large $z=0$};
				\node at (-0.6, 0.8) {$z>0,\; t>0$};
				\node at (-0.6, -0.8) {$z>0,\; t<0$};
				
				\draw[dashed,thick] (-2,0) -- (2,0);	
				
				\draw[purple,ultra thick] (-2,-3.5) -- (-2,3.5);						
				
			\end{tikzpicture}
		\end{minipage}
		\caption[Representation of the AdS$_5$ spacetime showing the bulk, conformal boundary, and Poincaré horizons, and the Penrose diagram of the Poincaré patch.]{(Left) Representation of the AdS$_5$ spacetime. The interior of the cylinder corresponds to the bulk and the lateral surface is the conformal boundary of the space. The inclined ellipses represent the Poincaré horizons, which delimit the Poincaré patch as the wedge between the horizons. (Right) Penrose diagram for the Poincaré patch of the AdS space.}
		\label{fig:poincarepatch}
	\end{figure}
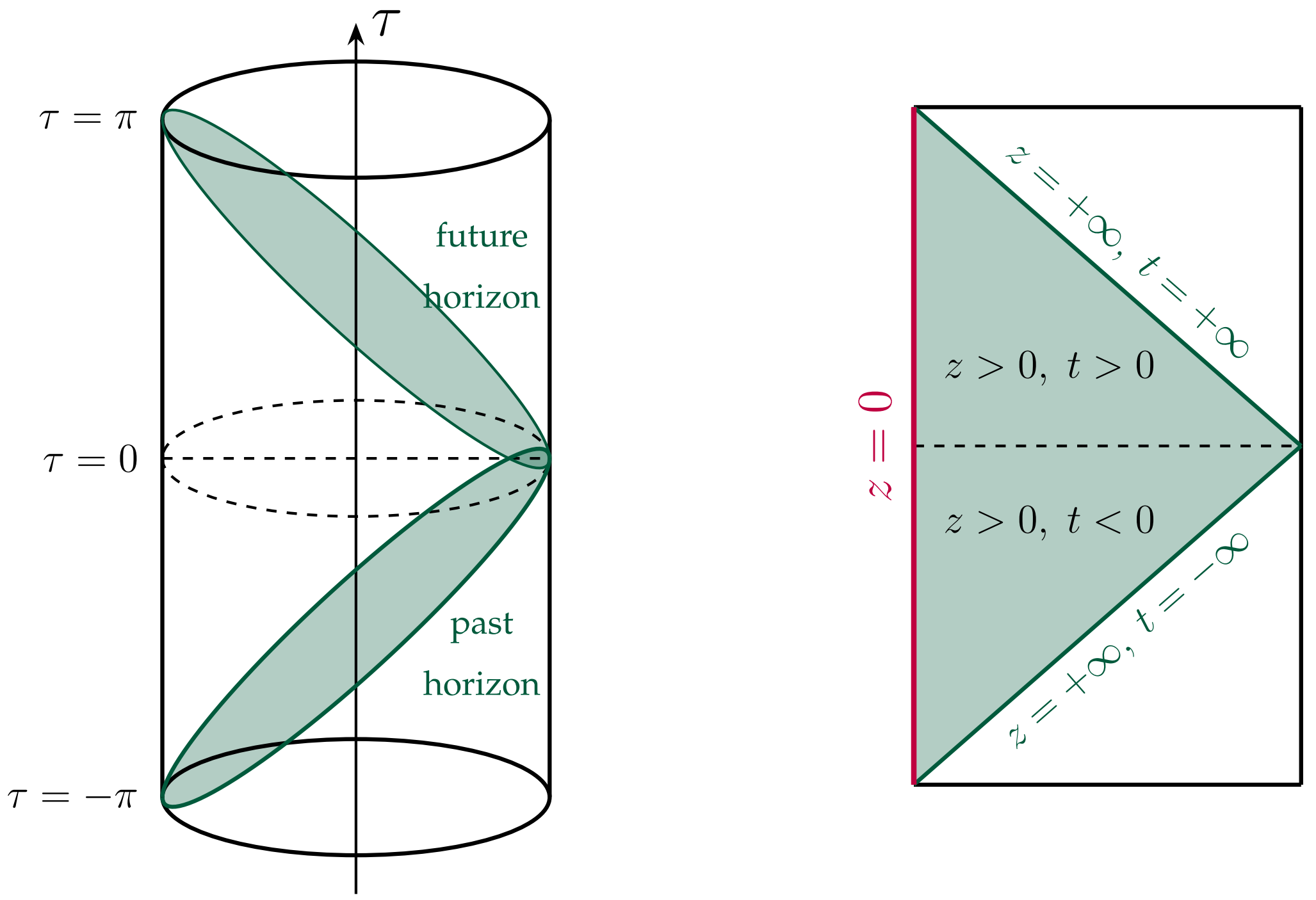
		
	According to Eq. (\ref{poincaremetric}), one can notice that the temporal term --- and, by consequence, the energy, since $E=i\hbar \partial/\partial t$ --- is accompanied by a  non-constant factor $\sqrt{-g_{00}}$. Because of this, the local proper energy $\tilde{E}$ for a state located at a specific depth $z$, when expressed in terms of the energy $E$ measured by an observer at the conformal boundary, is given by
		\begin{equation}
			\tilde{E} \sim \frac{1}{\sqrt{-g_{00}}}\frac{\partial}{\partial t} = \frac{1}{\sqrt{-g_{00}}}E = \frac{z}{L}E.
		\end{equation}
	Inverting this relation, it becomes clear that a process with a constant local energy $\tilde{E}$ deep inside the bulk will be perceived as having a redshifted, lower energy $E = (L/z)\tilde{E}$ by the observer at the boundary. So, the deeper the object is placed into the bulk (larger $z$), the lower its apparent energy becomes from the perspective of the boundary theory.
		
	This intrinsic feature turns out to be fundamentally relevant within the context of the AdS/CFT correspondence, since it maps the radial depth of the AdS bulk directly to the energy scale of the dual field theory living on the boundary. Consequently, small values of $z$ (approaching the boundary at $z=0$) correspond to high energies (the ultraviolet regime), while large values of $z$ (moving into the interior towards $z\to\infty$) correspond to low energies (the infrared regime) \cite{Polchinski:2001tt}.\\

	One last thing that is important to note in this section is that both the definition (\ref{ads}) and the metric (\ref{adsmetric}) are invariant under the transformations of SO$(2,d)$ group,  which acts on coordinates as $X^\mu \mapsto X^{\prime \mu}=\tensor{\Lambda}{^\mu_\nu} X^\nu$, where $\tensor{\Lambda}{^\mu_\nu} \in$ SO$(2,d)$. This group is analogous to the Lorentz group SO$(1,d)$ of rotations and boosts in the $(d+1)$-dimensional Minkowski spacetime.
		
	\section{Symmetries and conformal field theories}
	\label{section:cft}
	
	Understanding the symmetries of a theory is essential for defining some of its properties and the scope of its application. For example, the search for the symmetries of electromagnetism led to the discovery of a set of linear transformations known as \textit{Lorentz transformations}, which soon after became one of the foundations of Einstein's theory of special relativity.
	
	Lorentz transformations \index{Lorentz transformations} describe the change from one inertial frame of reference to another, with the invariance of physics under this transformation being one of the postulates of special relativity. This is evidenced by the invariance of the line element $ds^2$. The transformations include \textit{rotations} and \textit{boosts} (changes in coordinates between reference frames due to relative velocity), represented by the matrix $\tensor{\Lambda}{^\mu_\nu}$, which satisfies $\tensor{\Lambda}{^\mu_\rho}\tensor{\Lambda}{^\nu_\sigma} g_{\mu\nu}= g_{\rho\sigma}$.
	
	Those two types of transformations together form a group called the Lorentz group, SO$(1,d)$, for a $(d+1)$-dimensional flat spacetime. This group has $M^{\mu\nu}$ as its generators and is associated with its own algebra.
	
	The Lorentz group is a subgroup of a larger set of symmetries known as the \textit{Poincaré group}, ISO$(1,d)$. In addition to boosts and rotations, the Poincaré group also includes \textit{translations}, a displacement in the coordinates of the system. So, the transformations ruled by the group are generally expressed as $x^\mu \mapsto x^{\prime\mu} = \Lambda^\mu_\nu x^\nu + a^\mu$, combining Lorentz transformations and translations, parametrized by $\Lambda^\mu_\nu$ and $a^\mu$, respectively. The generator of translations is represented by $P^\mu$.
	
	\textit{Conformal transformations}\index{conformal field theory!conformal transformations}, on the other hand, are those that represent coordinate changes $x\mapsto x^\prime$ that locally rescale the metric as
	\begin{flalign}
		g_{\mu\nu} \mapsto g^\prime_{\mu\nu} (x)= \Omega^2(x)g_{\mu\nu},
		\label{conformal}
	\end{flalign}
	where angles (i.e., scalar products) between tangent vectors are preserved. The conformal factor $\Omega(x)$ is a positive function that depends on the coordinate $x$. By this definition, a theory in flat space is said to be \textit{conformal invariant} if it transforms the metric by a conformal factor without changing the line element $ds^2$ \cite{Nastase:2015wjb}, expressed as
	\begin{flalign}
		ds^2 = dx^\prime_\mu dx^{\prime\mu} = \left[\Omega(x)\right]^{-2}dx_\mu dx^\mu,
	\end{flalign}

	Therefore, a \textit{conformal field theory} \index{conformal field theory} is a quantum field theory that is invariant under conformal transformations. This definition leads to an important conclusion: since CFTs remain unchanged under these transformations, all scales must appear identical. Consequently, the theory cannot include scale-dependent parameters, such as mass, as these would introduce preferred scales (for instance, the Compton wavelength of a particle). This means that fields and states in these theories are massless, a property whose implications will be explored throughout this work.
		
	One of the conformal transformations is \textit{dilation}, a scale transformation that changes the coordinates by a constant factor as
	\begin{flalign}	
		x^\mu \mapsto x^{\prime\mu} = \lambda x^\mu,
		\label{conformalx}
	\end{flalign}
	which rescales the metric by a factor $\Omega(x)=\lambda$.
	
	The second type of transformation is the \textit{special conformal transformation} (SCT), a more complex coordinate change given by:
	\begin{flalign}
		x^\mu \mapsto x^{\prime \mu} = \frac{x^\mu - b^\mu x^2}{1-2(b\cdot x) + b^2x^2},
	\end{flalign}
	with $b^\mu$ being an arbitrary vector. In this case, the metric is rescaled by a factor $\Omega(x)=[1+2(b\cdot x)+b^2x^2]^{-1}$.
	
	It is straightforward to verify that rotations, boosts and translations are also conformal transformations. These can be combined with dilations and SCT to form a new group called the \textit{conformal group}. Thus, the Poincaré group is a subgroup of the conformal group by choosing $\Omega(x)=1$. The generators $D$ and $K_\mu$ represents dilations and SCT, respectively. 	\autoref{tab:conformal} summarize all the transformations and the explicit form of their generators.
	
	\begingroup	
	\renewcommand{\arraystretch}{1.15}
	\begin{table}[H]
		\centering
		\begin{tabular}{ccc}
			\multicolumn{3}{c}{\cellcolor{ufabc!50}\textbf{Conformal transformations}}      \\
			{\cellcolor{ufabc!30}}Translations & $x^\mu+a^\mu$ & $P_\mu=i\partial_\mu$  \\
			{\cellcolor{ufabc!10}} Lorentz (rotations and boosts) & $\tensor{\Lambda}{^\mu_\nu}x^\nu$ & $M^{\mu\nu}=i(x^\mu\partial^\nu - x^\nu\partial^\mu)$   \\
			{\cellcolor{ufabc!30}} Dilation                                                                   & $\lambda x$ & $D=ix^\mu\partial_\mu$   \\
			{\cellcolor{ufabc!10}} Special conformal transformation & $\frac{x^\mu - b^\mu x^2}{1-2b_\alpha x^\alpha + b^2x^2}$ & $K_\mu = i(x^2\partial_\mu - x_\mu x^\nu\partial_\nu) $
		\end{tabular}
		\caption{Summary of conformal transformations}
		\label{tab:conformal}	
	\end{table}
	\endgroup
	
	With these generators it is possible to construct the algebra of the conformal group and get the commutation relations:
	\begin{flalign}
		[D,P_\mu] &= -iP_\mu \nonumber\\
		[D,K_\mu] &= iK_\mu \nonumber\\
		[P_\mu,K_\nu] &=2i(M_{\mu\nu} - \eta_{\mu\nu}D) 	\nonumber\\
		[M_{\mu\nu},P_\rho] &=  -i(\eta_{\mu\rho}P_\nu -\eta_{\nu\rho}P_\mu) \nonumber \\
		[M_{\mu\nu},K_\rho] &= -i(\eta_{\mu\rho}K_\nu - 	\eta_{\nu\rho}K_\mu) \nonumber \\
		[M_{\mu\nu},M_{\rho\sigma}] &=  i(\eta_{\nu\rho}M_{\mu\sigma} + \eta_{\mu\sigma}M_{\nu\rho}
		+ \eta_{\mu\rho}M_{\nu\sigma} - \eta_{\nu\sigma}M_{\mu\rho}	),
		\label{conformalgenerators}
	\end{flalign}
	where the remaining commutators are zero. There is also representations defined in terms of the dilation operator $D$ and its eigenvalue $-i\Delta$, where $\Delta$ is the \textit{conformal dimension}\index{conformal field theory!conformal dimension} (also called scaling dimension) \cite{Nastase:2015wjb}.		
	Under (\ref{conformalx}) any quantum field $\phi$ transforms as
	\begin{flalign}
		\phi \mapsto \lambda^\Delta \phi,
	\end{flalign}
	and using (\ref{conformalgenerators}) the following is obtained:
	\begin{flalign}
		[D,P_\mu]\phi = -iP_\mu\phi	\; \mapsto \; D(P_\mu\phi) = P_\mu(D\phi) - iP_\mu\phi =-i(\Delta+1)(P_\mu\phi).
	\end{flalign}
	Analogously, this holds for $K_\mu$, so the generators $P_\mu$ and $K_\mu$ act as raising (creation) and lowering (annihilation) operators, respectively.
	
	Anticipating future developments in this thesis, it is relevant to extract new and significant information from the conformal group.
	
	First, a new set of generators $J_{MN}$ in ($d+2$) dimensions (with $M,N=0,\dots,d+1$) is defined, encompassing all four generators in $d$ dimensions of conformal group as follows \cite{Aharony:1999ti}
	\begin{flalign}
		J_{\mu\nu} = M_{\mu\nu}, \qquad 
		J_{(d+1)d} = D, \qquad
		J_{\mu d} = \frac{1}{2}(K_\mu-P_\mu), \qquad
		J_{\mu (d+1)} = \frac{1}{2}(K_\mu+P_\mu).
	\end{flalign}
	
	These generators are the same as those of the group SO$(2,d)$. Thus, the conformal group in $d$-dimensions is isomorphic to the group SO$(2,d)$. As discussed in \Cref{section:adsspace}, this is the same symmetry of a ($d+1$)-dimensional anti-de Sitter spacetime, suggesting a similarity between conformal field theories and a gravity theory with an additional dimension.
	
	\section{The AdS/CFT correspondence}
		\label{section:adscftcorrespondence}
		
		\subsection{$\mathcal{N}=4$ super Yang-Mills theory}
		\label{section:yangmills}
	
		As previously mentioned, from the gauge side of Maldacena conjecture resides an $\mathcal{N}=4$ super Yang-Mills (SYM) theory, and the best way to introduce it in this review is by dissecting each part of its name.
		
		The original Yang-Mills theory \index{Yang-Mills theory} was initially proposed in 1953 by Chen Ning Yang and Robert Mills in the context of nuclear interactions \cite{Yang:1954ek}. However, over time the name has been more used to identify a set of gauge theories of the same class as the original proposition, those based on non-Abelian groups. A group G is said to be Abelian when all its elements commute under the group operation, i.e., $a\ast b=b\ast a$, so a non-Abelian group is one that has at least one pair of elements that do not commute.
		
		Electromagnetism is the simplest example of an Abelian gauge theory. It has a global U(1) symmetry, which means that its action remains invariant under a transformation of the type $\Psi\mapsto \Psi^\prime= e^{iq\alpha}\Psi$, where $\Psi=\Psi(x)$ is a fermion field, $q\in\mathbb{R}$ is the electric charge and $\alpha$ is a real constant. However, for the case of a local U(1) transformation, when the parameter depends on the spacetime coordinates ($\alpha\equiv\alpha(x)$), the invariance does not hold, since the derivative $\partial_\mu\Psi$ does not transform in the same form (namely, $\partial_\mu\Psi\mapsto \partial_\mu\Psi'\neq e^{iq\alpha(x)}\partial_\mu\Psi$), thanks to the additional term that appears with the derivative acting on the exponential of $\alpha(x)$.
		
		To maintain invariance, it is necessary to insert a gauge field $A_\mu(x)$ and a covariant derivative $D_\mu=\partial_\mu - iqA_\mu(x)$, so that the transformation takes the same form as $\Psi$, that is $D_\mu\Psi\mapsto D_\mu\Psi'= e^{iq\alpha(x)}D_\mu\Psi$. For this to be true, the gauge field also needs to transform according to $A_\mu(x)\mapsto A_\mu(x)-\partial_\mu\alpha(x)$. As a result of these modifications, now terms of $A_\mu(x)$ and $\Psi$ interact, which in electromagnetism represents the interaction between their respective corresponding particles, photons and electrons.		
		With $A_\mu$ we can construct the field strength tensor $F_{\mu\nu}=\partial_\mu A_\nu - \partial_\nu A_\mu$, which appears in the action of electromagnetism as an $\sim F^2$ term and generates the kinetic terms $\sim(\partial A)^2$ of the photons. This format is characteristic of Abelian gauge theories \cite{Weinberg:1996kr}.
		
		The formulation for a non-Abelian theory is not very different, just a bit more complicated. 
		A good way to approach this topic is by using the SU($N_c$) group\footnote{The subscript in $N_c$ is not necessary for now, but we choose to keep it for easier identification when it gains a physical meaning.} as an example, as it is of great interest in theories such as quantum chromodynamics. For now, the intricacies of this group will not be discussed in detail, which will be done in \Cref{section:qcd} within the context of QCD, where its properties will gain a much more tangible physical meaning.
		
		The global SU($N_c$) transformations are of the type $\Psi(x) \mapsto U \Psi(x)$, where $U=e^{ig\alpha^at^a}$ represents the transformation matrix. In this expression $g$ is the coupling constant (whose significance will be clarified shortly), $\alpha^a$ are real constants, and $t^a$ are the generators of SU($N_c$). The indices $a=1,\dots,N_c^2-1$ identify each of the distinct generators of the group. 
		Again, the invariance does not hold if a local transformation is considered, so the process from the Abelian case can be repeated. However, there is a difference, since now a gauge field for each generator, $A^a_\mu$, needs to be considered in addition to the covariant derivative $D_\mu$. 
		
		When constructing the field strength tensor $F_{\mu\nu}=\partial_\mu A_\nu - \partial_\nu A_\mu-i[A_\mu,A_\nu]$, the commutator term that is not zero in the non-Abelian case causes the action's $\sim F^2$ part to generate not only the kinetic terms but also self-interaction terms --- $\sim g(\partial A)A^2$ and $\sim g^2A^4$ ---, which means that gluons, unlike photons, can interact with each other. The constant $g$, henceforth designated as $g\equiv g_\text{YM}$, is the                                                                                                                                                                                                                                                                                                                                                                                                                                                                                                                                                                                                                                                                                                 Yang-Mills coupling responsible for mediating these interactions.
		
		It is easy to verify how we can obtain the results of the Abelian theory from the non-Abelian one, which is why this is considered a generalization of both cases. Yang-Mills theories are at the heart of the standard model, serving as the foundation for describing electromagnetic, strong and weak interactions, and the configurations presented bring particular properties beyond those already shown, as we will see when discussing QCD in more detail.\\

		Having done that, we can now understand where the "super" comes from. In \Cref{section:cft}, we saw the importance of spacetime symmetries in the description of physical theories. However, there are other equally essential symmetries that are related to the intrinsic degrees of freedom of the fields, such as parity and charge conjugation. These are called \textit{internal symmetries}, defined as symmetries that are not due to transformations at points in spacetime.
		
		It seems a compelling idea to find a single symmetry group capable of encompassing both types. Mathematically, this means having at hand the generators of the Poincaré group responsible for parameterizing spacetime transformations ($P$ for translations and $M$ for rotations and boosts) and the generators representing the internal symmetries (which we will call $T$ here), and obtaining non-zero commutation relations. Unfortunately, since the generators of internal symmetries are always Lorentz scalars, the commutation relations we obtain are $[T_s,P_\mu]=[T_s,J_{\mu\nu}]=0$.
		
		In addition to this, the Coleman-Mandula theorem proved in 1967 \cite{Coleman:1967ad} establishes that the symmetry group of a QFT must be the direct product of a Poincaré group and an internal symmetry group, meaning there is no way to combine the two types of symmetry into a single group except in a trivial way. The solution to circumvent this impossibility is to look for loopholes in the premises of the theorem. One of the premises states that the theory has only massive particles, so a theory with conformal invariance would be suitable. However, the loophole that interests us in this review is the premise that all symmetries considered in the theorem are characterized by a Lie algebra. The idea, then, is to expand the problem to a symmetry characterized by a \textit{graded }Lie algebra \cite{Haag:1974qh}.
		
		A graded Lie algebra (also called a \textit{superalgebra})\index{superalgebra} encompasses not only commutators but also anticommutators, and it has new generators $Q$ known as \textit{supercharges}\index{supersymmetry!supercharges}. Roughly speaking, these supercharges are responsible for transforming bosons into fermions and fermions into bosons, which is the most notable characteristic of the so-called \textit{supersymmetry}\index{supersymmetry}. Indeed, its proposition as presented arose in the context of string theory, which in its original version had only bosons in its spectrum, and supersymmetry served to insert fermions into the theory.
		
		A theory can have more than one supersymmetry (an \textit{extended supersymmetry}), and from this, the supercharges gain an index, $Q^A$, such that $A=1,\dots,\mathcal{N}$, where the calligraphic $\mathcal{N}$ represents the number of independent supersymmetries of the theory. Thus, a boson can be transformed into an $\mathcal{N}$ number of fermions, its supersymmetric partners. The same applies to the transformation of fermions into bosons. The case of $\mathcal{N}=4$ is interesting because this is the maximum number of supersymmetries in a four-dimensional spacetime, which is why it is said that $\mathcal{N}=4$ SYM is the maximally supersymmetric theory in 4 dimensions. 
		
		There is an additional symmetry responsible for acting on the supercharges, the \textit{R-symmetry}\index{supersymmetry!R-symmetry}. The group that represents it depends on the value of $\mathcal{N}$. For $\mathcal{N}=1$, it is the U(1) group (usually written as U(1)$_R$), which only alters the phase of the supercharge. For $\mathcal{N}>1$ (that is, in extended supersymmetries), the symmetry acquires a global character and rotates the charges among themselves. For $\mathcal{N}=4$, this group is SU(4)$_R \simeq$ SO(6)$_R$.
			
		Although it will not be used directly in the subsequent calculations of this work, it is highly instructive for future comparisons to present the explicit form of the $\mathcal{N}=4$ SYM action, which is given by \cite{Ammon:2015wua}:
		\begin{equation}
			S_{\text{SYM}} = -\frac{1}{g_{\text{YM}}^2}\int d^4x\,
			\mathrm{Tr}\left(\frac{1}{2}F_{\mu\nu}F^{\mu\nu}
			+ D_\mu\phi^i D^\mu\phi^i
			- \frac{1}{2}[\phi^i,\phi^j]^2
			+ \text{fermionic terms}\right),
			\label{eq:SYM_action}
		\end{equation}
		where $g_{\text{YM}}$ is the Yang-Mills coupling constant, $F_{\mu\nu}$ is the non-Abelian field strength tensor, $D_\mu$ is the covariant derivative, and $\phi^i$ (with $i = 1, \dots, 6$) are real scalar fields. The trace is taken over the fundamental representation of the $SU(N_c)$ gauge group. The fermionic terms include the kinetic and interaction parts for four Weyl spinors $\lambda^A$ (with $A = 1, \dots, 4$), which are omitted here for brevity.		
		Through this explicit action, the field content of the maximally supersymmetric theory becomes evident: there is one gauge field $A_\mu$, six scalar fields $\phi^i$, and four Weyl spinors $\lambda^A$.		
		
		Finally, $\mathcal{N}=4$ SYM is a conformal theory, and is therefore invariant
		under the conformal group $SO(2,4)$, which is precisely the isometry group of
		AdS$_5$.\footnote{In fact, the theory is \textit{superconformal}, since conformal
			symmetry combines with supersymmetry and R-symmetry into the supergroup
			PSU(2,2|4), whose bosonic subgroup is SO(2,4)$\times$ SO(6). Only the
			bosonic conformal part is used in this work.}		
		Therefore, in summary,
		
		\begin{tcolorbox}[title={},colback=white,colframe=ufabc!50,colback=ufabc!10,coltitle=black]
			\centering
			$\mathcal{N}=4$ Super-Yang-Mills is a superconformal field theory with a \\ global SO(2,4)$\times$SO(6) symmetry.
		\end{tcolorbox}
	
		\subsection{Type IIB superstring theory}
		\label{section:stringtheory}
		
		On the gravitational side of the conjecture, there is one of the types of string theory. The subject is very broad and would require an entire course to fully address it. In this subsection, we will only cover a brief history, highlighting points that are necessary for the general understanding of the motivation and foundations of the conjecture. For further reading, references can be found in \cite{Zwiebach:2004tj, Green:1987sp, Becker:2006dvp, Polchinski:1998rq, Polchinski:1998rr}.
		
		\textit{String theory}\index{string theory} (ST) emerged as part of the effort to describe a quantum theory of the strong interactions. Today, as we well know, this task has been superseded by quantum chromodynamics, but there were still valuable tools in ST. In particular, with the addition of supersymmetry, string theory was promoted to \textit{superstring theory}\index{string theory!superstring theory} and its models began to encompass both bosons and fermions without the presence of anomalies (when a classical symmetry is violated in the quantum regime) and without the appearance of infinite values, which are common problems of quantum field theories. All of this served as an indication that superstrings would be an excellent candidate for a quantum gravity theory, which led to a significant increase in the number of works in the area, an event that became known as the first superstring revolution in the 1980s.
	
		The foundation of the theory lies in studying elementary units not as pointlike particles, but as one-dimensional objects called \textit{strings}. Strings have a characteristic length $l_s$ and can be open, where their two endpoints are distinct, or closed, where the endpoints meet to form a loop.
		Unlike the numerous elementary particles present in QFT models, the string is a unique object, and the variety of particles and their properties observed are manifestations of the oscillations of this single type of fundamental object.		
		The oscillation possibilities of open strings are reflected in degrees of freedom, so open strings represent a gauge theory by themselves, while the vibration states of closed strings represent a spin-$2$ particle, the graviton \cite{Natsuume:2014sfa}.
		
		Just as particles trace worldlines in spacetime, a string sweeps a two-dimensional surface called a worldsheet, whose embedding into the target spacetime (let us assume a D-dimensional Minkowski spacetime with metric $\eta_{MN}$) is parameterized by the function $X^M(\sigma^a)$, where $\sigma^a=\sigma^0,\sigma^1$ are the coordinates of the worldsheet. Through this description, strings are governed by the \textit{Nambu-Goto action},
		\begin{flalign}
			\mathcal{S}_\text{NG} = -\frac{1}{2\pi\alpha'} \int_{\Sigma} d^2\sigma\sqrt{\partial_\alpha X^M \partial_\beta X^N \eta_{MN}},
			\label{eq:nambugoto}
		\end{flalign}
		where $\alpha'$ is a parameter linked to the string length by $\alpha'=l_s^2$. Nevertheless, when the goal is to quantize the theory, the action can make the task more complicated, for example, because of the square root. To circumvent this adversity, it is possible to move to an equivalent action at the classical level, but much more convenient in quantization simply by inserting an auxiliary field $h_{\alpha\beta}(\sigma)$, the metric of the worldsheet. Thus, the \textit{Polyakov action} is written as
		\begin{flalign}
			\mathcal{S}_\text{P} = -\frac{1}{4\pi\alpha'} \int_{\Sigma} d^2\sigma \sqrt{-h} h^{\alpha\beta}\partial_\alpha X^M \partial_\beta X^N \eta_{MN},
		\end{flalign}
		with $h=\det(h_{\alpha\beta})$. An important feature of the action in this form is that it is invariant under Poincaré transformations (Lorentz and translations). It is also invariant under Weyl transformations (rescaling of the metric tensor), $h_{\alpha\beta}(\sigma)\mapsto \Omega(\sigma)^2h_{\alpha\beta}(\sigma)$. This allows the choice of a gauge in which the worldsheet metric is diagonal and flat, still maintaining Weyl invariance, $h_{\alpha\beta}(\sigma)\mapsto \Omega(\sigma)^2\eta_{\alpha\beta}(\sigma)$, which implies that string theory is a two-dimensional CFT in flat space.
		
\begin{figure}[h!]
			\centering
			\includegraphics[width=11cm]{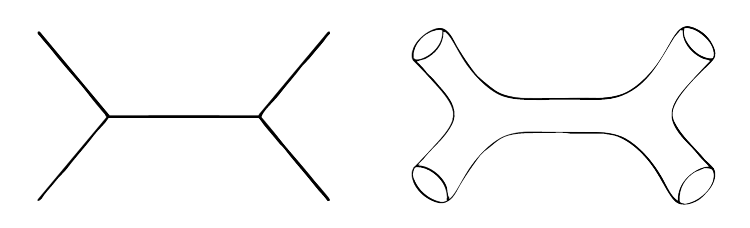}
			\caption[A tree-level Feynman diagram depicting a particle scattering process, and its corresponding string theory worldsheet representation]{A tree-level Feynman diagram (left) depicting a particle scattering process, and its corresponding string theory worldsheet representation (right).}
			\label{fig:strings}
		\end{figure}

		With these basic principles, strings can be quantized in a way very similar to what is done with point particles. In fact, many of the processes observed in QFTs can be reproduced in string theory with their due adjustments. This includes the interaction between strings, mediated by a coupling constant $g_s$, allowing the construction of an equivalent to Feynman diagrams to represent these interactions, but instead of lines, the topology of the worldsheets is used (\autoref{fig:strings}).
		
		However, the theory has notable setbacks, such as the presence of tachyons (particles with imaginary mass that are faster than light) and the need for many extra dimensions to maintain consistency with quantum mechanics. For instance, the purely bosonic theory requires 26 dimensions. With the advent of supersymmetry, this number drops to 10 dimensions. Supersymmetry also resulted in the disappearance of tachyons from the theory, making superstrings somewhat more promising.
		
		Currently, there are five effective superstring theories, but in 1995, Edward Witten \cite{Witten:1995ex} showed that they were all just different limits of the same 11-dimensional theory, \textit{M-theory}\index{string theory!M-theory}, interconnected by dualities (this discovery was the second superstring revolution). Two of these five theories, the type II (which are divided into type IIA and IIB), will be of greater interest in this section, as the IIB theory is crucial in one half of Maldacena conjecture.
		
\begin{figure}[h]
			\centering
			\includegraphics[width=11cm]{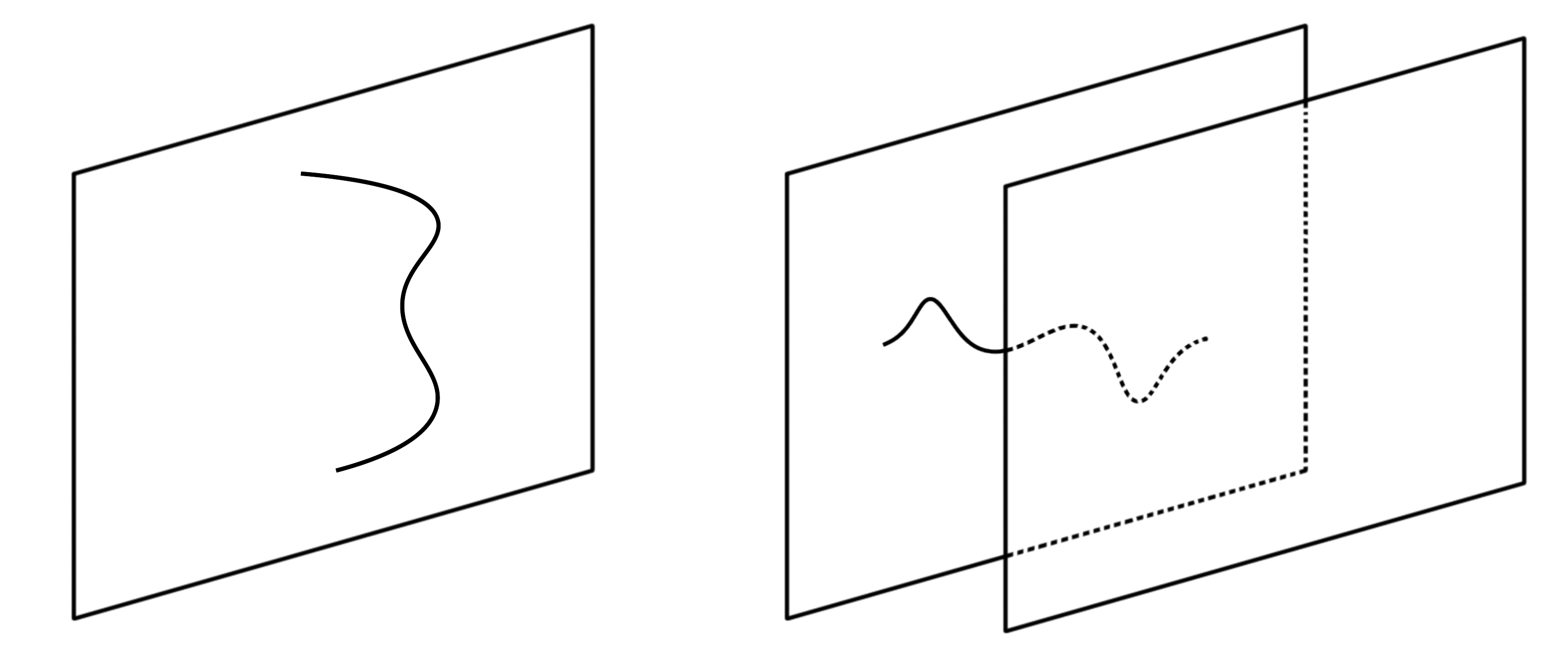}
			\caption[Different configurations of strings and branes]{Different configurations of strings and branes. Strings can have their endpoints attached to the same brane (left) or stretch between two different branes (right). Although the figure shows the branes as separated for visualization purposes, they are considered coincident.}
			\label{fig:branes}
		\end{figure}

		In addition to strings, another fundamental element that naturally appears in type II string theories are \textit{branes}\index{string theory!branes}, surfaces where the endpoints of strings connect, so that different endpoints of each string can end up on the same brane or on different branes (\autoref{fig:branes}). These objects are usually referred to as D$p$-branes, the D is an abbreviation for Dirichlet and has to do with boundary conditions, and $p$ represents the dimension of the brane, with a D0-brane being a particle, a D1-brane a string, a D2-brane a membrane, and so on. In the same way as a string, a brane sweeps a $(p+1)$-dimensional worldvolume as it moves in spacetime.
		
		To ensure the conservation of string properties, oscillations cannot end at their endpoints, so they must have effects propagated on the branes. This issue is resolved in the process of quantizing the strings, where oscillations transverse to the brane can be interpreted as fluctuations of the brane itself in transverse directions, showing that branes are objects with their own dynamics. On the other hand, oscillations longitudinal to the brane give rise to gauge fields living in the worldvolume of the brane. This leads to a major conclusion: a D$p$-brane describes a gauge theory in $(p+1)$-dimensions. 
		
		This work focuses on four-dimensional gauge theories, so D3-branes are a natural choice. It can be shown that the stability of the branes depends on their number of dimensions, with type IIA string theory maintaining stable branes if $p$ is even, while branes in type IIB theory are stable for odd $p$. Since $p=3$ is of interest, the IIB theory is preferred in the conjecture.
		
		If two parallel branes are considered, it is trivial to observe that a string has two possibilities of connection: it can start and end on the same brane, or it can stretch so that each endpoint is on a different brane.
		Generalizing for the case of $N_c$ branes, each endpoint of the string will have $N_c$ possibilities of connection.
		The situation becomes more interesting when a stack of branes located at the same position in spacetime is considered, so there are $N_c$ possible starting points and $N_c$ possible endpoints for the open strings. This leads to $N_c^2$ configurations, which can be interpreted as the number of gauge fields living in the worldvolume of this stack of branes, therefore representing a U($N_c$) gauge theory.
		In fact, the homomorphism between U($N_c$) and SU($N_c$)$\times$U(1) can be used to decouple the U(1) part, which describes the center of mass motion of the branes, leading to a SU($N_c$) gauge symmetry living in stack.\\
	
		So far, only the idea of branes as hypersurfaces where open strings end has been addressed, but there is another important manifestation of these objects. This second view emerges when considering a specific low-energy limit in superstring theory.
		As will be explored in more detail in \Cref{sec:hadronspectroscopy}, the modes generated by the vibration of the strings manifest themselves in spacetime as distinct particles.		
		The mass of these particles is directly determined by their excitation level $n$ (with $n = 0, 1, 2, \dots$), following a relation proportional to $m^2 \sim n/\alpha'$. The ground state ($n=0$) produces completely massless particles, which notably include the graviton. On the other hand, the excited states ($n \ge 1$) correspond to a tower of massive particles. 
		
		If we restrict our analysis to low-energy phenomena --- meaning energy scales $E$ much smaller than the string scale ($E \ll 1/\sqrt{\alpha'}$) --- the system does not possess enough energy to excite these massive modes. Mathematically, this regime is formalized by taking the limit $\alpha' \to 0$ while keeping the string coupling $g_s$ fixed. As $\alpha'$ approaches zero, two things happen simultaneously: the physical size of the string shrinks to a point ($l_s \to 0$), and the masses of all excited states approach infinity. 
		
		In quantum theory, particles with infinite mass require infinite energy to be created. Consequently, these heavy modes decouple completely from the physical spectrum, and the only states that survive this limit are the massless ones. Since the extended nature of the strings has been erased and only interacting point-like particles remain, the full superstring theory effectively reduces to a classical field theory of gravity: a 10-dimensional \textit{supergravity}\index{supergravity}. 
		
	For future reference when establishing the gravitational side of the duality, the bosonic sector of the type IIB supergravity action in the Einstein frame is given by \cite{Ammon:2015wua}:
		\begin{equation}
			S_{\text{SUGRA}} = \frac{1}{16\pi G_{10}} \int d^{10}x \sqrt{-g} \left( R - \frac{1}{2} \partial_M \Phi \partial^M \Phi - \dots \right),
			\label{eq:SUGRA_action}
		\end{equation}
	where $G_{10}$ is the 10-dimensional Newton's constant, $R$ is the Ricci scalar, and $\Phi$ is the dilaton field, a scalar field whose vacuum expectation value sets the string coupling constant, $g_s = e^{\langle \Phi \rangle}$, thereby controlling the strength of string interactions.\footnote{This ten-dimensional field should not be confused with the five-dimensional dilaton $\Phi(z)$ introduced in the bottom-up models of		\Cref{chapter:cims}, although the latter is motivated by it.}
		From this, one can obtain solutions for massive gravitational objects, such as black holes, by solving the supergravity equations of motion derived from this action, just as it is done in standard general relativity.
					
		In 1995, Joseph Polchinski demonstrated that branes were also solutions of supergravity, elevating them to objects capable of deforming spacetime \cite{Polchinski:1995mt}.
		From this, it is possible to think about what happens with the background spacetime in the presence of the stack of $N_c$ coincident D3-branes. The solution for the metric in this case is given by \cite{Ammon:2015wua}
		\begin{flalign}
			ds^2 = \left(1+\frac{L^4}{r^4}\right)^{-1/2}\left[-dt^2+dx_1^2 + dx_2^2 + dx_3^2\right] + \left(1+\frac{L^4}{r^4}\right)^{1/2}\left[dr^2+r^2d\Omega_5^2\right],
		\end{flalign}
		where $L^4=4\pi g_sN_c{\alpha'}^2$ is a length parameter \cite{Aharony:1999ti}. Depending on the value of the radial coordinate $r$, two different regions can be observed. For $r\gg L$ the result is a flat spacetime in 10 dimensions. Meanwhile, closer to the origin (the near-horizon region), with $r\ll L$, the solution reduces to
		\begin{flalign}
			ds^2 =  \frac{r^2}{L^2}\left[-dt^2+dx_1^2 + dx_2^2 + dx_3^2\right] + \frac{L^2}{r^2}dr^2+L^2d\Omega_5^2.
		\end{flalign}
		The last term represents a 5-sphere $S^5$ with radius $L$, while the remaining terms correspond to the AdS$_5$ spacetime with the same radius $L$. Therefore, the solution for the stack of D3-branes near the origin is an AdS$_5\times S^5$ spacetime.
		As presented in \Cref{section:adsspace}, the AdS$_5$ spacetime is invariant under the SO(2,4) group, while the symmetry group for the 5-sphere is SO(6).
		
		Therefore, in summary,
		
		\begin{tcolorbox}[title={},colback=white,colframe=ufabc!50,colback=ufabc!10,coltitle=black]
			\centering
			Type IIB superstring theory in AdS$_5\times S^5$ spacetime has a\\global SO(2,4)$\times$SO(6) symmetry.
		\end{tcolorbox}
	
		\subsection{Development of the correspondence}
		\label{section:conjecture}
		
		The first major step towards the AdS/CFT correspondence occurred in 1974, two decades before Maldacena’s proposition, when 't Hooft \cite{tHooft:1973alw} noticed that non-Abelian gauge theories became considerably simpler in the large $N_c$ limit.\index{Large $N_c$ limit} This might seem counterintuitive, as these theories are typically represented by groups like SU($N_c$) and SO($N_c$) with gauge fields $(A_\mu)_i^j$ described by $N_c \times N_c$ matrices. So, it was expected that the infinite increase of $N_c$ would make the theory more complicated, but the opposite happens. The result is that properties of the theory can be expanded in powers of 1/$N_c$ and treated perturbatively thanks to the adopted limit.
		
		In fact, the so-called 't Hooft limit is a bit more elaborate, being described as
		\begin{flalign}
			N_c\rightarrow\infty \quad \text{with} \quad \lambda\equiv g^2N_c \quad \text{kept fixed}.
		\end{flalign}
		with $\lambda$ being the \textit{'t Hooft coupling}.\index{Large $N_c$ limit!'t Hooft coupling}
			
		To visualize the implications of this limit, we start by adopting a new notation for Feynman diagrams. Instead of a single line, two lines oriented in opposite directions are used, each representing one of the indices of the gauge fields $(A_\mu)_i^j$. Now, propagators, vertices, and loops scales by a different factor of $N_c$. Using $V$, $E$, and $F$ as notation for the number of vertices, edges (propagators), and loops, respectively, the general rule for the diagram is given by
		\begin{flalign}
			\text{diagram} \sim	\left(\frac{N_c}{\lambda}\right)^V \left(\frac{\lambda}{N_c}\right)^E N_c^F = \lambda^{E-V} N_c^{V-E+F} \equiv \lambda^{E-V}N_c^{\chi}
		\end{flalign}
		where $\chi=V-E+F$ is the \textit{Euler characteristic}, the parameter that defines the power of $N_c$. For closed oriented surfaces, $\chi$ is given by the number of handles of the surface  (the genus $h$): $\chi=2-2h$.
		It is worth remembering that the rule applies to both planar and non-planar diagrams, with planar ones being those that can be drawn without crossing lines on a plane, while in non-planar ones the only way to avoid crossing lines is to draw them on higher-dimensional surfaces, such as a torus.
		
		With this, any physical quantity obtained from Feynman diagrams can also be expanded in terms of $N_c$ and $h$, and that can be done through the partition function (which is equivalent to the sum of all diagrams). This expansion results in an expression of the form
		\begin{flalign}
			\ln Z=\sum_{h=0}^{\infty} N^{2-2h}f_h(\lambda)
			\label{diagramsexpanded}
		\end{flalign}
		where $f_h(\lambda)=\sum_{i=0}^{\infty} c_{h,i}\lambda^i$ is a polynomial function of $\lambda$. It is easy to visualize that, in the $N_c\rightarrow\infty$ limit, the largest term is the one with the minimum genus (thus, maximum $\chi$), which corresponds to the planar diagrams that, scaling in the order of $N_c^2$, dominate above the other diagrams, which scale in powers of $1/N_c$.
		
		Therefore, in the 't Hooft limit, only the planar diagrams predominate, and this ultimately simplifies the theory. As the analysis was general, we can say that this result is valid for any gauge theory with adjoint matter fields, which includes $\mathcal{N}=4$ SYM theory, not just its pure version \cite{Aharony:1999ti}.
		Another great advantage of this idea is that Eq.~\eqref{diagramsexpanded} is the same as found in theories with oriented closed strings in the perturbative regime, assuming that the string coupling constant is of the order $g_s \sim 1/N_c$. The conclusion is that a weakly coupled string theory, that is, a quantum gravity theory, is very close to a gauge field theory.	
		
		To transform this abstract mathematical resemblance into a precise duality, we must recover the physical system described at the end of the previous section: the stack of $N_c$ coincident D3-branes. The formalization of the AdS/CFT correspondence relies on taking a specific low-energy limit on this system, known as Maldacena's \textit{decoupling limit} \cite{Maldacena:1997re}\index{AdS/CFT correspondence!decoupling limit}.
		
		This approach consists of analyzing the system at energy scales much lower than the string tension ($E \ll 1/\sqrt{\alpha'}$). As briefly introduced when discussing supergravity, mathematically this is achieved by taking the limit $\alpha' \to 0$. In this regime, the energy required to excite the massive string modes diverges, meaning only the massless modes of the strings survive.
		
		To see how this decoupling happens, let us first look at the system from the perspective of string theory. The full dynamics of the system can be described by an effective action that sums the contributions of closed strings (propagating freely in the 10D bulk), open strings, and the interaction between them:
		\begin{equation}
			\mathcal{S} = \mathcal{S}_{\text{closed}} + \mathcal{S}_{\text{open}} + \mathcal{S}_{\text{int}}.
		\end{equation}
		The full derivation of these actions is beyond the scope of this thesis but can be found in detail in Ref.~\cite{Ammon:2015wua}.

		The term $\mathcal{S}_{\text{closed}}$ governs the closed strings in the bulk. In the low-energy limit, it reduces to the action of free type IIB supergravity in a flat 10-dimensional Minkowski spacetime, which was presented in Eq. \eqref{eq:SUGRA_action}.
		The term $\mathcal{S}_{\text{open}}$ represents the dynamics of the open strings attached to the branes, which is described by the action:
		\begin{equation}
			\mathcal{S}_\text{open} = -\frac{1}{2\pi g_s} \int d^4x\left(\frac{1}{4}F_{\mu\nu}F^{\mu\nu} + \frac{1}{2}\partial_\mu\Phi^i \partial^\mu \Phi^i + \mathcal{O}(\alpha')\right)
		\end{equation}
		When expanding this expression for small $\alpha'$, it reduces exactly to the bosonic sector of the 4-dimensional $\mathcal{N}=4$ SYM action shown in Eq. \eqref{eq:SYM_action}, provided that we identify the coupling constants as
		\begin{equation}
			2\pi g_s = g^2_{\text{YM}}.
			\label{eq:coupling_relation}
		\end{equation}
		
		Finally, $\mathcal{S}_{\text{int}}$ describes the interaction between the bulk gravity and the fields on the brane, with terms of the form $\sim \Phi F^2$. This interaction is proportional to $g_s (\alpha')^2$, hence, taking the limit $\alpha' \to 0$ yields $\mathcal{S}_{\text{int}} \to 0$, causing the open strings on the brane and the closed strings in the bulk to decouple completely. Therefore, the system is split into two independent sectors: free supergravity in the bulk and $\mathcal{N}=4$ SYM on the branes, effectively establishing the correspondence.
			
		For this duality to be complete, the free parameters of both theories must be fully matched. On the CFT side, the free parameters are the Yang-Mills coupling constant $g_{\text{YM}}$ and the rank $N_c$. On the AdS side, the parameters are the string coupling $g_s$ and the characteristic radius $L/\sqrt{\alpha'}$. We have already seen from the brane action that $g^2_{\text{YM}} = 2\pi g_s$. By incorporating the geometric relation for the AdS radius $L^4 = 4\pi g_s N_c (\alpha')^2$, the complete map is established \cite{Natsuume:2014sfa}:
		\begin{equation}
			g^2_{\text{YM}} = 2\pi g_s \quad \text{and} \quad 2g^2_{\text{YM}}N_c = \frac{L^4}{{\alpha'}^{2}}.
			\label{adscftparameters}
		\end{equation}
				
		As none of these parameters have any restrictions initially, it is said that the way they are related is the "strongest version" of the correspondence. Because of this, it would be the ideal framework to continue with the studies. However, the calculations may prove to be very complicated if the parameters remain completely arbitrary. In addition, it is in the perturbative regime that string theory is better understood, so the ideal is to proceed assuming some limits and restrictions.
		
		Starting from the insight about string theory in the perturbative regime, the "strong version" of the correspondence assumes that $g_s\rightarrow0$ while keeping $L/\sqrt{\alpha'}$ constant and arbitrary, which, according to Eq. (\ref{adscftparameters}), implies $g_\text{YM}\rightarrow0$ and $N_c\rightarrow\infty$, but with $\lambda=g^2_\text{YM}N_c$ kept fixed. This is the 't Hooft limit that, as discussed earlier, simplifies the $\mathcal{N}=4$ SYM theory by making only the planar diagrams predominate.
		On the AdS side, the string theory is reduced to its classical limit. The only free parameters on each side are related by
		\begin{flalign}
			\lambda = \frac{L^4}{2{\alpha'}^2}.
		\end{flalign}
	
		Directly from the results above, one can imagine a new limit that simplifies the correspondence even further, leading to its "weak version", which consists of assuming $\lambda\rightarrow\infty$. This corresponds to ${\alpha'}^2/L^2 \rightarrow0$, which means that the string length $l_s=\sqrt{\alpha'}$ needs to be much smaller than the curvature radius $L$.
		
		What remains, then, is an $\mathcal{N}=4$ SYM mapped onto a classical type IIB supergravity in AdS$_5\times S^5$ spacetime. Not only that, the limit $\lambda\rightarrow\infty$, called the strong 't Hooft coupling, implies a strongly coupled QFT, which is quite difficult to work with, as perturbative methods are not applicable. At the same time, on the other side, there is a weakly curved AdS$_5\times S^5$ space. This is known as the weak/strong property of the correspondence, one of its most beneficial features, after all, a complicated theory can have its properties unraveled in the dual theory with weak coupling, where perturbative methods can be used, and then mapped back.
		
		\begingroup
		\renewcommand{\arraystretch}{1.6}
		\begin{table}[h!]
			\centering
			\setlength{\tabcolsep}{3mm}
			\begin{tabular}{ccc}
				\rowcolor{ufabc!50}	\bf Version 		&\bf Gravity side 		& \bf Gauge side  \\
				\rowcolor{ufabc!30}	Strongest		& \multicolumn{1}{c}{\begin{tabular}[c]{@{}c@{}}Type IIB string theory\\ on AdS$_5\times S^5$,\\ $\forall \, g_s,L/\sqrt{\alpha'}$\end{tabular}}	& \multicolumn{1}{c}{\begin{tabular}[c]{@{}c@{}}$\mathcal{N}=4$ SU($N_c$) \\SYM theory,\\ $\forall \, N_c, g_\text{YM}$ \end{tabular}} \\
				\rowcolor{ufabc!10}	Strong			&  		\multicolumn{1}{c}{\begin{tabular}[c]{@{}c@{}}Classical type IIB string theory\\ on AdS$_5\times S^5$,\\ $g_s\rightarrow0,L/\sqrt{\alpha'}$ constant \end{tabular}}				& \multicolumn{1}{c}{\begin{tabular}[c]{@{}c@{}}'t Hooft large $N_c$ limit of \\ $\mathcal{N}=4$ SU($N_c$) SYM theory,\\ $N_c\rightarrow\infty, \lambda=g_\text{YM}^2N_c$ fixed\end{tabular}}\\
				\rowcolor{ufabc!30}	Weak			&  \multicolumn{1}{c}{\begin{tabular}[c]{@{}c@{}}Classical type IIB supergravity\\ on AdS$_5\times S^5$,\\ $g_s\rightarrow0,L/\sqrt{\alpha'}\rightarrow\infty$ \end{tabular}} 					& \multicolumn{1}{c}{\begin{tabular}[c]{@{}c@{}}  Strong 't Hooft coupling of \\ $\mathcal{N}=4$ SU($N_c$) SYM theory,\\ $N_c\rightarrow\infty, \lambda\rightarrow\infty$\end{tabular}}
			\end{tabular}
			\caption{Summary of the different limits for AdS/CFT correspondence}
			\label{table:a}	
		\end{table}
		\endgroup	
		
		Consequently, the weak version of the correspondence is the one that finds more applications and possibilities for new insights. It remains to obtain the rules for mapping objects and properties between theories, which will be done next.\\	
		
		The objective now is to generalize the description, making explicit the rules that govern the correspondence.
				
		Duality implies an equivalence between the essential properties, such as symmetries and degrees of freedom, so that there is a one-to-one correspondence between each of these elements in the two theories. Symmetries are a good start. The isometry group of AdS$_d$ is SO(2,$d$), which also acts as the symmetry group for conformal theories on the boundary. This implies that the isometries of the bulk are mapped as global spacetime symmetries on the boundary.		
		At the same time, the SO(6) group, which is the symmetry group with all proper rotations of an $S^n$ sphere, is associated with the R-symmetry present in the boundary theory, suggesting an equivalence between the local gauge symmetry in the bulk with a global internal symmetry of the theory on the boundary
		
		In addition to symmetries, it is also necessary to verify the correspondence between observables in both theories. Soon after Maldacena's initial proposition, works by Gubser, Klebanov, and Polyakov \cite{Gubser:1998bc} and Witten \cite{Witten:1998qj} derived protocols for calculating the relations between these observables within the specific case provided by Maldacena, which was the primary verification of the conjecture. The formulation they presented provides the rules for translating quantities from the bulk to the boundary and vice versa, what became known as the \textit{holographic dictionary}\index{AdS/CFT correspondence!holographic dictionary}. 
			
		Fields propagating in the five-dimensional spacetime are referred to as \textit{bulk fields}, while the CFT operators are known as \textit{boundary fields}. According to the holographic dictionary, for each local, gauge-invariant boundary operator $\mathcal{O}(x^\mu)$, there exists a corresponding bulk field $\phi(x^\mu, z)$.
		The challenge is to find a one-to-one correspondence between them.
		
		Let us consider a scalar field $\phi(x^\mu, z)$ in the AdS$_{d+1}$ bulk with mass $m$. Its dynamics are governed by the Klein-Gordon equation in the curved AdS background:
		\begin{equation}
			\frac{1}{\sqrt{-g}} \partial_M \left( \sqrt{-g} g^{MN} \partial_N \phi \right) - m^2 \phi = 0,
			\label{eq:KG_AdS}
		\end{equation}
		where $g_{MN}$ is the AdS metric and $g = \det(g_{MN})=-(L/z)^{2d+2}$.
		Near the boundary $z \to 0$, the Eq. (\ref{eq:KG_AdS}) for $\phi(z)$ (assuming dependence only on $z$ for simplicity, or focusing on the $z$-dependence for fixed $x^\mu$) simplifies to 		
		\begin{flalign}
			\frac{1}{\sqrt{-g}} \partial_M \left( \sqrt{-g} g^{MN} \partial_N \phi \right) - m^2 \phi \\
			&=\left(\frac{z}{L}\right)^{d+1} \partial_z \left\{\left(\frac{L}{z}\right)^{d+1} g^{zz} \partial_z \phi \right\} - m^2\phi \nonumber\\
			&=z^{d+1} \partial_z \left\{z^{-(d-1)} \partial_z \phi \right\} - m^2L^2\phi \nonumber\\
			&= z^2\partial_z^2\phi - (d-1)z\partial_z\phi -m^2L^2\phi = 0,
			\label{eom_phi}
		\end{flalign}
		which has the form of a Cauchy-Euler differential equation. The general solution to such equation is of the form of a power-law, $\phi\sim z^\Delta$, where in this case the exponent is identified as the scaling dimension of the dual operator $\mathcal{O}(x^\mu)$.
		
		Plugging this ansatz into (\ref{eom_phi}) yields the indicial equation
			\begin{flalign}
				\Delta^2 - (d-1)\Delta -m^2L^2	=0,
			\end{flalign}
		with solutions given by the roots
		\begin{flalign}
			\Delta_{\pm} = \frac{d}{2} \pm \sqrt{\frac{d^2}{4}+m^2L^2}.
			\label{eq:scalindimension}
		\end{flalign}
		Thus, there are two independent solutions, $\phi(z \to 0) \sim z^{\Delta_+}$ and $\phi(z \to 0) \sim z^{\Delta_-}$. Defining $\Delta_+=d$ and $\Delta_-=d-\Delta$, the general asymptotic behavior of the bulk field near the boundary can be written as a linear combination:
		\begin{equation}
			\phi(z \to 0) \sim A(x^\mu) z^{d-\Delta} + B(x^\mu)z^{\Delta},
			\label{eq:general_asymptotic}
		\end{equation}
		where $A(x^\mu)$ and $B(x^\mu)$ are arbitrary functions of the coordinates of the boundary.		
		 
		From Eq. (\ref{eq:scalindimension}), the mass of the bulk field can be expressed in terms of the spacetime dimension $d$ and the scaling dimension $\Delta$ of the operator,
		\begin{flalign}
			m^2L^2 = \Delta(\Delta-d).
		\end{flalign}
		This relation holds for the scalar field, but similar steps can be followed to derive analogous relations for other types of fields.	
		More generally, the relation between the bulk mass and the scaling dimension can be extended to higher-rank tensor fields. In particular, for a $p$-form field (a totally antisymmetric tensor field of rank $p$) in the AdS bulk, which is dual to an antisymmetric tensor operator in the boundary CFT, the relation generalizes to
		\begin{flalign}
			m^2 L^2 = (\Delta - p)(\Delta + p - d).
			\label{massscalingrelation}
		\end{flalign}
		In this context, $p$ represents the degree of the differential form, which corresponds exactly to the number of antisymmetric Lorentz indices of the field. Its value is determined directly by observing the tensorial nature of the bulk field or its dual operator:
		for a scalar field, there are no spacetime indices, meaning $p=0$, which correctly recovers the previous relation; for a vector field (such as a gauge field $A_\mu$), there is a single index, yielding $p=1$, and so forth.
	
		\autoref{table:massscalingrelation} summarizes some of these correspondences for different types of fields.

		\begingroup
		\renewcommand{\arraystretch}{1.8}
		\begin{table}[h!]
			\centering
			\setlength{\tabcolsep}{3mm}
			\begin{tabular}{cc}
				\rowcolor{ufabc!50}	\bf Field & \bf Relation \\
				\rowcolor{ufabc!30}	$p$-form				& $m^2L^2=(\Delta-p)(\Delta+p-d)$ \\
				\rowcolor{ufabc!10}	Scalars, massive spin-$2$				& $m^2L^2=\Delta(\Delta-d)$\\
				\rowcolor{ufabc!30}	Massless spin-$2$		&  $m^2L^2=0$ \\
				\rowcolor{ufabc!10}	Spin-$1/2$, spin-$3/2$		& 	$\abs{m}L=\Delta-d/2$ \\
				\rowcolor{ufabc!30}	Rank $s$ symmetric traceless tensor		& $m^2L^2=(\Delta+s-2)(\Delta-s+2-d)$
			\end{tabular}
			\caption{Relation between mass and conformal dimension for different types of fields.}
			\label{table:massscalingrelation}	
		\end{table}
		\endgroup	
		
		A careful reader may have noticed that the relation between mass and conformal dimension for a $p$-form would require a lower bound such as $\Delta>p+d$ to prevent $m^2<0$, which would imply the existence of tachyons in the theory. This would be a highly undesirable instability, especially since $p$-forms represent many useful bulk fields. However, as discovered in \cite{Breitenlohner:1982bm}, the theory remains stable even for some negative values of $m^2$. In fact, the actual stability bound is given by 
		\begin{flalign}
			m^2 \ge - \frac{d^2}{4},
		\end{flalign}
		which is known as the \textit{Breitenlohner-Freedman bound} (see \cite{Nastase:2015wjb} for a more detailed discussion).\index{Breitenlohner-Freedman bound}\\
	
		Returning to the asymptotic solution near the conformal boundary (\ref{eq:general_asymptotic}), it is important to note that each of the two terms plays a distinct role in the holographic correspondence. The only exception occurs when $\Delta=d/2$, which makes the solutions degenerate. We then assume the limit $\Delta \ge d/2$.
		
		The term proportional to $z^{d-\Delta}$ is referred to as the non-normalizable mode. This nomenclature arises because its norm, when computed using the standard Klein-Gordon inner product in the $AdS_{d+1}$ background, diverges near $z=0$. This divergence occurs because the Klein-Gordon action used for normalization includes a factor $\sqrt{-g} \sim 1/z^{d+1}$, which grows faster near the boundary than the decay provided by $z^{d-\Delta}$, leading to a divergent integral.
	
		Furthermore, since the conformal dimension $\Delta$ satisfies the bound $\Delta \ge d/2$, the exponent $d - \Delta$ is less than or equal to $\Delta$. As a result, the term proportional to $z^{d - \Delta}$ decays more slowly than the $z^{\Delta}$ term in the limit $z \to 0$. Consequently, the non-normalizable mode dominates the asymptotic behavior of the bulk field near the boundary.
		To determine the coefficient $A(x^\mu)$, a boundary condition must be imposed, which is taken to be		
		\begin{equation}
			A(x^\mu) = \lim_{z\to 0}  z^{-(d-\Delta)} \phi(x^\mu, z) \equiv \phi_0.
			\label{eq:boundary_condition_A}
		\end{equation}
	This allows us to write $\phi(x^\mu,z\rightarrow0)=z^{d-\Delta}\phi_0(x^\mu)$ at leading order. The coefficient $\phi_0$ is then identified as the source of the bulk field $\phi$.
	
	In quantum field theory, an external source $J(x^\mu)$ is a function that is freely specified and coupled to an operator $\mathcal{O}(x^\mu)$ (typically via a term $\int \dd^d x \, J(x^\mu) \mathcal{O}(x^\mu)$ in the action) to probe the theory's dynamics.	
	Knowing the source is essential for obtaining the theory's observables, because the generating functional $Z[J]$, which encodes all observables (such as correlation functions), is by definition a functional of this source. Observables are then systematically extracted via functional derivatives of $Z[J]=\langle e^{-\int\mathcal{L}_J}\rangle$ (or its logarithm $W[J] = \ln Z[J]$) with respect to the source.
	
	Based on the identification established by the holographic dictionary, we can directly substitute the bulk field's boundary value $\phi_0$ identified with the source $J(x^\mu)$ into the expression for the CFT generating functional \cite{Zaanen:2015oix}:
	\begin{equation}
		Z_\text{CFT}[J = \phi_0] = \Bigg\langle \exp\left( \int \dd^d x\, \phi_0(x^\mu) \mathcal{O}(x^\mu) \right) \Bigg\rangle_\text{CFT}.
		\label{eq:gen_func}
	\end{equation}
	Then, the $n$-point correlation function for the operator $\mathcal{O}$ is given by successive functional derivatives of $Z_{\text{CFT}}$ with respect to the source, evaluated at $\phi_0=0$:
	\begin{flalign}
		\left\langle \mathcal{O}(x_1^\mu)\dots\mathcal{O}(x_n^\mu)\right\rangle = \left. \frac{\delta^n Z_{\text{CFT}}[\phi_0]}{\delta\phi_0(x^\mu_1)\dots\delta\phi_0(x^\mu_n)} \right|_{\phi_0=0}.
	\end{flalign}
	This explicitly shows how the generating functional, and thus all CFT observables derivable from it, can be viewed as depending on the boundary condition imposed in the bulk AdS theory.
	
	The idea that the boundary value of a bulk field in a higher-dimensional gravitational theory acts precisely as an external source for a quantum operator in the lower-dimensional field theory is a central and powerful concept within the AdS/CFT correspondence. It forms the primary operational link enabling calculations of CFT phenomena through gravitational computations.
	
	This connection is formalized by the \textit{Gubser-Klebanov-Polyakov-Witten} (GKPW) \textit{relation} \cite{Gubser:1998bc, Witten:1998qj}\index{AdS/CFT correspondence!GKPW relation}, which conjectures an equivalence between the partition functions of the two theories:
	\begin{flalign}
		Z_\text{CFT}[\phi_0] = Z_\text{AdS}[\phi_0]
	\end{flalign}
	Here, $Z_\text{AdS}[\phi_0]=e^{-\mathcal{S}_\text{AdS}[\phi_0]}$ represents the partition function of the gravitational theory in the AdS bulk. Consequently, the correlation functions of operators in the boundary field theory can be derived from the action of the higher-dimensional gravitational theory as follows:
	\begin{equation}
		\left\langle \mathcal{O}(x_1^\mu)\dots\mathcal{O}(x_n^\mu)\right\rangle = \left. \frac{\delta^n e^{-i\mathcal{S}_\text{AdS}[\phi_0]}}{\delta\phi_0(x^\mu_1)\dots\delta\phi_0(x^\mu_n)}\right|_{\phi_0=0}
		\label{eq:holographic_n_point}
	\end{equation}
		This remarkable result provides a concrete computational tool: CFT correlation functions, which can be notoriously difficult to calculate at strong coupling, can be obtained by solving classical field equations in the higher-dimensional AdS space, computing the corresponding action, and then taking functional derivatives with respect to the boundary value.
			
		\autoref{table:dictionaryadscft} presents some of the most important equivalences in the context of this work. Among them, we find, for instance, the correspondence between operators and fields, as well as the dual relations between symmetries. Anticipating the discussion presented in \Cref{chapter:thermalqcd}, the last two rows of the table illustrate how black holes in AdS space are related to the temperature on the boundary.

		\begingroup
		\renewcommand{\arraystretch}{1.8}
		\begin{table}[h!]
			\centering
			\setlength{\tabcolsep}{3mm}
			\begin{tabular}{cc}
				\rowcolor{ufabc!50}
				\bf Boundary (field theory) & \bf Bulk (gravity) \\
				\rowcolor{ufabc!30}	Partition function				& Partition function 	\\
				\rowcolor{ufabc!10}	Scalar operator	$\mathcal{O}_b$	& 	Scalar field $\phi$ \\
				\rowcolor{ufabc!30}	Fermionic operator $\mathcal{O}_f$		& Dirac field $\psi$	 \\
				\rowcolor{ufabc!10}	Global internal symmetry current $J^a$& Maxwell field $A^a$ \\
				\rowcolor{ufabc!30}	Global spacetime symmetry		& 	Local isometry \\
				\rowcolor{ufabc!10}	Global internal symmetry		& Local gauge symmetry 	\\
				\rowcolor{ufabc!30}	Energy-momentum tensor $T^{ab}$	&  Metric field $g^{ab}$ \\
				\rowcolor{ufabc!10} Entropy	& Area of black-hole horizon \\
				\rowcolor{ufabc!30}	Finite temperature & Black-hole Hawking temperature 
			\end{tabular}
			\caption{Dictionary for AdS/CFT correspondence}
			\label{table:dictionaryadscft}	
		\end{table}
		\endgroup	
		
		Before concluding this section, it is worth briefly commenting on the coefficient $B(x^\mu)$ from the asymptotic expansion (\ref{eq:general_asymptotic}), which has not been determined thus far. When we add a source $J(x^\mu)$ to the CFT, which we have identified with the bulk field's boundary value $\phi_0(x^\mu)$, the theory provides a response. A primary measure of this response is the vacuum expectation value (VEV) of the operator $\mathcal{O}(x^\mu)$ coupled to the source, denoted $\langle \mathcal{O}(x^\mu) \rangle_{\phi_0}$.

		Computing this relation explicitly requires taking functional derivative of the action with respect to the source, as indicated by Eq. (\ref{eq:holographic_n_point}), derived from the GKPW prescription. However, this procedure  demands additional formalism that is not covered here. A detailed derivation can be found in the seminal work by Klebanov and Witten \cite{Klebanov:1999tb}, which leads to
		\begin{equation}
			\langle \mathcal{O}(x^\mu) \rangle_{\phi_0} = (2\Delta - d) B(x^\mu),
		\end{equation}
		and thus we can finally write
		\begin{flalign}
			\phi = z^{d-\Delta} \phi_0 + z^\Delta \frac{\mean{\mathcal{O}}}{2\Delta-4}.
			\label{eq:phi_asymptotic_standard}
		\end{flalign}
			
		This concludes the essential discussion of the AdS/CFT correspondence, offering the conceptual groundwork needed for the analysis that follows in this thesis.

	\section{Quantum chromodynamics}
	\label{section:qcd}
	
	Quantum chromodynamics \index{quantum chromodynamics} is a gauge theory responsible for describing the behavior of the strong nuclear force, characterizing the binding of quarks to form hadrons (such as protons and neutrons), and ensuring that particles of different electric charges remain united in the atomic nucleus, among other tasks. 
	
	It is almost natural to make a comparison of QCD with quantum electrodynamics \index{quantum electrodynamics}, since both are quantum field theories that describe fundamental interactions mediated by spin-1 particles. While QED describes the electromagnetic interaction mediated by the photon, QCD describes the strong nuclear force mediated by gluons. Additionally, both theories are Yang-Mills theories, as briefly introduced in \Cref{section:yangmills}. However, despite these similarities, there are significant differences between the two theories.
	
	In most cases, QED behaves like a much simpler theory: it is a gauge theory based on the U(1) group, which results in only one type of charge, namely the electric charge. Also, it is described in terms of the gauge field $A_\mu$, which satisfies a set of linear equations, the Maxwell's equations:	
	\begin{flalign}
		\partial_\mu F^{\mu\nu} = J^\nu, \qquad
		\epsilon^{\mu\nu\rho\sigma}\partial_\nu F_{\rho\sigma} = 0,
	\end{flalign}
	where $\epsilon^{\mu\nu\rho\sigma}$ is the totally antisymmetric Levi-Civita tensor, $J^\nu$ is the electromagnetic current and $F_{\mu\nu}=\partial_\mu A_\nu - \partial_\nu A_\mu$ is the field strength tensor. With these objects it is possible to describe the behavior of electromagnetic fields that interact with charged particles mediated by photons.
	
	QCD is based on the $SU(N_c)$ group, a symmetry group that, with $N_c=3$ (we will soon see the meaning of this quantity), has eight generators $t^a$ $(a=1,2,\dots,8)$ defined by	
	\begin{flalign}
		t^a = \frac{1}{2}\lambda^a,
	\end{flalign}
	with $\lambda^a$ being the Gell-Mann matrices. 
	Sometimes, for simplicity, $t^a$ are also called Gell-Mann matrices. These generators obey the following algebra:	
	\begin{flalign}
		[t^a,t^b] = if^{abc}t^c,
	\end{flalign}
	where $f^{abc}$ are the totally anti-symmetric structure constants. Each of these generators corresponds to a type of gluon, resulting in a total of 8 independent types of gluons. This is a significant difference compared to the single photon of QED. These gluons are represented by the gauge field $A_\mu^a$, where now the index $a$ gains a physical meaning when representing each of the eight types. We can then construct the gluon field strength tensor
	\begin{flalign}
		G_{\mu\nu}^a = \partial_\mu A_\nu^a - \partial_\nu A_\mu^a + g_sf^{abc}A_\mu^bA_\nu^c ,
		\label{gluonfieldtensor}
	\end{flalign}
	with which we write the Yang-Mills equations of motion that govern the theory and are a generalization of Maxwell's equations, described as	
	\begin{flalign}
		\mathcal{D}_\mu G^{\mu\nu} =J^\nu,
	\end{flalign}
	where $g_s$ (or $\alpha_s=g_s^2/4\pi$)\footnote{In the literature, both $g_s$ and $\alpha_s$ are called the strong coupling constant, and we will alternate between them to use the most convenient notation.} is the strong coupling constant and $\mathcal{D}_\mu$ is the covariant derivative. It is easy to check that these equations are not linear, which means that there are terms of self-interaction, and consequently, the gluons are able to interact with each other, evidencing QCD as being a non-Abelian theory.
	
	Another fundamental difference between QED and QCD is the coupling constant that quantifies the strength of interaction. In QED, the intensity of the coupling between an electric charge and an electromagnetic field is characterized by a dimensionless constant called the fine structure constant, denoted by $\alpha$. In QCD, the equivalent is the coupling constant $\alpha_s$.
	
	Despite its name, the coupling constant $\alpha_s$ varies with the energy scale and has been observed to decrease as the energy increases. In fact, for high energies we have $\alpha_s\rightarrow0$, as discovered by Gross and Wilczek \cite{Gross:1973id} and independently by Politzer \cite{Politzer:1973fx} both in 1973. Thus, the strong interaction effectively disappears for high energy scales.
	This phenomenon is known as \textit{asymptotic freedom}\index{quantum chromodynamics!asymptotic freedom} and it is an essential property that needs to be included and respected in any AdS/QCD model.
	
	But there is an energy scale called the \textit{strong coupling scale}, also known as the QCD scale, where the coupling constant takes on a value $\alpha_s\approx1$. At this order, the strong force indeed lives up to its name as $\alpha_s$ becomes over 100 times greater than the fine structure constant $\alpha$ of QED. This characteristic energy scale is denoted by $\Lambda_\text{QCD}$ and its existence is one of the reasons why conformal invariance is broken in QCD.
	
	As the energy decreases below $\Lambda_\text{QCD}$, the coupling constant increases and eventually becomes so strong that perturbative methods break down. In this context, AdS/QCD models emerge as a possible solution to extract information at such low scales without depending on the values of certain parameters.
	
	Another way to understand the scale-dependent nature of the strong interaction is by examining its role in binding quarks together.
	First, we need to introduce a characteristic length defined in natural units as
	\begin{flalign}
		R_\text{QCD} = \frac{1}{\Lambda_\text{QCD}}.
	\end{flalign}
	
	The binding energy between a quark and an antiquark is a function of the distance $r$ that separates them. For distances smaller than $R_\text{QCD}$, it has a Coulomb-like form,
	\begin{flalign}
		V(r) \sim -\frac{1}{r}, \quad \text{ when} \quad r\ll R_\text{QCD}.
	\end{flalign}
	
	As the distance increases, the binding energy takes on a linear profile
	\begin{flalign}
		V(r) \sim r, \quad \text{ when} \quad r\gg R_\text{QCD}.
	\end{flalign}
	
	With this, we can observe that as we try to increase the distance between quarks, the required energy also increases. In fact, if we attempt to separate the quarks by an infinite distance, an infinite amount of energy would be needed to do so.		
	As the energy increases, there comes a point where it becomes more energetically favorable for a new quark-antiquark pair to be created through a fluctuation, with each one bonding to one of the initial two quarks.	
	Because of this, quarks never appear alone, they are always forming bound states. This phenomenon is known as \textit{confinement}\index{quantum chromodynamics!confinement}, one of the main ingredients of QCD. Incorporating confinement into holographic models, as we will see later, is an essential task.
	
	Confinement also applies to gluons, indicating that they share a common property with quarks. This property is called \textit{color}\index{quantum chromodynamics!color}, which arises from the fundamental representation of the SU(3) group. It is analogous to the QED electric charge in that it defines how quarks and gluons interact with the strong force, and is sometimes referred to as the \textit{color charge}.
	
	Usually, we consider three types of colors: red, green and blue (along with the corresponding anticolors: antired, antigreen and antiblue). A pair composed of color and its anticolor generates a color-neutral total charge. The same happens in the composition of three different colors. The number of colors is denoted by $N_c$ and this is the meaning of $N_c=3$ for QCD mentioned earlier in this section. However, it is also common to leave $N_c$ as a free parameter in order to explore and expand the properties of the theory.
	
	Confinement, in addition to preventing the existence of solitary quarks or gluons, is responsible for keeping them together in color-neutral bonded states called \textit{hadrons}. Despite being in agreement with experimental observations, confinement lacks rigorous analytical proof and is normally derived qualitatively from known results.
	
	Experiments involving hadrons played a crucial role in postulating and subsequently discovering a total of six  types of quarks. To distinguish these quarks, a new quantity known as \textit{flavor} was introduced.\index{quantum chromodynamics!flavor} The properties of each quark flavor are listed in \autoref{table:quarks}.
	
	\begingroup
	\renewcommand{\arraystretch}{1.8}
	\begin{table}[h!]
		\centering
		\setlength{\tabcolsep}{3mm}
		\begin{tabular}{ccccc}
			\rowcolor{ufabc!50}
			\bf Flavor & \bf Symbol &  \bf Mass (MeV) & \bf Constituent mass (MeV)& \bf Charge \\
			\rowcolor{ufabc!30}	Quark up 		& $u$	& $2$		& $336$		& $+2/3$	\\
			\rowcolor{ufabc!10}	Quark down		& $d$	& $4$		& $340$		& $-1/3$	\\
			\rowcolor{ufabc!30}	Quark strange	& $s$	& $95$		& $486$		& $-1/3$	\\
			\rowcolor{ufabc!10}	Quark charm		& $c$ 	& $1250$	& $1550$	& $+2/3$ 	\\
			\rowcolor{ufabc!30}	Quark bottom	& $b$ 	& $4200$	& $4730$	& $-1/3$ 	\\
			\rowcolor{ufabc!10}	Quark top		& $t$ 	& $174000$ 	& $177000$	& $+2/3$
		\end{tabular}
		\caption{Properties of quarks.}
		\label{table:quarks}	
	\end{table}
	\endgroup
	
	A significant disparity in the masses of quarks can be observed depending on their flavor. The reason for this discrepancy remains a mystery, but its implications are of great importance. For comparison purposes, it is worth noting that the scale at which the strong
	coupling becomes large, $\Lambda_{\text{QCD}}$, is of order
	$\Lambda_{\text{QCD}}\approx 300$ MeV.\footnote{The precise value of
		$\Lambda_{\text{QCD}}$ depends on the renormalization scheme and on the number of
		active flavors, ranging roughly from $200$ to $340$ MeV in the commonly used schemes.} Hence, we observe that the masses of the charm, bottom, and top quarks far exceed $\Lambda_\text{QCD}$, effectively excluding them from low-energy physics. Consequently, the number of flavors $N_f$ can be kept as a free parameter, offering flexibility in addressing specific problems.	
	Due to properties of symmetry between up and down quarks, which we will explore in detail in \Cref{section:hadrons}, it is common to work with $N_f=2$. It is also common to choose $N_f=3$ when working with systems that consider only the lighter quarks.\\
	
	For the sake of completeness, we will conclude this section by formalizing the concepts we have discussed thus far and introducing some essential topics for the next steps. 
	
	Quarks are spin-1/2 fermions represented by the quark field $q^a$ (with $a=1,\dots,N_f$ being the flavor index, the color index will be kept implicit) and therefore obey the Dirac equation		
	\begin{flalign}
		\left(i\slashed{D}-M\right)q^a = 0,
	\end{flalign}
	where $\slashed{D}=\gamma^\mu D_\mu$ is a derivative contracted with the Dirac gamma matrix, and $M=m^a_q\mathds{1}$ is a diagonal $N_f\cross N_f$ matrix that combines the quark masses for each flavor. In order to ensure the local gauge invariance of QCD, $D_\mu$ needs to be a covariant derivative of the form:
	\begin{flalign}
		D_\mu = \partial_\mu-i g_s A^a_\mu t^a.
	\end{flalign}
	
	The QCD Lagrangian is constructed by combining the quark part with the part responsible for the gluon dynamics, which is defined by the gluon field strength tensor (\ref{gluonfieldtensor}). This results in:	
	\begin{flalign}
		\mathcal{L}_\text{QCD} = \sum_{a=1}^{N_f}\bar{q}^{\;a}\left(i\slashed{D}-m^a\right)q^a -\frac{1}{2}\Tr\left[G_{\mu\nu} G^{\mu\nu}\right].
		\label{qcdlagrangian}
	\end{flalign}
	The generators of the theory are normalized according to $\Tr(t^a t^b) = \frac{1}{2}\delta^{ab}$.
	
	A general property of fields taking values in this Lie algebra, like the gauge potential $A_\mu$, is that they can be expanded in terms of these generators:	
	\begin{flalign}
		A_\mu = A_\mu^a t^a.
	\end{flalign}
	This decomposition, combined with the normalization property, is essential for evaluating traces over the matrix indices (e.g., color indices) that appear frequently.
	For instance, consider the gluon sector in the QCD Lagrangian (\ref{qcdlagrangian}), which involves the field strength tensor $G_{\mu\nu}$. Since $G_{\mu\nu}$ can be expanded as $G_{\mu\nu}=G_{\mu\nu}^a t^a$, the trace in its quadratic term simplifies as follows	
	\begin{flalign}
		\Tr\left[G_{\mu\nu} G^{\mu\nu}\right] = \Tr\left[G_{\mu\nu}^a t^a G^{\mu\nu}_b t^b\right] = G_{\mu\nu}^a G^{\mu\nu}_b \Tr[t^a t^b] = \frac{1}{2} G_{\mu\nu}^a G^{\mu\nu}_b \delta^{ab} = \frac{1}{2} G_{\mu\nu}^a G_a^{\mu\nu}. \label{eq:trace_gluon_sector}
	\end{flalign}
	This method will be used repeatedly throughout this thesis, sometimes implicitly.
	
	\subsection{Chiral symmetry breaking}
	\label{sec:chiral}
	\index{chiral symmetry breaking}	
	Confinement is not the only characteristic phenomenon of QCD that appears in the low-energy limits. Chiral symmetry, and even more importantly, its breaking, add a significant level of complexity to the theory and are crucial elements that need to be reproduced in a corresponding holographic theory.
		
	The starting point is to assume that all quark masses are zero, which is known as the chiral limit. By expanding the quark fields into their left- and right-handed components using the Dirac matrices, $q_L = (1-\gamma_5)q/2$ and $q_R = (1+\gamma_5)q/2$, the quark sector of the Lagrangian (\ref{qcdlagrangian}) is rewritten as 
		\begin{flalign}
			\mathcal{L}_{\text{quarks}} = i\bar{q}\slashed{D}q = 
			\bar{q}_L i\slashed{D}q_L + \bar{q}_R i\slashed{D}q_R,
			\label{chirallagrangian}
		\end{flalign}
		where flavor and color indices have been omitted (along with summations), as they will not be relevant in this analysis.	
		This particular form of the Lagrangian obeys a group of symmetries described as	
		\begin{flalign}
			\text{U}(N_f)_L \cross \text{U}(N_f)_R,
		\end{flalign}
		which is known as \textit{flavor chiral symmetry} or just \textit{chiral symmetry}\index{chiral symmetry breaking!chiral symmetry}.	
		It is possible to decompose these symmetries into
		\begin{flalign}
			\text{U}(1)_V \cross \text{U}(1)_A \cross \text{SU}(N_f)_L \cross \text{SU}(N_f)_R.
		\end{flalign}
		
		The U$(1)_V$ symmetry corresponds to a phase transformation that is applied in the same way for $q_L$ and $q_R$, i.e., $q_{L,R} = e^{i\theta}q_{L,R}$. It is associated with the conservation of baryon number, which represents the count of quarks and antiquarks before and after a strong interaction.
		The axial symmetry U$(1)_A$, on the other hand,  transforms fermions through a phase of the opposite sign, i.e., $q_{L,R} = e^{\pm i\theta}q_{L,R}$, and is not associated with a conserved quantity due to quantum effects, which is known as the \textit{axial anomaly}.
		These two Abelian symmetries will not be of major interest for this work and, therefore, will not be covered in more detail.
		
		For practical reasons, since we already know the outcomes of the $U(1)_V$ and $U(1)_A$ groups, we will focus on the two non-Abelian groups, namely
		\begin{flalign}
			\text{SU}(N_f)_L \cross \text{SU}(N_f)_R,
		\end{flalign}
		which acts on the quarks fields as
		\begin{flalign}
			\text{SU}(N_f)_L:\; q_L\mapsto Lq_L, \qquad \text{SU}(N_f)_R:\; q_R\mapsto Rq_R.
		\end{flalign}
		Here, $L$ and $R$ are $N_f\cross N_f$ unitary matrices with determinant 1. This implies that the $\text{SU}(N_f)_L \cross \text{SU}(N_f)_R$ transformations act independently on the left- and right-handed components of the quark fields, respectively.
		According to Noether's theorem, the invariance under the separate $\text{SU}(N_f)_L$ and $\text{SU}(N_f)_R$ transformations leads to two sets of $N_f^2 - 1$ conserved currents, one for each symmetry group, given by
		\begin{flalign}
			\left(j_{R,L}\right)_\mu^a = -i\frac{\partial\mathcal{L}}{\partial(\partial q_{R,L})}\left(t^a\right)q_{R,L}  = \bar{q}_{R,L}\gamma_\mu t^a q_{R,L}.
		\end{flalign}
		
		If we depart from the idealized scenario of massless quarks and reintroduce the mass term in (\ref{chirallagrangian}), the Lagrangian is no longer invariant under chiral symmetry, leading to \textit{explicit symmetry breaking}\index{chiral symmetry breaking!explicit chiral symmetry breaking}. 
		The quark mass term in the Lagrangian, 
		\begin{equation}
			\mathcal{L}_{\text{mass}} = - \bar{q} m q = - m(\bar{q}_L q_R + \bar{q}_R q_L )
			\label{eq:mass_term}
		\end{equation}
	explicitly mixes left- and right-handed components. It is evident that this scenario aligns with real-world observations, where quarks possess mass, as discussed earlier.
	In a scenario where only the up, down, and strange quarks are present, since their masses are small compared to the QCD scale, chiral symmetry can still be regarded as an approximate symmetry of QCD.
			
	Even in the chiral limit, there is also a \textit{spontaneous symmetry breaking} manifested by the existence of an non-vanishing expectation value for the bilinear operator $\bar{q}_Rq_L$ in the vacuum, 
		\begin{flalign}
			\bra{0}\bar{q}_R q_L\ket{0} = \mean{\bar{q}_R q_L} = \Sigma \neq 0,
		\end{flalign}
	known as a \textit{chiral condensate} (or \textit{quark condensate}) 	\index{chiral symmetry breaking!chiral condensate}, generated by the dynamics of the strong force in the vacuum.			
	With this, the chiral symmetry group is spontaneously broken to its diagonal subgroup,
		\begin{flalign}
			\text{SU}(N_f)_L \cross \text{SU}(N_f)_R \mapsto \text{SU}(N_f)_V,
			\label{breakinggroup}
		\end{flalign}
	which is also the same pattern for the explicit breaking. Therefore, due to the presence of quark masses and non-vanishing condensates in the vacuum, it is essential to emphasize that chiral symmetry is not an exact symmetry of QCD.
		
	According to Goldstone's theorem, the spontaneous breaking of a continuous global symmetry implies the existence of massless scalar bosons, known as Goldstone bosons. In the case of QCD with $N_f$ flavors, the $\text{SU}(N_f)_L \cross \text{SU}(N_f)_R \mapsto \text{SU}(N_f)_V$ breaking pattern involves $N_f^2 - 1$ broken axial generators. Consequently, $N_f^2 - 1$ massless Goldstone bosons should appear in the spectrum.
	In the present context, these bosons are bound states of quarks known as mesons.
	For $N_f = 2$, these are the three pions ($\pi^+, \pi^-, \pi^0$). For $N_f = 3$, these correspond to the pseudoscalar octet: the pions ($\pi$), kaons ($K$), and the eta ($\eta$) meson. Because the explicit symmetry breaking due to non-zero quark masses is small, these particles are observed not as massless, but as light \textit{pseudo-Goldstone bosons}.
		
	The masses of these pseudo-Goldstone bosons are formally related to the parameters governing both the explicit and spontaneous breaking of chiral symmetry, the quark masses and the chiral condensate, through the Gell-Mann-Oakes-Renner relation. For the pion, this relation is given by
	\begin{flalign}
		m_\pi^2 f_\pi^2 = -\frac{1}{2}(m_u + m_d)\bra{0}\bar{u}u + \bar{d}d\ket{0} + \mathcal{O}(m_{u,d}^2),
		\label{gmor}
	\end{flalign}
	where $f_\pi$ is the pion decay constant. Analogous expressions exist for the other mesons in the pseudoscalar octet, taking into account their respective quark compositions \cite{Weinberg:1996kr}.
	These mesons, as well as hadrons in general, will be covered in the next subsection.	
	
	\subsection{Hadrons: Mesons and baryons}
	\label{section:hadrons}
	
	With the advances in particle detectors and accelerators during the 1950s and 1960s, a huge number of particles began to be discovered even before a theory capable of explaining their existence had been formulated. While these discoveries opened up a wide range of possibilities, they also posed a major challenge: bringing order to this "particle zoo" \cite{Johnson:1999am}.
	
	One of the first attempts with this objective had already been made before the particle zoo, when there were only three particles discovered to deal with.
	In 1932, shortly after the discovery of the neutron, Werner Heisenberg noticed similarities of this new particle with the proton (which had already been discovered in 1919). Both have very similar masses (see \autoref{table:spin12baryons}) and behave indistinguishably under the force that keeps them bound in the nucleus (in modern terms: they interact in the same way with the strong force).
	Thus, Heisenberg introduced the concept of \textit{isospin}, an apparent symmetry that places the proton and neutron as different states of the same particle, the nucleon, distinguished only by their charges.

	In reality, isospin is not an exact symmetry because the interchangeability between protons and neutrons is only valid in the presence of the strong force. As we know well, protons have a positive charge and neutrons have no charge, so from the perspective of electromagnetic interactions, we are talking about a vast difference. The difference still exists in processes involving the weak force.

	Although isospin symmetry is conserved only in the strong interaction, it is common to take it into consideration in some analyses. This is because, for $N_f=2$, the physics reduces to two lightest quarks, with $m_u\approx m_d$, which makes the isospin subgroup SU$(2)_V$ much closer to being an exact symmetry than its larger group SU$(3)_V$. Indeed, the mathematics behind isospin is based on the concept of spin from quantum mechanics, which provides valuable tools for various studies that are still concerned with the organization of hadrons.
	
	After a series of attempts with partially successful models, Gell-Mann \cite{Gell-Mann:1964ewy} and Zweig \cite{Zweig:1964jf} independently proposed the \textit{quark model}\index{quark model} in 1964.
	The key ideas proposed are basically what we have seen in \Cref{section:qcd}, such as quarks as constituents of the particles observed in nature and the concepts of color and flavor. However, it is the organization of bound states of quarks that we will address in more detail in this section.
	
	In \Cref{section:qcd}, we discussed how the phenomenon of color confinement leads to the manifestation of quarks in nature as bound states, ensuring their overall color neutrality. These bound states are called \textit{hadrons} (from the Greek "\textit{hadrós}", "robust"), and they can be organized according to the quark model into multiplets following specific criteria.

	Among the main quantum numbers to consider for now are those used in the literature, particularly in particle catalogues, to identify hadrons. Typically, the notation used follows this format:
	\begin{flalign}
		I(J^{PC}).
	\end{flalign}
	The first of these quantum numbers, $I$, is the isospin, named after the symmetry proposed by Heisenberg, although it does not have much resemblance to the original idea. Its third component, $I_3$, is also relevant when distinguishing particles. With the advent of the quark model, the up and down quarks are assigned the isospin $I=1/2$, but they differ by having $I_3=1/2$ for the up quark and $I_3=-1/2$ for the down quark, which extends to the hadrons they form. The other quark flavors have $I=I_3=0$ \cite{ParticleDataGroup:2020ssz}.
	
	The total angular momentum $J$ combines the intrinsic angular momentum (spin) $S$ and the orbital angular momentum $L$, taking values in the range $\abs{L-S} \leq J \leq \abs{L+S}$ in integer steps. 
	
	The index $P$ represents parity and indicates how a quantity behaves under spatial inversion: it takes the value "$+$" for quantities that remain unchanged and "$-$" for those that are altered. By convention of the quark model, quarks have positive intrinsic parity and antiquarks have negative intrinsic parity \cite{ParticleDataGroup:2020ssz}. 
	Lastly, $C$ represents charge conjugation parity (C-parity), which describes the behavior of a system under the charge conjugation operation, an operation that transforms a particle into its antiparticle. A value of "$+$" indicates the state is even under charge conjugation, while "$-$" indicates it is odd.
	
	For the purposes of this thesis, it will often be more common to use a shorter notation, $J^P$, which includes the most important quantum numbers for the analyses to be developed.
	
	Other important quantum numbers within the quark model which will be detailed later in their respective contexts.
	
	\subsubsection{Mesons}
	
	In 1935, Yukawa proposed a new particle capable of carrying the nuclear force and calculated its mass \cite{Yukawa:1935xg}. This particle would be responsible for keeping protons and neutrons together, overcoming their electromagnetic repulsion and ensuring the cohesion of atomic nuclei. These particles were named \textit{mesons}\index{mesons} (from the Greek \textit{mesos}, "intermediate"). Today, we understand mesons as a type of hadron that achieves color neutrality through the formation of quark-antiquark pairs or linear combinations thereof. This structure results in mesons having integer spins, which classifies them as bosons.

	The simplest combinations we can observe are in a scenario with only up and down quarks and their respective antiquarks, joined with antiparallel spins to form hadrons with spin $J=0$. These are the mesons $\pi^+$, $\pi^-$, and $\pi^0$, now known as pions. Their compositions and properties are listed in \autoref{table:pseudoscalarmesons}. Pions are the lightest hadrons that exist and were the first mesons to be discovered in experiments with cosmic rays in 1947.
	
	The three types of pions arise from the spontaneous breaking of chiral symmetry $\text{SU}(2)_L \cross \text{SU}(2)_R \mapsto \text{SU}(2)_V$. According to the Goldstone theorem, the bosons resulting from this breaking would be massless. However, due to the non-zero quark masses, which break the chiral symmetry, the pions are considered pseudo-Goldstone bosons with a small mass.
	
	The pions are not alone among the lightest mesons. In fact, they are some examples of \textit{light mesons}\index{mesons!light mesons}, a term used in the literature to describe mesons composed of light quarks -- up, down, and strange. And it is by adding the strange quark to the scenario that new possibilities of hadrons, such as \textit{strange mesons}\index{mesons!strange mesons}, and new phenomena emerge.
	
	Still within the family of spin-0 mesons, there is a group of strange mesons collectively known as kaons, which can be divided into two isospin $I=1/2$ doublets: one composed of the kaons $K^+(u\bar{s})$ and $K^0(d\bar{s})$, and the other composed of their respective antiparticles, $K^-(\bar{u}s)$ and $\bar{K}^0(\bar{d}s)$. The fact that kaons are almost three times more massive than pions highlights the mass difference between the quarks $u$ and $d$ when compared to the quark $s$.
	In general, kaons are full of peculiarities that make them very useful in CP violation and leptonic decay studies \cite{Griffiths:2008zz}, to name a few examples.
	
	An eighth spin-0 meson, heavier than the previous ones, is called the eta ($\eta$) meson, which is composed of a linear combination of quark-antiquark pairs of the same flavor, resulting in a neutral particle.
	
	This octet of mesons formed by three pions, four kaons and the eta meson represents the pseudo-Goldstone bosons generated in the spontaneous breaking of chiral symmetry with three flavors of quark, $\text{SU}(3)_L \cross \text{SU}(3)_R \mapsto \text{SU}(3)_V$. Therefore, mesons can be bound states of quarks and, at the same time, pseudo-Goldstone bosons.
	
	There is a ninth light spin-0 meson that forms a singlet, the eta prime ($\eta^\prime$) meson. Although it is not part of the octet mentioned earlier, it is commonly grouped together to form a nonet since they share properties enough to be classified in the same family. All these mesons are summarized in \autoref{table:pseudoscalarmesons}.
		
	An interesting fact was observed regarding strange hadrons: some of them, such as kaons, are created through particle collision processes with no particular peculiarity, but their decay time is much slower than other hadrons, even those with very similar masses.
	
	\begingroup
	\renewcommand{\arraystretch}{1.2}
	\begin{table}[H]
		\centering
		\setlength{\tabcolsep}{3mm}
		\begin{tabular}{cccccc}
			\rowcolor{ufabc!50}
			\bf Meson & \bf Composition & \bf Mass (MeV) & $I$ & $I_3$ & \bf Charge\\
			\rowcolor{ufabc!30} $\pi^+$ &$u\bar{d}$ &$139$&$1$&$1$&$+1$ \\
			\rowcolor{ufabc!10} $\pi^-$ &$\bar{u}d$ &$139$&$1$&$-1$&$-1$\\
			\rowcolor{ufabc!30} $\pi^0$ &$\frac{1}{\sqrt{2}}(u\bar{u}-d\bar{d})$&$135$&$1$&$0$&$0$\\
			\rowcolor{ufabc!10} $K^+$ 	&$u\bar{s}$&$494$&$1/2$&$1/2$&$+1$\\
			\rowcolor{ufabc!30} $K^0$ 	&$d\bar{s}$&$498$&$1/2$&$-1/2$&$0$\\
			\rowcolor{ufabc!10} $K^-$ 	&$\bar{u}s$&$494$&$1/2$&$1$/$2$&$-1$\\
			\rowcolor{ufabc!30} $\bar{K}^0$ &$\bar{d}s$&$498$&$1/2$&$-1/2$&$0$\\
			\rowcolor{ufabc!10} $\eta$ 	&$\frac{1}{\sqrt{6}}(u\bar{u}-d\bar{d}-2s\bar{s})$&$548$&$0$&$0$&$0$\\
			\rowcolor{ufabc!30} $\eta^\prime$ &$\frac{1}{\sqrt{3}}(u\bar{u}-d\bar{d}+s\bar{s})$&$958$&$0$&$0$&$0$
		\end{tabular}
		\caption{Spin-0 mesons.}
		\label{table:pseudoscalarmesons}
	\end{table}	
	\endgroup
	
	The solution to this disparity was independently proposed by Gell-Mann, Nakano, and Nishijima in 1953 \cite{Gell-Mann:1964ewy,Nakano:1953zz}, with the introduction of a new conserved quantity: \textit{strangeness} \index{strangeness}. This also led to the introduction of a new quantum number, denoted as $\mathtt{S}$, which sums to -1 for each strange quark in the hadron's composition and +1 for each strange antiquark.
	
	The main idea behind the modern concept of strangeness is that the strong interaction cannot change the number of strange quarks, as shown in the process:
	\begin{flalign}
		\pi^- + p &\rightarrow K^+ + K^- + n \\
		\mathtt{S}: (0) + (0) &\rightarrow (+1) + (-1) + (0), \nonumber
	\end{flalign}	
	where the pion $\pi^-$ and a proton $p$ decay into a positive kaon $K^+$, a negative kaon $K^-$, and a neutron $n$, keeping the total strangeness $(\mathtt{S}=0)$ unchanged.
	But hadrons can also decay through electromagnetic and weak interactions, with only the weak interaction not conserving strangeness, as in the process:
	\begin{flalign}
		K^0 &\rightarrow \pi^+ + \pi^0 \\
		\mathtt{S}: (+1) &\rightarrow (0) + (0),
	\end{flalign}
	which represents the weak decay of a neutral kaon $K^0$ into two pions, violating strangeness conservation. As the weak decay duration range is over 10 orders of magnitude greater than that of the strong interaction, processes that do not preserve strangeness proceed through this pathway, resulting in much slower decay of the hadrons.
	
	The inclusion of strangeness allows hadrons to be organized in an elegant scheme known as the \textit{eightfold way}, which groups multiplets according to their properties.
	Each hadron is positioned at a vertex, where those in the same horizontal line share the same strangeness value, and those in the same left-leaning diagonals share the same electric charge.
	Diametrically opposite particles are antiparticles of each other.
	
\begin{figure}[H]
		\centering
		\begin{tikzpicture}
			\filldraw [black] (0,0) circle (3pt) node[label=$\pi^0\text{, } \eta \text{, } \eta^\prime$,xshift=0.6cm]{};
			\filldraw [black] (3.5,0) circle (3pt) node[label=$\pi^+$,xshift=0.4cm]{};
			\filldraw [black] (-3.5,0) circle (3pt) node[label=$\pi^-$,xshift=0.5cm]{};
			\filldraw [black] (-1.75,3.08) circle (3pt) node[label=$K^0$,xshift=0.5cm]{};
			\filldraw [black] (1.75,3.08) circle (3pt) node[label=$K^+$,xshift=0.5cm]{};
			\filldraw [black] (-1.75,-3.08) circle (3pt) node[label=$K^-$,xshift=0.5cm]{};
			\filldraw [black] (1.75,-3.08) circle (3pt) node[label=$\bar{K}^0$,xshift=0.5cm]{};
			
			\node at (-5.2,0) {$\mathtt{S}=0$};
			\node at (-5.2,3.08) {$\mathtt{S}=+1$};
			\node at (-5.2,-3.08) {$\mathtt{S}=-1$};
			
			\node at (5.82,-4.3) {$q=+1$};
			\node at (-1.18,-4.3) {$q=-1$};
			\node at (2.32,-4.3) {$q=0$};
			
			\draw [dashed] (-1.75,3.08) -- (2.32,-4.08);
			\draw [dashed] (-3.5,0) -- (-1.18,-4.08);
			\draw [dashed] (1.75,3.08) -- (5.82,-4.08);
			\draw [dashed] (1.75,3.08) -- (3.5,0);
			\draw [dashed] (3.5,0) -- (1.75,-3.08);
			\draw [dashed] (-3.5,0) -- (-1.75,3.08);
			
			\draw [dashed] (-4.4,0) -- (3.5,0);
			\draw [dashed] (-4.4,-3.08) -- (1.75,-3.08);
			\draw [dashed] (-4.4,3.08) -- (1.75,3.08);	
		\end{tikzpicture}
		\caption{Nonet of spin-0 mesons.}
		\label{fig:eightfoldway}
	\end{figure}
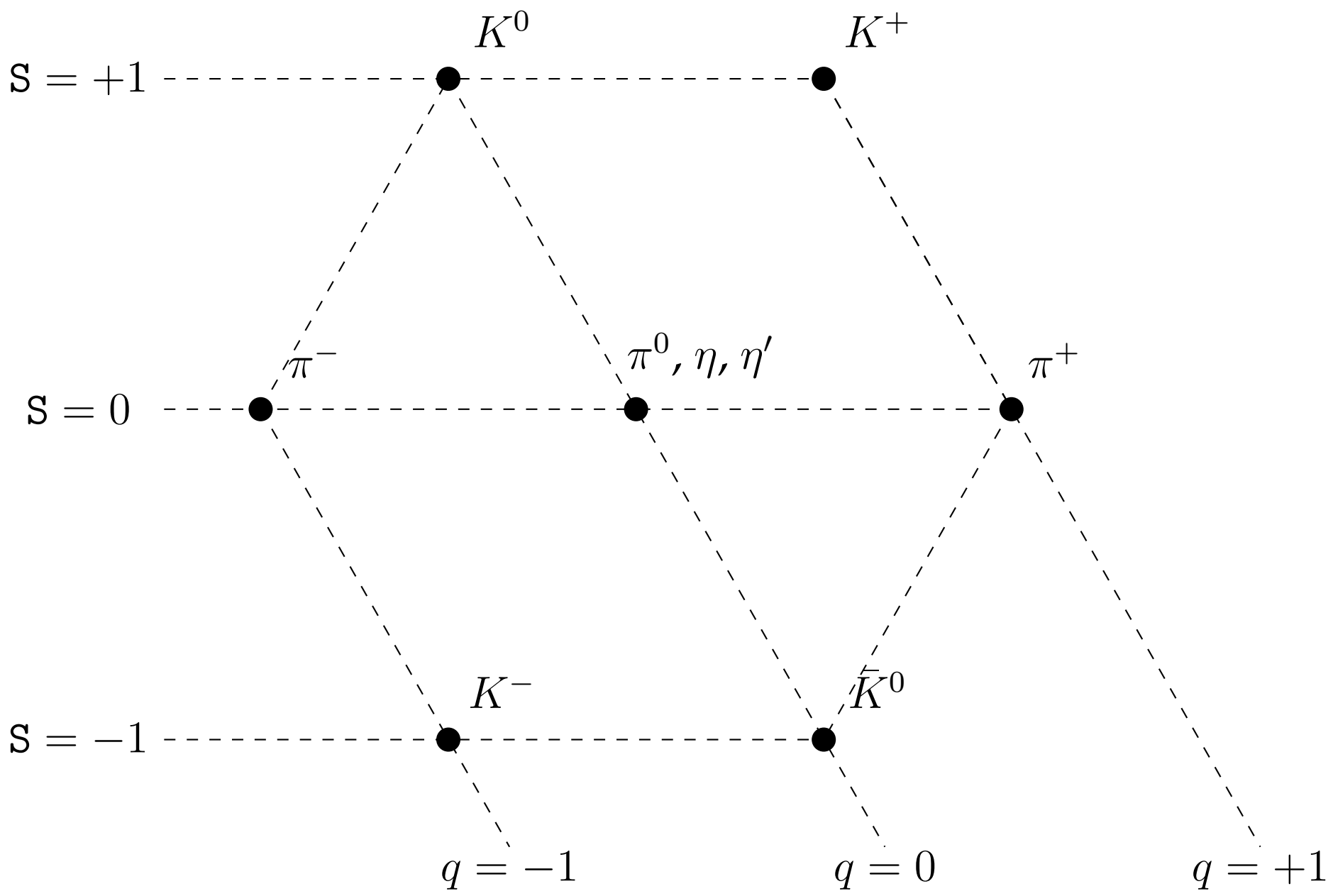

	Later, the eightfold way would be incorporated into the quark model in a more accurate version with respect to experimental results.
	For example, for the spin-0 meson nonet, we have the scheme shown in \autoref{fig:eightfoldway}.
	
	Analogous to strangeness, other quantum numbers are associated with heavier quark flavors discovered later, such as charm, topness, and bottomness. Consequently, extended versions of the eightfold way can include this other quantum numbers, requiring a three-dimensional representation of the scheme.
	Since the mesons analyzed in this thesis do not include these heavier quarks, we will not address these properties in detail.

	Mesons can be grouped into families depending on their principal quantum numbers, such as spin and parity. For example, mesons with spin $J=1$ belong to the class of vector mesons, which, in turn, can be of two types depending on their parity. This is because a vector can be either polar, meaning its sign is reversed under a coordinate inversion, or axial, where its sign remains unchanged under the same transformation. Mathematically, this can be observed by applying the parity operator $P$ to arbitrary polar vectors $\boldsymbol{u}$ and $\boldsymbol{v}$, which results in $P\boldsymbol{u}=-\boldsymbol{u}$ and $P\boldsymbol{v}=-\boldsymbol{v}$. However, the cross product of these vectors yields a third vector, $\boldsymbol{w}=\boldsymbol{u}\cross\boldsymbol{v}$, for which the application of $P$ does not change the sign, since $P\boldsymbol{w}=P(\boldsymbol{u}\cross\boldsymbol{v})=(-\boldsymbol{u})\cross(-\boldsymbol{v})=\boldsymbol{w}$. Therefore, since $\boldsymbol{u}$ and $\boldsymbol{v}$ are polar vectors, then $\boldsymbol{w}$ is an axial vector, also known as a pseudovector \cite{Griffiths:2008zz}.
	
	Due to this property of parity inversion, mesons with $J^P=1^-$ are called vector mesons, while those with $J^P=1^+$ are called axial-vector or pseudovector mesons. The same logic applies to distinguishing scalar ($J^P=0^+$) and pseudoscalar ($J^P=0^-$) mesons. This indicates that the mesons presented in \autoref{table:pseudoscalarmesons} are of the pseudoscalar type. 
	
	The analysis developed in \Cref{section:lightmesons} will address four of these types:
	the $f_{0}$ family representing the scalars, the pion $(\pi)$ family for the pseudoscalar mesons, the $a_{1}$ family representing the axial-vector mesons, and the rho $(\rho)$ mesons representing the vector mesons.	
	 
	In addition to the previously discussed pseudoscalar kaons, there are also vector kaons cataloged by the PDG, formed by a combination of an up or down quark and a strange antiquark (or their respective antiparticles). Conventionally, an asterisk is added to vector kaons (e.g., $K^*$) to differentiate them from their pseudoscalar counterparts. They are the main subject of study in \Cref{section:strangekaons}, where their properties will be presented in more detail.

\subsubsection{Baryons}

	The other way in which quarks can arrange themselves to achieve a color-neutral configuration is in sets of three, each with a different color state. Particles formed by this arrangement of three quarks are called \textit{baryons} \index{baryons} (from the Greek "\textit{barýs}", "heavy"). Since all quarks are spin-1/2 particles, baryons also have half-integer spin, making them fermions.
	\footnote{Exotic baryons, those with compositions different from the usual $qqq$ quark combination, have been discovered in experiments over the last decade. Examples of these exotic baryons include tetraquarks and pentaquarks. Since these baryons will not be addressed in this thesis, we will continue with the most simple explanation.}
	
	The three-quark configuration requires additional care when dealing with baryons within the quark model. For example, it is common for baryons with the same quark composition to represent different particles. Therefore, it is necessary to consider other quantum numbers during the analysis. 
	As this section aims to provide only a brief overview of the objects of study for this thesis, we will not adopt the more fundamental approach that involves starting from the representation of baryons in group theory.
	Interested readers may consult Refs. \cite{Griffiths:2008zz,Fayyazuddin:1994wh} for a more in-depth approach.
		
	The most well-known baryons are the nucleons, namely the proton $p$ and the neutron $n$, with quark compositions of ($uud$) and ($udd$), respectively. Since quarks are fermions, their total wavefunction must be antisymmetric, according to the Pauli exclusion principle in its generalized form. For the ground state nucleon, this constraint leads to a total spin configuration of $J=1/2$. From the perspective of isospin, the proton and neutron are two different states of the nucleon, a hadron with isospin $I=1/2$, where the state $I_3=+1/2$ corresponds to the proton and $I_3=-1/2$ to the neutron.
	
	The grouping of spin-1/2 baryons forms an octet composed of the nucleons and six other baryons from three different families:
	the Sigma ($\Sigma$), Xi ($\Xi$), and Lambda ($\Lambda^0$) baryons. Their compositions and other properties are summarized in \autoref{table:spin12baryons}.
	In representations that include heavier flavors, other spin-1/2 baryons of the Sigma and Xi types can be included, such as the charmed Sigma ($\Sigma_c$), bottom Sigma ($\Sigma_b$), charmed Xi ($\Xi_c$), and even the double-charmed Xi ($\Xi_{cc}$).
		
	\begingroup
	\renewcommand{\arraystretch}{1.8}
	\begin{table}[h!]
		\centering
		\setlength{\tabcolsep}{3mm}
		\begin{tabular}{cccccc}
			\rowcolor{ufabc!50}
			Baryon & Composition & Mass (MeV) & $I$ & $I_3$ & Charge\\
			\rowcolor{ufabc!30} $p$ 		& $uud$ &$938$	&$1/2$	&$1/2$	&$+1$ \\
			\rowcolor{ufabc!10} $n$ 		& $udd$ &$939$	&$1/2$	&$-1/2$	&$0$\\
			\rowcolor{ufabc!30} $\Sigma^+$	& $uus$ &$1189$	&$1$	&$1$	&$+1$\\
			\rowcolor{ufabc!10} $\Sigma^0$	& $uds$ &$1192$	&$1$	&$0$	&$0$\\
			\rowcolor{ufabc!30} $\Sigma^-$	& $dds$ &$1197$	&$1$	&$-1$	&$-1$\\
			\rowcolor{ufabc!10} $\Lambda^0$	& $uds$ &$1115$	&$0$	&$0$	&$0$\\
			\rowcolor{ufabc!30} $\Xi^0$		& $uss$ &$1314$	&$1/2$	&$1/2$	&$0$\\
			\rowcolor{ufabc!10} $\Xi^-$ 	& $dss$ &$1321$	&$1/2$	&$-1/2$	&$-1$
		\end{tabular}
		\caption{Spin-1/2 baryons.}
		\label{table:spin12baryons}
	\end{table}	
	\endgroup
	
	To highlight the importance of other quantum numbers for baryons, we can examine the baryons that form the spin-3/2 decuplet. Among them is the family of Delta ($\Delta$) baryons, which are composed solely of up and down quarks, this time with their spins aligned, granting them a total spin of $J=3/2$. They are represented as $\Delta^{++}$, $\Delta^+$, $\Delta^0$, and $\Delta^-$.
	
	The baryons $p$ and $\Delta^+$ share the same quark composition and electric charge, as do the pair $n$ and $\Delta^0$. However, the delta baryons have higher mass and spin, so they can be considered excited states of the proton and neutron.	
	This correspondence as excited states does not hold for $\Delta^{++}$ and $\Delta^-$. Even though $\Delta^-$ shares the same charge as the antiproton ($\bar{p}$), they do not possess the same quark composition, as $\bar{p}$ is composed of the same quarks as the proton but with particles exchanged for their corresponding antiparticles.
	
	This feature reappears in the decuplet. In addition to the Delta baryons, there are Sigma and Xi baryons with the same quark compositions as their octet counterparts, but with a higher spin ($J=3/2$) and slightly greater mass. They are distinguished by an asterisk (e.g., $\Sigma^*, \Xi^*$) or by the explicit addition of their mass in parentheses (e.g., $\Sigma(1385),\Xi(1530)$) and are also considered excited states of their spin-1/2 counterparts.
	
	Concluding the decuplet is the Omega baryon ($\Omega^-$), which is composed of three strange quarks. By definition, baryons in this family do not contain up or down quarks. The experimental detection of this baryon is regarded as a major success for the quark model, as it took place after Gell-Mann had predicted its existence and properties, thereby consolidating the predictive capability of the eightfold way multiplets \cite{Griffiths:2008zz}.	
	
	Many other particles, compositions, and arrangements could be discussed. However, for the scope of this thesis, this brief overview is sufficient, as it provides an appropriate introduction to the objects of study in this work.

	\subsection{Mass generation in QCD}
	\index{quantum chromodynamics!mass generation}
	
	At this point, one may have noticed that the mass of hadrons is not equal to the sum of the masses of the quarks that compose them. The proton, for instance, has a mass of approximately $938$ MeV, while the three quarks in its composition barely add up to $10$ MeV. Understanding where the remaining $\sim 95\%$ of the mass comes from involves some of the most subtle aspects of QCD, which will be addressed here, as their description will be relevant throughout this thesis.
	
	In the chiral limit, the classical QCD Lagrangian contains no dimensionful parameters. This includes the coupling constant, which is dimensionless in natural units. As discussed in \Cref{section:cft}, this implies that the classical theory possesses a conformal symmetry.	
	The Noether current associated with scale invariance is the dilatation current,
	\begin{equation}
		J^\mu_\text{D} = x_\nu T^{\mu\nu},
	\end{equation}
	where $T^{\mu\nu}$ is the energy-momentum tensor \cite{Donoghue:1992dd}. The divergence of this current is given by the trace of the energy-momentum tensor,
	\begin{equation}
		\partial_\mu J^\mu_\text{D} = T^\mu_{\ \mu}.
	\end{equation}
	Classically, for a massless gauge theory, this trace vanishes, $T^\mu_\mu = 0$, ensuring scale invariance.
	However, this symmetry is broken at the quantum level, a phenomenon known as the \textit{conformal anomaly} (or \textit{trace anomaly})\index{conformal field theory!conformal anomaly}\footnote{The term \textit{anomaly} denotes the breaking of a classical symmetry by quantum effects, just like the axial anomaly seen previously.} \cite{Chanowitz:1972da,Crewther:1972kn,Collins:1976yq,Aguilar:2019teb}. The breaking arises because the quantization and renormalization procedures introduce a mass scale, the renormalization scale $\mu$. This leads to the description given by the beta function\index{quantum chromodynamics!beta function}, which determines how the coupling constant $g$ depends on the renormalization scale,
	\begin{equation}
		\beta(g) = \mu \frac{dg}{d\mu},
		\label{betafunction}
	\end{equation}
	effectively describing the running of the coupling constant.
	
	The quantum correction to the trace of the energy-momentum tensor is therefore non-zero. In the presence of massive quarks, it takes the form \cite{Donoghue:1992dd}
	\begin{flalign}
		T^\mu_\mu = \frac{\beta(g)}{2g} G^a_{\mu\nu}G^{a\mu\nu} + \sum_a m_a (1+\gamma_m)\bar{q}_a q_a,
		\label{traceanomaly}
	\end{flalign}
	where
	\begin{flalign}
		\gamma_m = \frac{\mu}{m}\frac{\partial m}{\partial\mu}
		\label{anomalousdimension}
	\end{flalign}
	is the anomalous dimension of the quark mass operator\index{quantum chromodynamics!anomalous dimension}, accounting for the scale dependence introduced by renormalization.
	
	It is interesting to note that the breaking of the conformal symmetry in QCD is due to 
	both the explicit breaking of chiral symmetry (via quark mass terms) and the non-perturbative fluctuations of the gluon fields. Thus, even in the chiral limit $m_q \to 0$, the first term in Eq. (\ref{traceanomaly}) survives and the anomaly remains present.
	
	The self-interaction term between gluon fields is intimately linked to another structure of the QCD vacuum. 
	Similar to the formation of the chiral condensate $\mean{\bar{q}q}$, the non-perturbative dynamics of QCD lead to a non-vanishing vacuum expectation value for the gluon operator appearing in the trace anomaly, i.e.,
	\begin{flalign}
		\Big\langle 0 \hspace{0.8mm} \Big| \frac{\alpha_s}{\pi} G^a_{\mu\nu}G_a^{\mu\nu} \Big | \hspace{0.6mm} 0 \Big\rangle = (350\pm30 \text{ MeV})^4,
		\label{gluoncondensate}
	\end{flalign}
	which is known as the \textit{gluon condensate}\index{gluon condensate}, representing the energy density of the QCD vacuum due to non-perturbative gluon fluctuations, and its numerical value presented here was calculated using sum rules \cite{Shifman:1978bx,Ji:1995sv}. 
	
	This self-interaction of the QCD gauge bosons generates a fundamental characteristic distinct from QED: the existence of a \textit{mass gap}. Even in the absence of quarks (pure Yang-Mills theory), the theory predicts that the lowest energy excitation above the vacuum has a finite, positive mass determined by the scale $\Lambda_{\text{QCD}}$. In other words, the lightest particles in the theory are not massless. A direct consequence of this color confinement and the mass gap is the theoretical prediction of bound states composed purely of gluons, called \textit{glueballs} \cite{Mathieu:2008me}. Thus, the gluon field does not merely serve as a force mediator but carries enough energy to form matter by itself.
	
	The energy stored in the gluon field is crucial for understanding the mass of visible matter. Although it is well known that fermions, such as quarks, and gauge bosons acquire mass through the Higgs field\footnote{Although the Higgs mechanism plays a fundamental role in generating the masses of gauge bosons, a detailed discussion is beyond the scope of this thesis and the context of dynamical mass generation in QCD. We refer the reader to Refs. \cite{Djouadi:2005gi} and \cite{Gunion:1989we} for a comprehensive review.}, this mechanism is responsible for only a tiny fraction of the mass of light hadrons, such as the proton and the neutron.
	
	To observe this mathematically, we utilize the mass decomposition derived from the matrix elements of the energy-momentum tensor. For a hadron represented by a state $\ket{P}$ with four-momentum $P^\mu$ and mass $M$, the matrix element of the energy-momentum tensor is \cite{Ji:1995sv}
	\begin{flalign}
		\bra{P}T^{\mu\nu}\ket{P} = 2P^\mu P^\nu,
	\end{flalign}	
	and, with a normalization given by $\braket{P}{P}=(E/M)(2\pi)^3\delta^3(\mathbf{0})$, taking the trace of this expression leads to the scalar relation
	\begin{flalign}
		\bra{P}T^{\mu}_\mu\ket{P} = 2M^2.
	\end{flalign}
	
	Therefore, by substituting the expression (\ref{traceanomaly}) into the left-hand side, we arrive at the mass decomposition formula \cite{Hoferichter:2025ubp}:
		\begin{flalign}
			M = \frac{1}{2M} \Big\langle P \hspace{0.8mm} \Big| \frac{\beta(g)}{2g} G^a_{\mu\nu}G_a^{\mu\nu} \Big | \hspace{0.6mm} P \Big\rangle 
			+ \frac{1}{2M} \Big\langle P \hspace{0.8mm} \Big| \sum_q m_q(1+\gamma_m)\bar{q}q \Big | \hspace{0.6mm} P \Big\rangle,
			\label{massdecomp}
		\end{flalign}
	where it was used that, to first-order approximation, $\beta(g)\sim g^3$.

	This equation allows us to quantify the contribution of each term to the total hadron mass. The second term on the right-hand side, known as the $\sigma$-term (sigma term), represents the contribution from the explicit breaking of chiral symmetry by the quark masses generated by the Higgs field. For the nucleon, phenomenological analyses and Lattice QCD calculations estimate this contribution to be small, approximately $45$ MeV \cite{Gasser:1990ap,Ji:1994av}. This corresponds to approximately 5\% of the nucleon's mass, as expected.
			
	Moreover, the first term accounts for the majority of the hadron mass. It arises from the gluon condensate, being a result of non-perturbative vacuum dynamics and the conformal anomaly. Thus, we can conclude that the bulk of hadronic mass emerges from gluon dynamics, encompassing the interaction energy and kinetic energy of these confined particles.
		
	It is worth noting that there are different ways to decompose the nucleon mass. The Hamiltonian decomposition, often referred to as the Ji decomposition \cite{Ji:1994av,Ji:1995sv}, explicitly separates the kinetic energies of quarks and gluons from the anomaly and mass terms. 	
	From the perspective of mass generation via symmetry breaking, the trace decomposition presented here is the most direct approach (despite having some flaws), as it demonstrates that without the conformal anomaly and the condensates, the nucleon would be nearly massless.\\
		
	Various phenomenological approaches exist to address the structure of hadrons. One such approach is the bag model, which groups quarks and gluons into spherical regions of space with radius $r\sim\Lambda_\text{QCD}^{-1}$.
	However, for the purposes of this work, we adopt the \textit{constituent quark model}\index{quark model!constituent quark model}. 
		
	In this context, according to the constituent quark model, a distinction must be made between two types of quarks. The quark fields appearing in the QCD Lagrangian, which are used to determine the quantum numbers of hadrons, are called \textit{current quarks}. Their masses are generated by the Higgs mechanism and are listed in the third column of \autoref{table:quarks}. They represent, essentially, a "bare" quark.
	
	On the other hand, \textit{constituent quarks} are quasiparticles that share the same quantum numbers as current quarks but are significantly heavier. This additional mass arises from the cloud of gluons and quark-antiquark pairs that dresses the current quark. Consequently, the mass of constituent quarks depends on the binding energy of the surrounding environment and their motion within this sea of particles. Hence, it is referred to as a \textit{dynamically} generated mass.
	In this sense, the constituent quark mass provides an effective way of representing hadron masses, and their values are shown in the fourth column of \autoref{table:quarks} for comparison with the current quark masses.
	
	A comparison between the current and constituent masses of the three heaviest quarks reveals that the discrepancy between them is minimal, especially when contrasted with the light quark sector. This arises because, for flavors heavier than the strange quark, the current mass significantly exceeds the characteristic QCD confinement scale, $\Lambda_{\text{QCD}}\approx300$ MeV.	
	Consequently, the relative influence of the cloud that dresses the quarks diminishes. In this regime, the explicit symmetry breaking induced by the Higgs mechanism dominates over the dynamical chiral symmetry breaking. As a result, unlike light hadrons, the mass of heavy hadrons is primarily determined by the sum of the intrinsic masses of their valence quarks, rather than by the energy stored in the gluonic field.
	This is why the energy-momentum tensor trace decomposition and the Ji decomposition are more efficient in light hadrons, such as the nucleon.

	Now that the topic of hadron masses in QCD has been covered, we proceed to the mass spectroscopy of hadrons.

	\subsection{Hadron spectroscopy}
	\label{sec:hadronspectroscopy}	
				
	Before the advent of QCD, some of the early attempts to explain the strong force followed alternative paths to quantum field theory, given the latter's difficulty in describing the relevant phenomena using perturbative methods. One such path revived the S-matrix theory proposed by Heisenberg in the 1940s, which relied exclusively on the scattering matrix.			
				
	In 1959, Tullio Regge introduced the study of scattering processes in terms of a generalization of angular momentum that allowed for complex values \cite{Regge:1959mz}. Later, Geoffrey Chew and Steven Frautschi extended Regge's classical description to a relativistic framework and proposed a possible connection with strong interaction phenomenology \cite{Chew:1961ev}. As a result, it was observed that many of the detected particles could be organized into families defined by so-called \textit{Regge trajectories}\index{Regge trajectories}, linear trajectories in the plane of angular momentum $J$ versus the squared mass $m^2$ \cite{Zwiebach:2004tj,Collins:1977jy}.
	This conclusion was central to the idea of "nuclear democracy," strongly advocated by Chew \cite{Schwarz:2012zc}. Around this concept emerged what became known as the "bootstrap model," the hypothesis that there is no distinction between elementary and composite particles, such that all particles and their interactions are self-consistently determined solely by the general principles that govern the S-matrix itself (therefore, without any sign of quantum field theory) \cite{Tang:2000tb}.

	\begin{figure}[H]
		\centering
		\includegraphics[width=11cm]{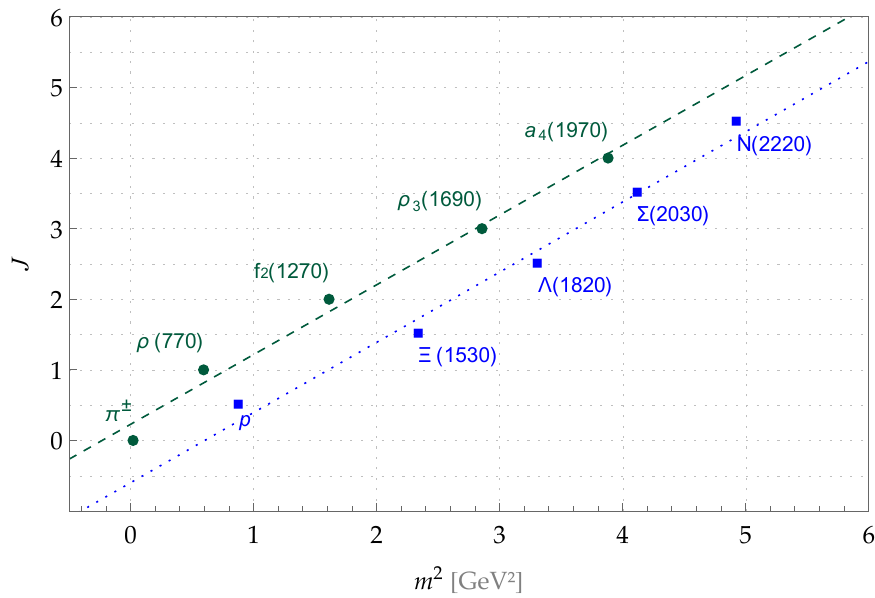}
		\caption[Regge trajectories for light-flavor mesons and baryons]{Regge trajectories for light-flavor mesons (green) and baryons (blue).}
		\label{fig:reggeexample}
	\end{figure}

	In 1968, Gabriele Veneziano mathematically proposed a scattering amplitude that proved consistent with experimental data on the strong interaction. The expression, now known as the Veneziano amplitude, satisfied the properties of S-matrix theory and properly incorporated Regge trajectories. Veneziano's result is widely regarded as the starting point of string theory \footnote{For a more in-depth exposition of Veneziano's groundbreaking work, see \cite{Zwiebach:2004tj,Schwarz:2012zc,Rickles:2014fha}.}.
	Around 1970, Yoichiro Nambu, Holger Bech Nielsen, and Leonard Susskind independently realized that the Veneziano amplitude could be given a physical interpretation by modeling hadrons as manifestations of vibrating, one-dimensional strings.

	This background explains how string theory emerged as an excellent candidate to explain the strong interaction. For instance, Regge trajectories emerge naturally within ST, since the linear relationship between angular momentum and squared mass can be derived by considering a rotating open string. In such a configuration, the angular momentum resulting from the rotation contributes to the total energy (and hence the squared mass) of the string, giving rise to the observed $J\sim m^2$ Regge behavior. The resulting expression for the trajectory, then takes a linear form,
	\begin{flalign}
		J = \alpha'm^2 + \alpha_0.
		\label{eq:regge}
	\end{flalign}
	Here, $\alpha_0$ is an intercept introduced by quantum corrections. The parameter $\alpha'\sim\text{(1 GeV)}^{-2}$ is the same that appears in the Nambu-Goto action (\ref{eq:nambugoto}) and is referred to as the Regge slope, due to its origin as the proportionality factor between $J$ and $m^2$. In addition to being related to the string length $\ell_s$, $\alpha'$ also represents the inverse of the string tension $\sigma_s$, according to \cite{Natsuume:2014sfa}
	\begin{flalign}
		\sigma_s = \frac{1}{2\pi\alpha'} = \frac{1}{2\pi\ell_s^2}.
	\end{flalign}

	String theory also offers a compelling picture of quark confinement. A meson can be represented as an open string with a quark at one endpoint and an antiquark at the other. The gluonic flux tube that forms between the quark and antiquark can thus be modeled as a string with a tension holding the particles together. As the quarks are pulled apart, the potential energy stored in the string increases with the distance until the string eventually breaks. The resulting two pieces, each with two endpoints, correspond to new quark-antiquark pairs rather than free quarks, which effectively represents the confinement.

	The description given by Eq. (\ref{eq:regge}) can be generalized to include other states from the same hadron family. These states, called \textit{resonances}\index{Regge trajectories!resonances}, are higher-mass excitations that share the same defining quantum numbers as the family's ground-state particle, such as spin and parity.
	The $\rho$ meson, for example, has the $\rho(770)$ as its ground state and, among its known resonances, the states $\rho(1450)$, $\rho(1700)$, $\rho(2150)$, to name a few. 
	Therefore, the trajectories shown in \autoref{fig:reggeexample} represent (and organize) only hadrons in their ground state.
	There are dozens of resonances listed in the Particle Data Group (PDG) \cite{ParticleDataGroup:2024cfk}, some with unknown quantum numbers and others whose families have not yet been identified.
	
	The generalized relation has the form:	
	\begin{flalign}
		m^2 \sim J + n.
	\end{flalign}	
	Thus, for a fixed $J^{PC}$, it is possible to construct Regge trajectories for a specific family of particles by organizing their resonances according to their \textit{radial quantum number} $n$.	
	In the context of string theory, this quantum number arises naturally, representing the contribution of radial excitations of the vibrating string to the total energy/mass of the state.
	Each vibrational level adds a discrete amount of energy to the ground-state configuration, resulting in an increase in the observed mass \cite{Collins:1977jy}.
	In QCD, the radial quantum number also emerges naturally from the solution of equations with confining potentials, as will be demonstrated in the next section.
	
	Following a period of intense attention in the 1960s, interest in Regge trajectories declined as QCD provided new insights into the composition and dynamics of hadrons. Nevertheless, with the vast increase in experimental data being produced continuously, Regge trajectories have proven to be an effective tool for organizing and classifying observed states, greatly benefiting hadron spectroscopy.
	
	After the discovery of particles containing heavy quarks, such as the $J/\psi$ charmonium, it was observed that Regge trajectories exhibit a non-linear behavior for these hadrons. This motivated the adoption of alternative approaches to study heavy hadrons \cite{DeRujula:1975qlm,Godfrey:1985xj}. The absence of a unified model capable of simultaneously addressing the different quark masses remains one of the weaknesses of hadron spectroscopy to this day.
	For a more detailed discussion, we refer the reader to Ref. \cite{Afonin:2014nya}.
		
	In this work, we will employ what we shall refer to as a \textit{Regge-like trajectories}, which retains the same structure as described in this section, but incorporates parameters beyond mass and angular momentum, such as configurational information and temperature.
	We will also consider non-linear Regge trajectories through a modified framework that explicitly incorporates the constituent quark mass.\\
		
	With this not-so-brief introduction to the underlying foundations completed, the next chapter will present in detail the main developments of this thesis.

\chapter{Configurational information of hadrons}
\label{chapter:cims}		
		
		Originally conceived in the context of communications, \textit{information theory} was formulated by Claude Shannon in the 1940s. It provided a mathematical framework to address questions concerning the quantification of previously abstract concepts such as information and data processing.	
		Over time, information theory has transcended its original scope of communication systems and has served as the foundation for studies in various fields and applications, ranging from cryptography and economics to biomedicine and even social sciences. Naturally, it has also found numerous applications in physics.
		
		Inspired by the concepts introduced by Shannon, Gleiser and Stamatopoulos \cite{Gleiser:2011di} proposed a new approach to analyze physical problems based on information extracted from the system configuration in momentum space. In subsequent works, this idea was expanded to form a toolkit known as \textit{configurational information measures} (CIMs)\index{configurational information measures}.
		
		In essence, CIMs are concerned with extracting information about systems from their shape, that is, from their field configuration in space.
		Physical systems with spatially localized configurations can be characterized by their structure in momentum space. Therefore, understanding the shape of a structure implies knowing its dynamic properties, including interactions and boundary conditions. Usually, the description is achieved through Fourier transforms that create an equivalent representation of the field configuration in terms of momentum modes	\cite{Gleiser:2018jpd,Gleiser:2018kbq}.
		Thus, CIMs -- namely the configurational entropy and the configurational complexity -- measure and quantify properties through their distinct approaches to counting these modes.
		
		In the early works in the field, the term "configurational entropy" was broadly applied, serving as a synonym for all information measurements in this context. Indeed, much of what was once referred to as configurational entropy is now more accurately understood as configurational complexity. In Ref. \cite{Gleiser:2018jpd}, Gleiser and Damian Sowinski provided a more detailed description that outlined the distinctions among the types of CIMs and we will always use these definitions throughout the present work.
		
		In Ref. \cite{Gleiser:2011di}, the CIMs were introduced as a way to measure the amount of information in systems with spatially localized energy like solitons and bounces in one dimension. It was demonstrated that these tools are capable of indicating the best approximation for soliton-like solutions, selecting among different trial functions with degenerate energies.
		The discriminatory nature observed by Gleiser and Stamatopoulos was also applied to investigate the properties of kink-like solitons in models with two interacting scalar fields \cite{Correa:2014boa} and to predict the spontaneous decay rates of hydrogenic atoms	with an accuracy better than $3.7\%$ when compared to established data in literature \cite{Gleiser:2017cro}.
		
		The CIMs were also employed to study the properties of travelling wave solutions (travelling solitons) in scenarios with violation of Lorentz and CPT symmetries, revealing the best energy density profile for the parameter that breaks the symmetries \cite{Correa:2015lla}. 
		The approach used consists of seeking which values for the essential parameters of the scenarios lead to a system in its most organized form, that is, which minimize the entropy and complexity values. 	Thus, it was possible to place strict bounds on the parameters of modified gravity theories, such as $f(R)$ \cite{Correa:2015ako}, $f(R,T)$ \cite{Correa:2015qma}, sine-Gordon \cite{Correa:2015vka,Barreto:2022mbx}, Weyl \cite{Correa:2016noh}, Bloch \cite{Cruz:2017ems}, among others \cite{Casadio:2016aum}, brane models; and in 6-dimensional braneworld models, such as topological Abelian string-vortex and the string-cigar scenarios \cite{Correa:2016pgr}.
		
		In Ref. \cite{Gleiser:2013mga}, it was demonstrated that the maximum and minimum values of these measures are linked to the boundary between stability and instability for spatially-bound objects. Moreover, in this study, the application of this property correctly predicted the stability of Q-balls and the Chandrasekhar limit for compact stars. Subsequently, these results were extended to investigate the stability of neutron and boson stars \cite{Gleiser:2015rwa}, and black holes in AdS space \cite{Braga:2016wzx}.
		
		Other applications arise in understanding the information dynamics of phase transitions, utilizing the ability of CIMs to measure information flow and determine the critical point where entropy is minimized \cite{Gleiser:2015aav, Sowinski:2016vxz}. There are also applications in the study of glueballs \cite{Bernardini:2016qit}, nonequilibrium dynamics of spontaneous symmetry-breaking \cite{Gleiser:2012tu} and in the context of events on cosmological scales \cite{Bernardini:2019stn}. Other applications include Refs. \cite{Fernandes-Silva:2019fez, Bernardini:2019stn, Barreto:2022len, Casadio:2022pla, daRocha:2021jzn, daRocha:2022bnk, Casadio:2023pmh}.	
		Finally, there are also applications in AdS/QCD models that we will cover in greater detail in the following sections as they will be the focus of this thesis.\\
		
		Following the presentation of the fundamental concepts in the previous chapter, \Cref{section:adsqcd} reviews the main bottom-up AdS/QCD models, namely the hard-wall and soft-wall approaches, with particular emphasis on the latter, as it will serve as the basis for analyzing the properties of hadrons in this work.

		In this chapter, \Cref{section:informationtheory} is devoted to a brief review of information theory, serving as a way to introduce fundamental concepts such as entropy, information, and complexity.
		
		\Cref{section:cims} provides an individual introduction to each of the CIMs, outlining the construction of their formalisms and highlighting their differences.
		
		\Cref{section:cimadsqcd} constitutes the theoretical core of this thesis, as it presents the main formalism employed in the subsequent analyses. The protocol for applying CIMs in conjunction with AdS/QCD models is introduced in detail.
		
		The following sections begin the applications, implementing the CIM-based protocol alongside AdS/QCD models. In \Cref{section:lightmesons}, CIMs are applied to the analysis of four families of light-flavor mesons within the dynamical soft-wall model. Subsequently, in \Cref{section:strangekaons}, the objects of study are the strange vector kaons, analyzed in the modified soft-wall model that incorporates constituent quark masses, thereby giving rise to non-linear Regge trajectories.	
		
		\pagebreak
		
	\section{AdS/QCD models}
	\label{section:adsqcd}
	
		So far, we have seen the construction of the AdS/CFT correspondence from its fundamental ingredients, rules, and limits so that a duality can be developed between two theories in different dimensions.
		However, as may have become clear in the previous chapter, QCD does not entirely satisfy these essential elements. In fact, it has few properties in common with the $\mathcal{N}=4$ SYM theory.
		
		Among the differences that can be listed based on what we have seen up to this point, the most prominent is the fact that $\mathcal{N}=4$ SYM is conformally invariant, while QCD has a characteristic energy scale. At low energies, QCD is strongly coupled and exhibits confinement, while it shows asymptotic freedom at high energies. On the other hand, $\mathcal{N}=4$ SYM, being conformal, is strongly coupled at all scales and does not have a confinement mechanism.
		The field content of $\mathcal{N}=4$ SYM (a gauge field, 4 Weyl spinors, and 6 scalar fields) transforms in the adjoint representation of the gauge group, while the quarks and gluons of QCD transform in the fundamental representation.
		Last but not least, $\mathcal{N}=4$ SYM is supersymmetric, while QCD is not.
		
		Some of these differences cease to exist or become irrelevant when a finite temperature is introduced to the system. But until the end of this chapter, all AdS/QCD models will be considered as being at zero temperature. 
		Only from \Cref{chapter:thermalqcd} onwards will we address in detail these same models at finite temperature.
		
		There are actually two fundamental ways to approach the desired holographic QCD.
		One of them is the \textit{top-down}\index{AdS/QCD!top-down approach} approach, a formulation that starts from a well-defined higher-dimensional gravitational theory (usually string theory) and seeks to find a background capable of reproducing a dual theory with the characteristics of QCD.
		
		The other one, known as the \textit{bottom-up}\index{AdS/QCD!bottom-up approach} approach, involves starting with a 4-dimensional QFT and constructing the dual theory by imposing constraints based on phenomenological observations or experimental data. This approach prioritizes the ability to reproduce known features of the QCD over the fundamental structure of the underlying gravitational theory.
	
		Nevertheless, each approach has its drawbacks. The top-down requires a robust mathematical framework and is usually based on complicated theories. On the other hand, the bottom-up is built to reproduce specific properties and may not provide the fundamental structure of the theory, which can hide some problems in unexpected places. Thus, each approach needs to be chosen where it is most suitable.
		
		Since we are more interested in phenomenological aspects, especially those related to the mass spectra of hadrons, the bottom-up approach is a better choice. Despite its drawbacks, this approach has shown accuracy in its phenomenological results and adequately incorporates essential features of QCD, such as confinement and chiral symmetry.

		There are two prominent bottom-up AdS/QCD models, the hard-wall and the soft-wall.	
		They differ in the way they introduce QCD properties into the holographic model and both will be described below.
				
		In the analysis of both models, only the light quarks will be considered, so $N_f=3$, and the quark field is given by
		\begin{equation}
			q_{L,R} = \begin{pmatrix}
				u \\ d \\ s
			\end{pmatrix}_{L,R}.
		\end{equation}
			
		The quark masses will be input parameters of the model, organized in a matrix given by
		\begin{flalign}
			M=\begin{pmatrix}
				m_u & 0 & 0 \\ 
				0 & m_d & 0 \\
				0 & 0 & m_s
			\end{pmatrix}.
		\end{flalign}
		Usually, isospin symmetry is assumed, so that $m_u=m_d$, especially in models with $N_f=2$. For simplicity, in this section dedicated to explaining the main properties of this AdS/QCD models, we will extend this symmetry to an exact SU$(3)$ flavor symmetry and consider the three light quarks as having the same mass, thus $m_q\coloneqq m_u=m_d=m_s$.
		
		The values of the chiral condensate $\Sigma = \langle \bar{q}q \rangle$ will also be treated as input parameters of the model \cite{Erlich:2005qh}. They are given by:
		\begin{flalign}
			\Sigma = \begin{pmatrix}
				\sigma_u & 0 & 0 \\ 
				0 & \sigma_d & 0 \\
				0 & 0 & \sigma_s
			\end{pmatrix}.
		\end{flalign}
		In this work, we also extend the symmetry to the three light quarks, so that $\sigma \coloneqq \sigma_u=\sigma_d=\sigma_s$. 		
		An analysis considering only isospin symmetry, with $\sigma_q \neq \sigma_s$, can be found in \cite{Abidin:2009aj}. In addition, chiral condensates were also studied in 4-flavor AdS/QCD, as in Refs. \cite{Ballon-Bayona:2017bwk,Chen:2021wzj,Karapetyan:2023sfo}.

		Although there are infinite operators in QCD, we will focus on those that are relevant to the physics studied in this thesis. These include the left- and right-handed currents $\left(j_L\right)_\mu^a=\tensor{\bar{q}}{_L}\gamma_\mu t^a \tensor{q}{_L}$ and $(j_R )^a_\mu=\tensor{\bar{q}}{_R}\gamma_\mu t^a \tensor{q}{_R}$, and the bilinear quark $\tensor{\bar{q}}{ _R} \tensor{q}{_L}$ (all introduced in \Cref{sec:chiral}), which correspond in $\text{AdS}_5$ to the gauge fields $L_M^a$, $R_M^a$ and to the scalar field $(2/z)X$, respectively.	
	
		The 5D masses $m_5$ of these fields are obtained using the relation (\ref{massscalingrelation}),
		\begin{flalign}
			m_5^2L^2 = (\Delta-p)(\Delta+p-4),
			\label{massdimension_relation}
		\end{flalign}
		where $\Delta$ is the conformal dimension and $p$ is the form degree of the operator ($p=0$ for scalar, $p=1$ for currents). The relations for all operators and fields are summarized in \autoref{table:dictionary}.	
		\begingroup
		\renewcommand{\arraystretch}{1.8}
		\begin{table}[h!]
			\centering
			\setlength{\tabcolsep}{3mm}
			\begin{tabular}{ccccc}
				\rowcolor{ufabc!50}
				\bf 4D: $\mathcal{O}(x)$ & \bf 5D: $\phi(x,z)$ & $p$ & $\Delta$ & $(m_5)^2$\\
				\rowcolor{ufabc!30} $\tensor{\bar{q}}{_L}\gamma_\mu t^a \tensor{q}{_L}$&$L_M^a$&$1$&$3$&$0$ \\
				\rowcolor{ufabc!10} $\tensor{\bar{q}}{_R}\gamma_\mu t^a \tensor{q}{_R}$&$R_M^a$&$1$&$3$&$0$\\
				\rowcolor{ufabc!30}$\bar{q}_R^{\; a} q_L^b$ 	&$(2/z)X^{ab}$&$0$&$3$&$-3$
			\end{tabular}
			\caption{Operators and fields with their respective masses of the model.}
			\label{table:dictionary}
		\end{table}	
		\endgroup
		
	 	At first glance, the negative mass for the field $X$ (which also leads to it being called a tachyonic scalar field) may seem odd. However, as discussed at the end of \Cref{section:conjecture}, the stability of scalar bulk fields in $\text{AdS}_{d+1}$ is ensured as long as the mass obeys the Breitenlohner-Freedman bound $m_5^2 \ge -d^2/4$.
	 
		These are the first preparations needed to approach the models.
	
		\subsection{Hard-wall model}
		\label{subsection:hardwall}
	
		The \textit{hard-wall} (HW) \textit{model}\index{AdS/QCD!hard-wall model} was initially proposed by Polchinski and Strassler \cite{Polchinski:2001tt} and used in the early days of AdS/QCD to study the glueball mass spectrum \cite{Boschi-Filho:2002wdj,Boschi-Filho:2002xih}. Subsequently, the model was modified by Erlich, Katz, Son, and Stephanov \cite{Erlich:2005qh}, as well as Da Rold and Pomarol \cite{DaRold:2005mxj,DaRold:2005vr}. While there have been later modifications to address specific cases, this section will primarily follow \cite{Polchinski:2001tt,Erlich:2005qh,DaRold:2005mxj,DaRold:2005vr}.	
			
		In order to emulate the phenomenon of confinement, the HW model constrains the energy scale by introducing two boundaries. One of them is positioned at the ultraviolet (UV) limit, specifically at $z=z_0$. This boundary is introduced to prevent divergences and maintain finite calculations, although it is important to note that the limit $z_0\rightarrow0$ is implicit. The other boundary is an infrared (IR) cutoff that serves as the QCD scale and is located at $z=z_m$, with $z_m\sim \Lambda_{\text{QCD}}^{-1}$. Thus, the fields are restricted to propagating only between these two limits.
				
		The hard-wall model is successful in reproducing and calculating certain QCD properties, such as form factors and mass spectra, as can be found in Refs. 		\cite{DaRold:2005vr,Brodsky:2006uqa,Evans:2005ip,Hirn:2005nr,Katz:2005ir,Erlich:2006hq,Hirn:2005vk,deTeramond:2005su,Ballon-Bayona:2014oma}. 
		\autoref{table:hardwall} summarizes the results obtained in Ref. \cite{Erlich:2005qh}, where the masses of the $\pi$, $\rho$, and $a_1$ mesons were computed using the HW model and compared with experimental data.
		However, the model presents a significant limitation when attempting to describe the mass spectra of resonances belonging to the same hadron family.
		In other words, the model accurately reproduces the $m^2 \sim J$ trajectories for ground-state hadrons (similar to the plot shown in \autoref{fig:reggeexample}), yet it does not reproduce the correct mass trajectories with respect to the radial quantum number, $m^2 \sim n$.
	
		\begingroup
		\renewcommand{\arraystretch}{1.8}
		\begin{table}[h!]
			\centering
			\setlength{\tabcolsep}{3mm}
			\begin{tabular}{ccc}
				\rowcolor{ufabc!50}	Observable & Experimental (MeV) & HW model\\
				\rowcolor{ufabc!30} $m_\pi$ & $139.57039 \pm 0.00017$ & $141$ \\
				\rowcolor{ufabc!10} $m_\rho$ & $ 775.8 \pm 0.5$& $832$ \\
				\rowcolor{ufabc!30} $m_{a_1}$ & $ 1230 \pm 40$ & $1220$
			\end{tabular}
			\caption[Comparison between experimental masses and those predicted by the hard-wall model for three mesons]{Comparison between experimental masses and those predicted by the hard-wall model for three mesons. Source: \cite{Erlich:2005qh}.}
			\label{table:hardwall}
		\end{table}	
		\endgroup

		This characteristic of the HW model to describe only a single type of trajectory is due to the fact that the normalized solutions of the equations of motion, denoted by $\psi(z)$, are expressed in terms of Bessel functions.
		Imposing a boundary condition in the UV limit, $\psi(z \to 0) \to 0$, the solution reduces to a Bessel function of the first kind, $\psi(z) \sim J_1(m_n z)$. 
		The expected discretization of particle masses arises naturally when infrared boundary conditions are imposed at the IR cutoff, $\partial_z \psi(z_m) = 0$, which restricts the allowed physical states to the discrete zeros (the roots) of these Bessel functions. However, because the $n$-th zero of a Bessel function grows asymptotically linear with $n$, the resulting mass spectrum yields a relation of the form $m_n^2 \sim n^2$.				
		This behavior is accurate for the ground state $n=1$, but inconsistent with experimental observations, which show that the squared mass spectrum grows linearly with $n$. Consequently, the hard-wall model fails to provide an accurate description of the particle mass spectrum for $n>1$.
		
		Since a significant part of this thesis focuses on studying the mass spectra of baryon and meson resonances as functions of $n$, the hard-wall model is not the most suitable holographic QCD framework for our purposes. For this reason, we choose not to elaborate on its details in this section.
	
		\subsection{Soft-wall model}
		\label{subsection:softwall}
		
		Proposed in \cite{Karch:2006pv}, the \textit{soft-wall} (SW) \textit{model} \index{AdS/QCD!soft-wall model} offers a solution to the problem encountered in the HW model in accurately describing the particle mass spectrum. This solution involves modifying the background of the theory, resulting in a better reproduction of the trajectories, and demonstrating that the issue was not an intrinsic feature of AdS/QCD, but rather of a specific model.
		
		What characterizes the SW model is a background parameterized by two $z$-dependent functions. The first one, denoted by $A(z)$, appears as the warp factor in the line element of the model, modifying the familiar AdS$_5$ background to
			\begin{flalign}
				ds^2=e^{A(z)}\left(\eta_{\mu\nu}dx^\mu dx^\nu +dz^2\right).
				\label{softwallmetric}
			\end{flalign}
		In order to recover the usual conformal factor of AdS space, the warp factor is commonly taken as $A(z)=\ln(L/z)$.
		
		The second parameter is a dilaton field $\Phi(z)$.		
		While phenomenologically it introduces a smooth cutoff in the AdS$_5$ bulk (replacing the abrupt cutoff of the hard-wall model), physically it is responsible for the breaking of conformal invariance. According to the holographic dictionary, this bulk scalar field maps to the gluon operator $G^a_{\mu\nu}G_a^{\mu\nu}$ on the boundary \cite{Csaki:2006ji}. Hence, its non-trivial profile encodes the effects of the gluon condensate (\ref{gluoncondensate}), providing the mass scale responsible for confinement.
	
		The dilaton appears explicitly in the action of the model, which is given by
			\begin{flalign}
				\mathcal{S}_\text{SW} = \int d^5x  \sqrt{-g} e^{-\Phi(z)}\Tr\left\{\abs{D_M X}^2+m_X^2\abs{X}^2-\frac{1}{4g_5^2}\left(F_L^2+F_R^2\right)\right\},
				\label{softwallaction}
			\end{flalign}
		The first term is the covariant derivative of the scalar field $X$, which involves all three fields selected in the correspondence:
		\begin{flalign}
			D_MX = \partial_M X - iL_MX + iXR_M.
		\end{flalign}
		Here, we have used the shorthand notation $L_N=L_N^a t^a$ and $R_N=R_N^a t^a$, where $L_N^a$ and $R_N^a$ are real numbers and $t^a$ are the SU($N_c$) generators. 
		The second term is the mass term of the theory and, as shown in \autoref{table:dictionary}, $m_X^2=-3$.
		
		In the last term, $F_L^2$ and $F_R^2$ stand for $F_{L}^2=F^{L}_{MN}F_{L}^{MN}$ and $F_{R}^2=F^{R}_{MN}F_{R}^{MN}$, where the field strength tensor is defined as
		\begin{flalign}
				F_{MN}^L = \partial_M L_N  - \partial_N L_M - i[L_M,L_N], \\
				F_{MN}^R = \partial_M R_N  - \partial_N R_M - i[R_M,R_N].
		\end{flalign}
		
		The five-dimensional gauge coupling $g_5$ is determined by comparing the two-point vector correlation function in the model with the leading term in the operator product expansion of QCD for the same function, and is fixed as $g_5^2=12\pi^2/N_c$ \cite{Erlich:2005qh}.
		
		A major advantage of the dilaton, and consequently the model, is that its profile can be calibrated to reproduce desired properties, as long as it satisfies the conditions at the two ends of the holographic coordinate. Near the boundary, the dilaton must vanish, so that the factor $e^{-\Phi}$ tends to one and both the geometry and the fields recover their AdS asymptotics, on which the holographic dictionary relies and which represents the ultraviolet regime of QCD, where the theory is asymptotically free. In the infrared, it must grow without bound, so that the same factor suppresses the fields at large $z$ and the spectrum becomes discrete, a role played by the cutoff in the hard-wall model. In particular, to obtain the Regge trajectories, the dilaton must have the asymptotic behavior
		\begin{flalign}
			\Phi(z\rightarrow0)=0 \quad\text{and}\quad
			\Phi(z\rightarrow\infty)=\kappa^2 z^2,
			\label{dilatonasymptotic}
		\end{flalign}
		where $\kappa$ plays the role of $\Lambda_{\text{QCD}}$ by introducing the QCD scale and the mass gap. Throughout this section, as the equations of motion are solved for each particle sector of the model, it will become evident how this dilaton profile is responsible for imposing a confining potential on the modes propagating along the holographic coordinate, which reproduces the mass spectrum with the form $m_n^2\sim n$ characteristic of Regge trajectories.\\
		
		In order for the action (\ref{softwallaction}) to describe mesons, the physical fields representing these particles must be made explicit. Thus, the $L_M$ and $R_M$ fields will be replaced by the vector field $V_M$ and the axial field $A_M$, related as follows
		\begin{flalign}
			V_M=\frac{L_M+R_M}{2} \quad \text{ and } \quad A_M=\frac{L_M-R_M}{2},
			\label{axialandvector}
		\end{flalign}
		which leads to the respective gauge field strength tensors
		\begin{flalign}
			F_{MN}^V &= \partial_M V_N  - \partial_N V_M - i[V_M,V_N], \\
			F_{MN}^A &= \partial_M A_N  - \partial_N A_M - i[A_M,A_N].
		\end{flalign}
		Accordingly, the covariant derivative is also rewritten:
		\begin{flalign}
			D_M X = \partial_M X - i\left\{A_M,X\right\} + i\left[V_M, X\right]	.	
		\end{flalign}		
	
	 	The field $X$ is slightly more complex, as it contains the scalar field $S$, the pseudoscalar field $\uppi$, and a vacuum expectation value $X_0=\mean{X}$. With flavor symmetry, the expression is given by:
	 	\begin{flalign}
	 		X = \left(X_0 + S \right)e^{2i\uppi}.
	 	\end{flalign}
	 	It is important to note that the flavor index is always implied, $S=S^at^a$ and $\uppi=\uppi^at^a$, as well for $V_M=V_M^a t^a$ and $A_M=A_M^a t^a$. 
	 	
		Before proceeding to the equations of motion for each meson type, it is necessary to determine the VEV of $X$. 
		Since $X$ is the holographic dual to the quark bilinear operator $\bar{q}_R q_L$, its dynamics are crucial because its vacuum expectation value implements the breaking of chiral symmetry in the dual theory. This calculation is done by turning off all other fields, assuming they do not manifest in the vacuum, and then solving for $X_0$.
		With this assumption, the model's action is written as
		\begin{flalign}
			\mathcal{S}_X = \int d^5x \sqrt{-g}e^{-\Phi} \Big(\abs{DX_0}^2 + m_X^2\abs{X_0}^2\Big),
		\end{flalign}
		and the resulting equation of motion in this case is
		\begin{flalign}
			\chi'' - \left(\Phi' + \frac{3}{z}\right)\chi' + \frac{3}{z^2}\chi =0,
			\label{eom_vev}
		\end{flalign}
		where we denote with a prime the derivatives with respect to the $z$ coordinate and, for convenience, we rewrite the VEV as $\chi(z)=2X_0(z)$.
		
		In the hard-wall model, the absence of a dilaton field simplifies the equation above, leading to the solution \cite{DaRold:2005vr}
		\begin{flalign}
			\chi(z) = c_1 z + c_2 z^3,
		\end{flalign}
		where $c_1$ and $c_2$ are integration constants. These coefficients can be obtained by analyzing the solution in the UV limit and comparing it with results from the holographic dictionary. In doing so, they can be identified (up to a normalization constant) as corresponding to the quark mass and the chiral condensate, $c_1\sim m_q$ and $c_2\sim \sigma$ \cite{Cherman:2008eh}. This identification is detailed in \autoref{appendix:fullsoftwall}.
				
		In the soft-wall model, with a dilaton in its simplest form $\Phi=\kappa^2z^2$, the solution for (\ref{eom_vev}) is more complicated:
		\begin{flalign}
			\chi(z) = C_1  U\left(\frac{1}{2};0;\kappa^2z^2\right) + C_2 \prescript{}{1}{F}_1 \left(\frac{3}{2};2;\kappa^2z^2\right),
		\end{flalign}
		where $C_1$ and $C_2$ are new constants, $U$ is the Kummer's function of the second kind (also known as Tricomi confluent hypergeometric function) and $\prescript{}{1}{F}_1$ is the Kummer's function of the first kind.
		However, the second term diverges when $z\rightarrow\infty$, meaning the action is not finite in the IR limit. Since this cannot correspond to a physical quantity, the second term must be discarded. Identifying the coefficient $C_1$ (see \autoref{appendix:fullsoftwall}), the final expression for the VEV solution in the soft-wall model is:
		\begin{flalign}
			\chi(z) = m_q z\Gamma\left(\frac{3}{2}\right) U\left(\frac{1}{2};0;\kappa^2z^2\right),
		\end{flalign}
		with $\Gamma$ being the gamma function.	
		
		Near the boundary, the function $U$ can be expanded, and the solution becomes  \cite{Colangelo:2008us}
		\begin{flalign}
			\lim_{z \to 0}\chi(z) \approx  m_q z - \frac{m_q \kappa^2 z^3}{2} \left(1 - 2\gamma_E - 2\ln(\kappa z ) - \Psi\left(\frac{3}{2}\right)\right),
			\label{eq:vacuum_chi}
		\end{flalign}
		where $\gamma_E$ is the Euler-Mascheroni constant and $\Psi(y)=\Gamma'(y)/\Gamma(y)$ is the digamma function. 		
		Comparing this with the standard AdS/CFT dictionary solution, we observe that the coefficients of both the $z$ and $z^3$ terms are proportional to the quark mass $m_q$, which leads to  $\sigma \propto m_q$.
		This result does not match physical expectations, as the mechanisms for spontaneous and explicit symmetry breaking must be distinct. In particular, spontaneous symmetry breaking must still occur in the chiral limit, where $m_q \rightarrow 0$.
				
		This issue can be addressed by introducing an additional quartic term $\sim|X|^4$ in the action, as discussed in Ref. \cite{Gherghetta:2009ac}, which makes the equation for $\chi$ nonlinear.	
		Another proposal, presented in Ref. \cite{Cherman:2008eh}, keeps Eq.~\eqref{eom_vev} unchanged and instead fixes the UV behavior of the solution as
		\begin{equation}
			\chi(z\to0)=m_q\zeta z+\frac{\sigma}{\zeta}z^3,
			\label{eq:uvbehavior}
		\end{equation}
		with $2\pi\zeta=\sqrt{N_c/N_f}$ a normalization constant, so that $m_q$ and $\sigma$ enter as independent parameters. We follow this second approach, which also allows the matching with the IR solution of the dynamical model discussed below.
			
		In order to obtain the large-$z$ behavior, Refs. \cite{Li:2012ay,Li:2013oda} considered an extension of the soft-wall model that includes a gravitational action coupled to the matter action (\ref{softwallaction}). This extended model will be discussed in detail in \Cref{sec:dynamical_soft_wall}, so we abstain from entering into further specifics here. For the present purposes, it suffices to note that the proposed solution in the IR limit takes the form:
		\begin{flalign}
			\chi(z \rightarrow \infty) = \sqrt{\frac{8}{\alpha}}\kappa e^{-\Phi/2},
			\label{eq:irsolution}
		\end{flalign}
		where $\alpha$ is a parameter that couples the simplest SW action and the gravitational one.
		
		To ensure that both asymptotic solutions, (\ref{eq:uv_condition}) and (\ref{eq:irsolution}), are simultaneously satisfied, the VEV $\chi$ must take a general form constrained by the following condition:
		\begin{equation}
			\chi'(z) = \sqrt{\frac{8}{\alpha}} \kappa e^{-\Phi/2} \left( 1 + a_1 e^{-\Phi} + a_2 e^{-2\Phi} \right),
			\label{eq:chi_condition}
		\end{equation}
		with the coefficients $a_1$ and $a_2$ determined by the following expressions:
		\begin{subequations}
			\begin{align}
				a_1 &= -2 + \frac{5\sqrt{2\alpha}m_q\zeta}{8\kappa} + \frac{3\sqrt{2\alpha}\sigma}{4\zeta\kappa^3}, \\
				a_2 &= 1 - \frac{3\sqrt{2\alpha}m_q\zeta}{8\kappa} - 	\frac{3\sqrt{2\alpha}\sigma}{4\zeta\kappa^3}. 
			\end{align}
			\label{eq:coefs}
		\end{subequations}
		These coefficients explicitly connect the parameters of the model, such as $m_q$, $\sigma$, $\zeta$, and $\kappa$, to the overall structure of the solution $\chi(z)$.\\
			
		With the VEV in hand and after all field redefinitions, the full soft-wall action takes the form:
		\begin{flalign}
			\mathcal{S}_\text{SW} &=\frac{1}{2}\int d^5x \sqrt{-g}e^{-\Phi} \bigg\{(\partial_M S)^2 - 3S^2  
			+ \chi^2(\partial_M\uppi - A_M)^2 \nonumber\\ 
			&\quad + \frac{1}{2g_5^2}\left[(\partial_MV_N - \partial_NV_M)^2 + (\partial_M A_N - \partial_N A_M)^2\right] \bigg\}.
			\label{fullsoftwallaction}
		\end{flalign}
		
		In the framework of AdS/QCD models, the connection between the continuous 5D fields postulated in the bulk and the discrete spectrum of 4D hadrons is established through the \textit{Kaluza-Klein decomposition}\index{Kaluza-Klein decomposition}. This mathematical technique allows a 5D field, denoted generically as $\uppsi(x^\mu, z)$, to be expressed as an infinite sum of products:
		\begin{flalign}
			\uppsi(x^\mu, z) = \sum_{n=0}^\infty \upphi_n(x^\mu) \psi_n(z)
		\end{flalign}
		Each term in this expansion consists of a 4D field $\upphi_n(x^\mu)$, which is identified with a specific hadron, and a corresponding eigenfunction $\psi_n(z)$ along the coordinate $z$. The discrete mass spectrum $m_n^2$ for these hadrons arises from the eigenvalue problem satisfied by the profile functions $\psi_n(z)$, which is derived from the 5D equations of motion and appropriate boundary conditions in the extra dimension. This procedure effectively translates the dynamics in the fifth dimension into a tower of massive 4D particles		
		\footnote{While the original Kaluza-Klein theory aimed to unify gravity and electromagnetism, its concepts of dimensional reduction and the emergence of effective lower-dimensional fields from higher-dimensional ones have found significant modern application, especially within string theory. Interested readers are directed to the Chapter 8 of Ref. \cite{Polchinski:1998rq} for a detailed discussion of these connections.}.
			
	To obtain physical results, these functions $\psi_n(z)$ must be regular near the UV boundary. This requires them to be well-behaved as $z \to 0$, often by imposing a condition such as
	\begin{equation}
		\lim_{z \to 0} \psi_n(z) = 0.
		\label{eq:uv_condition}
	\end{equation}
	Another necessary condition for the functions $\psi_n(z)$ to represent physical states is that they must be normalizable. Therefore, the integral of their absolute value squared must be finite, that is,
	\begin{equation} \label{eq:normalizability_integral}
		\int_0^\infty dz \, \sqrt{-g}e^{-\Phi} \abs{\psi_n(z)}^2 < \infty
	\end{equation}
	This is usually achieved by imposing the boundary condition at the IR region:
	\begin{equation}
		\lim_{z \to \infty} \psi_n(z) = 0.
		\label{eq:ir_condition}
	\end{equation}
	These conditions, applied to the equation of motion for $\psi_n(z)$, lead to a discrete mass spectrum, $m_n$.
	Next, each sector of this action will be analyzed separately.
	
		\subsubsection{Scalar sector}
		
		The scalar sector of the soft-wall model is responsible for describing the properties and spectrum of scalar mesons. The dynamics of this sector are governed by the terms depending on the scalar field $S$ in the overall action (\ref{fullsoftwallaction}),
		\begin{flalign}
			\mathcal{S}_\text{S} = \int d^5x \sqrt{-g} e^{-\Phi} \Big((\partial_M S)^2 -3S^2 \Big).
			\label{eq:action_scalar}
		\end{flalign}
		
		From this action, the equation of motion for the scalar field is obtained.	After performing a Fourier transform for the 4D spacetime coordinates $x^\mu \to k^\mu$, such that
		\begin{flalign}
			S(x,z) = \int \frac{d^4k}{(2\pi)^4} e^{-ik \cdot x} \widehat{S}(k,z),
		\end{flalign}
		the equation of motion in momentum space is written as
		\begin{flalign}
			z^3 e^{\Phi}\partial_z \left[\frac{e^{-\Phi}}{z^3}\partial_z \widehat{S}\right] - k^2 \widehat{S} + \frac{3}{z^2}\widehat{S} =0.
			\label{eom_scalar}
		\end{flalign}
		This equation can be simplified by rewriting the field as $\widehat{S}\mapsto e^{B/2}\widehat{S}$, with $B\equiv B(z)=\Phi-3A=\Phi + 3\ln(z)$.
		To simulate the discrete spectrum of mesonic particles, the scalar field undergoes a Kaluza-Klein expansion, 
		\begin{flalign}
			\widehat{S}(x,z) = \sum_{n=0}^\infty S_n(x)s_n(z),
		\end{flalign}
		where $S_n(k)$ represent the 4D scalar meson fields with mass $m_{s_n}$, and $s_n(z)$ are their corresponding wavefunctions. Substituting the KK expansion into (\ref{eom_scalar}) and identifying $k^2=-m_{s_n}^2$, the equation for the profiles $s_n(z)$ can be transformed into a Schrödinger-like equation,
		\begin{flalign}
			\left[-\partial_z^2 + U_S(z)\right]s_n = m_{s_n}^2 s_n,
			\label{eq:scalar_eom}
		\end{flalign}
		with the effective potential $U_S(z)$ given by
		\begin{flalign}
			U_S(z) = -\frac{B''}{2} + \frac{(B')^2}{4} +\frac{3}{z^2}.
		\end{flalign}
		In the case of the simplest soft-wall model, with the dilaton given by $\Phi(z)=\kappa^2z^2$ and the warp factor written as $A(z)=-\ln(z)$, the potential is equal to
		\begin{flalign}
			U_S(z) = \kappa^4z^2 + \frac{3}{4z^2} + 2\kappa^2.
		\end{flalign}
	
		With the boundary conditions (\ref{eq:uv_condition}) and (\ref{eq:ir_condition}), which ensure that $s_n$ is regular and normalizable, Eq. (\ref{eq:scalar_eom}) can be solved analytically, yielding an expression given by
		\begin{flalign}
			s_n (z) = \sqrt{\frac{2}{n+1}}\kappa^2z^{3/2}e^{-\kappa^2z^2/2}L_n^1(\kappa^2z^2),
		\end{flalign}	
		which is written in terms of the generalized Laguerre polynomials,
		\begin{flalign}
			L_n^\mu(x) = \frac{x^{-\mu} e^x}{n!}\frac{d^n}{dx^n}(e^{-x} x^{\mu+n}).
		\end{flalign}	
		With this solution, the corresponding mass spectrum is given by			
		\begin{flalign}
			m_{s_n}^2 = 4\kappa^2\left(n+\frac{1}{2}\right).
		\end{flalign}
	
		It is therefore evident that, in the soft-wall model, the mass spectrum of the scalar mesons follows a relation $m_n^2\sim n$, which is consistent with the experimentally confirmed behavior of the Regge trajectories.
		
		\subsubsection{Vector sector}
		
		Following the scalar sector, we now explore the vector sector of the soft-wall model, which is employed to describe vector mesons. The dynamics of this sector are governed by the action:
		\begin{flalign}
			\mathcal{S}_\text{V} = -\frac{1}{4g_5^2}\int d^5x \sqrt{-g}e^{\Phi(z)}\left(\partial_M V_N - \partial_N V_M\right)^2.
			\label{eq:action_vector}
		\end{flalign}
	
		One can see that the vector field is invariant under a gauge transformation of the type $V_N\mapsto V_N' = V_N-\partial_N \lambda$, by noting that
		\begin{flalign}
			\partial_M V_N' - \partial_N V_M' &=  \partial_M (V_N-\partial_N \lambda) - \partial_N (V_M-\partial_M \lambda) \nonumber\\
			&=\partial_M V_N - \partial_M\partial_N \lambda - \partial_N V_M + \partial_N\partial_M \lambda \nonumber\\
			&=\partial_M V_N  - \partial_N V_M.
			\label{gaugetest}
		\end{flalign}						
		This invariance allows us to fix the axial gauge by choosing a parameter $\lambda$ such that $\partial_z \lambda = V_z$, which allows us to set the fifth component of the vector field to zero, $V_z=0$.
				
		The next steps are analogous to what was done in the scalar sector.		
		Starting from the action (\ref{eq:action_vector}), one can derive the equation of motion for the vector field which, in momentum space, takes the form:		
		\begin{flalign}
			ze^{\Phi} \partial_z \left[\frac{e^{-\Phi}}{z}\partial_z \widehat{V}\right]  + k^2 \widehat{V} =0.
		\end{flalign}
		
		A transformation $\widehat{V}\mapsto e^{B/2}\widehat{V}$, with $B(z)=\Phi(z)-A(z)=\Phi(z)+\ln(z)$, and a Kaluza-Klein expansion of the type
		\begin{flalign}
			\widehat{V}(x,z) = \sum_{n=0}^{\infty} V_n(x)v_n(z)
		\end{flalign}
		allows the equation of motion to be rewritten in terms of a function $v_n$ associated with a discrete mass spectrum $m_{v_n}$. The resulting EOM becomes:
		\begin{flalign}
			\bigg[-\partial_z^2 + U_V(z)\bigg]v_n = m_{v_n}^2v_n,
		\end{flalign}
		where the effective potential is given by
		\begin{flalign}
			U_V(z) = -\frac{B''}{2} + \frac{(B')^2}{4},
		\end{flalign}
	
		For the specific case of the dilaton $\Phi(z)=\kappa^2z^2$ and the warp factor given by $A(z)=-\ln(z)$, the potential is written as
		\begin{flalign}
			U_V(z) = \kappa^4z^2 + \frac{3}{4z^2}.
		\end{flalign}
	
		The boundary conditions (\ref{eq:uv_condition}) and (\ref{eq:ir_condition}) ensure regularity and normalizability of the solutions, which are given by
		\begin{flalign}
			v_n (z) = \sqrt{2}\kappa z^{2}e^{-\kappa^2z^2/2}L_n^0(\kappa^2z^2),
		\end{flalign}
		with a corresponding mass spectrum
		\begin{flalign}
			m_{v_n}^2 = 4\kappa^2\left(n+1\right).
		\end{flalign}
		The above expression also follows the experimentally detected profile $m_n^2\sim n$.
		
		\subsubsection{Axial sector}
		
		We now proceed to the axial-vector sector of the soft-wall model, responsible for the description of axial-vector mesons. The dynamics within this sector are characterized by the corresponding part of the action,
		\begin{flalign}
			\mathcal{S}_\text{A} &=\int d^5x \sqrt{-g}e^{-\Phi} \bigg\{\chi^2(\partial_M\uppi - A_M)^2 + \frac{1}{2g_5^2}(\partial_M A_N - \partial_N A_M)^2 \bigg\}.
			\label{eq:action_axial}
		\end{flalign}
			
		Just like the vector field, the axial field also allows for a gauge transformation. It is evident in the second term, which has the same form as the term shown in (\ref{gaugetest}).	
		However, the axial sector presents an additional challenge, as its main field is coupled to the pseudoscalar field, as can be seen in the first term:
		\begin{flalign}
				(\partial_M\uppi - A_M)^2 = g^{MN}(\partial_M\uppi\partial_N\uppi - 2\partial_M\uppi A_N + A_MA_N),
		\end{flalign}
				
		To decouple them, it is necessary to perform a Helmholtz decomposition, which separates the field into its transverse and longitudinal components, $A_M = A_M^\perp + A_M^\parallel = A_M^\perp + \partial_M \varphi$, with $\varphi$ being an auxiliary longitudinal field.
			
		Axial mesons are described only by the transverse components, the longitudinal ones will compose the representation of pseudoscalar mesons. Therefore, we focus solely on the transverse terms and write the equation of motion in momentum space as:
		\begin{flalign}
			ze^\Phi\partial_z \left[\frac{e^{-\Phi}}{z}\partial_z \widehat{A}\right]
			- k^2 \widehat{A} + \frac{\chi^2g_5^2}{z^2}\widehat{A} = 0,
		\end{flalign}
		where we wrote the transverse field $A^\perp$ as $\widehat{A}$ after the Fourier transformation.
		After applying a transformation of the type $\widehat{A}\mapsto e^{B/2}\widehat{A}$, with $B(z)=\Phi(z)-A(z)=\Phi(z)+\ln(z)$, and performing a Kaluza-Klein decomposition given by
		\begin{flalign}
			\widehat{A}(x,z) = \sum_{n=0}^{\infty} A_n(x)a_n(z),
		\end{flalign}
		the equation of motion can be rewritten in terms of the functions $a_n$ as
			\begin{flalign}
				\bigg[-\partial_z^2 + U_A(z)\bigg]a_n = m_{a_n}^2a_n,
			\end{flalign}
		with an effective potential given by
			\begin{flalign}
			U_A(z) = \frac{B''}{2} - \frac{(B')^2}{4} + \frac{\chi^2g_5^2}{z^2}.
		\end{flalign}
		
		In the specific case of the dilaton given by $\Phi=\kappa^2z^2$ and the warp factor $A(z)=-\ln(z)$, the potential is
			\begin{flalign}
				U_A(z) = \kappa^4z^2 + \frac{3}{4z^2} + \frac{\chi^2g_5^2}{z^2}.
			\end{flalign}
				
		The potential for the axial field has a similar structure to that of the vector field, except for a $z$-dependent mass term that arises from the coupling between the axial field and the VEV of the scalar field \cite{Kapusta:2009ab}.
		This difference does not modify the formulation of the solution, which, following the boundary conditions (\ref{eq:uv_condition}) and (\ref{eq:ir_condition}), is written as
		\begin{flalign}
			a_n(z) = \sqrt{\frac{2}{n+1}}\kappa^2z^{3/2}e^{-\kappa^2z^2/2}L_n^1(\kappa^2z^2)
		\end{flalign}
		with a corresponding eigenvalue equation given by
		\begin{equation}
			m_{a_n}^2 = 4\kappa^2 (n+1)
		\end{equation}
	
		Once again, the resulting mass spectrum follows $m_n^2\sim n$, as expected in agreement with experimental data.
		
		\subsubsection{Pseudoscalar sector}
		
	The pseudoscalar sector, responsible for describing pseudoscalar mesons, starts from the same action used in the axial sector (\ref{eq:action_axial}).	
	This time, we turn our attention to the pseudoscalar field $\uppi$. 
	
	As previously seen, the axial field undergoes a decomposition into a transverse and a longitudinal component, which is responsible for decoupling the terms that describe the axial mesons. On the other hand, after this decomposition, the auxiliary longitudinal field $\varphi$ couples to the field $\uppi$.
	Thus, the dynamics of the pseudoscalar sector is governed by two coupled equations of motion, which in momentum space can be written as:
	\begin{subequations}
		\begin{align}
			\begin{split}
				e^{-\Phi}k^2 \partial_z \widehat{\varphi} + \beta\partial_z\widehat{\uppi} &= 0,
			\end{split}\\
			\begin{split}
				\frac{z}{\beta}\partial_z \left[\frac{e^{-\Phi}}{z}\partial_z \widehat{\varphi}\right] 
				+ \widehat{\varphi}-\widehat{\uppi} &=0.
			\end{split}
		\end{align}
	\end{subequations}
	By defining a new field $\widehat{Y}=\frac{e^{-\Phi}}{z}\partial_z\widehat{\varphi}$, these two equations can be combined into a single one: 
	\begin{flalign}
	\widehat{Y}'' + \left(\frac{1}{z} - \frac{\beta'}{\beta}\right)\widehat{Y}'+  \left(\frac{\beta}{e^{-\Phi}}+ k^2\right)\widehat{Y} = 0,
	\end{flalign}
	where $\beta(z)=\chi^2g_5^2z^{-2}e^{-\Phi}$.
	
	The procedure then follows the usual steps. A transformation of the form $\widehat{Y}\mapsto e^{-B/2}\widehat{Y}$, with $B(z)=\Phi(z)-3A(z)=\Phi(z)+3\ln(z)$, along with a Kaluza-Klein decomposition given by
	\begin{flalign}
		\widehat{Y} = \sum_{n=0}^{\infty} Y_n(x)y_n(z),
	\end{flalign}
	allows the equation of motion to be rewritten in a Schrödinger-like form
	\begin{flalign}
		\bigg[-\partial_z^2 + U_\uppi(z) \bigg]y_n = m_{\uppi_n}^2y_n,
	\end{flalign}
	whose effective potential is given by
	\begin{flalign}
		U_\uppi(z) = - \frac{B''}{2} +  \frac{3(B')^2}{4} - \frac{\chi^2g_5^2}{z^2}.
	\end{flalign}
	
	As in the axial sector, the potential also contains a term dependent on the scalar VEV. However, this conclusion does not affect the results, which follow the same pattern as in the other sectors, with solutions of the form 
	\begin{flalign}
		y_n(z) = \sqrt{2}\kappa z^{2}e^{-\kappa^2z^2/2}L_n^0(\kappa^2z^2),
	\end{flalign}
	and a corresponding mass spectrum given by 
		\begin{flalign}
		m_{\uppi_n}^2 = 4\kappa^2 n
		\label{eq:pseudoscalar_spectrum}
	\end{flalign}	
	also consistent with experimental results.

		\subsection{Dynamical soft-wall model}
		\label{sec:dynamical_soft_wall}
		
		While phenomenologically successful in certain aspects, a notable limitation of the standard soft-wall model is the \textit{ad hoc} nature of the dilaton profile. It is imposed by hand to break conformal invariance and generate confinement, rather than emerging from the underlying dynamics of the 5D gravitational theory itself.
		This means the model neglects backreaction, the fact that the dilaton's own energy-momentum should deform the assumed AdS$_5$ geometry, but is instead treated within a fixed, non-dynamic background.
		
		This issue had already been recognized shortly after the first works involving the hard-wall model, which also exhibited the same limitations regarding backreaction (among others), as pointed out in Ref. \cite{Csaki:2006ji} , where a possible solution to this problem is also proposed.
		
		The \textit{dynamical soft-wall} (DSW) \textit{model}\index{AdS/QCD!dynamical soft-wall model} aims to address these limitations by deriving the confining background dynamically.
		In this approach, the dilaton field $\Phi$, along with the metric of the 5D spacetime, is determined self-consistently by solving the coupled Einstein-scalar equations of motion \cite{Batell:2008zm,dePaula:2008ph,Ballon-Bayona:2023zal}.

		The action that defines the model is divided into two parts: one responsible for the background, and another that describes the matter content. The coupling can be implemented as proposed in \cite{Li:2012ay,Li:2013oda}, taking the following form:
		\begin{flalign}
			\mathcal{S}_\text{DSW} = \mathcal{S}_\text{G} + \frac{N_f}{N_c}\mathcal{S}_\text{M}.
			\label{eq:actioncoupling}
		\end{flalign}			
		Here, $\mathcal{S}_\text{M}$ represents the action for additional matter fields, such as gauge fields dual to QCD operators, namely, the action (\ref{softwallaction}) from the simplest soft-wall model. The action $\mathcal{S}_\text{G}$ describes the gravitational background and takes the standard graviton-dilaton form:	
		\begin{flalign}
			\mathcal{S}^s_\text{G} = \frac{1}{16\pi G_5} \int d^5x \sqrt{-g^s} e^{-2\Phi^s} \left( R^s + 4\partial_M\Phi^s \partial^M\Phi^s - V_\Phi^s\right),
			\label{action_stringframe}
		\end{flalign}
		where $R^s$ is the Ricci scalar, $G_5$ is the 5D Newton's constant, $\Phi^s$ is the dilaton field, and $V_\Phi^s$ is the dilaton potential. The superscript $s$ indicates that this fields, and thus the action (\ref{action_stringframe}), are written in what is known as the \textit{string frame}, which is more natural in the context of string theory. However, when dealing with spacetime structure from a general relativity perspective, the so-called \textit{Einstein frame} is more convenient, as it recovers the standard Einstein-Hilbert form for the action \cite{Polchinski:1998rq}.
		
		Transitioning between frames involves a Weyl rescaling that absorbs the dilaton field into the metric,
		\begin{flalign}
			g^E_{MN} = e^{2\omega}g^s_{MN},
 		\end{flalign}
		where $\omega=2(\Phi_0-\Phi)/(D-2)$	and the coefficient $\Phi_0$ is chosen for convenience. In the case studied here, we choose $\Phi_0=0$, and, with $D=5$ dimensions, this leads to the transformation:
		\begin{flalign}
			g^E_{MN} = e^{-\frac{4}{3}\Phi}g^s_{MN}.
		\end{flalign}
		This conformal transformation also redefines the fields accordingly:
		\begin{flalign}
			\Phi^E = \sqrt{\frac{8}{3}}\Phi^s, \quad \text{and} \quad V_\Phi^E = e^{\frac{4}{3}\Phi}V_\Phi^s.
		\end{flalign}
		
		For simplicity, we will omit the superscript $E$ from now on. With this, the background action in the Einstein frame becomes \cite{dePaula:2008fp}
		\begin{flalign}
			\mathcal{S}_\text{G} =  \frac{1}{16\pi G_5} \int d^5x \sqrt{-g} \left(R - \frac{1}{2}\partial_M\Phi \partial^M \Phi - V_\Phi\right)
			\label{action_einsteinframe}
		\end{flalign}
		
		The metric is usually taken to be of the form:
		\begin{equation}
			ds^2 = e^{2A(z)} (\eta_{\mu\nu}dx^\mu dx^\nu + dz^2),
			\label{eq:dsw_metric}
		\end{equation}
		where $A(z)=A^s(z) - \frac{2}{3}\Phi(z)$ is the warp factor in the Einstein frame.
		Some formulations may omit the conformal factor only along the holographic coordinate, but here we choose to follow a more general approach.
		
		Another aspect that makes the DSW model attractive concerns the Wilson loop, whose expectation value for a rectangular contour provides the static potential between a quark and an antiquark. In the large-$N_c$ limit, this observable can be computed holographically from the on-shell Nambu--Goto action of a fundamental string whose worldsheet extends into the bulk \cite{Maldacena:1998im}. For a metric of the form (\ref{eq:dsw_metric}) written in the string frame, the holographic confinement criterion states that the quark-antiquark potential grows linearly with their separation when the function $e^{2A^s(z)}=\sqrt{G_{00}(z)G_{xx}(z)}$ has a finite, positive minimum at an accessible radial position $z=z_*$. In this case, the string tension is given by $\sigma=e^{2A^s(z_*)}/(2\pi\alpha')$ \cite{Kinar:1998vq}. In the standard SW model, the AdS$_5$ metric remains undeformed, and the corresponding string-frame warp factor does not develop such a minimum; consequently, the standard implementation does not yield a linearly rising quark--antiquark potential from the fundamental string worldsheet, even though the dilaton is used to reproduce important features of the hadronic spectrum. In the DSW model, by contrast, the string-frame warp factor obtained from the coupled Einstein-dilaton equations develops a finite, positive minimum \cite{Ballon-Bayona:2023zal}, allowing the background geometry to produce a confining quark-antiquark potential.
		
		Disregarding $\mathcal{S}_\text{M}$ for now, we can use the action $\mathcal{S}_\text{G}$ to obtain the fields directly from the background dynamics of the system.
		First, the equations of motion can be derived by solving the Einstein field equations and varying Eq. (\ref{eq:dsw_metric}) with respect to $\Phi$.		As a result, one obtains a set of coupled equations:	
		\begin{subequations}
			\begin{align}
				\begin{split}
					\Phi'' + 3A'\Phi' - e^{2A} \frac{\partial V_\Phi}{\partial \Phi} &= 0
				\end{split}\\
				\begin{split}
					6(A')^2 - \frac{1}{2}(\Phi')^2 + e^{2A}V_\Phi &= 0
				\end{split}\\
				\begin{split}
					3A'' + 3(A')^2 + \frac{1}{2}(\Phi')^2 + e^{2A}V_\Phi &= 0
				\end{split}
			\end{align}
			\label{eq:eom_dsw}
		\end{subequations}
		This set of equations is overdetermined and can be conveniently rearranged to express, for example, the derivative of the dilaton in terms of the warp factor. Then, it suffices to know the geometry of the spacetime to determine the profile of the dilaton and its potential, up to integration constants that can be fixed by boundary conditions.
		
		In \Cref{section:lightmesons}, the dynamical soft-wall model will be employed to analyze light mesons. In addition to revisiting some concepts introduced here, the complete equations of motion for the DSW model -- including the mesonic action -- will also be presented.
		
		All derivations discussed here and throughout the analysis of light mesons consider the model at zero temperature. For a finite-temperature treatment, Refs. \cite{He:2013qq,Zhang:2021itx} are recommended, as they provide a general analysis including the derivation of the corresponding equations of motion.	
			
		\section{Information theory}
		\label{section:informationtheory}
		
		In his groundbreaking work of 1948 \cite{Shannon:1948zz}, Claude Shannon introduced the fundamental concepts of information theory\index{information theory}. In this work, he described the general functioning of a communication system along with its essential components -- an information source, a transmission channel and a receiver -- and presented methods for quantifying information.		
		In this way, Shannon transformed the seemingly abstract concept of information into quantifiable numerical values. 
		
		To formalize this idea, consider a discrete system described by an alphabet $\mathcal{A}=\{a_1,a_2,\dots,a_N\}$. This alphabet represents the set of all possible symbols or events, where each individual element $a_i$ has a corresponding probability of occurrence $p_i = p(a_i)$. 
		
		In this framework, the occurrence of an event itself carries a measurable amount of information. Typically, this quantity is referred to as the \textit{information content}\index{information theory!information content} (also known as Shannon information, self-information, or surprisal). Mathematically, the information content of a specific event $a$ is given by
		\begin{flalign}
			I(a)=-\log p_a,
			\label{shannoninformation}
		\end{flalign}
		which represents the minimum number of units required to encode and transmit that particular outcome.
		The base of the logarithm in this equation determines the unit of measurement. With a base 2, information is measured in bits (a familiar unit in computer science); with a base 10, the unit is the hartley (also known as dit or ban); and for many cases in physics, the natural base $e$ is commonly used, yielding entropy measured in nats.
		For example, when rolling a fair 4-sided die, a binary system needs two digits (bits) to transmit the result (e.g., 00, 01, 10, 11). We can arrive at this exact value using Eq. (\ref{shannoninformation}): since each outcome has an equal probability of $p_a=1/4$, calculating $-\log_2(1/4)$ yields an information content of exactly 2 bits.
		
		While the information content evaluates the surprise of a single isolated event, analyzing a complete system requires understanding its overall behavior. Therefore, to evaluate the system as a whole, we must calculate the expected value -- or the average -- of the information content across all possible events in the alphabet. 
		This is a core element of information theory called \textit{Shannon entropy}\index{information theory!Shannon entropy}, mathematically represented by weighting the information content by its respective probability:
		\begin{flalign}
			H = -\sum_{a\in \mathcal{A}} p_a \log p_a.
			\label{shannonentropy}
		\end{flalign}
		{The unit of measurement for entropy is the same as the quantity of information, depending on the base of the logarithm, such as bits for base 2 and nats for base $e$.}
			
		Conceptually, Shannon entropy quantifies the degree of uncertainty in a system before an event occurs. If a system is highly predictable (for instance, one outcome is almost certain to happen), its entropy is low, meaning little information is gained when the event actually takes place. If all outcomes in the alphabet are equally likely, the uncertainty is maximized. In such a state, observing the actual outcome provides the maximum possible amount of information.	

		From these definitions, the Shannon entropy being the expected value of the information content becomes explicitly clear through the relation:
		\begin{flalign}
			H = -\sum_{a\in \mathcal{A}} p_a \log p_a = \sum_{a\in \mathcal{A}} p_a I(a) = \langle I \rangle.
		\end{flalign}
	
		For an event that is absolutely certain to occur ($p_a=1$), its information content evaluates to zero, as there is no surprise associated with a fully expected outcome. Consequently, if a system is completely deterministic -- meaning one state has $p=1$ and all others have $p=0$ -- its overall entropy vanishes.
		
		Conversely, for an event that is impossible ($p_a=0$), the logarithm in the equation is formally undefined. However, taking the limit as $p_a\to 0$ reveals that the information gained from observing such an event tends towards infinity, reflecting maximum surprise. Despite this infinite individual surprisal, its contribution to the total entropy tends to zero, since $\lim_{p_a \to 0} p_a \log p_a = 0$. This mathematically ensures that events that practically never occur do not artificially inflate the system's average uncertainty.\\

		We can extend these descriptions to the continuous limit, which would be equivalent to adopting an alphabet with infinite symbols. Since this is the case for most physical problems, having a differential version of entropy is essential. A direct way to achieve this is by introducing a probability density $p_a=\rho(a)da$ and making $da \to 0$, then rewriting (\ref{shannonentropy}) as
		\begin{flalign}
			H = -\lim\limits_{da\rightarrow0}\sum_{a \in \mathcal{A}} \rho(a)da \log\left[\rho(a)da\right] 
			= -\lim\limits_{da\rightarrow0}\sum_{a\in \mathcal{A}} \left[\rho(a)da \log\rho(a) + \log(\frac{1}{da})\right].
		\end{flalign}
		
		Visualizing the integral in the first term is straightforward, but the second term presents a divergence. Initially, we can safely disregard the second term, bearing in mind that physical applications usually occur in scenarios with certain cutoff scales that will address this issue \cite{Gleiser:2018kbq}.
		Therefore, the differential entropy is written as
		\begin{flalign}
			H=\int da \rho(a) \log\rho(a).
			\label{differentialentropy}
		\end{flalign}
		Another issue that can be raised about the above equation is that it is not invariant under a coordinate transformation. In information theory, this problem is solved with the Kullback-Leibler divergence, which measures the relative entropy between two different distributions \cite{Sowinski}. Later we will see how this issue is treated within the approach that underlies this thesis.\\
		
		Finally, it is necessary to provide a brief introduction to the notion of \textit{complexity}\index{information theory!complexity}. The concept emerged in the 1960s and is defined as the minimal amount of information required to describe an object completely and precisely in any possible language \cite{Kolmogorov:1963,Kolmogorov:1968,Kolmogorov:1968a}. Essentially, it measures how incompressible a given piece of data is, making it one of the foundations of algorithmic information theory, a branch of information theory that focuses on the computational treatment of information.
		
		A classical example illustrates this idea: consider the two symbol sequences "\textit{aaaaaaaaaa}" and "\textit{4c1j5b3p0c}." The first can easily be described as "ten letters \textit{a}," a short and straightforward definition, implying low complexity. By contrast, the second sequence cannot be expressed through such a simple rule, as no obvious pattern governs its structure. The only option is to list each symbol individually, which corresponds to high complexity. Thus, an object is considered complex if it cannot be represented more succinctly than by the object itself.
		
		Complexity is related to entropy, but they quantify different properties. This can be clarified with another classical example. A cup of coffee may be regarded as a system with both low entropy and low complexity, since its state is simple and highly predictable. When milk is poured into the cup, turbulence develops, and the mixing of the two liquids leads simultaneously to an increase in entropy and in complexity, as new patterns and structures rich in information emerge. A full description of this state is long and essentially incompressible, as it requires specifying the intricate vortices and the boundaries between the fluids. After some time, however, the liquids settle into a homogeneous mixture. At this stage, the complexity decreases again, since the system is uniform and easily described, but the entropy remains high, as the information that distinguished the two liquids has been irreversibly lost. The uncertainty about the initial configuration of the system is maximal, and the final description, "coffee with milk," though simple, does not allow reconstruction of the previous state.
		
		Therefore, complexity peaks between perfect order and complete disorder, that is, between minimum and maximum entropy. A system is not complex when it is fully ordered nor when it is entirely random, but rather when it displays structured patterns that require a large amount of information to be faithfully described.
		
		With these introductory remarks, we are now prepared to move on to the configurational approach.	
			
		\section{Configurational entropy and complexity}
		\label{section:cims}
		
		As indicated in the introduction to this chapter, CIMs are concerned with extracting information from the shape of a system. In this section, we can finally examine these concepts in detail.
		
		To understand the fundamental ideas behind CIMs, one may start by considering a square-integrable scalar field configuration $\phi(\mathbf{r})$. Such a configuration specifies how the scalar field is distributed within a given volume. The field may be uniformly spread or display localized fluctuations at different positions, and regardless of the specific arrangement, this complete spatial structure is what we refer to as the \textit{shape}.
		
		Fluctuations in the field can be characterized by analyzing how the value of the field at one point relates to its value at another point. This relationship is quantified by the two-point correlation function, defined as
		\begin{flalign}
			\Pi(\mathbf{r}) = \frac{1}{V} \int_V d^d \mathbf{r}' \phi(\textbf{r}') \phi(\textbf{r}'+\textbf{r}).
			\label{eq:2pointcorrelation}
		\end{flalign}
		In this expression, the integral accounts for covering all correlations between points separated by a distance vector $\mathbf{r}$ within the volume. By comparing the field with a shifted version of itself, this function is also referred to as the autocorrelation, denoted $(\phi\star\phi)(\mathbf{r})$, making $\Pi(\mathbf{r})$ its volume-averaged value.
		
		To analyze the configuration in momentum space, it is first essential to recall a property of correlation functions that is analogous to the convolution theorem. It states that the Fourier transform of the correlation of two functions is the product of the complex conjugate of their individual Fourier transforms, i.e., $\mathcal{F}[f\star g]=\overline{\mathcal{F}[f]}\cdot\mathcal{F}[g]$.
		Using this property, the Fourier transform of (\ref{eq:2pointcorrelation}) can be written as	
		\begin{flalign}
			\mathcal{F}\big[\Pi\big] = \mathcal{F}\bigg[\frac{1}{V}(\phi\star\phi)\bigg] = \frac{1}{V} \overline{\mathcal{F}[\phi]}\cdot\mathcal{F}[\phi] = \frac{1}{V} \big|\mathcal{F}[\phi] \big|^2.
		\end{flalign}
	
		Although the field $\phi(\mathbf{r})$ is defined over a continuous volume, for the purpose of performing a spectral analysis it is convenient, at least at the beginning of this section, to adopt a representation in a discrete momentum space. In doing so, an equivalence between the continuous spatial average, defined by an integral over the volume, $(1/V)\int$, and its discrete counterpart, a sum over all $N$ points, $(1/N)\sum$, is established.	
		With this equivalence in mind, the Fourier transform of the field can be written as
		\begin{flalign}
			\mathcal{F}[\phi(\mathbf{r})] = \widehat{\phi}_\mathbf{k} = \frac{1}{V} \sum_\mathbf{r} \phi(\mathbf{r}) e^{-i\mathbf{k}\cdot\mathbf{r}},
		\end{flalign}
		whose inverse transform is given by
		\begin{flalign}
			\phi(\mathbf{r}) = \sum_\mathbf{k} 	\widehat{\phi}_\mathbf{k} e^{i\mathbf{k}\cdot\mathbf{r}},
		\end{flalign}
		which, from a physical point of view, represents the decomposition of the field into Fourier modes, where $\widehat{\phi}_\mathbf{k}$ denotes the amplitude of each mode.  
		This interpretation is central to the analysis of CIMs.
		
		With this, it follows that $\mathcal{F}[\Pi] \sim \abs{\widehat{\phi}_\mathbf{k}}^2$, and consequently, applying the inverse transform to both sides yields
		\begin{flalign}
			\frac{1}{V} \int d^d \mathbf{r}' \phi(\textbf{r}') \phi(\textbf{r}'+\textbf{r}) = \sum_\mathbf{k} 	\abs{\widehat{\phi}_\mathbf{k}}^2 e^{i\mathbf{k}\cdot\mathbf{r}}
			\label{eq:wienerkinchin}
		\end{flalign}
		Therefore, we can conclude that the Fourier transform of the two-point correlation function is the power spectrum of the field.
		This is a well-known result in information theory and signal processing referred to as the Wiener-Khinchin theorem \cite{Cover:2012}.
		
		This result establishes a direct connection between the physical properties of the system and the information measures.  
		Since the power spectrum is a positive-definite and normalizable function, it can be treated as a probability distribution.  
		Moreover, as it captures the contributions from all modes composing the configuration, measures performed on this distribution provide direct information about the physical system.
	
		From this conclusion, the probability of measuring the contribution of a specific mode is given by:
		\begin{flalign}
			p_\mathbf{k} = 	\frac{\abs{\widehat{\phi}_\mathbf{k}}^2}{\sum_{\mathbf{k}^\prime}\abs{\widehat{\phi}_{\mathbf{k}^\prime}}^2}.
			\label{eq:modal_disc_ce}
		\end{flalign}
		This object is one of the main ingredients in the calculation of CIMs and is referred to as the \textit{modal fraction}.
	
		In the current definition, the \textit{configurational entropy}\index{configurational information measures!configurational entropy} (CE) quantifies how many wave modes, and, effectively, the amount of information, are required to construct the field configuration.
		This measure is obtained by counting the contributions of the modes. Therefore, CE is defined in the same way as Shannon entropy, but applied to the modal fraction:
			\begin{flalign}
			\ce[\phi] = -\sum_{\mathbf{k}}p_\mathbf{k} \log p_\mathbf{k}.
		\end{flalign}
		For this reason, CE is also referred to as Shannon entropy for correlations \cite{Braga:2018fyc}.
		As in information theory, systems with uniformly distributed probabilities lead to maximum uncertainty (since all outcomes are equally possible), thus the entropy is also at its maximum. For a system with $N$ modes carrying equal weight, the modal fraction $p_\mathbf{k}=1/N$ results in the maximum configurational entropy $\ce=\log N$.	
		This is the case, for instance, with uncorrelated noise, which has a uniform power spectrum. Conversely, for a system described by a single mode, such as a single plane wave, the configuration is perfectly determined and the entropy is zero.
	
		While CE counts the units needed to construct a configuration, how intricately it is structured can be quantified by another measure, the \textit{configurational complexity}\index{configurational information measures!configurational complexity} (CC). 
		The modal fraction associated with CE accounts for the contribution of a given mode $\abs{\widehat{\phi}_{\mathbf{k}^\prime}}^2$ with respect to the total power of the system, $\sum_{\mathbf{k}^\prime}\abs{\widehat{\phi}_{\mathbf{k}^\prime}}^2$.
		In contrast, CC requires a new modal fraction, constructed by comparing a specific mode to the one with the highest contribution in the system, $\abs{\widehat{\phi}^{\text{max}}_{\mathbf{k}}}^2$.
		This difference in approach is sufficient for CE and CC to measure distinct properties.
		The modal fraction for complexity is written as:
		\begin{flalign}
			f_\mathbf{k} = \frac{\abs{\widehat{\phi}_\mathbf{k}}^2}{\abs{\widehat{\phi}^{\text{max}}_{\mathbf{k}}}^2},
			\label{modal_discrete_cc}
		\end{flalign}
		whose normalization with respect to the mode of highest contribution implies $f_\mathbf{k} \le 1$.
		Thus, the expression for the configurational complexity is given by:
		\begin{flalign}
			\cc[\phi] = -\sum_\mathbf{k} f_\mathbf{k} \log f_\mathbf{k},
		\end{flalign}
		that is always positive due to the normalization made in (\ref{modal_discrete_cc}). 
		It is easy to visualize that a uniform distribution represents such a simple pattern that it implies no complexity. As the distribution deviates from uniformity , i.e., the modes begin to differ from the one with the largest contribution, complexity increases. However, for a plane wave, both the CC and the CE are zero.
		Therefore, the complexity measure has its own properties and demands its own definition.
			
		An important point is that high entropy does not necessarily imply high complexity. Consider again the example of uncorrelated noise, which has maximal configurational entropy. In this case, the modal fraction for complexity is $f_\mathbf{k} = 1$, since all modes have the same power, which results in zero complexity. Conceptually, we can interpret uncorrelated noise as having a high degree of uncertainty. However, its complexity is zero because the description needed to generate it contains no specific features beyond purely random signals, with no interconnections between them.
		
		Other formulations based on information-theoretic quantities have also been employed in the study of hadrons; Ref.~\cite{daRocha:2025zth}, for instance, applies the Shannon entropy and related divergences to the distribution of leading digits of the hadron masses listed by the PDG, linking its deviation from a scale-invariant profile to the QCD scale.
		\\
					
		Both entropy and complexity can be obtained in the continuous limit of momentum space.
		In this scenario, the Fourier transform in $d$ spatial dimensions is
		\begin{flalign}
			\widehat{\phi}(\mathbf{k}) =  \int \frac{d^d\mathbf{r}}{(2\pi)^{d/2}} \phi(\mathbf{r}) e^{-i\mathbf{k}\cdot\mathbf{r}},
			\label{fourier}
		\end{flalign}
		and the continuous probability distribution $\rho(\mathbf{k})$ that acts as the modal fraction for the entropy is
		\begin{flalign}
			\rho(\mathbf{k}) =
			\frac{\abs{\widehat{\phi}(\mathbf{k})}^2}{\int d^d\mathbf{k}^\prime\abs{\widehat{\phi}(\mathbf{k}^\prime)}^2}.
			\label{modal_continuum_dce}
		\end{flalign}
		
		Therefore, the continuous version of the CE, formally known as the \textit{differential configurational entropy}\index{configurational information measures!differential configurational entropy} (DCE), is given by \cite{Gleiser:2018kbq}:
		\begin{flalign}
			\dce[\phi] = -\int d^d\mathbf{k}\:  \rho(\mathbf{k})\log \rho(\mathbf{k}).
			\label{dce}
		\end{flalign}		
		Just like its discrete counterpart, the DCE is maximal for systems with a uniform-type power spectrum, such as uncorrelated noise, and it is zero for systems with a single mode, like a single plane wave.
		
		As we observed when discussing entropy in continuous form (\ref{differentialentropy}) for the first time, the expression for DCE (\ref{dce}) is also not invariant under a coordinate transformation. From the perspective of CIMs, this issue is expected. This is because we know that information can change depending on the system used: the word $\varphi\nu\sigma\iota\kappa\acute{\eta}$ is constructed within a set of symbols that we recognize as the Greek alphabet; if we choose to use another alphabet, the format of the information will change. Hence, the shape of a field demanding a different set of data for its description depending on the coordinate system is, in fact, treated as a characteristic of CIMs.
		
		The modal fraction associated with the measure of complexity is also constructed from the probability distribution. Just like in the discrete case, the modal fraction for a mode $\mathbf{k}$ is normalized with respect to the mode of maximum contribution,
		\begin{flalign}
			f(\mathbf{k}) = \frac{\abs{\widehat{\phi}(\mathbf{k})}^2}{\abs{\widehat{\phi}^{\text{max}}(\mathbf{k}')}^2},
			\label{modal_continuum_dcc}
		\end{flalign}
		with which we define the \textit{differential configurational complexity}\index{configurational information measures!differential configurational complexity} (DCC):
		\begin{flalign}
			\dcc[\phi] = -\int d^d\mathbf{k} \: f(\mathbf{k})\log f(\mathbf{k}).
		\end{flalign}
		Since the modal fraction is constrained by $f(\mathbf{k}) \leq 1$, it is guaranteed that the DCC will always yield a positive value.
		And, just like the CC, the DCC vanishes for systems with straightforward descriptions, such as a single plane wave or a uniformly distributed power spectrum. This description makes it simple to imagine how the complexity of these systems increases as the power distribution deviates from a uniform profile.
		
		At first glance, it may appear that the modal fractions associated with entropy and complexity, respectively (\ref{modal_continuum_dce}) and (\ref{modal_continuum_dcc}), are different only due to a normalization factor, which would imply that DCE and DCC are essentially measuring the same information.
		However, we can observe that the normalization in (\ref{modal_continuum_dce}) is carried out with respect to a constant real number computed from an integration across the entire momentum space, $\int d^d\mathbf{k}|\widehat{\phi}(\mathbf{k})|^2$.
		On the other hand, the normalization of the modal fraction (\ref{modal_continuum_dcc}) is done relative to the absolute square of a function that depends on the value of a specific mode $\mathbf{k}$, $\abs{\widehat{\phi}^{\text{max}}}^2$, which even ensures the positivity of DCC. Consequently, this difference is sufficient for DCE and DCC to measure distinct properties.
		
		Nevertheless, there is another point that deserves careful attention. From a physical standpoint, there is no need to ensure that the DCE remains positive, as is the case with the DCC. This is because negative values of the DCE arise only from probability distributions that do not correspond to any physical system \cite{Gleiser:2018jpd}.
		
		Considering that the DCE equation is defined as an integral over the entire space, it is convenient that the integrand be well-behaved in order to avoid divergences. To guarantee convergence, one of the proposed solutions includes introducing a second normalization in the probability distribution 		\cite{Gleiser:2013mga}, 
		\begin{flalign}
			\check{\rho}(\mathbf{k}) = \frac{\rho(\mathbf{k})}{\rho^{\text{max}}(\mathbf{k}')},
		\end{flalign}
		where $\rho^{\text{max}}$ represents the value of the distribution at the mode of largest contribution. This additional normalization recasts the distribution as a dimensionless quantity bounded between 0 and 1, avoiding potential problems. As a consequence, the DCE is rewritten as
		\begin{flalign}
			\dce = -\int d^d\mathbf{k} \;  \check{\rho}(\mathbf{k})\log\check{\rho}(\mathbf{k}).
		\end{flalign}
		
		Another proposed solution involves a change of variables that rearranges the limits of integration, as will be presented in the next section, where the context renders its explanation more illustrative.
		
		\section{CIM-based AdS/QCD hybrid method}
		\label{section:cimadsqcd}
		
		In addition to the various applications highlighted in the introduction of this chapter, CIMs were also used to investigate some aspects of AdS/CFT \cite{Barreto:2022ohl,Braga:2019jqg,Cruz:2018qby,Bazeia:2018uyg,Bazeia:2021stz,Alves:2017ljt,Alves:2014ksa,Braga:2020opg,Lee:2018zmp}.		Applications in QCD phenomenology have found fruitful results with new insights into topics such as nuclides \cite{MartinContreras:2022lxl}, 
		color-glass condensate \cite{Karapetyan:2016fai,Karapetyan:2017edu} and several properties of meson families, including scalar, exotic, light- and heavy-flavor mesons \cite{dePaula:2009za,Frederico:2014bva,Colangelo:2018mrt}, quarkonia in quark-gluon plasma with a magnetic field \cite{Braga:2021fey,Braga:2020hhs}, and glueballs with their finite-temperature description \cite{Braga:2021zyi}, in addition to other applications \cite{GoncalvesdaSilva:2017bvk, Barbosa-Cendejas:2018mng,Ma:2018wtw,LeThien:2019lxh,Karapetyan:2020yhs,daRocha:2021xwq,Karapetyan:2021vyh,Karapetyan:2021crv,Karapetyan:2023sfo}.
		
		Furthermore, the use of CIMs has proven to be highly efficient in scrutinizing the mass spectroscopy of hadrons in holographic QCD.
		Bernardini and da Rocha \cite{Bernardini:2016hvx} were the first to propose DCE as a means of analyzing mesons in AdS/QCD, establishing a direct relationship between the configurational entropy and the spin values of mesons. This enabled the estimation of the stability of higher resonance states, indicating a possible explanation for the difficulty of detecting and measuring them.
		Later, the same procedure was applied to other families of mesons \cite{Bernardini:2018uuy,Ferreira:2019inu,daRocha:2021imz,Bernardini:2025wkb}, in addition to tensor \cite{Ferreira:2019nkz} and isovector mesons \cite{daRocha:2021ntm}, bottomonium and charmonium \cite{Braga:2017fsb}, odd- and even-spin glueballs \cite{MarinhoRodrigues:2020yzh, MarinhoRodrigues:2020ssq}, others exotic mesons \cite{Karapetyan:2021ufz}, and baryons \cite{Ferreira:2020iry,Guo:2024nrf},  with many results in accordance with the further state data in Particle Data Group \cite{ParticleDataGroup:2024cfk}. 
		
		To apply CIMs to AdS/QCD models, the first step consists in determining the configurational profile of the system. In the case of hadrons, the representation can be carried out through the energy density, which corresponds to the temporal component of the energy-momentum tensor.  
		As discussed at the beginning of \Cref{sec:dynamical_soft_wall}, the simplest soft-wall model is not fully consistent with general relativity due to the introduction of background fields without accounting for backreaction. Nevertheless, we may assume that the energy-momentum tensor can still be constructed in the same way as in the Einstein field equations, namely, as
		\begin{flalign}
			T_{00}(r) = \frac{2}{\sqrt{-g}}\left[\frac{\partial(\sqrt{-g}\mathcal{L})}{\partial g^{00}}
			- \frac{\partial}{\partial x^\beta}\frac{\partial(\sqrt{-g}\mathcal{L})}{\partial(\partial_\beta g^{00})}
			\right],
			\label{energymomentumtensor}
		\end{flalign}	
		where $\mathcal{L}$ is the Lagrangian density. Thus, obtaining the CIMs of the system requires only the Lagrangian function, which can be directly obtained from the model's action.
		
		The Fourier transform acts on this tensor as  
		\begin{flalign}
			T_{00}(k) = \int_{-\infty}^{+\infty} d^d r \frac{\sqrt{-g}}{(2\pi)^{d/2}} T_{00}(r) e^{ikr},
		\end{flalign}
		which decomposes the field into momentum space, generating the configuration in terms of the momentum	$k=k_M=(k_\mu,k_z)$. The modal fraction associated with the entropy is then obtained as  
		\begin{flalign}
			\ltau(k) = \frac{\abs{T_{00}(k)}^2}{\displaystyle\int_{-\infty}^{+\infty} d^dk\abs{T_{00}(k)}^2},
		\end{flalign}
		and consequently, the DCE can be written as  
		\begin{flalign}
			\dce = - \int_{-\infty}^{+\infty} d^d k \; \ltau(k) \log \ltau(k).
		\end{flalign}
		
		The challenge is then to evaluate this integral. Although the computation is performed numerically, it requires additional procedures to make the resolution computationally feasible. Since the relevant fields considered here, such as the dilaton, the warp factor, and the VEV, depend exclusively on the holographic coordinate $z$, we can fix $d=1$ and restrict the integration to this coordinate only.  
		Moreover, since $z$ represents the energy scale, this integration runs from the AdS boundary into the bulk, which translates into an integration over the non-negative real axis, $[0,+\infty)$, thus excluding negative values of $z$.  
		
		As discussed at the end of the previous section, the modal fraction as originally defined does not guarantee convergence for the DCE. Instead of introducing the second normalization, an alternative approach consists in performing a change of variables, which is the method adopted from now on. A transformation of the type $z\mapsto\tilde{z}/(1-\tilde{z})$ maps the integration interval $z\in [0,+\infty)$ into a finite integration interval $\tilde{z}\in[0,1]$.
		
		Finally, with these definitions, the DCE is rewritten as		
		\begin{flalign}
			\dce = \int^1_0  \frac{dk}{(1-k)^2} \; \ltau\left(\frac{k}{1-k}\right) \ln\left[\ltau\left(\frac{k}{1-k}\right)\right].
		\end{flalign}

		The procedure to compute the DCC is considerably more straightforward and can be carried out under the same arguments discussed above. In this case, the modal fraction is expressed as
		\begin{flalign}
			f_\tau(k) = \frac{\abs{T_{00}(k)}^2}{\abs{T_{00}^{\text{max}}}^2},
		\end{flalign}
		where $T_{00}^{\text{max}}$ denotes the dominant mode with which the modal fraction is normalized, ensuring that the integral is properly evaluated. The next step is to carry out the change of variables in order to redefine the limits of integration. With this, the DCC is given by:
		\begin{flalign}
			\dcc = -\int_0^1\frac{dk}{(1-k)^2} \; f_\tau \left(\frac{k}{1-k}\right) \log f_\tau \left(\frac{k}{1-k}\right).
			\label{eq:dcc_expression}
		\end{flalign}

		Next, these expressions are employed to calculate the DCE and DCC of hadronic resonances and to construct what we shall refer to as \textit{Regge-like trajectories}, due to their similarities with Regge trajectories without necessarily incurring their properties or even derivation method.
		The first type of Regge-like trajectory plots the values of the DCE/DCC as a function of the radial quantum number $n$ of each resonance. The expressions that give rise to these trajectories, denoted as $\dce(n)$ and $\dcc(n)$, are obtained by interpolating the CIM values corresponding to each quantum number $n$.
		
		According to the interpolation theorem, if $\mathsf{p}$ resonances are catalogued for a given family of particles, then the expression for the DCE that passes through all data points must be a polynomial of maximum degree $\mathsf{p}-1$, which takes the following general form:
		\begin{flalign}
			\dce(n) = a_{\mathsf{p}-1}n^{\mathsf{p}-1} + a_{\mathsf{p}-2}n^{\mathsf{p}-2} + \dots + a_2n^2 + a_1 n + a_0,
			\label{eq:generaldcen}
		\end{flalign}
		and similarly for the DCC.
		
		The second type of Regge-like trajectory consists of the CIM values as a function of the squared mass of each resonance.
		The expression for these trajectories, denoted as $\dce(m^2)$ and $\dcc(m^2)$, is obtained through an interpolation based on the experimentally detected masses of the resonances, yielding an expression of the form
		\begin{flalign}
			\dce(m^2) = b_{\mathsf{p}-1}(m^2)^{\mathsf{p}-1} + b_{\mathsf{p}-2}(m^2)^{\mathsf{p}-2} + \dots + b_1m^2 + b_0,
			\label{eq:generaldcem}
		\end{flalign}
		and the DCC follows the same form.		
		By using experimentally catalogued data for the interpolation, we ensure that these curves maintain a high degree of accuracy with respect to what is observed in nature. This empirical foundation represents one of the main advantages of the method presented here.
		
		The trajectories themselves already carry physical information about the system. As observed in \cite{Bernardini:2016hvx}, the DCE tends to increase with both $n$ and $m$. This rise in entropy reflects a corresponding enhancement in the uncertainty and unpredictability of the information encoded in the configuration, which can be interpreted as a manifestation of growing system instability. Such behavior offers a reasonable explanation for the much more frequent detection of hadrons in lower excited states, characterized by smaller radial quantum numbers and masses, as these states are expected to be considerably more stable.
		
		Thus, we obtain the same informational measures as functions of two independent particle properties. This duality allows the estimation of the masses of new states, which will represent one of the central conclusions of the analyses presented in this work. Once an expression for the CIMs as a function of the radial number has been obtained, as in Eq. (\ref{eq:generaldcen}), one can insert any higher value of $n$ to estimate the corresponding DCE/DCC value for that resonance level. This CIM value is then substituted into the left-hand side of the mass-dependent equation (\ref{eq:generaldcem}), which can subsequently be solved algebraically to find the allowed solutions for $m$. The estimated mass of the extrapolated state is selected from these solutions after discarding complex and unphysical results (such as decreasing masses associated with higher resonances).
		
		Finally, these estimated masses can be compared with the uncertainty ranges of further states detected in experiments and cataloged in the PDG, but still lacking identification. In this way, extrapolated states that exhibit possible correspondences may be regarded as candidates for these unidentified further states, thereby suggesting a potential direction for future experimental investigations.
		
		Although such an extrapolation could in principle be performed using the eigenvalue equations derived from the Schrödinger-like formalism for each particle type, the fact that the CIM-based AdS/QCD hybrid method employs an interpolation grounded in experimental data makes its predictions substantially more robust and firmly anchored to reality.
		Moreover, the insights it provides regarding system stability are particularly valuable, opening the door to further analyses such as those developed in Ref. \cite{Braga:2018fyc} and in \Cref{chapter:thermalqcd}, where CIMs are employed to investigate stability with respect to temperature.\\		

		A pertinent issue that may be raised concerns the application of CIMs -- quantities intrinsically connected to probability distributions -- to a well-defined function in space such as the energy-momentum tensor. If that were the case, one would essentially be attempting to extract information measures such as uncertainty from a deterministic system, which would be logically inconsistent.		
		The resolution of this point lies in the fact that the formalism underlying CIMs does not require the original field to be probabilistic. Instead, a probability density is constructed from the Fourier-mode decomposition of the system's spatial structure. Since these modes can exhibit different intensities, ranging from uniform to completely distinct, the informational measures remain valid and applicable.
		The source of uncertainty, therefore, emerges from the transformation of the original function into a configurational profile.
		
		With this issue clarified, and after the necessary presentation of the main method of this thesis, we can now proceed to its practical applications in the analysis of hadrons.

		\section{CIMs of light-flavor mesons}
		\label{section:lightmesons}
		
		A reasonable starting point for applying CIMs in conjunction with AdS/QCD models is the study of light-flavor mesons, given their relatively simple description and the abundance of experimental results available. However, these advantages have also made light mesons extensively researched in the context of AdS/QCD and even CIMs, as evidenced by works such as \cite{Karch:2006pv,Bernardini:2018uuy,Ferreira:2019inu}.
		
		Given the considerable number of detected states for light mesons, the first analysis in this thesis focuses on using the DCE to search for new, heavier resonances beyond those already identified. To broaden this search, the dynamical soft-wall model will be employed to analyze the families of pseudoscalar, axial vector, scalar, and vector mesons.
			
		In addition to this setup, the fields will incorporate an anomalous mass correction. In bottom-up AdS/QCD models, it is usually assumed that the holographic dictionary remains unchanged despite modifications to the AdS background. While the conformal dimension is well-defined in a CFT, perturbative calculations in QCD suggest a correction to the conformal dimension in the form of $\Delta-\gamma(z)$, where $\gamma(z)$ is the anomalous dimension that depends on the energy scale $\sim 1/z$. Consequently, Eq. (\ref{massdimension_relation}) is rewritten as			
		\begin{flalign}
			m^2 = \big[\Delta -\gamma(z) \big] \big[ \Delta - \gamma(z) - 4\big].
			\label{masscorrected}
		\end{flalign}
		
		The asymptotic scalar VEV $\chi$ in (\ref{eq:uvbehavior}) is also modified by these corrections, acquiring the following asymptotic form:
		\begin{eqnarray}
			\chi(z\rightarrow 0) = m_q \upzeta z^{1+\gamma(z)} + \frac{\sigma}{\upzeta} z^{3-\gamma(z)},
			\label{chi_anomalous}
		\end{eqnarray}
		where $\upzeta = \zeta z^{\gamma(z)}$ was redefined to make the mass dimension of the terms on the right-hand side of Eq. (\ref{chi_anomalous}) compatible.
		With this, the condition (\ref{eq:chi_condition}) is now written as
		\begin{flalign}\label{fas1}
			\chi'(z) = \sqrt{\frac{8}{\alpha}}\kappa e^{-\Phi(z)/2} \left(1+a_1(z)e^{-\Phi(z)} + a_2(z)e^{-2\Phi(z)}\right),
		\end{flalign}
		with
		\begin{eqnarray}
			a_1(z) &=& -2 + \frac{5\sqrt{2\alpha}m_q e^2\upzeta}{8\kappa} + \frac{3\sqrt{2\alpha}\sigma }{4e^2\upzeta \kappa^3}, \\
			a_2(z) &=& 1 - \frac{3\sqrt{2\alpha}m_q e^2\upzeta}{8\kappa} - \frac{3\sqrt{2\alpha}\sigma }{4e^2\upzeta \kappa^3}, 
		\end{eqnarray}
		where the factor $e^2$ is inserted here to ensure that the dimension is recovered in the UV limit.
	
		There is more than one way to describe the anomalous dimension. Ref. \cite{Gubser:2008yx} demonstrates that it is possible to relate the dynamic evolution of the anomalous dimension to the beta function (\ref{betafunction}). In perturbative QCD, the beta function is commonly represented as a expansion in terms of the powers of the coupling constant, with its terms derived from loop corrections at different orders.
		To fit the soft-wall model, the 't Hooft coupling  is considered and the renormalization scale $\mu$ is expressed as a function of the inverse of the $z$ coordinate. This reformulation enables the beta function to be rewritten in a modified form as
		\begin{flalign}
			\beta(\lambda(z)) = -z\frac{d\lambda(z)}{dz}.
		\end{flalign}
			
		According to \cite{FolcoCapossoli:2016uns}, in the one-loop approximation, the beta function correctly reproduces the perturbative behavior of the UV regime by assuming the form $\beta(\lambda)\sim b_0\lambda^2$, with $b_0=\frac{1}{8\pi^2}(\frac{11}{3}-\frac{2}{9}N_f)$. To reproduce the non-perturbative behavior of the IR regime, it is necessary to modify the beta function by a squared correction factor, which is given by
		\begin{flalign}
			\beta(\lambda) = -b_0 \lambda^2 \bigg(1-\frac{\lambda}{\lambda_0}\bigg)^2.
			\label{modified_beta}
		\end{flalign}
		Thus, $\lambda_0$, defined at a fixed point in the IR, becomes an input parameter in the model.
		Also, based on the one-loop correction results, the anomalous dimension can be expressed in terms of the coupling constant as
		\begin{flalign}
			\gamma = \frac{(N_c^2-1)}{32\pi^2 N_c^2}\lambda,
			\label{eq:gamma1}
		\end{flalign}
		which leads to
		\begin{flalign}
			\frac{d\gamma}{dz} = -\frac{(N_c^2-1)}{32\pi^2 N_c^2 z}\beta(\lambda),
			\label{eq:gamma2}
		\end{flalign}
		whose solution provides the expression for $\gamma$ as a function of the beta function, offering one possible approach to determining the anomalous dimension.
			
		To further expand the analysis, the four cases of anomalous dimension presented in \cite{Chen:2022pgo} will be considered. Case I will serve as a control and does not account for the correction, meaning $\gamma(z)=0$. Case II incorporates the correction in the form of
		\begin{flalign}
			\gamma(z)=-\frac{3}{2}\left(e^{-\Phi^2(z)/2}-1\right),
			\label{caseii}
		\end{flalign}
		leading to corresponding adjustments in the input parameters, which are now written as
		\begin{eqnarray}
			m_q&\mapsto&m_qz^{\gamma(z)}\left[\frac{1}{2}(1+\gamma(z))+\ln(z)\gamma'(z)\right],\\
			\sigma&\mapsto&\sigma z^{-\gamma(z)}\left[\frac{1}{2}(3-\gamma(z))-\ln(z)\gamma'(z)\right],
		\end{eqnarray}
		to test the dominance of the non-perturbative regime over the UV limit.
		
		Case III employs the dynamic evolution of $\gamma(z)$ based on the beta function, as described by Eqs. (\ref{eq:gamma1}, \ref{eq:gamma2}), whose inverse function is given by
		\begin{eqnarray}
			\gamma^{-1}(z)=  \frac{32 \pi ^2 b_0
				N_c^2 }{N_c^2-1}\left[\frac{1}{z}+\frac{1}{2} \ln\frac{1-z}{1+z}\right]\ln (z). 
			\label{caseiii}
		\end{eqnarray}
		
		Finally, case IV considers a constant correction, where $\gamma(z)=\gamma$ will be treated as an input parameter.
		
		For this analysis, the input parameters will be $m_q=5$ MeV, $\sigma = (240\text{ MeV})^3$, and $\kappa=0.43$ GeV. Newton's constant will be fixed as $G_5=3L^3/4=0.750$ \cite{Li:2012ay}. In case III, the coupling at the IR fixed point, $\lambda_0$, is set to 1 and the initial condition used is $\lim_{z \to 1/M_Z}\lambda(z) = 0.1184$, where $M_Z$ is the $Z$ boson mass, frequently used as a mass scale in the analysis of the running of the QCD coupling constant. And for case IV, the constant value of the anomalous dimension will be $\gamma=0.45$ \cite{Chen:2022pgo}.\\
		
		The action considered is the DSW action, which follows from the coupling between the gravitational and matter parts as in Eq. (\ref{eq:actioncoupling}). In its complete form, this action is written as
		\begin{flalign}
			\mathcal{S}_{\text{DSW}} = \frac{1}{16\pi G_5} \int {\rm d}^5x\sqrt{-g} \left[ e^{-2\Phi} \bigg( R - 4\partial_M\Phi \partial^M\Phi - V_\Phi  \bigg) 
			- \alpha e^{-\Phi} \left( \frac{1}{2}\partial_M \chi \partial^M \chi -\frac{3}{4}\abs{\chi}^2
			\right)\right],
		\end{flalign}
		where we define the coupling parameter as $\alpha=16\pi G_5 N_f/N_c$.		
		From this complete form, we obtain the equations of motion analogous to Eqs. (\ref{eq:eom_dsw}), but now supplemented by the fields arising from $\mathcal{S}_\text{M}$.
				
		Following the same procedure presented in \Cref{subsection:softwall}, it is possible to obtain the Schrödinger-like equations for the fields representing the mesons. Here, we denote $\uppi_n$ as the function describing the pseudoscalar mesons, $a_n$ for the axial vector mesons, $s_n$ for the scalar mesons, $v_n$ for the vector mesons, and $\varphi_n$ is an auxiliary function coupled to the pseudoscalar sector.
		
		The equations are as follows:
		
		\begin{subequations}
			\setlength{\abovedisplayskip}{2pt}
			\setlength{\belowdisplayskip}{2pt}
			\begin{alignat}{4}
				\bigg(-\partial^2_z + U_s\bigg)s_n &= m_n^2s_n,\\
				\bigg(-\partial^2_z + U_v\bigg)v_n &= m_n^2 v_n,\\
				\bigg(-\partial^2_z + U_a\bigg)a_n &= m_n^2 a_n,\\
				\bigg(-\partial^2_z + U_{\uppi,\varphi}\bigg)\uppi_n &= m_n^2 (\uppi_n - e^{A} \chi \varphi_n),\\
				\bigg(-\partial^2_z + U_\varphi\bigg)\varphi_n &= g_5^2 e^{A} \chi (\uppi_n - e^{A} \chi \varphi_n),
			\end{alignat}
			\label{schrodingerlike}
		\end{subequations}
		with the potentials 
		\begin{subequations}
			\begin{eqnarray}
				U_s &=& \frac{1}{2}(3A'' - \Phi'') + \frac{1}{4}(3A' - \Phi')^2 -3e^{2A},\\
				U_v &=&\frac{1}{2}(A''-\Phi'') + \frac{1}{4}(A'-\Phi')^2,\\
				U_a &=& \frac{1}{2}(A''-\Phi'') + \frac{1}{4}(A'-\Phi')^2 + g_5^2 e^{2A}\chi^2, \\
				U_{\uppi,\varphi} &=& \frac{1}{2}\bigg( 3A'' - \Phi'' + \frac{2\chi''}{\chi} - \frac{2{\chi'}^2}{\chi^2}\bigg) + \frac{1}{4}\bigg(3A'-\Phi'+\frac{2\chi'}{\chi}\bigg)^2,\\
				U_\varphi &=& \frac{1}{2}(A''-\Phi'') + \frac{1}{4}(A'-\Phi')^2.
			\end{eqnarray}
			\label{schrodingerlikepotentials}
		\end{subequations}
		
		The $n=1,2,3,\dots$ solutions to these equations thus form the families of mesons studied in this section. 
		For scalar mesons, we will analyze the $f_0$ meson family, which includes nine cataloged resonances. The pseudoscalar mesons chosen for the analysis belong to the pion family, which includes its four other cataloged resonances. The chosen axial-vector mesons are from the $a_1$ family, which has five cataloged resonances. Finally, the vector mesons analyzed in this section will be the rho ($\rho$) mesons, which have seven cataloged resonances.
		
		The results were obtained numerically with the input parameters presented previously and compiled in the tables \ref{table:spectra_pseudoscalar}, \ref{table:spectra_axial}, \ref{table:spectra_scalar}, and \ref{table:spectra_vector}, with distinction for the four cases analyzed.
			
		\begingroup
		\renewcommand{\arraystretch}{1.8}
		\begin{table}[h!]
			\centering
			\setlength{\tabcolsep}{1.2mm}
			\begin{tabular}{cc!{\color{white}\vrule width 2pt}ccccc}
				\rowcolor{ufabc!50} 
				\cellcolor{ufabc!50} $n$ & \cellcolor{ufabc!50} \bf State & \bf $M_\text{exp}$ (MeV) & \multicolumn{1}{c}{\bf $M_\text{I}$ (MeV)} & \multicolumn{1}{c}{\bf $M_\text{II}$ (MeV)} & \multicolumn{1}{c}{\bf $M_\text{III}$ (MeV)} & \multicolumn{1}{c}{\bf $M_\text{IV}$ (MeV)} \\
				\rowcolor{ufabc!30} 1 & $\uppi^\pm$ & \small $139.57039 \pm 0.00017$ & \small $140 \pm 0.000178$ &\small $146 \pm 0.000192$ & \small $158 \pm 0.000173$ &\small $109 \pm 0.000171$ \\
				\rowcolor{ufabc!10} 2 & $\uppi(1300)$ & $1300 \pm 100$ & $1600 \pm 123$ & $1408 \pm 108$ & $1410 \pm 108$ & $1415 \pm 109$ \\
				\rowcolor{ufabc!30} 3 & $\uppi(1800)$ & $1810^{+9}_{-11}$ & $1897\pm12$ & $1873\pm 11$ & $1821\pm11$ & $1841\pm11$ \\
				\rowcolor{ufabc!10} 4* & $\uppi(2070)$ & $2070 \pm 35$ & $2116 \pm 36$ & $2123 \pm 36$ & $2070 \pm 35$ & $2096 \pm 36$ \\
				\rowcolor{ufabc!30} 5* & $\uppi(2360)$ & $2360 \pm 25$ & $2299 \pm 25$ & $2313 \pm 25$ & $2264 \pm 25$ & $2289 \pm 25$ 
			\end{tabular}
			\caption[Experimental and predicted mass spectra for $\uppi$ pseudoscalar mesons mass]{Experimental and predicted mass spectra for $\uppi$ pseudoscalar mesons mass. Meson states marked with the "*" are omitted from the summary table in the PDG.}
			\label{table:spectra_pseudoscalar}
		\end{table}
		\endgroup
				
		\begingroup
		\renewcommand{\arraystretch}{1.8}
		\begin{table}[h!]
			\centering
			\setlength{\tabcolsep}{3mm}
			\begin{tabular}{cc!{\color{white}\vrule width 2pt}ccccc}
				\rowcolor{ufabc!50} 
				\cellcolor{ufabc!50} $n$ & \cellcolor{ufabc!50} \bf State & \bf $M_\text{exp}$ (MeV) & \multicolumn{1}{c}{\bf $M_\text{I}$ (MeV)} & \multicolumn{1}{c}{\bf $M_\text{II}$ (MeV)} & \multicolumn{1}{c}{\bf $M_\text{III}$ (MeV)} & \multicolumn{1}{c}{\bf $M_\text{IV}$ (MeV)} \\
				\rowcolor{ufabc!30} 1 & $a_1(1260)$ & $1230 \pm 40$ & $1316 \pm 43$ & $1222 \pm 40$ & $1163 \pm 40$ & $1232 \pm 40$ \\
				\rowcolor{ufabc!10} 2 & $a_1(1640)$ & $1655 \pm 16$ & $1735 \pm 17$ & $1676 \pm 16$ & $1649 \pm 16$ & $1707 \pm 17$ \\
				\rowcolor{ufabc!30} 3* & $a_1(1930)$ & $1930\pm71$ & $1969\pm71$ & $1971\pm71$ & $1926\pm71$ & $1980\pm72$ \\
				\rowcolor{ufabc!10} 4* & $a_1(2095)$ & $2096 \pm 41 \pm 81$ & $2163 \pm 126$ & $2178 \pm 127$ & $2132 \pm 124$ & $2181 \pm 127$ \\
				\rowcolor{ufabc!30} 5* & $a_1(2270)$ & $2270\pm55$ & $2336\pm57$ & $2356\pm57$ & $2311\pm56$ & $2358\pm57$ 
			\end{tabular}
			\caption[Experimental and predicted mass spectra for $a_1$ axial vector mesons]{Experimental and predicted mass spectra for $a_1$ axial vector mesons. Meson states marked with the "*" are omitted from the summary table in the PDG.}
			\label{table:spectra_axial}
		\end{table}
		\endgroup
		
		\begingroup
		\renewcommand{\arraystretch}{1.8}
		\begin{table}[h!]
			\centering
			\setlength{\tabcolsep}{3mm}
			\begin{tabular}{cc!{\color{white}\vrule width 2pt}ccccc}
				\rowcolor{ufabc!50} 
				\cellcolor{ufabc!50} $n$ & \cellcolor{ufabc!50} \bf State & \bf $M_\text{exp}$ (MeV) & \multicolumn{1}{c}{\bf $M_\text{I}$ (MeV)} & \multicolumn{1}{c}{\bf $M_\text{II}$ (MeV)} & \multicolumn{1}{c}{\bf $M_\text{III}$ (MeV)} & \multicolumn{1}{c}{\bf $M_\text{IV}$ (MeV)} \\
				\rowcolor{ufabc!30} 1 & $f_0(500)$ & $550^{+250}_{-180}$ & $231 \pm 212$ & $488 \pm 222$ & $539 \pm 245$ & $609 \pm 277$ \\
				\rowcolor{ufabc!10} 2 & $f_0(980)$ & $990 \pm 20$ & $1106 \pm 22$ & $1000 \pm 20$ & $1043 \pm 21$ & $1040 \pm 21$ \\
				\rowcolor{ufabc!30} 3 & $f_0(1370)$ & $1370^{+218}_{-156}$ & $1395 \pm 222$ & $1267 \pm 202$ & $1366 \pm 217$ & $1357 \pm 216$ \\
				\rowcolor{ufabc!10} 4 & $f_0(1500)$ & $1522 \pm 25$ & $1632 \pm 27$ & $1591 \pm 26$ & $1620 \pm 27$ & $1608 \pm 26$ \\
				\rowcolor{ufabc!30} 5 & $f_0(1770)$ & $1784 \pm 16$ & $1846 \pm 17$ & $1848 \pm 17$ & $1837 \pm 16$ & $1824 \pm 16$ \\
				\rowcolor{ufabc!10} 6 & $f_0(2020)$ & $1982 \pm 3^{+54}_{-0}$ & $2039 \pm 59$ & $2063 \pm 59$ & $2030 \pm 58$ & $2016 \pm 58$ \\
				\rowcolor{ufabc!30} 7* & $f_0(2100)$ & $2095 \pm 19$ & $2215 \pm 20$ & $2249 \pm 20$ & $2206 \pm 20$ & $2192 \pm 20$ \\
				\rowcolor{ufabc!10} 8* & $f_0(2330)$ & $2330 \pm 153$ & $2376 \pm 156$ & $2410 \pm 158$ & $2367 \pm 155$ & $2355 \pm 155$ \\
				\rowcolor{ufabc!30} 9* & $f_0(2470)$ & $2470 \pm 4^{+4}_{-6}$ & $2520 \pm 8$ & $2544 \pm 8$ & $2516 \pm 8$ & $2504 \pm 8$
			\end{tabular}
			\caption[Experimental and predicted mass spectra for $f_0$ scalar mesons]{Experimental and predicted mass spectra for $f_0$ scalar mesons. Meson states marked with the "*" are omitted from the summary table in the PDG.}
			\label{table:spectra_scalar}
		\end{table}
		\endgroup

	\begingroup
	\renewcommand{\arraystretch}{1.8}
	\begin{table}[h!]
		\centering
		\setlength{\tabcolsep}{3mm}
		\begin{tabular}{cc!{\color{white}\vrule width 2pt}ccccc}
			\rowcolor{ufabc!50} 
			\cellcolor{ufabc!50} $n$ & \cellcolor{ufabc!50} \bf State & \bf $M_\text{exp}$ (MeV) & \multicolumn{1}{c}{\bf $M_\text{I}$ (MeV)} & \multicolumn{1}{c}{\bf $M_\text{II}$ (MeV)} & \multicolumn{1}{c}{\bf $M_\text{III}$ (MeV)} & \multicolumn{1}{c}{\bf $M_\text{IV}$ (MeV)} \\
			\rowcolor{ufabc!30} 1 & $\rho(770)$ & $775.26 \pm 0.23$ & $771 \pm 0.23$ & $729 \pm 0.23$ & $737 \pm 0.23$ & $741 \pm 0.23$ \\
			\rowcolor{ufabc!10} 2 & $\rho'(1450)$ & $1350 \pm 20^{+20}_{-30}$ & $1143 \pm 34$ & $1135 \pm 34$ & $1136 \pm 34$ & $1137 \pm 34$ \\
			\rowcolor{ufabc!30} 3 & $\rho(1450)$ & $1465 \pm 25$ & $1431 \pm 25$ & $1423 \pm 25$ & $1425 \pm 25$ & $1426 \pm 25$ \\
			\rowcolor{ufabc!10} 4 & $\rho(1700)$ & $1720 \pm 20$ & $1670 \pm 20$ & $1663 \pm 20$ & $1665 \pm 20$ & $1666 \pm 20$ \\
			\rowcolor{ufabc!30} 5* & $\rho(1900)$ & $1900^{+51}_{-60}$ & $1878 \pm 51$ & $1873 \pm 51$ & $1874 \pm 51$ & $1875 \pm 51$ \\
			\rowcolor{ufabc!10} 6* & $\rho(2150)$ & $2150 \pm 172$ & $2065 \pm 172$ & $2061 \pm 172$ & $2062 \pm 172$ & $2062 \pm 172$ \\
			\rowcolor{ufabc!30} 7* & $\rho(2270)$ & $2270 \pm 60$ & $2237 \pm 60$ & $2234 \pm 60$ & $2234 \pm 60$ & $2235 \pm 60$
		\end{tabular}
		\caption[Experimental and predicted mass spectra for $\rho$ vector mesons]{Experimental and predicted mass spectra for $\rho$ vector mesons. Meson states marked with the "*" are omitted from the summary table in the PDG.}
		\label{table:spectra_vector}
	\end{table}
	\endgroup
	
	\pagebreak
	
\begin{figure}[H]
		\centering
		\includegraphics[width=11cm]{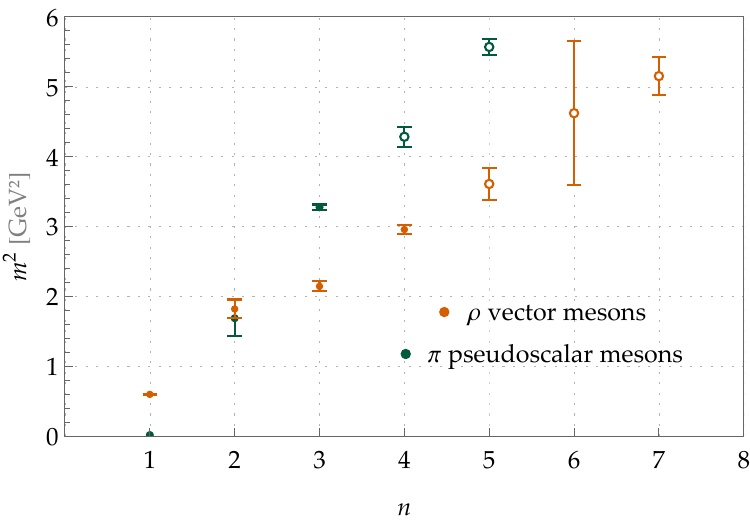}
			\caption[Mass spectra for the $\uppi$ and $\rho$ mesons from experimentally obtained masses as a function of the radial quantum number $n$]{Mass spectra for the $\uppi$ and $\rho$ mesons from experimentally obtained masses as a function of the radial quantum number $n$.	Meson states represented by open circles are omitted from the summary table in PDG.}
	\label{fig:masses1}
	\end{figure}

\begin{figure}[H]
				\centering
				\includegraphics[width=11cm]{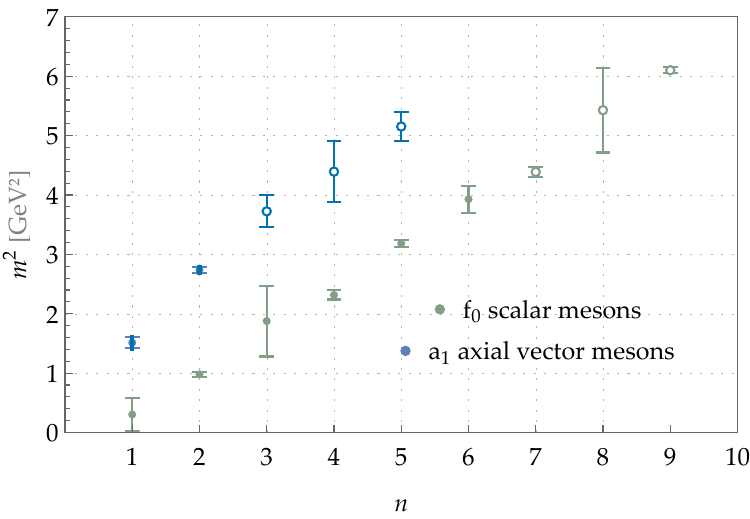}
				\caption[Mass spectra for the $a_1$ and $f_0$ mesons from experimentally obtained masses as a function of the radial quantum number $n$]{Mass spectra for the $a_1$ and $f_0$ mesons from experimentally obtained masses as a function of the radial quantum number $n$. States represented by open circles are omitted from the summary table in PDG.}
			\label{fig:masses2}
		\end{figure}

	The mass spectra obtained by this method are also presented in \Cref{fig:masses1,fig:masses2}, where a linear pattern in the Regge trajectories can be observed.
	
	In the following subsections, the DCE will be used to quantify the information of these mass spectra, with which expressions will be constructed to estimate the heavier resonances beyond those recorded in the PDG.

		\subsection{Pseudoscalar mesons}
			
		Using the protocol described in \Cref{section:cimadsqcd}, the DCE was calculated for the family of $\uppi$ mesons across all four cases of anomalous mass correction. The results for each resonance are presented in \autoref{table:DCE_pion}.
				
		\begingroup
		\renewcommand{\arraystretch}{1.8}
		\begin{table}[h!]
			\centering
			\setlength{\tabcolsep}{3mm}
			\begin{tabular}{cc!{\color{white}\vrule  width 2pt}cccc}
				\rowcolor{ufabc!50} 
				\cellcolor{ufabc!50} & \cellcolor{ufabc!50} & \multicolumn{4}{c}{\cellcolor{ufabc!50}\textbf{DCE (nat)}} \\ 
				\rowcolor{ufabc!50} 
				\rowcolor{ufabc!50} \multirow{-2}{*}{$n$} & \multirow{-2}{*}{\cellcolor{ufabc!50} \bf State} & \bf Case I & \multicolumn{1}{c}{\bf Case II} & \multicolumn{1}{c}{\bf Case III} & \multicolumn{1}{c}{\bf Case IV} \\
				\rowcolor{ufabc!30} 1 & $\pi^\pm$ & 4.5837 & 4.8583 & 5.1290 & 4.6988 \\ 
				\rowcolor{ufabc!10} 2 & \;$\pi(1300)$\; & 6.5613 & 6.8907 & 7.0809 & 6.6787 \\ 
				\rowcolor{ufabc!30} 3 & $\pi(1800)$ & 8.6321 & 8.9870 & 9.2464 & 8.8342 \\ 
				\rowcolor{ufabc!10} 4 & $\pi(2070)$ & 10.1594 & 10.7347 & 10.9556 & 10.4175 \\ 
				\rowcolor{ufabc!30} 5 & $\pi(2360)$ &\; 11.9119 \; &\; 12.4429 \;&\; 12.8191 \;&\; 12.1823
			\end{tabular}
			\caption{DCE of the radial resonances of the $\uppi$ meson family.}
			\label{table:DCE_pion}
		\end{table}
		\endgroup
		
\begin{figure}[H]
			\centering
			\includegraphics[width=11cm]{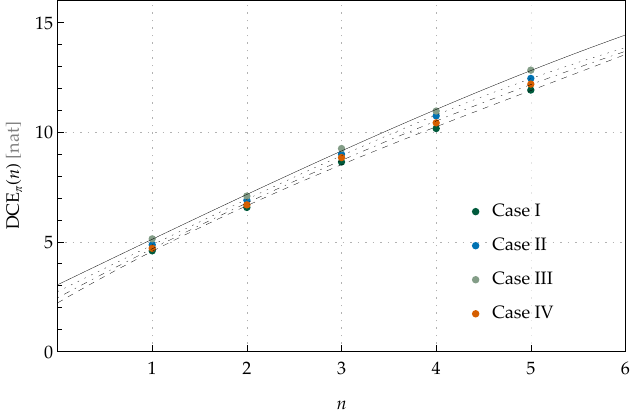}
			\caption[DCE of the $\uppi$ resonances as a function of the $n$ radial excitation level]{DCE of the $\uppi$ resonances as a function of the $n$ radial excitation level. The first type of DCE-Regge-like trajectories are depicted following Eqs. (\ref{dce_pion_n1}) - (\ref{dce_pion_n4}), and are  respectively plotted as continuous, dotted, dot-dashed, and dashed lines.}
			\label{fig:dce_pion_n}
		\end{figure}

		From these values, two types of Regge-like trajectories can be constructed. The first type is obtained by expressing the DCE as a function of the radial quantum number $n$, leading to the trajectories shown in \autoref{fig:dce_pion_n}. A quadratic polynomial interpolation of these data yields the following expressions:
		\begin{subequations}
			\setlength{\abovedisplayskip}{2pt}
			\setlength{\belowdisplayskip}{2pt}
				\begin{alignat}{4}
					{{\rm DCE}}_{\uppi, \text{I}}(n) &= -7.109 \cdot 10^{-2} n^2 + 2.252 n + 2.395, \label{dce_pion_n1} \\
					{{\rm DCE}}_{\uppi, \text{II}}(n) &= -4.523 \cdot 10^{-2} n^2 + 2.197 n + 2.953, \label{dce_pion_n2} \\
					{{\rm DCE}}_{\uppi, \text{III}}(n) &= -7.161 \cdot 10^{-2} n^2 + 2.300 n + 2.449, \label{dce_pion_n3} \\
					{{\rm DCE}}_{\uppi, \text{IV}}(n) &= -7.121 \cdot 10^{-2} n^2 + 2.329 n + 2.580, \label{dce_pion_n4},
				\end{alignat}
				\label{dce_pion_n}
			\end{subequations}
		where the quadratic interpolation keeps the root mean square deviation (RMSD) within $10^{-3}$.
		
\begin{figure}[H]
				\centering
				\includegraphics[width=11cm]{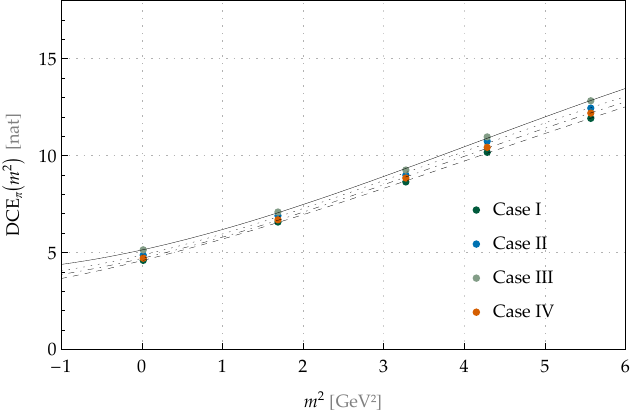}
				\caption[DCE of the $\uppi$ family as a function of their squared mass]{DCE of the $\uppi$ family as a function of their squared mass, for  $n=1,\dots,5$.	The second type of DCE-Regge-like trajectories is depicted following Eqs. (\ref{dce_pion_m1}) - (\ref{dce_pion_m4}). They are respectively plotted as continuous, dotted, dot-dashed, and dashed lines.}
				\label{fig:dce_pion_m}
			\end{figure}

		The second type of Regge-like trajectory is constructed by expressing the DCE values as a function of the experimental meson masses cataloged in the PDG and presented in \autoref{table:spectra_pseudoscalar}. The results are shown in \autoref{fig:dce_pion_m}, and the interpolation of the data produced the following expressions for the trajectories:
			\begin{subequations}
				\setlength{\abovedisplayskip}{2pt}
				\setlength{\belowdisplayskip}{2pt}
				\begin{alignat}{4}
					{{\rm DCE}}_{\uppi, \text{I}}(m^2) &= 3.049 \cdot 10^{-2} m^4 + 1.159 m^2 + 4.5449, \label{dce_pion_m1} \\
					{{\rm DCE}}_{\uppi, \text{II}}(m^2) &= 3.343 \cdot 10^{-2} m^4 + 1.197 m^2 + 4.8094, \label{dce_pion_m2} \\
					{{\rm DCE}}_{\uppi, \text{III}}(m^2) &= 4.796 \cdot 10^{-2} m^4 + 1.133 m^2 + 5.0798, \label{dce_pion_m3} \\
					{{\rm DCE}}_{\uppi, \text{IV}}(m^2) &= 3.236 \cdot 10^{-2} m^4 + 1.182 m^2 + 4.6479. \label{dce_pion_m4}
				\end{alignat}
				\label{dce_pion_m}
			\end{subequations}
			within $0.023\%$ RMSD.
			
		The following steps proceed as presented in \autoref{section:cimadsqcd}. To estimate the masses of resonances for $n>5$, the extrapolated values of $n$ are first inserted into Eqs. (\ref{dce_pion_n}), obtaining the corresponding DCE value for each new resonance. Then, these DCE values are inserted into the left-hand side of Eqs. (\ref{dce_pion_m}), which, when solved for $m$, provide an estimated mass as the solution.
		
		The first extrapolated state of the $\uppi$ mesons is $n=6$, which will be referred to as $\uppi^\star_6$. To estimate its mass, the quantum number $n=6$ is inserted into (\ref{dce_pion_n}), resulting in DCE values of $13.3487$ nat, $13.9882$ nat, $14.5061$ nat, and $13.6728$ nat for cases I, II, III, and IV, respectively. Then, by substituting these DCE values into the left-hand side of Eq. (\ref{dce_pion_m}), the estimated mass values for each case are obtained.
		
		The same procedure can be applied to obtain the state $\uppi^\star_7$ by inserting $n=7$ into Eqs. (\ref{dce_pion_n}), resulting in DCE values of $14.6766$ nat, $15.3910$ nat, $16.115$ nat, and $15.0421$ nat. For the state $\uppi^\star_8$, the same is done with $n=8$, resulting in $17.0573$ nat, $16.6514$ nat, $17.6335$ nat, and $16.2682$ nat for cases I-IV. Inserting the DCE values into Eqs. (\ref{dce_pion_m}) and solving for $m$, the estimated masses are obtained again. All these results are summarized in \autoref{table:pion_masses}.
		
		\begingroup
		\renewcommand{\arraystretch}{1.2}
		\begin{table}[H]
			\centering
			\setlength{\tabcolsep}{1.2mm}
			\begin{tabular}{cc!{\color{white}\vrule width 2pt}ccccc}
				\rowcolor{ufabc!50} 
				\cellcolor{ufabc!50} $n$ & \cellcolor{ufabc!50} \bf State & \bf $M_\text{exp}$ (MeV) & \multicolumn{1}{c}{\bf $M_\text{I}$ (MeV)} & \multicolumn{1}{c}{\bf $M_\text{II}$ (MeV)} & \multicolumn{1}{c}{\bf $M_\text{III}$ (MeV)} & \multicolumn{1}{c}{\bf $M_\text{IV}$ (MeV)} \\
				\rowcolor{ufabc!30} 1 & $\uppi^\pm$ & \small $139.57039 \pm 0.00017$ & \small $140 \pm 0.000178$ & \small $146 \pm 0.000192$ & \small $158 \pm 0.000173$ & \small $109 \pm 0.000171$ \\
				\rowcolor{ufabc!10} 2 & $\uppi(1300)$ & $1300 \pm 100$ & $1600 \pm 123$ & $1408 \pm 108$ & $1410 \pm 108$ & $1415 \pm 109$ \\
				\rowcolor{ufabc!30} 3 & $\uppi(1800)$ & $1810^{+9}_{-11}$ & $1897\pm12$ & $1873\pm 11$ & $1821\pm11$ & $1841\pm11$ \\
				\rowcolor{ufabc!10} 4* & $\uppi(2070)$ & $2070 \pm 35$ & $2116 \pm 36$ & $2123 \pm 36$ & $2070 \pm 35$ & $2096 \pm 36$ \\
				\rowcolor{ufabc!30} 5* & $\uppi(2360)$ & $2360 \pm 25$ & $2299 \pm 24$ & $2313 \pm 25$ & $2264 \pm 24$ & $2289 \pm 24$ \\
				\rowcolor{ufabc!10} 6$^\star$ & $\uppi^\star_6$ & --- & $2547 \pm 56$ & $2547 \pm 56$ & $2553 \pm 56$ & $2546 \pm 56$ \\
				\rowcolor{ufabc!30} 7$^\star$ & $\uppi^\star_7$ & --- & $2706 \pm 59$ & $2708 \pm 59$ & $2772 \pm 61$ & $2706 \pm 59$ \\
				\rowcolor{ufabc!10} 8$^\star$ & $\uppi^\star_8$ & --- & $2838 \pm 62$ & $2841 \pm 62$ & $2867 \pm 63$ & $2838 \pm 62$ 
			\end{tabular}
			\caption[Extended mass spectrum of the $\uppi$ pseudoscalar meson family]{Extended mass spectrum of the $\uppi$ pseudoscalar meson family, including higher $n$ resonances. The masses for $n=6, 7, 8$ indicated with the "$\,^\star$" denote the values extrapolated by the concomitant use of the DCE-Regge-like trajectories (\ref{dce_pion_n}, \ref{dce_pion_m}). Error propagation is included in the model predictions.}
			\label{table:pion_masses}
		\end{table}
		\endgroup
		
		Considering the data available in the PDG, the $\uppi^\star_6$ and $\uppi^\star_8$ resonances have not found any possible matches. Only the $\uppi^\star_7$ resonance, except for case III, found a potential match with the further meson state $X(2680)$ in PDG \cite{ParticleDataGroup:2024cfk}. The predicted mass for this resonance in cases I, II, and IV are, respectively, $2706.92 \pm 59$ MeV, $2708.76 \pm 59$ MeV, and $2706.80 \pm 59$ MeV, which fall within the detected measured mass $2676\pm 27$ MeV for the $X(2680)$ state.
		
		\subsection{Axial vector mesons}
		
		The same DCE protocol applied to the pseudoscalar mesons was also applied to the axial-vector mesons. The configurational entropy values obtained are presented in \autoref{table:DCE_a1} for the five detected resonances of the $a_1$ meson.
		
			\begingroup
			\renewcommand{\arraystretch}{1.2}
			\begin{table}[H]
				\centering
				\setlength{\tabcolsep}{3mm}
				\begin{tabular}{cc!{\color{white}\vrule width 2pt}cccc}
					\rowcolor{ufabc!50} 
					\cellcolor{ufabc!50} & \cellcolor{ufabc!50} & \multicolumn{4}{c}{\cellcolor{ufabc!50}\textbf{DCE (nat)}} \\ 
					\rowcolor{ufabc!50} 
					\rowcolor{ufabc!50} \multirow{-2}{*}{$n$} & \multirow{-2}{*}{\cellcolor{ufabc!50} \bf State} & \bf Case I & \multicolumn{1}{c}{\bf Case II} & \multicolumn{1}{c}{\bf Case III} & \multicolumn{1}{c}{\bf Case IV} \\
					\rowcolor{ufabc!30} 1 & $a_1(1260)$ & 5.8237 & 6.4485 & 6.8135 & 6.0852 \\ 
					\rowcolor{ufabc!10} 2 & $a_1(1640)$ & 7.3888 & 8.2860 & 8.6997 & 7.7314 \\ 
					\rowcolor{ufabc!30} 3 & $a_1(1930)$ & 9.2468 & 10.1272 & 10.6694 & 9.4873 \\ 
					\rowcolor{ufabc!10} 4 & $a_1(2095)$ & 10.9983 & 12.2035 & 12.8673 & 11.3531 \\ 
					\rowcolor{ufabc!30} 5 & \;$a_1(2270)$\; &\; 12.6515 \;&\; 14.0436 \;&\; 14.8783 \;&\; 13.1214 \;
				\end{tabular}
				\caption{DCE of the radial resonances of the $a_1$ meson family.}
				\label{table:DCE_a1}
			\end{table}
			\endgroup
			
\begin{figure}[H]
				\centering
				\includegraphics[width=11cm]{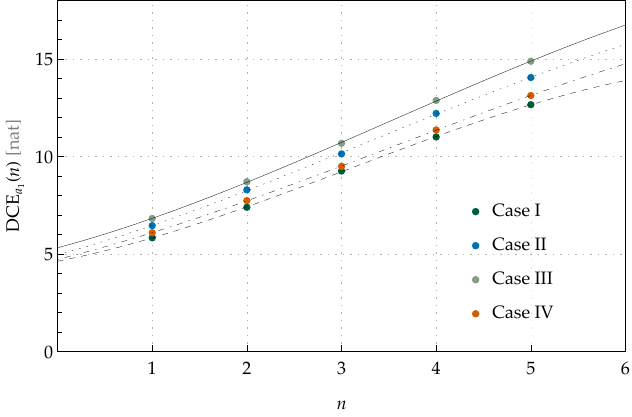}
				\caption[DCE of the $a_1$ resonances as a function of the $n$ radial excitation level]{DCE of the $a_1$ resonances  as a function of the $n$ radial excitation level.	The first form of DCE-Regge-like trajectories is depicted following Eqs. (\ref{dce_a1_n1}) - 
					(\ref{dce_a1_n4}), and is respectively plotted as continuous, dotted, dot-dashed, and dashed lines.}
				\label{fig:dce_a1_n}
			\end{figure}

		By interpolating these values as a function of the quantum number $n$, it is possible to construct the first type of Regge-like trajectories, presented in \autoref{fig:dce_a1_n}. The expressions for these trajectories are given by
			\begin{subequations}
				\setlength{\abovedisplayskip}{2pt}
				\setlength{\belowdisplayskip}{2pt}
				\begin{alignat}{4}
					\text{DCE}_{a_1,\text{I}}(n) &= 4.949 \cdot 10^{-3} n^2 + 1.697 n + 4.077, \label{dce_a1_n1} \\
					\text{DCE}_{a_1,\text{II}}(n) &= 3.414 \cdot 10^{-2} n^2 + 1.825 n + 4.935, \label{dce_a1_n2} \\
					\text{DCE}_{a_1,\text{III}}(n) &= 2.528 \cdot 10^{-2} n^2 + 1.618 n + 4.424, \label{dce_a1_n3} \\
					\text{DCE}_{a_1,\text{IV}}(n) &= 1.718 \cdot 10^{-2} n^2 + 1.808 n + 4.610, \label{dce_a1_n4}
				\end{alignat}
				\label{dce_a1_n}
			\end{subequations}
			within $0.002\%$ RMSD.
			
\begin{figure}[H]
				\centering
				\includegraphics[width=11cm]{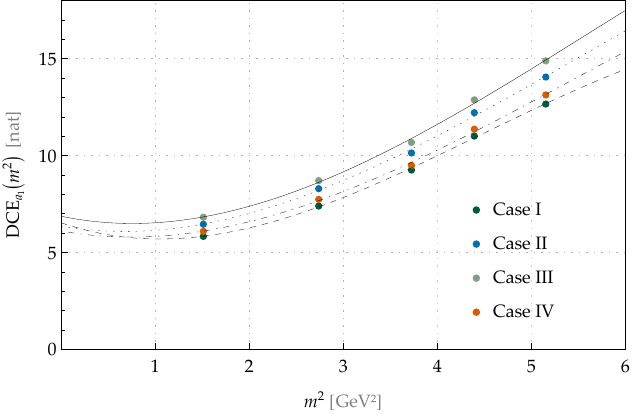}
				\caption[DCE of the $a_1$  family  as a function of their squared mass]{DCE of the $a_1$  family  as a function of their squared mass, for  $n=1,\dots,5$.		The second form of DCE-Regge-like trajectories is depicted following Eqs. (\ref{dce_a1_m1}) - (\ref{dce_a1_m4}), and is respectively plotted as continuous, dotted, dot-dashed, and dashed lines.}
				\label{fig:dce_a1_m}
			\end{figure}

			A similar interpolation can be performed using the DCE with respect to masses. The experimentally obtained mass values listed in \autoref{table:spectra_axial} were used to construct the second type of Regge-like trajectories, which is described by the following equations:
			\begin{subequations}
				\setlength{\abovedisplayskip}{2pt}
				\setlength{\belowdisplayskip}{2pt}
				\begin{alignat}{4}
					\text{DCE}_{a_1,\text{I}}(m^2) &= 2.235 \cdot 10^{-1} m^4 + 4.206 \cdot 10^{-1} m^2 + 4.6398, \label{dce_a1_m1} \\
					\text{DCE}_{a_1,\text{II}}(m^2) &= 2.553 \cdot 10^{-1} m^4 + 4.111 \cdot 10^{-1} m^2 + 5.2311, \label{dce_a1_m2} \\
					\text{DCE}_{a_1,\text{III}}(m^2) &= 2.906 \cdot 10^{-1} m^4 + 3.061 \cdot 10^{-1} m^2 + 5.6727, \label{dce_a1_m3} \\
					\text{DCE}_{a_1,\text{IV}}(m^2) &= 2.483 \cdot 10^{-1} m^4 + 3.011 \cdot 10^{-1} m^2 + 5.0481, \label{dce_a1_m4}
				\end{alignat}
				\label{dce_a1_m}
			\end{subequations}
			within $0.038\%$ RMSD, and presented in \autoref{fig:dce_a1_m}.
			
			The first heavier resonance in the $a_1$ family corresponds to $n = 6$ and is denoted as $(a_1)_6^\star$. Substituting this quantum number into Eqs. (\ref{dce_a1_n}), the DCE values obtained for cases I, II, III, and IV are $14.4359$ nat, $16.0743$ nat, $17.1138$ nat, and $15.0409$ nat, respectively. By inserting these values into the left-hand side of Eqs. (\ref{dce_a1_m}), we obtain algebraic equations that can be solved to determine the corresponding mass spectrum values, $m$.  
			
			The same procedure is applied for $n = 7$, corresponding to the state $(a_1)_7^\star$. The DCE values obtained from Eqs. (\ref{dce_a1_n}) for the four studied cases are $16.1970$ nat, $18.1054$ nat, $19.3825$ nat, and $16.9873$ nat, respectively. These values enable the estimation of the mass for $(a_1)_7^\star$.  
			
			Finally, for $n = 8$, following the same procedure, the DCE values obtained are $17.9680$ nat, $20.1708$ nat, $21.7195$ nat, and $18.9842$ nat. The resulting mass values for $(a_1)_8^\star$, along with those for $(a_1)_6^\star$ and $(a_1)_7^\star$, are summarized in \autoref{table:a1_extended}.  
			
			\begingroup
			\renewcommand{\arraystretch}{1.2}
			\begin{table}[H]
				\centering
				\setlength{\tabcolsep}{1.2mm}
				\begin{tabular}{cc!{\color{white}\vrule width 2pt}ccccc}
					\rowcolor{ufabc!50} 
					\cellcolor{ufabc!50} $n$ & \cellcolor{ufabc!50} \bf State & \bf $M_\text{exp}$ (MeV) & \multicolumn{1}{c}{\bf $M_\text{I}$ (MeV)} & \multicolumn{1}{c}{\bf $M_\text{II}$ (MeV)} & \multicolumn{1}{c}{\bf $M_\text{III}$ (MeV)} & \multicolumn{1}{c}{\bf $M_\text{IV}$ (MeV)} \\
					\rowcolor{ufabc!30} 1 & $a_1(1260)$ & $1230 \pm 40$ & $1316 \pm 43$ & $1222 \pm 40$ & $1163 \pm 40$ & $1232 \pm 40$ \\
					\rowcolor{ufabc!10} 2 & $a_1(1640)$ & $1655 \pm 16$ & $1735 \pm 17$ & $1676 \pm 16$ & $1649 \pm 16$ & $1707 \pm 17$ \\
					\rowcolor{ufabc!30} 3* & $a_1(1930)$ & $1930\pm71$ & $1969\pm71$ & $1971\pm71$ & $1926\pm71$ & $1980\pm72$ \\
					\rowcolor{ufabc!10} 4* & $a_1(2095)$ & $2096 \pm 41 \pm 81$ & $2163 \pm 126$ & $2178 \pm 127$ & $2132 \pm 124$ & $2181 \pm 127$ \\
					\rowcolor{ufabc!30} 5* & $a_1(2270)$ & $2270\pm55$ & $2336\pm57$ & $2356\pm57$ & $2311\pm56$ & $2358\pm57$ \\
					\rowcolor{ufabc!10} 6$^\star$ & $({a_1})^\star_6$ & --- & $2397 \pm 70$ & $2400 \pm 71$ & $2402 \pm 71$ & $2401 \pm 71$ \\
					\rowcolor{ufabc!30} 7$^\star$ & $({a_1})^\star_7$ & --- & $2512 \pm 74$ & $2518 \pm 74$ & $2522 \pm 74$ & $2520 \pm 74$ \\
					\rowcolor{ufabc!10} 8$^\star$ & $({a_1})^\star_8$ & --- & $2614 \pm 77$ & $2624 \pm 77$ & $2631 \pm 77$ & $2628 \pm 77$
				\end{tabular}
				\caption[Extended mass spectrum of the $a_1$ axial vector meson family]{Extended mass spectrum of the $a_1$ axial vector meson family, including higher $n$ resonances. The extrapolated masses for $n=6, 7, 8$, denoted by "$^\star$", were obtained through DCE-Regge-like trajectories (\ref{dce_a1_n}, \ref{dce_a1_m}), interpolating the experimental masses for $n=1,\dots, 5$. Error propagation is included in the model predictions.}
				\label{table:a1_extended}
			\end{table}
			\endgroup
			
			According to the data from PDG \cite{ParticleDataGroup:2024cfk}, the axial vector meson resonance $({a_1})^\star_6$ exhibits mass spectrum compatibility, across all four cases of the anomalous dimension, with the higher meson state $X(2340)$, which has an experimentally measured mass of $2340 \pm 20$ MeV.  
			The resonance $({a_1})^\star_7$ does not correspond to any higher meson states cataloged in the PDG.  
			Finally, for the resonance $({a_1})^\star_8$, potential matches are found with the higher states $X(2600)$, $X(2632)$, and $X(2680)$ across all four cases. These states have masses of $2618.3 \pm 2.0_{-1.4}^{+16.3}$ MeV, $2631.6 \pm 2.1$ MeV, and $2676 \pm 27$ MeV, respectively, encompassing all the predicted mass values in cases I, II, III, and IV.  

			\subsection{Scalar mesons}
			
			The third particle group analyzed consists of the $f_0$ scalar meson family. The DCE values obtained from applying the protocol are listed in \autoref{table:DCE_f0}.					
			\begingroup
			\renewcommand{\arraystretch}{1.8}
			\begin{table}[h!]
				\centering
				\setlength{\tabcolsep}{3mm}
				\begin{tabular}{cc!{\color{white}\vrule width 2pt}cccc}
					\rowcolor{ufabc!50} 
					\cellcolor{ufabc!50} & \cellcolor{ufabc!50} & \multicolumn{4}{c}{\cellcolor{ufabc!50}\textbf{DCE (nat)}} \\ 
					\rowcolor{ufabc!50} 
					\rowcolor{ufabc!50} \multirow{-2}{*}{$n$} & \multirow{-2}{*}{\cellcolor{ufabc!50} \bf State} & \bf Case I & \multicolumn{1}{c}{\bf Case II} & \multicolumn{1}{c}{\bf Case III} & \multicolumn{1}{c}{\bf Case IV} \\
					\rowcolor{ufabc!30} 1 & $f_0(500)$ & 1.1851 & 1.6410 & 1.8817 & 1.3796 \\ 
					\rowcolor{ufabc!10} 2 & $f_0(980)$ & 3.5514 & 3.8472 & 4.3234 & 3.7181 \\ 
					\rowcolor{ufabc!30} 3 & $f_0(1370)$ & 5.4049 & 5.7932 & 6.5052 & 5.6311 \\ 
					\rowcolor{ufabc!10} 4 & $f_0(1500)$ & 7.1802 & 7.7611 & 8.7170 & 7.5041 \\ 
					\rowcolor{ufabc!30} 5 & $f_0(1770)$ & 8.8281 & 9.4346 & 10.6263 & 9.0468 \\ 
					\rowcolor{ufabc!10} 6 & $f_0(2020)$ & 10.2014 & 10.9801 & 12.4143 & 10.7022 \\ 
					\rowcolor{ufabc!30} 7 & $f_0(2100)$ & 11.6711 & 12.5994 & 14.2711 & 12.2893 \\ 
					\rowcolor{ufabc!10} 8 & $f_0(2330)$ & 13.1114 & 14.0281 & 15.9391 & 13.6493 \\ 
					\rowcolor{ufabc!30} 9 & $f_0(2470)$ & 14.4653 & 15.5772 & 17.7300 & 15.1908 \\ 
				\end{tabular}
				\caption{DCE of the radial resonances of the $f_0$ meson family.}
				\label{table:DCE_f0}
			\end{table}
			\endgroup
			
\begin{figure}[H]
				\centering
				\includegraphics[width=11cm]{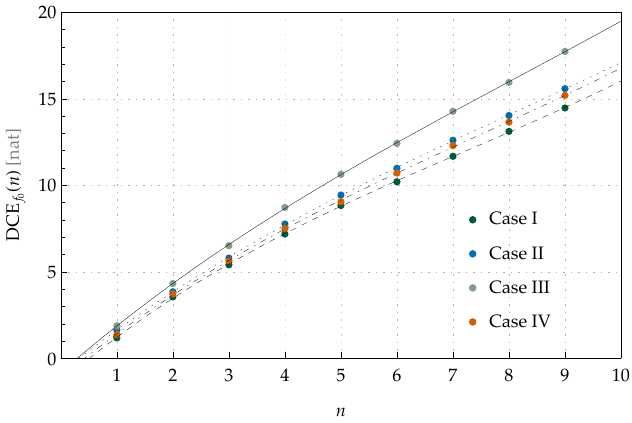}
				\caption[DCE of the $f_0$ resonances as a function of the $n$ radial excitation level]{DCE of the $f_0$ resonances as a function of the $n$ radial excitation level.	The first form of DCE-Regge-like trajectories is depicted following Eqs. (\ref{dce_f0_n1}) -  (\ref{dce_f0_n4}), and is respectively plotted as continuous, dotted, dot-dashed, and dashed lines.}
				\label{fig:dce_f0_n}
			\end{figure}

			The first type of Regge-like trajectories is obtained by interpolating the DCE values, which depend on the radial excitation level $n$ of the resonances, using the data presented in \autoref{table:spectra_scalar}. These results are shown in \autoref{fig:dce_f0_n}, and the curves representing the interpolations are given by the following equations:			
			\begin{subequations}
				\setlength{\abovedisplayskip}{2pt}
				\setlength{\belowdisplayskip}{2pt}
				\begin{alignat}{4}
					\text{DCE}_{f_0,\text{I}}(n) &= 8.7296 \cdot 10^{-3} n^3 - 1.8928 \cdot 10^{-1} n^2 + 2.7579 n - 1.3609, \label{dce_f0_n1} \\
					\text{DCE}_{f_0,\text{II}}(n) &= 5.3675 \cdot 10^{-3} n^3 - 1.3152 \cdot 10^{-1} n^2 + 2.5684 n - 0.8109, \label{dce_f0_n2} \\
					\text{DCE}_{f_0,\text{III}}(n) &= 5.3554 \cdot 10^{-3} n^3 - 1.3131 \cdot 10^{-1} n^2 + 2.8062 n - 0.8090, \label{dce_f0_n3} \\
					\text{DCE}_{f_0,\text{IV}}(n) &= 7.2202 \cdot 10^{-3} n^3 - 1.6056 \cdot 10^{-1} n^2 + 2.6715 n - 1.1077, \label{dce_f0_n4}
				\end{alignat}
				\label{dce_f0_n}
			\end{subequations}
			within $0.001\%$ RMSD.
			
\begin{figure}[H]
				\centering
				\includegraphics[width=11cm]{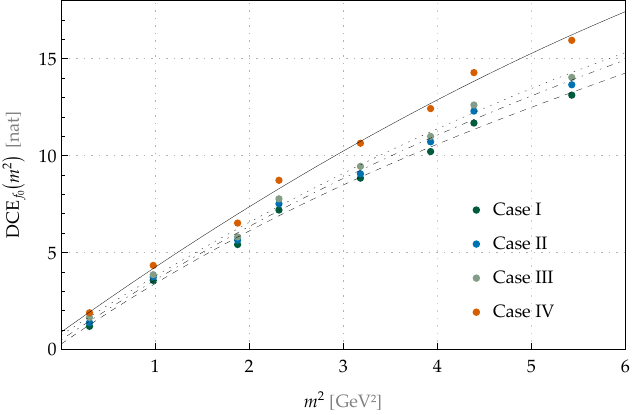}
				\caption[DCE of the $f_0$  family  as a function of their squared mass]{DCE of the $f_0$  family  as a function of their squared mass, for  $n=1,\dots,9$.			The second form of DCE-Regge-like trajectories is depicted following Eqs. (\ref{dce_f0_m1})  - (\ref{dce_f0_m4}), and is respectively plotted as continuous, dotted, dot-dashed, and dashed lines.}
				\label{fig:dce_f0_m}
			\end{figure}

			The DCE values can also be expressed as a function of the experimental mass of each resonance, as presented in \autoref{table:spectra_scalar}. These values are plotted in \autoref{fig:dce_f0_m} alongside the mass values for the four cases of the anomalous dimension studied. The functions interpolating these points, which form the second type of Regge-like trajectories, are defined by		
			\begin{subequations}
				\setlength{\abovedisplayskip}{2pt}
				\setlength{\belowdisplayskip}{2pt}
				\begin{alignat}{4}
					\text{DCE}_{f_0,\text{I}}(m^2) &= 1.1377 \cdot 10^{-2} m^6 - 2.3773 \cdot 10^{-1} m^4 + 3.3474 m^2 + 0.2658, \label{dce_f0_m1} \\
					\text{DCE}_{f_0,\text{II}}(m^2) &= -2.3917 \cdot 10^{-4} m^6 - 1.1738 \cdot 10^{-1} m^4 + 3.4705 m^2 + 0.8764, \label{dce_f0_m2} \\
					\text{DCE}_{f_0,\text{III}}(m^2) &= 6.6807 \cdot 10^{-3} m^6 - 1.8313 \cdot 10^{-1} m^4 + 3.2675 m^2 + 0.4780, \label{dce_f0_m3} \\
					\text{DCE}_{f_0,\text{IV}}(m^2) &= 1.7533 \cdot 10^{-3} m^6 - 1.3350 \cdot 10^{-1} m^4 + 3.1689 m^2 + 0.7245. \label{dce_f0_m4}
				\end{alignat}
				\label{dce_f0_m}
			\end{subequations}
			The RMSD are 0.72\% for Eq. (\ref{dce_f0_m1}), 0.38\% for Eq. (\ref{dce_f0_m2}), 0.31\% for Eq. (\ref{dce_f0_m3}), and 0.28\% for Eq. (\ref{dce_f0_m4}). 
			
			Eqs. (\ref{dce_f0_n}, \ref{dce_f0_m}) are used to estimate the $f_0$ states for $n>9$. By substituting the desired $n$ value into the DCE expressions (\ref{dce_f0_n}), the corresponding configurational entropy value for each case is obtained. This value is then inserted into the mass relation (\ref{dce_f0_m}) to solve for the mass $m$.
			
			For $n=10$, corresponding to the state $({f_0})^\star_{10}$, Eqs. (\ref{dce_f0_n}) provide the following DCE values for cases I, II, III, and IV: $16.0197$ nat, $17.0886$ nat, $19.4774$ nat, and $16.7715$ nat. These values are then substituted into (\ref{dce_f0_m}), and solving the resulting algebraic equations yields the mass solutions, which are summarized in \autoref{table:f0_extended}.
			
			The same method is applied for $n=11$, corresponding to the state $({f_0})^\star_{11}$. In this case, the DCE values are $17.6922$ nat, $18.6717$ nat, $21.2987$ nat, and $18.4611$ nat, respectively for cases I-IV. By inserting these values into Eqs. (\ref{dce_f0_m}) and solving for $m$, the mass solutions are obtained and summarized in \autoref{table:f0_extended}.
			
			\begingroup
			\renewcommand{\arraystretch}{1.8}
			\begin{table}[h!]
				\centering
				\setlength{\tabcolsep}{1.2mm}
				\begin{tabular}{cc!{\color{white}\vrule width 2pt}ccccc}
					\rowcolor{ufabc!50} 
					\cellcolor{ufabc!50} $n$ & \cellcolor{ufabc!50} \bf State & \bf $M_\text{exp}$ (MeV) & \multicolumn{1}{c}{\bf $M_\text{I}$ (MeV)} & \multicolumn{1}{c}{\bf $M_\text{II}$ (MeV)} & \multicolumn{1}{c}{\bf $M_\text{III}$ (MeV)} & \multicolumn{1}{c}{\bf $M_\text{IV}$ (MeV)} \\
					\rowcolor{ufabc!30} 1 & $f_0(500)$ & $550^{+250}_{-180}$ & $231\pm 212$ & $488\pm222$ & $539\pm245$ & $609\pm277$ \\
					\rowcolor{ufabc!10} 2 & $f_0(980)$ & $990 \pm 20$ & $1106 \pm 22$ & $1000\pm 20$ & $1043 \pm 21$ & $1040 \pm 21$ \\
					\rowcolor{ufabc!30} 3 & $f_0(1370)$ & $1370^{+218}_{-156}$ & $1395\pm222$ & $1267\pm202$ & $1366\pm217$ & $1357\pm216$ \\
					\rowcolor{ufabc!10} 4 & $f_0(1500)$ & $1522 \pm 25$ & $1632 \pm 27$ & $1591 \pm 26$ & $1620 \pm 27$ & $1608 \pm 26$ \\
					\rowcolor{ufabc!30} 5 & $f_0(1770)$ & $1784\pm16$ & $1846\pm17$ & $1848\pm17$ & $1837\pm16$ & $1824\pm16$ \\
					\rowcolor{ufabc!10} 6 & $f_0(2020)$ & $1982 \pm 3^{+54}_{-0}$ & $2039\pm59$ & $2063\pm59$ & $2030\pm58$ & $2016\pm58$ \\
					\rowcolor{ufabc!30} 7* & $f_0(2100)$ & $2095\pm19$ & $2215\pm20$ & $2249\pm20$ & $2206\pm20$ & $2192\pm20$ \\
					\rowcolor{ufabc!10} 8* & $f_0(2330)$ & $2330\pm153$ & $2376\pm156$ & $2410\pm158$ & $2367\pm155$ & $2355\pm155$ \\
					\rowcolor{ufabc!30} 9* & $f_0(2470)$ & $2470 \pm 4^{+4}_{-6}$ & $2520\pm8$ & $2544\pm8$ & $2516\pm8$ & $2504\pm8$ \\
					\rowcolor{ufabc!10} 10$^\star$ & $({f_0})^\star_{10}$ & --- & $2653 \pm 192$ & $2660 \pm 192$ & $2661 \pm 192$ & $2657 \pm 192$ \\
					\rowcolor{ufabc!30} 11$^\star$ & $({f_0})^\star_{11}$ & --- & $2832 \pm 204$ & $2860 \pm 206$ & $2862 \pm 206$ & $2844 \pm 205$
				\end{tabular}
				\caption[Extended mass spectrum of the $f_0$ scalar meson family]{Extended mass spectrum of the $f_0$ scalar meson family, including higher $n$ resonances. The extrapolated masses for $n=10, 11$, denoted by "$^\star$", were obtained through DCE-Regge-like trajectories (\ref{dce_f0_n}, \ref{dce_f0_m}), interpolating the experimental masses for $n=1,\dots,9$. Error propagation is included in the model predictions.}
				\label{table:f0_extended}
			\end{table}
			\endgroup
			
			The estimated masses for the $f_0^\star$ scalar meson resonances exhibit a significant margin of error, which permits a broader range of potential matches. The predicted mass values for the $({f_0})^\star_{10}$ resonance encompass all of the experimentally observed mass ranges for the further meson states $X(2600)$, $X(2632)$, and $X(2680)$, which are $2618.3 \pm 2.0^{+16.3}{-1.4}$ MeV, $2631.6 \pm 2.1$ MeV, and $2676 \pm 27$ MeV, respectively.
			Regarding the $({f_0})^\star_{11}$ resonance, while it also displays a wide range in the predicted masses, it only matches the detected mass of the further meson state $X(2680)$ across all four cases of the anomalous dimension. A slight exception is found in case I, where with a mass of $2832 \pm 204$ MeV, it also fits within the correspondence range for the $X(2632)$ state.
			
			\subsection{Vector mesons}
			
			Finally, the $\rho$ vector meson family is analyzed using the DCE protocol. The obtained DCE values are summarized in \autoref{table:DCE_rho}.

			\begingroup
			\renewcommand{\arraystretch}{1.8}
			\begin{table}[h!]
				\centering
				\setlength{\tabcolsep}{3mm}
				\begin{tabular}{cc!{\color{white}\vrule width 2pt}cccc}
					\rowcolor{ufabc!50} 
					\cellcolor{ufabc!50} & \cellcolor{ufabc!50} & \multicolumn{4}{c}{\cellcolor{ufabc!50}\textbf{DCE (nat)}} \\ 
					\rowcolor{ufabc!50} 
					\rowcolor{ufabc!50} \multirow{-2}{*}{$n$} & \multirow{-2}{*}{\cellcolor{ufabc!50} \bf State} & \bf Case I & \multicolumn{1}{c}{\bf Case II} & \multicolumn{1}{c}{\bf Case III} & \multicolumn{1}{c}{\bf Case IV} \\
					\rowcolor{ufabc!30} 1 & $\rho(770)$ & 3.4904 & 3.9357 & 4.4100 & 3.6840 \\ 
					\rowcolor{ufabc!10} 2 & $\rho'(1450)$ & 5.7290 & 6.4184 & 6.8178 & 6.0210 \\ 
					\rowcolor{ufabc!30} 3 & $\rho(1450)$ & 7.6540 & 8.4414 & 8.7933 & 8.0035 \\ 
					\rowcolor{ufabc!10} 4 & $\rho(1700)$ & 10.3740 & 10.9784 & 11.3980 & 10.8100 \\ 
					\rowcolor{ufabc!30} 5 & $\rho(1900)$ & 12.0950 & 12.7642 & 13.4660 & 12.3370 \\ 
					\rowcolor{ufabc!10} 6 & $\rho(2150)$ & 13.5550 & 14.7020 & 15.3020 & 14.0280 \\ 
					\rowcolor{ufabc!30} 7 & $\rho(2270)$ & 15.0450 & 16.2040 & 16.8760 & 15.4860 \\ 
				\end{tabular}
				\caption{DCE of the radial resonances of the $\rho$ meson family.}
				\label{table:DCE_rho}
			\end{table}
			\endgroup
			
\begin{figure}[H]
				\centering
				\includegraphics[width=11cm]{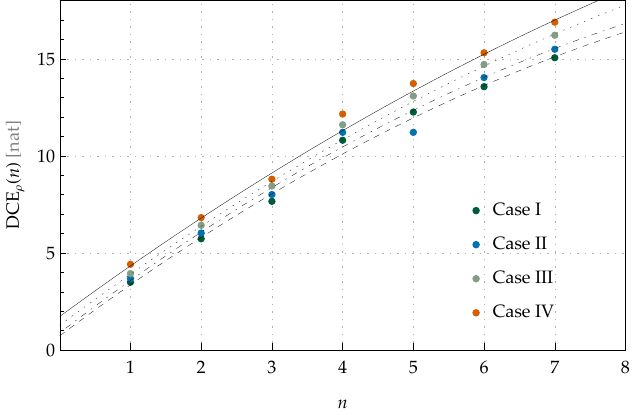}
				\caption[DCE of the $\rho$ resonances as a function of the $n$ radial excitation level]{DCE of the $\rho$ resonances as a function of the $n$ radial excitation level. 	The first form of DCE-Regge-like trajectories is depicted following Eqs. (\ref{dce_rho_n1}) - (\ref{dce_rho_n4}), and is respectively plotted as continuous, dotted, dot-dashed, and dashed lines.}
				\label{fig:dce_rho_n}
			\end{figure}

			The DCE values as a function of the radial quantum number $n$ are shown in \autoref{fig:dce_rho_n}. By interpolating these values, the first type of DCE-Regge-like trajectories is obtained, as described by
			\begin{subequations}
				\setlength{\abovedisplayskip}{2pt}
				\setlength{\belowdisplayskip}{2pt}
				\begin{alignat}{4}
					\text{DCE}_{\rho,\text{I}}(n) &= -0.0960 n^2 + 2.7240 n + 0.7306, \label{dce_rho_n1} \\
					\text{DCE}_{\rho,\text{II}}(n) &= -0.0813 n^2 + 2.7113 n + 1.2736, \label{dce_rho_n2} \\
					\text{DCE}_{\rho,\text{III}}(n) &= -0.0707 n^2 + 2.6743 n + 1.7269, \label{dce_rho_n3} \\
					\text{DCE}_{\rho,\text{IV}}(n) &= -0.1004 n^2 + 2.7939 n + 0.8848, \label{dce_rho_n4}
				\end{alignat}
				\label{dce_rho_n}
			\end{subequations}
			within $0.002\%$ RMSD.
			
\begin{figure}[H]
				\centering
				\includegraphics[width=11cm]{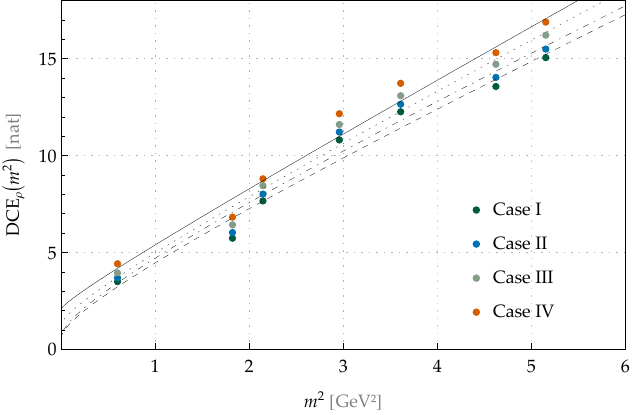}
				\caption[DCE of the $\rho$  family  as a function of their squared mass]{DCE of the $\rho$  family  as a function of their squared mass, for  $n=1,\dots,7$.		The second form of DCE-Regge-like trajectories is depicted following Eqs. (\ref{dce_rho_m1}) - (\ref{dce_rho_m4}), and is respectively plotted as continuous, dotted, dot-dashed, and dashed lines.}
				\label{fig:dce_rho_m}
			\end{figure}

			The DCE values can also be expressed as a function of the experimental mass for each resonance, as shown in \autoref{table:spectra_vector}. These DCE values are plotted alongside the corresponding mass values for the four cases considered in \autoref{fig:dce_rho_m}. The functions that interpolate these data points, forming the second type of DCE-Regge-like trajectories, are given by
			\begin{subequations}
				\setlength{\abovedisplayskip}{2pt}
				\setlength{\belowdisplayskip}{2pt}
				\begin{alignat}{4}
					\text{DCE}_{\rho,\text{I}}(m^2) &= -1.6132 \cdot 10^{-1} m^6 + 1.2135 m^4 + 3.3776 \cdot 10^{-1} m^2 + 2.7424, \label{dce_rho_m1} \\
					\text{DCE}_{\rho,\text{II}}(m^2) &= -1.6831 \cdot 10^{-1} m^6 + 1.2868 m^4 + 3.2691 \cdot 10^{-1} m^2 + 3.6457, \label{dce_rho_m2} \\
					\text{DCE}_{\rho,\text{III}}(m^2) &= -1.6161 \cdot 10^{-1} m^6 + 1.2037 m^4 + 4.5905 \cdot 10^{-1} m^2 + 2.8653, \label{dce_rho_m3} \\
					\text{DCE}_{\rho,\text{IV}}(m^2) &= -1.4731 \cdot 10^{-1} m^6 + 1.0957 m^4 + 7.6060 \cdot 10^{-1} m^2 + 2.9784. \label{dce_rho_m4}
				\end{alignat}
				\label{dce_rho_m}
			\end{subequations}
			within $1.52\%$ RMSD. 
			
			To estimate the $\rho$ meson states for $n>7$, Eqs. (\ref{dce_rho_n}, \ref{dce_rho_m}) are applied. This process is similar to previous ones, where the value of $n$ is inserted into the DCE expressions in (\ref{dce_rho_n}). The obtained DCE values are then used in the left-hand side of Eqs. (\ref{dce_rho_m}), which are solved to derive the corresponding mass values.
			
			For the resonance ${\rho}^\star_8$, substituting $n=8$ into Eqs. (\ref{dce_rho_n}) provides the DCE values for each of the four cases. These values are $16.1674$ nat, $17.4496$ nat, $18.2374$ nat, and $16.5997$ nat for cases I, II, III, and IV, respectively. Using these values and applying Eqs. (\ref{dce_rho_m}), the mass of this resonance is determined for the four cases.
			
			A similar procedure is followed for $n=9$, leading to the DCE values for the resonance ${\rho}^\star_9$ in the four cases: $17.0310$ nat, $18.4236$ nat, $19.3162$ nat, and $17.4322$ nat. The resulting mass values are presented in \autoref{table:rho_extended}, alongside the masses of ${\rho}^\star_8$.
			
			\begingroup
			\renewcommand{\arraystretch}{1.2	}
			\begin{table}[H]
				\centering
				\setlength{\tabcolsep}{1.2mm}
				\begin{tabular}{cc!{\color{white}\vrule width 2pt}ccccc}
					\rowcolor{ufabc!50} 
					\cellcolor{ufabc!50} $n$ & \cellcolor{ufabc!50} \bf State & \bf $M_\text{exp}$ (MeV) & \multicolumn{1}{c}{\bf $M_\text{I}$ (MeV)} & \multicolumn{1}{c}{\bf $M_\text{II}$ (MeV)} & \multicolumn{1}{c}{\bf $M_\text{III}$ (MeV)} & \multicolumn{1}{c}{\bf $M_\text{IV}$ (MeV)} \\
					\rowcolor{ufabc!30} 1 & $\rho(770)$ & $775.26 \pm 0.23$ & $771 \pm 0.23$ & $729 \pm 0.22$ & $737 \pm 0.23$ & $741 \pm 0.22$ \\
					\rowcolor{ufabc!10} 2 & $\rho'(1450)$ & $1350 \pm 20^{+20}_{-30}$ & $1143\pm34$ & $1135\pm34$ & $1136\pm34$ & $1137\pm34$ \\
					\rowcolor{ufabc!30} 3 & $\rho(1450)$ & $1465 \pm 25$ & $1431 \pm 25$ & $1423 \pm 25$ & $1425 \pm 25$ & $1426 \pm 25$ \\
					\rowcolor{ufabc!10} 4 & $\rho(1700)$ & $1720 \pm 20$ & $1670 \pm 20$ & $1663 \pm 20$ & $1665 \pm 20$ & $1666 \pm 20$ \\
					\rowcolor{ufabc!30} 5* & $\rho(1900)$ & $1900^{+51}_{-60}$ & $1878\pm51$ & $1873\pm 51$ & $1874\pm 51$ & $1875\pm 51$ \\
					\rowcolor{ufabc!10} 6* & $\rho(2150)$ & $2150\pm172$ & $2065\pm 172$ & $2061\pm 172$ & $2062\pm 172$ & $2062\pm 172$ \\
					\rowcolor{ufabc!30} 7* & $\rho(2270)$ & $2270\pm 60$ & $2237\pm 60$ & $2234\pm 60$ & $2234\pm 60$ & $2235\pm 60$ \\
					\rowcolor{ufabc!10} 8$^\star$ & $\rho^\star_8$ & --- & $2315 \pm 67$ & $2347 \pm 68$ & $2336 \pm 68$ & $2314 \pm 67$ \\
					\rowcolor{ufabc!30} 9$^\star$ & $\rho^\star_9$ & --- & $2341 \pm 68$ & $2378 \pm 69$ & $2368 \pm 68$ & $2339 \pm 68$ \\
				\end{tabular}
				\caption[Extended mass spectrum of the $\rho$ vector meson family]{Extended mass spectrum of the $\rho$ vector meson family, including higher $n$ resonances. The extrapolated masses for $n=8,9$, denoted by "$^\star$", were obtained through DCE-Regge-like trajectories (\ref{dce_rho_n}, \ref{dce_rho_m}), interpolating the experimental masses for $n=1,\dots,7$. Error propagation is included in the model predictions.}
				\label{table:rho_extended}
			\end{table}
			\endgroup
			
			The mass values for the ${\rho}^\star_8$ and ${\rho}^\star_9$ resonances were very close, leading to all representatives of both states falling within the error range of the same additional mesonic state in the PDG, $X(2340)$. This state has an experimentally measured mass of $2340 \pm 20$ MeV, which falls within the predicted ranges for the four cases of both the ${\rho}^\star_8$ and ${\rho}^\star_9$ states, as shown in \autoref{table:rho_extended}.
			
			\subsection{Concluding remarks}
			
			This section was devoted to the DCE applied to the dynamical soft-wall model modified by an anomalous mass correction, with the aim of investigating four meson families.
			By computing the DCE and constructing the Regge-like trajectories, it was possible to estimate the masses of higher excited states within these families. The predicted masses were subsequently compared with experimentally detected but unidentified meson states listed in the Particle Data Group to identify potential correspondences.
			The search for possible matches among further states of these long-studied particles, with a high number of detections, revealed some margin for the existence of new resonances.
						
			For the pseudoscalar mesons, the estimated mass for the $n=6$ state, denoted as $\pi_6^\star$, found a possible match with the $X(2680)$ state in three of the four scenarios (cases I, II, and IV). 
			
			In the axial-vector meson sector, the $(a_1)_6^\star$ resonance showed a unique correspondence with the $X(2340)$ state across all four anomalous dimension cases, representing the only instance of a single, unambiguous match in this study. The higher excitation, $(a_1)_8^\star$, corresponded with the $X(2600)$, $X(2632)$, and $X(2680)$ states in all four cases. 
			
			For the scalar $f_0$ mesons, the predicted mass for the $(f_0)_{10}^\star$ state also fell within the range of the $X(2600)$, $X(2632)$, and $X(2680)$ states for all scenarios. The subsequent $(f_0)_{11}^\star$ state was matched with $X(2632)$ and $X(2680)$ in case I, and exclusively with $X(2680)$ in the other three cases. 
			
			Finally, for the vector $\rho$ mesons, the predicted masses for both the $\rho_8^\star$ and $\rho_9^\star$ resonances were found to be very close, with both states pointing to a possible correspondence with the single further meson state $X(2340)$ in all scenarios. 
			
			As expected, it was observed that the DCE increases monotonically with the radial quantum number $n$ for all meson families and scenarios, indicating that configurational instability grows with the excitation level. This finding reinforces the hypothesis that the increasing instability is reflected in the greater experimental difficulty of detecting heavier resonances.

			These results were published in	\cite{daRocha:2024sjn}. 
			
			\pagebreak
		
		\section{CIMs of the strange vector kaon family}
		\label{section:strangekaons}
		
			The strange vector kaons belong to the spin-1 meson nonet, which also includes the $\rho$ meson studied in the previous section.
			They are composed of a strange quark or antiquark and a lighter up or down antiquark or quark. This composition gives rise to four distinct types: the charged $K^*(892)^+=(u\bar{s})$ and $K^*(892)^-=(s\bar{u})$, and the neutral $K^*(892)^0=(d\bar{s})$ and its antiparticle $\bar{K}^*(892)^0=(s\bar{d})$ \cite{greiner1996gauge}.			
			
			Unlike the more stable pseudoscalar kaons, $K^{*}$ mesons are resonances with a very short lifetime, decaying rapidly via the strong interaction into a pion and a kaon ($K^* \rightarrow \pi + K$). Their detection is performed by reconstructing these decay products. However, since pions and pseudoscalar kaons continue to interact, either through absorption or rescattering, the measurement of vector kaons becomes a rather challenging task, especially at the LHC energies, which ranges from 8 to 13 TeV \cite{Ilner:2016xqr}. In addition, vector kaons are also produced in heavy-ion collisions at relativistic energies, where they serve as valuable probes for studying the quark-gluon plasma \cite{Rafelski:2001hp,Markert:2002rw,ALICE:2014jbq,ALICE:2021xyh}.
					
			Together with the $K^*(892)$, which correspond to the ground state with $n=1$, the family of strange vector kaons is completed by two additional experimentally detected resonances, $K^*(1410)$ and $K^*(1680)$, corresponding to $n=2$ and $n=3$, respectively.
			These resonances have been experimentally observed in the decays of strange $B$ mesons. Specifically, the neutral $K^*(1410)^0$ and $K^*(1680)^0$ states are produced in $B^0_s$ decays alongside charmonia; the former appears with both $J/\psi$ and $\psi(2S)$, while the latter has only been reported with $J/\psi$ thus far \cite{Li:2019pzx}. Furthermore, charged resonances like $K^*(1410)^\pm$ can be produced from the decay of either $B^0_s$ or $B^0$ mesons, although the corresponding production of the $K^*(1680)^\pm$ is noted to be a much rarer event \cite{Li:2018psm}.\\
			
			The vector kaons can be described by the vector sector of the SW model, resulting in a relatively simple action:			
			\begin{flalign}
				\mathcal{S}_{K^*} = \frac{1}{4g_5^2} \int \sqrt{-g} e^{-\Phi}F^{MN}F_{MN}.
				\label{eq:kaonaction}
			\end{flalign}
			Following the same procedure presented in \Cref{subsection:softwall}, we can reduce the equation of motion to the Schrödinger-like form
			\begin{flalign}
				\left[-\partial_z^2 + U(z)\right]\xi_n = m_n^2\xi_n,
				\label{eq:schrodinger-like_kaon}
			\end{flalign}
			where $\xi_n\equiv\xi_n(z)$ is the function that represents the kaons. The effective potential has the general form given by
			\begin{flalign}
				U(z) = -\frac{B''}{2} + \frac{(B')^2}{4},
				\label{eq:general_potential}
			\end{flalign}
			with $B(z)=\Phi(z)-A(z)$.
				
			Before proceeding to the CIM-based AdS/QCD hybrid method, it is interesting to have a glimpse of the results for the mass spectrum obtained by the solution of Eq. (\ref{eq:schrodinger-like_kaon}), within the simplest SW model with $\Phi(z)=(\kappa z)^2$ and $A(z)=-\ln(z)$.			
			
			\autoref{table:mass_neutral} presents the mass values for the neutral strange vector kaons, comparing the masses detected experimentally \cite{ParticleDataGroup:2024cfk} with the results predicted by the AdS/QCD model, as well as the relative error (RE) between them.
			\begingroup
			\renewcommand{\arraystretch}{1.2}
			\begin{table}[H]
				\centering
				\setlength{\tabcolsep}{3mm}
				\begin{tabular}{ccccc}
					\rowcolor{ufabc!50}
					\bf $n$					& \bf State 	&  $M_\text{exp}$(MeV)	& $M_\text{AdS/QCD}$ (MeV) 	&	RE (\%)  \\
					\rowcolor{ufabc!30}	1 	& $K^*(892)$ 	& $895.55 \pm 0.2$ 		& $1038.4 \pm 0.3 $					& 13.8	\\
					\rowcolor{ufabc!10}	2	& $K^*(1410)$ 	& $1414 \pm 15$ 			& $1451 \pm 15 $					& 2.6	\\
					\rowcolor{ufabc!30}	3	& $K^*(1680)$ 	& $1718 \pm 18$ 			& $1755 \pm 18 $					& 2.1
				\end{tabular}
				\caption{Mass spectrum of neutral strange vector kaons resonances.}
				\label{table:mass_neutral}	
			\end{table}
			\endgroup
			
\begin{figure}[H]
				\centering
				\includegraphics[width=11cm]{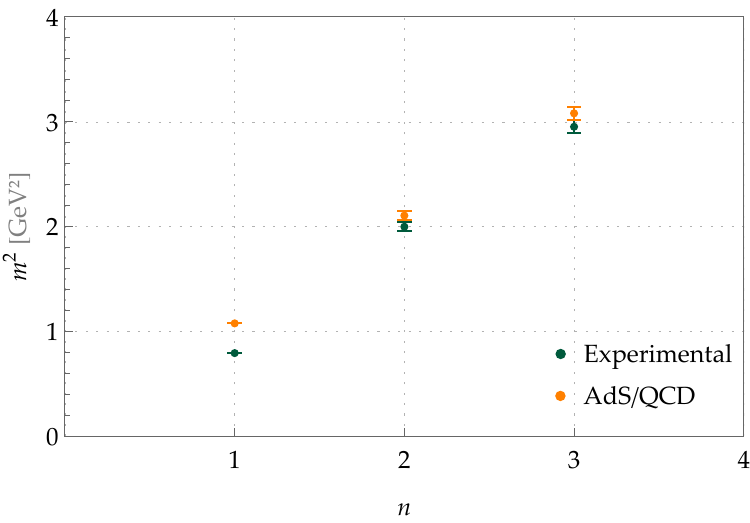}
				\caption[Mass spectrum of neutral strange vector kaons radial resonances]{Mass spectrum of neutral strange vector kaons radial resonances: experimental values and AdS/QCD prediction, for $n=1,2,3$.}
				\label{fig:masses_neutral}
			\end{figure}

			By interpolating this data, it is possible to write the linear trajectory that relates the squared mass to the level $n$ as
			\begin{flalign}
				m_n^2 = 1.0747 n - 0.2319,
			\end{flalign}
			which recreates the mass spectrum for the neutral strange vector kaons with a 5.8\% RMSD, proving to be a precise expression to reproduce the experimental values. \autoref{fig:masses_neutral} presents the plot of these results with a linear Regge trajectory.	
			
			For the charged strange vector kaons, \autoref{table:mass_charged} compiles the experimentally obtained mass values and the results predicted by the AdS/QCD model, with their respective RE.
			
			\begingroup
			\renewcommand{\arraystretch}{1.2}
			\begin{table}[H]
				\centering
				\setlength{\tabcolsep}{3mm}
				\begin{tabular}{ccccc}
					\rowcolor{ufabc!50}
					\bf $n$					& State 		&  $M_\text{exp}$(MeV)	& $M_\text{AdS/QCD}$ (MeV) 	& RE (\%)  \\
					\rowcolor{ufabc!30}	1 	& $K^*(892)$ 	& $891.67 \pm 0.26$ & $1035.2 \pm 0.3$ 				& 13.9	\\
					\rowcolor{ufabc!10}	2	& $K^*(1410)$ 	& $1414 \pm 15$ 	& $1447 \pm 15$ 					& 2.3	\\
					\rowcolor{ufabc!30}	3	& $K^*(1680)$ 	& $1718 \pm 18$ 	& $1750 \pm 18$ 					& 1.8
				\end{tabular}
				\caption{Mass spectrum of charged strange vector kaons resonances.}
				\label{table:mass_charged}	
			\end{table}
			\endgroup
			
\begin{figure}[H]
				\centering
				\includegraphics[width=11cm]{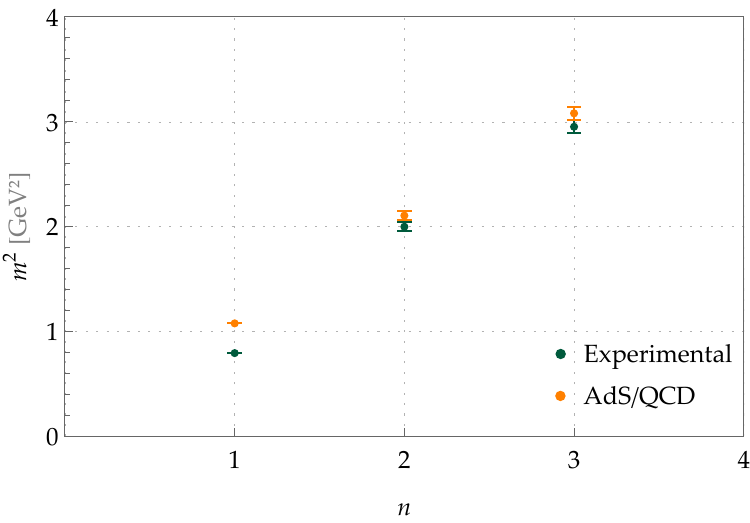}
				\caption[Mass spectrum of charged strange vector kaons radial resonances]{Mass spectrum of charged strange vector kaons radial resonances: experimental values and AdS/QCD prediction, for $n=1,2,3$.}
				\label{fig:masses_charged}
			\end{figure}

			The linear Regge trajectory that corresponds to the mass spectrum is presented in the \autoref{fig:masses_charged}. 
			And the expression for this trajectory is given by
			\begin{flalign}
				m_n^2 = 1.0782 n - 0.2411,
			\end{flalign}
			which reproduces the experimental mass spectrum within a 5.7\% RMSD, proving to be a very precise approximation for the results.\\

			Although remarkably accurate in organizing light-flavor hadrons, Regge trajectories become much less reliable once heavy quarks, or even the strange quark, are involved in the composition of the particles. In such cases, the trajectories acquire a non-linear profile, significantly different from what was originally proposed 		\cite{Chen:2018nnr,Chen:2018bbr,Chen:2018hnx,Gershtein:2006ng}.
			From this observation, it is reasonable to infer that the linearity of Regge trajectories depends on the constituent mass of the quarks forming the hadrons.
			
			Indeed, AdS/QCD models are not constructed with the explicit quark content of hadrons in mind. An immediate consequence of this limitation can be seen in the results obtained from the soft-wall model discussed earlier. The holographic QCD approach to meson spectroscopy relies on two parameters: the bulk mass, which encodes information about the dual operators at the boundary, and an energy scale -- represented by the parameter $\kappa$ in the SW dilaton profile -- that determines the slope of Regge trajectories. Consequently, when writing, for instance, Eq. (\ref{eq:pseudoscalar_spectrum}) for the pseudoscalar meson mass spectrum, one cannot distinguish whether it refers to a pion, a kaon, or any other member of the pseudoscalar nonet.
			
			Since the vector kaons studied in this section contain strange quarks in their composition, it is important to analyze their spectrum from this perspective. Although the strange quark is not heavy compared to charm and bottom, for example, its constituent mass is still considerably larger than that of the up and down quarks, which justifies a change of approach. Ref. \cite{MartinContreras:2020cyg} proposes a method to incorporate the constituent quark mass into the description of non-linear Regge trajectories, and we adopt this approach as the foundation for the AdS/QCD model to which the DCE and DCC will be applied.

			A meson, understood as a bound quark-antiquark pair, has the average constituent quark mass defined as
			\begin{flalign}
				\bar{m}(q_1,q_2) = \frac{1}{2}(m_{q_1}+m_{q_2}),
			\end{flalign}
			with the constituent quark mass values found in the \autoref{table:quarks}.
			
			We will proceed with the soft-wall model using (\ref{softwallmetric}) as the metric and a warp factor given by $A(z)=-\ln(z)$.
			Since the linear patterns were obtained with the dilaton in its usual form, $\Phi(z)=(\kappa z)^2$, a deviation from this format can introduce a non-linear pattern in the trajectories. In Ref. \cite{MartinContreras:2020cyg}, the proposed deviation is caused by a shift in the quadratic exponent:
			\begin{flalign}
				\Phi(z) = (\kappa z)^{2-\alpha},
				\label{shifted_dilaton}
			\end{flalign}
			and the dependence on the constituent quark mass, as suggested in the initial hypothesis, is made through the parameters $\kappa$ and $\alpha$, defined as	
			\begin{flalign}
				\alpha(\bar{m})&= 0.8454 - 0.8485 e^{-0.4233\bar{m}^2}, \\
				\kappa(\bar{m})&= 15.2085 - 14.8082 e^{-0.0524\bar{m}^2},
			\end{flalign}
				where the average constituent quark mass $\bar{m}$ is in GeV. With the dilaton given by (\ref{shifted_dilaton}) and inserted into the Eq. (\ref{eq:general_potential}), the potential takes the form of
			\begin{flalign}
				U(\kappa,\alpha,z) = \left(1-\frac{\alpha}{2}\right)\frac{\kappa}{z}(\kappa z)^{1-\alpha}
				+ (\alpha-2)^2\frac{\kappa^2}{4}(\kappa z)^{2-2\alpha}
				+ \frac{3}{4z^2}
				+\left(\frac{3\alpha}{2}-\frac{\alpha^2}{2}-1\right)\kappa^2(\kappa z)^{-\alpha}.
			\end{flalign}
			Looking at the dilaton and the potential, it is possible to notice that in the limit $\alpha\rightarrow0$ we recover the usual model for light-flavor mesons.

			The calculation of the CIMs requires the energy density $T_{00}$, which can be obtained from the Lagrangian density of the action (\ref{eq:kaonaction}) with the new dilaton, namely,  
			\begin{equation}
				\mathcal{L} = \frac{1}{4} e^{-(\kappa z)^{2-\alpha}} F^{MN}F_{MN}
			\end{equation}
			leading to the following expression:  
			\begin{equation}
				T_{00} = \frac{1}{4g_5^2} e^{-(\kappa z)^{2-\alpha}}g_{00}F^{MN}F_{MN} - F^{0M}F_{0M}.
			\end{equation}
			
			In order to simplify the evaluation of this term of the energy-momentum tensor, the axial gauge $V_z=0$ is adopted, and the remaining components of the vector field are assumed to follow a plane-wave solution, $V_\mu = \epsilon_\mu e^{-imt}\xi_n$,
			where $\epsilon_\mu=(0,1,0,0)^\mathsf{T}$ denotes a transverse polarization vector.  
			
			Under these assumptions, the energy density becomes 
			\begin{flalign}
				T_{00} = \frac{z^2e^{-(\kappa z)^{2-\alpha}}}{2g_5^2}\left[-(\partial_z\xi)^2 + m_n^2\xi^2\right].
			\end{flalign}
			
			With this result, the DCE and DCC can be directly evaluated for both charged and neutral strange vector kaons.

			\pagebreak

			\subsection{DCE of neutral strange vector kaons}
				
				In this first part, the DCE will be calculated for the neutral strange vector kaons.				
				
				The first resonance, given by $n=1$, represents the $K^*(892)^0=d\bar{s}$ and its associated antikaon $\bar{K}^*(892)^0=(s\bar{d})$. The resonance $n=2$ represents $K^*(1410)^0$ and $n=3$ represents $K^*(1680)^0$, with their respective antiparticles and the same quark composition.
				The combination of $d$ and $s$ quarks results in an average constituent mass of $\bar{m}=413$ MeV, and consequently, $\alpha=0.056$ and $\kappa=532.06$ MeV.
				
				The procedure for calculating DCE and DCC is the same as that presented in \Cref{section:cimadsqcd}, with the specifications described in the beginning of this section. The values obtained for DCE are presented in \Cref{table:dce_neutral}.
				\begingroup
				\renewcommand{\arraystretch}{1.2}
				\begin{table}[H]
					\centering
					\setlength{\tabcolsep}{3mm}
					\begin{tabular}{ccc}
						\rowcolor{ufabc!50}
						\bf $n$ & \bf State &  {\bf DCE} (nat)	  \\
						\rowcolor{ufabc!30}	1 	& $K^*(892)^0$ 		& $104.79$ 	\\
						\rowcolor{ufabc!10}	2	& $K^*(1410)^0$ 	& $176.05$	\\
						\rowcolor{ufabc!30}	3	& $K^*(1680)^0$ 	& $253.21$
					\end{tabular}
					\caption{DCE of the radial resonances of the neutral $K^*$ meson family}
					\label{table:dce_neutral}	
				\end{table}
				\endgroup
				
\begin{figure}[H]
					\centering
					\includegraphics[width=11cm]{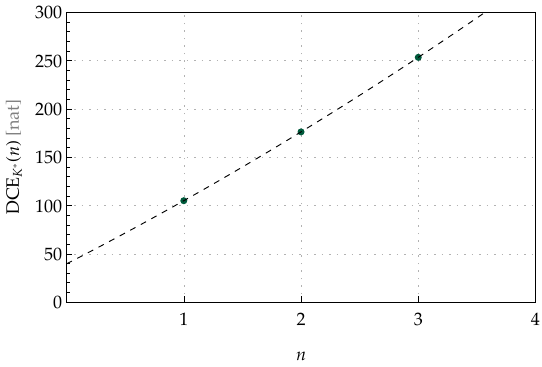}
					\caption{DCE of the neutral $K^*$ family as a function of the $n$ radial excitation level.}
					\label{fig:dce_neutral_n}
				\end{figure}

				The DCE-Regge-like trajectories constructed with these results are presented in \autoref{fig:dce_neutral_n}, with the DCE given as a function of the $n$ radial excitation level. The quadratic polynomial interpolation of these data follows the expression
				\begin{flalign}
					\dce_{K^*}(n) = 2.95 n^2 + 62.41 n + 39.43
					\label{eq_dce_neutral_n}
				\end{flalign}
				within 0.001\% RMSD. From this expression it is possible to infer the value of DCE for any other higher excitation level.
				
\begin{figure}[H]
					\centering
					\includegraphics[width=11cm]{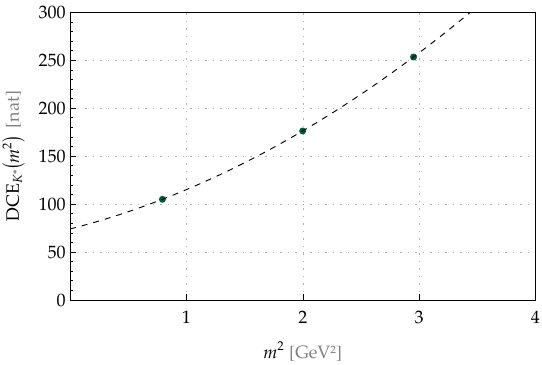}
					\caption[DCE of the neutral $K^*$ family as a function of their squared mass]{DCE of the neutral $K^*$ family as a function of their squared mass, for $n=1,2,3$.}
					\label{fig:dce_neutral_m}
				\end{figure}

				The second type of DCE-Regge-like trajectories can be constructed by plotting the previously obtained DCE values (available in \autoref{table:dce_neutral}) against the experimentally detected mass of each resonance (available in \cite{ParticleDataGroup:2024cfk} and listed in \autoref{table:mass_neutral}). The results are presented in the \autoref{fig:dce_neutral_m}, and a polynomial interpolation of these data is expressed by	
				\begin{flalign}
					\dce_{K^*} (m^2) = 10.1151 m^4 + 30.8854 m^2 + 73.8901,
					\label{eq_dce_neutral_m}
				\end{flalign}
				within 0.001\% RMSD.
				
			Expressions (\ref{eq_dce_neutral_n}) and (\ref{eq_dce_neutral_m}) provide a way to calculate DCE based on both radial resonance level $n$ and kaon mass $m$, the core properties of the neutral strange vector kaon family. This establishes a direct relationship between these two quantities, which is one of CIMs main properties. By extrapolating $n$ values, the mass of further neutral kaon states can be estimated.
			Here, the procedure consists of first inserting the value of $n$ into (\ref{eq_dce_neutral_n}), obtaining the DCE value for the resonance. Then this value is inserted on the left-hand side of (\ref{eq_dce_neutral_m}), and by solving the resulting equation for $m$ the mass value corresponding to that resonance state is obtained.
			
			This approach uses experimental values for the kaon mass and DCE, making it a more realistic approximation than standard AdS/QCD methods that rely on solving Schrödinger-like equations to predict mass spectra. The resulting new states can be compared with cataloged experimental results to identify potential candidates for members of the strange vector kaon family.
			
			In order to obtain the mass of the $K^*_4$ strange vector kaon resonance, the radial number $n=4$ is inserted into Eq. (\ref{eq_dce_neutral_n}), which results in a DCE equal to $329.24$ nat. By applying this value to the left-hand side of (\ref{eq_dce_neutral_m}) and solving the equation, the mass $m_{K^*_4}=1933.81$ MeV is obtained. The same procedure is applied to the following states. For $n=5$, the DCE is equal to $397.14$ nat, which results in a mass for the $K^*_5$ resonance equal to $m_{K^*_5}=2092.71$ MeV. For the $K^*_6$ resonance, with $n=6$, the DCE is $449.87$ nat and the mass obtained from this value is $m_{K^*_6}=2203.43$ MeV.	
			The results obtained, adjusted with the error bar from the experimental data, are summarized in \autoref{table:dce_neutral_preview}.
			
			\begingroup
			\renewcommand{\arraystretch}{1.2}
			\begin{table}[H]
				\centering
				\setlength{\tabcolsep}{3mm}
				\begin{tabular}{cccc}
					\rowcolor{ufabc!50}
					\bf $n$ 				& \bf State 		&  $M_\text{exp}$ (MeV)	& 	$M_\text{th}$ (MeV)	\\
					\rowcolor{ufabc!30}	1 	& $K^*(892)$ 		& $895.55 \pm 0.26 $	& $1038.4 \pm 0.3$				\\
					\rowcolor{ufabc!10}	2	& $K^*(1410)$ 		& $1414 \pm 15$ 		& $1451 \pm 15$				\\
					\rowcolor{ufabc!30}	3	& $K^*(1680)$ 		& $1718 \pm 18$ 		& $1755 \pm 18$				\\
					\rowcolor{ufabc!10}	4	& $(K^*_4)^\star$ 	& ---					& $1934 \pm 19$				\\
					\rowcolor{ufabc!30}	5	& $(K^*_5)^\star$	& --- 					& $2093 \pm 21$				\\
					\rowcolor{ufabc!10}	4	& $(K^*_6)^\star$	& ---					& $2203 \pm 22$
				\end{tabular}
				\caption[Extended mass spectrum of the neutral $K^*$ meson family using DCE]{Extended mass spectrum of the neutral $K^*$ meson family, with theoretical values estimated for $n=4,5,6$. The states with an "$\star$" indicate that their estimated masses were obtained through the extrapolation of DCE-Regge like trajectories that interpolate the experimental values.}
				\label{table:dce_neutral_preview}	
			\end{table}
			\endgroup
			
			From these results, further mesonic states in the PDG \cite{ParticleDataGroup:2024cfk} can be consulted for potential candidates that correspond to the found masses.
			The $K_5^*$ strange vector kaon resonance has a mass predicted by the model using DCE as being $m_{K^*_5}=2093 \pm 21$ MeV, a value that fits the experimental mass of the $X(2075)$ state, which is $2075\pm12\pm5$ MeV, therefore these two might match.	
			Considering the error bar, $K_5^*$ might also match the $X(2080)$ state, which has a mass of $2080\pm10$ MeV.
			In addition, it may also correspond to the $X(2100)$ state, whose mass is $2100\pm40$ MeV.
			As for the $K_6^*$ resonance, with a predicted mass $m_{K^*_6}=2203 \pm 22$ MeV, it might match the $X(2210)$ state, whose experimental mass is $2210^{+79}_{-15}$ MeV.
						
			\subsection{DCC of neutral strange vector kaons}
				
				Complementing the DCE analysis of neutral kaons, this subsection will also present an approach using DCC.
				In the case of neutral strange vector kaons, the construction of the constituent quark model is similar to that done for DCE: $K^*(892)^0=d\bar{s}$ and its antikaon $\bar{K}^*(892)^0=s\bar{d}$ correspond to level $n=1$, while $K^*(1410)^0$ and $K^*(1680)^0$ (with their respective antikaons) correspond to states $n=2$ and $n=3$. The combination of $d$ and $s$ quarks results in an average constituent mass of $\bar{m}=413$ MeV, which yields $\alpha=0.056$ and $\kappa=532.06$ MeV.
								
			The values for DCC obtained through the already known procedure are presented in \Cref{table:dcc_neutral}.
			\begingroup
				\renewcommand{\arraystretch}{1.8}
				\begin{table}[h!]
					\centering
					\setlength{\tabcolsep}{3mm}
					\begin{tabular}{ccc}
						\rowcolor{ufabc!50}
						\bf $n$ & \bf State &  {\bf DCC} 	  \\
						\rowcolor{ufabc!30}	1 	& $K^*(892)$ 	& $279.12$ 	\\
						\rowcolor{ufabc!10}	2	& $K^*(1410)$ 	& $510.92$	\\
						\rowcolor{ufabc!30}	3	& $K^*(1680)$ 	& $770.20$
					\end{tabular}
					\caption{DCC of the radial resonances in the neutral $K^*$ meson family}
					\label{table:dcc_neutral}	
				\end{table}
				\endgroup
				
\begin{figure}[H]
					\centering
					\includegraphics[width=11cm]{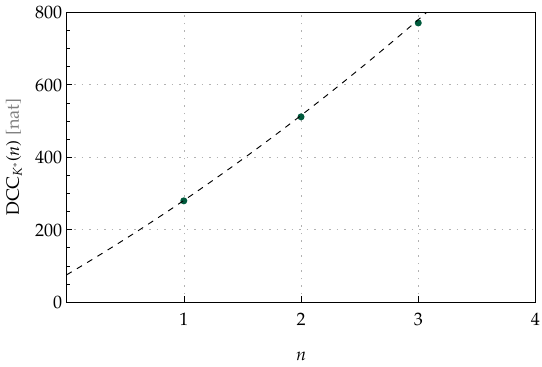}
					\caption{DCC of the neutral $K^*$ family as a function of the $n$ radial excitation level.}
					\label{fig:dcc_neutral_n}
				\end{figure}

				These results are plotted in \autoref{fig:dcc_neutral_n}, where the first type of DCC-Regge-like trajectory is presented, with DCC given as a function of the radial resonance level $n$.
				The polynomial interpolation of this trajectory is given by the expression	
				\begin{flalign}
					\dcc(n)_{K^*} = 14.74n^2 + 190.58n + 74.80,
					\label{eq_dcc_neutral_n}
				\end{flalign}
				within $0.001\%$ RMSD.
				
\begin{figure}[H]
				\centering
				\includegraphics[width=11cm]{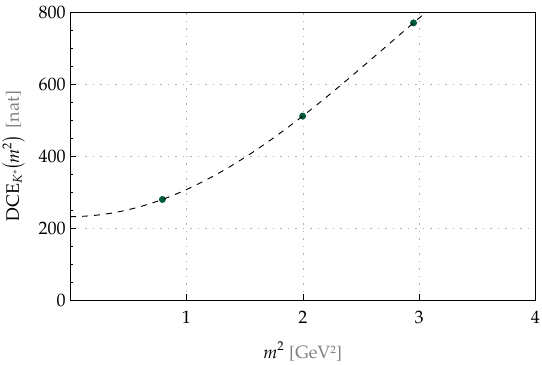}
				\caption[DCC of the neutral $K^*$ family as a function of their squared mass]{DCC of the neutral $K^*$ family as a function of their squared mass, for $n=1,2,3$.}
				\label{fig:dcc_neutral_m}
			\end{figure}

			The second type of DCC-Regge-like trajectory is obtained by plotting the values collected from DCC as a function of the experimentally detected mass spectrum (available in \autoref{table:mass_neutral}). The plot is presented in \autoref{fig:dcc_neutral_m} and the polynomial interpolation that generates this trajectory is given by
			\begin{flalign}
				\dcc_{K^*}(m^2) =  - 1.7222 m^4 + 76.8417 m^2 + 231.3610,
				\label{eq_dcc_neutral_m}
			\end{flalign}
			within $0.001\%$ RMSD.
			
			Following the same procedure as before, the predicted mass of the $K^*_4$ strange vector resonance can be obtained by inserting $n=4$ into (\ref{eq_dcc_charged_n}), obtaining a DCC equal to $1062.16$ nat. Applying this value to the left-hand side of (\ref{eq_dcc_charged_m}) and solving the equation yields a mass $m_{K^*_4}=1937.67$ MeV. For $n=5$, the DCC results in $1350.01$ nat, which leads to a mass for the $K^*_5$ state equal to $m_{K^*_5}=2107.73$ MeV. Finally, for $n=6$, the $K^*_6$ strange vector kaon resonance has a DCC equal to $1624.96$ nat and leads to a mass $m_{K^*_6}=2242.16$ MeV. The results obtained, adjusted with the error bar from the experimental data, are summarized in \autoref{table:dcc_neutral_preview}.
			
			\begingroup
			\renewcommand{\arraystretch}{1.2}
			\begin{table}[H]
				\centering
				\setlength{\tabcolsep}{3mm}
				\begin{tabular}{cccc}
					\rowcolor{ufabc!50}
					\bf $n$ 				& \bf State 		&  $M_\text{exp}$ (MeV)	& 	$M_\text{th}$ (MeV)	\\
					\rowcolor{ufabc!30}	1 	& $K^*(892)$ 		& $895.55 \pm 0.2$ 	& 	$1038.4 \pm 0.3$			\\
					\rowcolor{ufabc!10}	2	& $K^*(1410)$ 		& $1414  \pm 15$ 		& 	$1451\pm15$			\\
					\rowcolor{ufabc!30}	3	& $K^*(1680)$ 		& $1718  \pm 18$		& 	$1755\pm18$			\\
					\rowcolor{ufabc!10}	4	& $(K^*_4)^\star$ 	& ---					& $1938\pm19$		\\
					\rowcolor{ufabc!30}	5	& $(K^*_5)^\star$ 	& --- 					& $2108\pm21$		\\
					\rowcolor{ufabc!10}	4	& $(K^*_6)^\star$ 	& ---					& $2242\pm22$
				\end{tabular}
				\caption[Extended mass spectrum of the neutral $K^*$ meson family using DCC]{Extended mass spectrum of the neutral $K^*$ meson family, with theoretical values estimated for $n=4,5,6$. The states with an "$\star$" indicate that their estimated masses were obtained through the extrapolation of DCC-Regge like trajectories that interpolate the experimental values.}
				\label{table:dcc_neutral_preview}	
			\end{table}
			\endgroup
			
			The $K^*_5$ strange vector kaon resonance, whose mass is predicted by the model as $m_{K^*_5}=2108\pm21$ MeV, might match the $X(2100)$ state in \cite{ParticleDataGroup:2024cfk}, which has the experimental mass given by $2100\pm40$ MeV. And just as obtained for DCE, the $K^*_6$ resonance, with predicted mass $m_{K^*_6}=2242\pm22$, might match the $X(2210)$ further mesonic state, which has an experimental mass $2210^{+79}_{-21}$ MeV.
									
			\subsection{DCE of charged strange vector kaons}
			
			For the charged strange vector kaons, the resonance $n=1$ represents both the $K^*(892)^+=u\bar{s}$ state and its antikaon, $K^*(892)^-=s\bar{u}$. Similarly, the $K^*(1410)^\pm$ and $K^*(1680)^\pm$ states, respectively for $n=2$ and $n=3$, are described using the same constituent quark model. The combination of $u$ and $s$ quarks results in an average constituent mass of $\bar{m}=411$ MeV, which leads to $\alpha=0.0554$ and $\kappa=530.79$ MeV.
			
			Using the same procedure as done for the neutral kaons, the DCE of the charged vector kaons could be calculated and the results are presented in \Cref{table:dce_charged}.
			\begingroup
			\renewcommand{\arraystretch}{1.2}
			\begin{table}[H]
				\centering
				\setlength{\tabcolsep}{3mm}
				\begin{tabular}{ccc}
					\rowcolor{ufabc!50}
					\bf $n$ & \bf State &  {\bf DCE} (nat)	  \\
					\rowcolor{ufabc!30}	1 	& $K^*(892)$ 	& $106.98$ 	\\
					\rowcolor{ufabc!10}	2	& $K^*(1410)$ 	& $179.11$	\\
					\rowcolor{ufabc!30}	3	& $K^*(1680)$ 	& $258.15$
				\end{tabular}
				\caption{DCE of the radial resonances of the charged $K^*$ meson family.}
				\label{table:dce_charged}	
			\end{table}
			\endgroup
			
\begin{figure}[H]
				\centering
				\includegraphics[width=11cm]{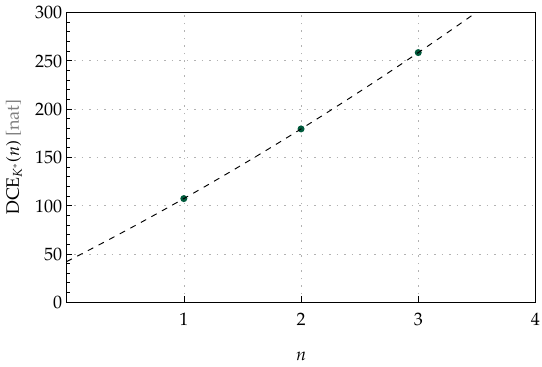}
				\caption{DCE of the charged $K^*$ family as a function of the $n$ radial excitation level.}
				\label{fig:dce_charged_n}
			\end{figure}

		The values of DCE are plotted in \autoref{fig:dce_charged_n}, from which it is possible to construct the first type of DCE-Regge-like trajectory in which DCE depends on the values of $n$. Interpolating the data for $n=1,2,3$ results in an expression given by
		\begin{flalign}
			\dce_{K^*}(n) = 3.45n^2 + 61.76n + 41.76,
			\label{eq_dce_charged_n}
		\end{flalign}
		within $0.001\%$ RMSD.
			
\begin{figure}[H]
			\centering
			\includegraphics[width=11cm]{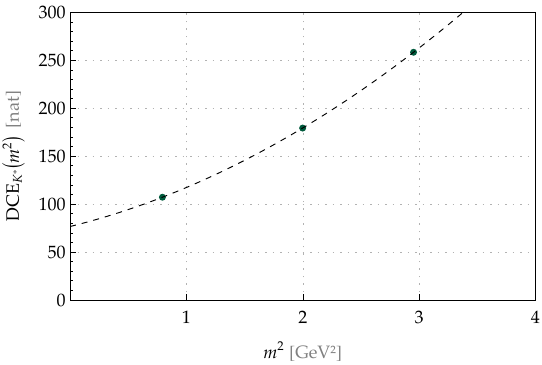}
			\caption[DCE of the charged $K^*$ family as a function of their squared mass]{DCE of the charged $K^*$ family as a function of their squared mass, for $n=1,2,3$.}
			\label{fig:dce_charged_m}
		\end{figure}

		The second type of DCE-Regge-like trajectory, in which DCE is plotted against the mass values of kaons, is presented in \autoref{fig:dce_charged_m}. Interpolating these data results in an equation given by
		\begin{flalign}
			\dce_{K^*}(m^2) = 10.6947 m^4 + 29.9886 m^2 + 76.4267,
			\label{eq_dce_charged_m}
		\end{flalign}
		within $0.001\%$ RMSD.
		
	Similarly to what was done for neutral strange vector kaons, it is possible to infer the mass of the next resonance states at $n>3$ using the same procedure described previously. The properties of the $K^*_4$ resonance are obtained when we put $n=4$ in (\ref{eq_dce_charged_n}), where DCE results in $337.51$ nat. Inserting this value into (\ref{eq_dce_charged_m}) and solving the equation leads to a mass of $m_{K^*_4}=1933.16$ MeV. Repeating the process for $n=5$, the DCE for the $K^*_5$ state is $410.59$ nat, which when inserted into (\ref{eq_dce_charged_m}) results in a mass of $m_{K^*_5}=2118.11$ MeV. Similarly, the DCE for the $n=6$ resonance is $470.81$ nat, which yields a mass for the $K^*_6$ state equal to $m_{K^*_5}=2239.02$ MeV.
	These results, adjusted with the error bar obtained from the experimental data, are summarized in \autoref{table:dce_charged_preview}, along with the experimental mass detected from the first three resonances.
	
	\begingroup
	\renewcommand{\arraystretch}{1.2}
	\begin{table}[H]
		\centering
		\setlength{\tabcolsep}{3mm}
		\begin{tabular}{cccc}
			\rowcolor{ufabc!50}
			\bf $n$ 				& \bf State 		&  $M_\text{exp}$ (MeV)	& 	$M_\text{th}$ (MeV)	\\
			\rowcolor{ufabc!30}	1 	& $K^*(892)$ 		& $891.67 \pm 0.26$ 	& $1035.2 \pm 0.3$				\\
			\rowcolor{ufabc!10}	2	& $K^*(1410)$ 		& $1414 \pm 15$			& $1447\pm15$			\\
			\rowcolor{ufabc!30}	3	& $K^*(1680)$ 		& $1718 \pm  18$ 		& $1750\pm18$				\\
			\rowcolor{ufabc!10}	4	& $(K^*_4)^\star$ 	& ---					& $1933\pm19$			\\
			\rowcolor{ufabc!30}	5	& $(K^*_5)^\star$	& --- 					& $2093\pm21$			\\
			\rowcolor{ufabc!10}	4	& $(K^*_6)^\star$	& ---					& $2207\pm22$
		\end{tabular}
		\caption[Extended mass spectrum of the charged $K^*$ meson family using DCE]{Extended mass spectrum of the charged $K^*$ meson family, with theoretical values estimated for $n=4,5,6$. The states with an "$\star$" indicate that their estimated masses were obtained through the extrapolation of DCE-Regge like trajectories that interpolate the experimental values.}
		\label{table:dce_charged_preview}	
	\end{table}
	\endgroup
	
	Based on the further mesonic states found in \cite{ParticleDataGroup:2024cfk}, the $K^*_5$ strange vector kaon resonance, whose predicted mass is $m_{K^*_5}=2093\pm21$ MeV, might match the $X(2075)$ state, which has experimental mass of $2075\pm12\pm5$ MeV, and the $X(2080)$, with a mass of $2080\pm10$ MeV. 
	In addition, it might also match another further mesonic state, $X(2100)$, which has an experimental mass equal to $2100\pm40$ MeV. Similarly, the $K^*_6$ strange vector kaon resonance, with mass $m_{K^*_6}=2207\pm22$ MeV, might match the $X(2210)$ state, which has an experimental mass equal to $2210^{+79}_{-15}$ MeV.
			
			\subsection{DCC of charged strange vector kaons}
		
			Complementing the work for the charged strange vector kaons, the DCC is calculated following the same constituent quark model for the DCE analysis: the $K^*(892)^+=u\bar{s}$ state and its antikaon, $K^*(892)^-=\bar{u}s$ represent the $n=1$ resonance, while the states $K^*(1410)^\pm$ and $K^*(1680)^\pm$ represent $n=2$ and $n=3$, respectively. The combination of $u$ and $s$ quarks results in an average constituent mass of $\bar{m}=411$ MeV, which leads to $\alpha=0.0554$ and $\kappa=530.79$ MeV.
		
			The procedure for calculating DCC presented in \Cref{section:cimadsqcd} was used again and the results are compiled in \Cref{table:dcc_charged}.
			\begingroup
			\renewcommand{\arraystretch}{1.8}
			\begin{table}[h!]
				\centering
				\setlength{\tabcolsep}{3mm}
				\begin{tabular}{ccc}
					\rowcolor{ufabc!50}
					\bf $n$ & \bf State &  {\bf DCC} 	  \\
					\rowcolor{ufabc!30}	1 	& $K^*(892)$ 	& $286.68$	\\
					\rowcolor{ufabc!10}	2	& $K^*(1410)$ 	& $454.60$	\\
					\rowcolor{ufabc!30}	3	& $K^*(1680)$ 	& $582.90$
				\end{tabular}
				\caption{DCC of the radial resonances of the charged $K^*$ meson family}
				\label{table:dcc_charged}	
			\end{table}
			\endgroup
			
\begin{figure}[H]
				\centering
				\includegraphics[width=11cm]{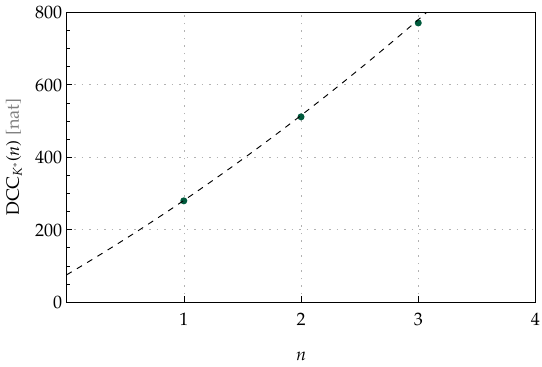}
				\caption{DCC of the charged $K^*$ family as a function of the $n$ radial excitation level.}
				\label{fig:dcc_charged_n}
			\end{figure}

			With the values calculated for the DCC, the plot shown in \autoref{fig:dcc_charged_n} is obtained, resulting in the first type of DCC-Regge-like trajectory when plotted as a function of the radial resonance level $n$. The equation that generates this trajectory is given by			
			\begin{flalign}
				\dcc_{K^*}(n) = -19.81n^2 + 227.35n + 79.14,
				\label{eq_dcc_charged_n}
			\end{flalign}
			within $0.001\%$ RMSD.
			
\begin{figure}[H]
				\centering
				\includegraphics[width=11cm]{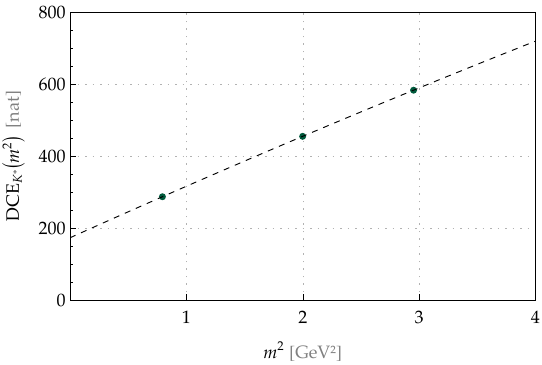}
				\caption[DCC of the charged $K^*$ family as a function of their squared mass]{DCC of the charged $K^*$ family as a function of their squared mass, for $n=1,2,3$.}
				\label{fig:dcc_charged_m}
			\end{figure}

			Similarly, placing the DCC as a function of the experimentally detected mass spectrum (available in \autoref{table:mass_charged}) results in the second type of DCC-Regge-like trajectory, presented in \autoref{fig:dcc_charged_m}. The polynomial interpolation for this trajectory is given by the equation
			\begin{flalign}
				\dcc_{K^*}(m^2) = -2.1896 m^4 + 145.4680 m^2 + 172.5590 
				\label{eq_dcc_charged_m}
			\end{flalign}
			within $0.001\%$ RMSD.	
			
		The predicted mass of the $K^*_4$ strange vector resonance is derived by inserting $n=4$ into Eq. (\ref{eq_dcc_charged_n}). This results in a DCC of $1056.35$ nat. This value is then applied to the left-hand side of  (\ref{eq_dcc_charged_m}) and the equation is solved to yield a mass of $m_{K^*_4}=1935.34$ MeV.
		When $n=5$, the DCC becomes $1315.70$ nat, leading to a mass of $m_{K^*_5}=2099.83$ MeV for the $K^*_5$ state. Lastly, for $n=6$, the $K^*_6$ strange vector kaon resonance has a DCC of $1543.43$ nat, resulting in a mass of $m_{K^*_6}=2223.76$ MeV. The obtained results, adjusted with the error bar from the experimental data, are summarized in \autoref{table:dcc_charged_preview}.
		
		\begingroup
		\renewcommand{\arraystretch}{1.2}
		\begin{table}[H]
			\centering
			\setlength{\tabcolsep}{3mm}
			\begin{tabular}{cccc}
				\rowcolor{ufabc!50}
				\bf $n$ 				& \bf State 	&  $M_\text{exp}$ (MeV)	& 	$M_\text{th}$ (MeV)	\\
				\rowcolor{ufabc!30}	1 	& $K^*(892)$ 	& $891.67 \pm 0.26$ 	& 	$1035.2 \pm 0.3$		\\
				\rowcolor{ufabc!10}	2	& $K^*(1410)$ 	& $1414 	 \pm 15 $	& 	$1447\pm15$		\\
				\rowcolor{ufabc!30}	3	& $K^*(1680)$ 	& $1718 	 \pm 18 $	& 	$1750\pm18$		\\
				\rowcolor{ufabc!10}	4	& $(K^*_4)^\star$ 		& ---			& 	$1935\pm19$	\\
				\rowcolor{ufabc!30}	5	& $(K^*_5)^\star$ 		& --- 					&	$2100\pm21$	\\
				\rowcolor{ufabc!10}	4	& $(K^*_6)^\star$ 		& ---					& 	$2224\pm22$
			\end{tabular}
		\caption[Extended mass spectrum of the charged $K^*$ meson family using DCC]{Extended mass spectrum of the charged $K^*$ meson family, with theoretical values estimated for $n=4,5,6$. The states with an "$\star$" indicate that their estimated masses were obtained through the extrapolation of DCC-Regge like trajectories that interpolate the experimental values.}
			\label{table:dcc_charged_preview}	
		\end{table}
		\endgroup
		
		As before, the $K^*_5$ strange vector kaon resonance, whose mass is predicted using DCC as $m_{K^*_5}=2100\pm21$ MeV, might match the $X(2100)$ further mesonic state in \cite{ParticleDataGroup:2024cfk}, whose mass is $2100\pm40$ MeV. Additionally, the $K^*_6$ resonance, with predicted mass $m_{K^*_6}=2224\pm22$, might match the $X(2210)$ further mesonic state, which has an experimental mass $2210^{+79}_{-21}$ MeV.

		\subsection{Concluding remarks}
		
		The analysis in this section employs Regge-like trajectories as functions of the radial quantum number $n$ and the mass $m$ to investigate the family of strange vector kaons. The soft-wall model with a deformed dilaton was adopted to incorporate the constituent quark mass. This choice is motivated by the presence in the composition of the vector kaons of the strange quark, which, although formally classified as a light quark, is considerably heavier than the up and down quarks. As experimentally anticipated in previous works, the inclusion of heavier quarks in meson composition leads to the emergence of non-linear Regge trajectories, an effect successfully reproduced here through the dilaton deformation.
		
		In addition to the DCE, the DCC was also employed to corroborate the predictions obtained with the method developed in this thesis. The formation of non-linear Regge-like trajectories in both approaches enabled the estimation of possible new resonances for the vector kaons.
		
		The DCE results indicate that both charged and neutral vector kaons may correspond to the states $X(2075)$, $X(2080)$, and $X(2100)$ at the $n=5$ resonance. For $n=6$, the analysis suggests a correspondence with the state $X(2210)$ listed in the PDG. Interestingly, for $n=4$, no candidate was identified among the further states currently cataloged.
		
		The application of the DCC reinforced the DCE findings for the $n=6$ resonance, supporting the correspondence with the $X(2210)$ state. However, for $n=5$, the DCC estimates pointed to both charged and neutral kaons corresponding solely to the $X(2100)$ state, thus highlighting it as the most robust candidate.
		
		Analogous to the case of light mesons, the analysis also revealed an increase in entropy and complexity with higher radial quantum numbers and, consequently, larger particle masses. This behavior suggests enhanced configurational instability and may indicate increasing experimental difficulty in detecting higher excitations of vector kaons beyond $n=6$.
		
		These results were published in \cite{daRocha:2023nsb}.
		
		\chapter{AdS/QCD at Finite Temperature}
		\label{chapter:thermalqcd}
		
		Applications of the AdS/CFT correspondence and holographic QCD at zero temperature provide a wide range of results, as we have seen in the previous chapters. However, it is at finite temperature that AdS/QCD shows greater potential for applications and results closer to the real world.
		
		At low temperature, strong interaction continues to exhibit the same properties well described by QCD at zero temperature. Quarks and gluons remain confined in hadrons, which form the nuclei of elements and allow for the existence of hadronic matter as we know it.
		But once we move to scenarios with high temperature and density, this entire structure collapses. For example, quarks and gluons are deconfined and form a \textit{quark-gluon plasma} (QGP) \index{quark-gluon plasma}. 
		The QCD phase diagram containing the transition to this state of matter is also not very well understood, both experimentally and theoretically \cite{Zaanen:2015oix}.
		
		In this chapter, we implement the configurational information measures together with AdS/QCD to study mass spectroscopy of hadrons at finite temperature.
		\Cref{section:adscft_finiteT} is dedicated to introducing one of the approaches for incorporating finite temperature into AdS/CFT, including a review of thermal quantum field theory and a brief overview of black holes in the context of AdS spacetime, introducing some aspects of phase transitions in QCD.
		
		\Cref{section:hadrons_termal} will address AdS/QCD at finite temperature and provide the foundations for working with mesons and baryons within this framework. This includes a brief discussion of the difficulties inherent to the treatment of baryons within the correspondence.
		
		Finally, \Cref{section:nucleons} presents the CIM-based AdS/QCD hybrid method applied to nucleons at finite temperature.
		
	\section{AdS/CFT correspondence at finite temperature}
		\label{section:adscft_finiteT}
		
		\subsection{Thermal field theory}
		\label{sec:thermalfieldtheory}
		
		Before understanding how finite temperature appears in the AdS/CFT correspondence, it is essential to review how temperature is introduced in a QFT when formulating a thermal field theory.
		
		Recalling the fundamental concepts of statistical thermodynamics, the density matrix of a system in thermal equilibrium is given by
		\begin{flalign}
			\rho= e^{-\beta\mathcal{H}},
		\end{flalign}
		where $\beta=1/k_B T$ is the inverse temperature, with $k_B$ being the Boltzmann constant (for simplicity, we will take $k_B\equiv1$ from now on). The Hamiltonian $\mathcal{H}$ must be chosen based on the most appropriate ensemble to describe the scenario. In the canonical ensemble, we have $\mathcal{H}=H$, where $H$ is the dynamic Hamiltonian of the system. In the grand canonical ensemble, the Hamiltonian is $\mathcal{H}=H-\mu N$, where $N$ is the particle number operator and $\mu$ is the chemical potential. For the purpose of this brief introduction, we will maintain a more general analysis.
		
		Regardless of the ensemble, having the density matrix allows us to obtain the partition function,
		\begin{flalign}
			Z = \Tr{\rho}=\Tr{e^{-\beta \mathcal{H}}},
		\end{flalign}
		where the trace can be seen as the sum of all expected values of the density matrix.
		
		In a QFT at zero temperature, the expected value of an operator $A$ is given by the familiar expression:
		\begin{flalign}
			\mean{A}_0 = \Tr{\rho A}.
		\end{flalign}
		Thus, when the system is in thermal contact with a heat bath at equilibrium temperature $T$, the expected value of the operator is defined as
		\begin{flalign}
			\mean{A}_\beta = \frac{1}{Z}\Tr{Ae^{-\beta H}} = \frac{\Tr{\rho A}}{\Tr {\rho}}.
		\end{flalign}
	
		With this, the correlation functions can be calculated. But before proceeding, the operators will be treated in the Heisenberg picture, where operators can incorporate time dependence according to
		\begin{flalign}
			A_H(t) = e^{i\mathcal{H}t}A e^{-i\mathcal{H}t},
		\end{flalign}
		Thus, the correlation function between any two operators $A$ and $B$ within the theory at finite temperature can be written as
		\begin{flalign}
			\mean{A_H(t)B_H(t')}_\beta &= Z^{-1} \Tr{\rho A_H (t) B_H(t')} \nonumber\\
			&=Z^{-1}
			\Tr{e^{-\beta\mathcal{H}}A_H (t) B_H(t')} \nonumber \\
			&=Z^{-1}
			\Tr{e^{-\beta\mathcal{H}}A_H (t) (e^{\beta\mathcal{H}}e^{-\beta\mathcal{H}}) B_H(t')} \nonumber \\
			&=Z^{-1}
			\Tr{e^{-\beta\mathcal{H}}e^{i\mathcal{H}t}A (t) e^{-i\mathcal{H}t} e^{\beta\mathcal{H}}e^{-\beta\mathcal{H}} e^{i\mathcal{H}t'}B(t')e^{-i\mathcal{H}t'}} \nonumber \\
			&=Z^{-1}
			\Tr{e^{i\mathcal{H}(t+i\beta)}A (t) e^{-i\mathcal{H}(t+i\beta)}e^{-\beta\mathcal{H}} e^{i\mathcal{H}t'}B(t')e^{-i\mathcal{H}t'}} \nonumber \\
			&=Z^{-1}
			\Tr{e^{-\beta\mathcal{H}} e^{i\mathcal{H}t'}B(t')e^{-i\mathcal{H}t'}e^{i\mathcal{H}(t+i\beta)}A (t) e^{-i\mathcal{H}(t+i\beta)}} \nonumber \\
			&=\mean{B_H(t')A_H(t+i\beta)}_\beta.
			\label{correlation_finiteT}
		\end{flalign}
		For visual clarity, spatial coordinates have been omitted. The manipulation above utilizes the property of cyclic permutation of the trace, ensuring that it is a general operation independent of the commutation properties of the operators. This is important because such operators can be fermionic or bosonic.
		
		One can directly notice that the operator $e^{-\beta\mathcal{H}}$ is essentially the time evolution operator $e^{-i\mathcal{H}t}$ if we consider $t=i\beta$. However, since $\beta=T^{-1}$ is real, the time variable $t$ needs to be imaginary.	
		This visualization is done formally through a Wick rotation to Euclidean space, redefining the time variable as $\tau=it$, effectively introducing an imaginary Euclidean time. Eq. (\ref{correlation_finiteT}) is known as the Kubo-Martin-Schwinger (KMS) relation, and from it, it is possible to show that imaginary time has a period $\beta$, i.e., $\tau\equiv\tau+\beta$. In fact, $\tau$ can be periodic or antiperiodic depending on whether the operators are fermionic or bosonic, respectively, but this is a more complex discussion that we will not cover here.
		
		Therefore, one way to formulate a quantum field theory at finite temperature is through the formalism of imaginary (Euclidean) time with a period of $\beta$.
		
		\subsection{Black hole thermodynamics}
		\label{sec:blackholethermo}
		
		At the beginning of this thesis, in \Cref{section:holographic}, we briefly touched upon some concepts related to black holes, but the topic of temperature was just mentioned. The notion that black holes emit radiation seems contrary to their original concept as astronomical objects from which nothing can escape. However, arguments for the existence of black hole radiation have existed since Hawking's proposition in 1974 \cite{Hawking:1974rv}.
		
		Hawking's demonstration succeeded in bringing quantum phenomena into an initially geometric problem.
		Roughly speaking, quantum fluctuations in the vacuum describe the rapid creation followed by annihilation of particle-antiparticle pairs. All of this happens on energy scales and time intervals that do not violate conservation laws thanks to the Heisenberg uncertainty principle.
		When these fluctuations occur near the event horizon, there is a possibility that they are disturbed by the gravitational effects of the black hole. As a result, one of the particles in the pair may be captured before annihilation, while the other is effectively "added" to the universe, propelled by the black hole. This process is analogous to a system losing mass through radiation emission, except that in this case, the radiation consists of particles that were originally not part of the black hole. Despite never being directly observed, this Hawking radiation has provided valuable insights into the thermodynamic properties of black holes, such as temperature and energy.
		
		The derivation of Hawking temperature for black holes, i.e., the temperature of Hawking radiation, can be approached in various ways, but here we will follow the formalism started in the preceding subsection.
		
		For a static black hole, the simplest metric solution takes the form
		\begin{flalign}
			ds^2 = -f(r)dt^2 + \frac{dr^2}{f(r)} +r^2 d\Omega^2 ,
			\label{staticblackhole}
		\end{flalign}
		where the primordial example is the Schwarzschild solution (\ref{schwarzschild}), which has the so-called \textit{blackening factor} given by $f(r)=1-2M/r$. A Wick rotation of this metric allows us to write it as
		\begin{flalign}
			ds^2 = f(r)d\tau^2 + \frac{dr^2}{f(r)} + r^2d\Omega^2 ,
		\end{flalign}
		which now has a Euclidean signature.
		When we examine the Wick-rotated Schwarzschild black hole,
		\begin{flalign}
			ds^2 = \left(1-\frac{2M}{r}\right)d\tau^2 + \left(1-\frac{2M}{r}\right)^{-1}dr^2 + r^2d\Omega^2,
		\end{flalign}
		it becomes easier to visualize that within the event horizon ($r<r_h=2M$), the signature ceases to be Euclidean. Consequently, it is impossible to traverse the horizon inward. Simultaneously, the horizon transitions from being singular to smooth. To ensure this, we must analyze the vicinity of the horizon.
		
		Given that the event horizon is defined as the point $r_h$ such that $f(r_h)=0$ and assuming that $f'(r_h)\neq 0$, we can expand the blackening factor as a first-order Taylor expansion $f(r)\simeq f'(r_h)(r-r_h)$, yielding
		\begin{flalign}
			ds^2 \simeq f'(r_h)(r-r_h)d\tau^2 + \frac{dr^2}{f'(r_h)(r-r_h)}.
		\end{flalign}
		
		By introducing the new coordinates $\rho=2\sqrt{(r-r_h)/f'(r_h)}$ and $\theta=f'(r_h)\tau/2$, we can rewrite the metric as
		\begin{flalign}
			ds^2=\rho^2 + \rho^2 d\theta^2.
		\end{flalign}
		This expression is analogous to that of a plane in polar coordinates, provided that $\theta$ is periodic with a period of $2\pi$, so $\theta\equiv\theta+2\pi$. If this condition is not satisfied, $\rho = 0$ (which corresponds to $r=r_h$) becomes a singular point known as conical singularity, as the resulting metric describes a cone.
		Since the Euclidean black hole must be smooth at $r_h$, the condition needs to be followed, avoiding the conical singularity. With this, the horizon becomes the origin of the polar coordinate system.
		
		Remembering that the time $\tau$ has $\beta=1/T$ as its period, the definition of $\theta$ and its periodicity condition lead us to write, finally, an equation for the Hawking temperature of static black holes:
		\begin{flalign}
			T = \frac{f'(r_h)}{4\pi}.
		\end{flalign}
		And, for the case of the Schwarzschild solution, the temperature is
		\begin{flalign}
			T= \frac{1}{4\pi r_h} = \frac{1}{8\pi M}.
			\label{schwarzschild_temperature}
		\end{flalign}
		
		Through this equation, one can observe that if a black hole radiates, losing energy in the form of Hawking radiation, it consequently loses mass and its temperature increases. Since Eq. (\ref{schwarzschild_temperature}) does not exhibit any upper limitations, we can imagine that the temperature grows infinitely. This process is called \textit{evaporation}, and it implies that if the black hole does not gain mass in any way, its fate is to evaporate until it disappears, releasing an enormous amount of energy at the end of the process.
		
		By rearranging the temperature expression to $M=1/8\pi T$, one can directly calculate the heat capacity of the black hole:
		\begin{flalign}
			C = \frac{\partial M}{\partial T} = -\frac{1}{8\pi T^2} <0.
		\end{flalign}
		That is, in this situation, black holes have a negative heat capacity. It means that black holes heat up as energy is removed from the system and cool down as energy is added. Intuitively, the opposite should occur. In fact, according to the second law of thermodynamics, systems are only in equilibrium if they possess a positive heat capacity. 
		
		To understand why this leads to instability, consider a black hole in thermal contact with a large heat reservoir (a heat bath) at the same temperature. If a small fluctuation causes the black hole to absorb a small amount of energy from the bath, its mass increases. Due to the negative specific heat, its temperature consequently decreases. Being now cooler than the bath, it absorbs even more energy, cooling further and growing indefinitely. Conversely, if it radiates a bit more energy than it absorbs, it loses mass, heats up, and radiates even more intensely, leading to a runaway evaporation process. Thus, any thermal equilibrium with a static heat bath is unstable \cite{Susskind:2005js}.
				
		Therefore, these black holes are thermodynamically unstable and do not represent an equilibrium state.
		
		This conclusion about equilibrium frustrates any idea of generalizing the formalism and placing a QFT at finite temperature by simply inserting a black hole into the system.
		The discussion regarding the implications of the obtained results extends and remains open to the present. For the purposes of this work, we content ourselves with investigating the problem in other configurations.
	
		\subsection{Schwarzschild-AdS black hole}
		\label{sec:sch-adsblackhole}
		
		The discussion can now be extended to the AdS spacetime. The Schwarzschild solution for a black hole in a ($d+1$)-dimensional AdS space is given by \cite{Nastase:2015wjb}
		\begin{flalign}
			ds^2 = \left(1+\frac{r^2}{L^2}-\frac{w_d M}{r^{d-2}}\right)dt^2 + \left(1+\frac{r^2}{L^2}-\frac{w_d M}{r^{d-2}}\right)^{-1}dr^2 + r^2d\Omega_{d-1}^2,
			\label{schwarzschild-adsbh}
		\end{flalign}
		where, according to the general form for static black holes (\ref{staticblackhole}), the blackening factor is given by:
		\begin{flalign}
			f(r)=1+\frac{r^2}{L^2}-\frac{w_d M}{r^{d-2}}.
			\label{f_schwarzschild-adsbh}
		\end{flalign}
		Here, $L$ is the AdS radius, which we will keep explicit for analysis, and the factor $w_d$ is defined as
		\begin{flalign}
			w_d \equiv \frac{8\pi}{(d-2)\Omega_{d-1}},
		\end{flalign}
		where $\Omega_{d-1}$ is the volume of a unit sphere in ($d-1$) dimensions. From this, we can recover some well-known results. For $d=3$, corresponding to AdS$_4$ spacetime, the volume is $\Omega_2=4\pi$, leading to $w_3 = 2$. 
		For $d=4$ (AdS$_5$) we have $\Omega_3=2\pi^2$ and $w_4 = 2/\pi$.
		
		Near the origin, $r\rightarrow0$, the blackening factor becomes $f(r)\sim 1 -w_dM/r^{d-2}$, approaching the asymptotically flat Schwarzschild solution. An exact way to obtain the Schwarzschild solution is by assuming $L\rightarrow \infty$, which restores the flat space.	
		As the radial coordinate moves away from the origin toward the boundary, $r \rightarrow \infty$, the solution (\ref{schwarzschild-adsbh}) reduces to the metric in static coordinates (\ref{staticcoordinates}), one of the coordinate systems mentioned in the \Cref{section:adsspace}. This implies that a Schwarzschild-AdS (SAdS) black hole is asymptotically an AdS spacetime.\\
		
		The procedure for calculating the Hawking temperature is analogous to what was done in the previous subsection, but with some particularities.
		
\begin{figure}[h!]
			\centering
			\includegraphics[width=11cm]{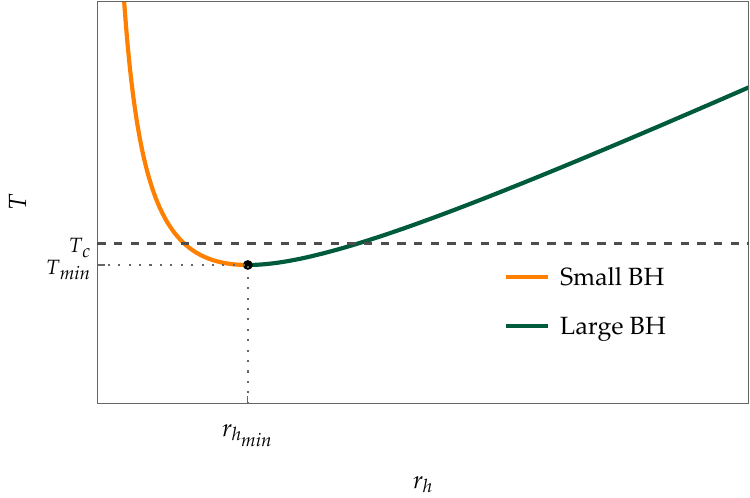}
			\caption[Hawking temperature as a function of the black hole horizon radius in AdS space]{Hawking temperature $T$ as a function of the black hole horizon radius $r_h$ in anti-de Sitter space with $L=1$. The orange curve represents the small black hole branch, while the evergreen curve corresponds to the large black hole branch. The minimum temperature required for black hole existence is $T_{\text{min}}$, corresponding to the radius $r_{h_\text{min}}$. The horizontal dotted line denotes the critical temperature $T_c$ of the Hawking-Page phase transition.}
			\label{fig:blackholes}
		\end{figure}

		The function (\ref{f_schwarzschild-adsbh}) has more than one solution for $f(r_h)=0$, so the analysis is performed with the outermost black hole horizon defined as the larger of these solutions.
		The calculation is quite complicated and the result can be expressed more clearly as
		\begin{flalign}
			w_d M = r^{d-2}_h \left(1+\frac{r_h^2}{L^2}\right).
			\label{outerhorizon}
		\end{flalign}
		Rewriting the metric with this, we again find that the time coordinate must be periodic to avoid a singularity. Consequently, we conclude that the temperature of an AdS black hole is given by:
		\begin{flalign}
			T= \frac{r^2_h d + (d-2)L^2}{4\pi L^2 r_h}.
			\label{sads_temperature}
		\end{flalign}
		It is common in the literature to treat two types of SAdS black holes based on their size relative to the AdS radius, remembering that the size of the event horizon and the mass are directly related. These types are: the large, defined by $r_h\gg L$, and the small, with $r_h\ll L$.
		Applying these limits to Eq. (\ref{sads_temperature}), we find that the temperature for small SAdS black holes scales as $T \sim 1/r_h$, while for large ones, it scales as $T \sim r_h$. 
		This implies that there exists a point where the temperature is minimal, marking the transition. \Cref{fig:blackholes} illustrates these curves.
		
		By calculating $dT/dr_h=0$, we can easily find the size of the horizon at the minimum point, $r_{h_\text{min}}=\sqrt{d-2}L/\sqrt{d}$, and subsequently determine the minimum temperature, which is given by the expression:
		\begin{flalign}
			T_\text{min} = \frac{\sqrt{d(d-2)}}{2\pi L}.
		\end{flalign}	
		
		Next, since mass does not explicitly appear in the temperature equations, it is convenient to rewrite the expression for the heat capacity as
		\begin{flalign}
			C = \frac{\partial M}{\partial T} = \frac{\partial M}{\partial r_h}\frac{\partial r_h}{\partial T} = \frac{\partial M}{\partial r_h}\left(\frac{\partial T}{\partial r_h}\right)^{-1}.
		\end{flalign}
		
		From Eq. (\ref{outerhorizon}), we know that $\partial M/\partial r_h$ is always positive. Therefore, the sign of the thermal capacity depends solely on the form of $T$.
		The temperature for small black holes is given by $T \sim 1/r_h$, leading to $\partial T/\partial r_h < 0$.
		Consequently, the heat capacity is always negative. Hence, they are also thermodynamically unstable.
		
		On the other hand, the temperature for large black holes is given by $T\sim r_h$, which results in positive heat capacity. However, this condition alone is not sufficient to determine whether large SAdS black holes are stable. We must also ensure that the black hole’s free energy $F_\text{BH}$ is lower than the free energy of AdS spacetime without the black hole $F_\text{AdS}$.	
		The complete calculation can be found in \cite{Natsuume:2014sfa}. The result is that the stability condition $\Delta F = F_\text{BH} - F_\text{AdS} < 0$ is fulfilled only if
		\begin{flalign}
			T>T_c \equiv \frac{d-1}{2\pi L} >T_\text{min},
		\end{flalign}
		where $T_c$ represents the critical temperature. In summary, considering a black hole solution with temperature $T$, we can highlight four key points:
		
		\begin{enumerate}[(I)]
			\item $T<T_\text{min}$: black holes cannot exist; the solution representing the globally stable state is the AdS spacetime with a Euclidean time of periodicity $\beta$, which is called the \textit{thermal AdS spacetime} (ThAdS).
			\item $T_\text{min}< T < T_c$: after the temperature increase, black holes can exist, but as the free energy condition is not satisfied, instead it is $\Delta F>0$, they are not stable. Consequently, the black hole decays via Hawking radiation, and the solution reduces to the thermal AdS spacetime.
			\item $T=T_c$: at this point, a transition from the thermal AdS spacetime to an SAdS black hole occurs, known as the \textit{Hawking-Page phase transition} \cite{Hawking:1982dh}.
			\item $T>T_c$: the solution describes a large black hole with $\Delta F<0$, therefore, it is stable and in thermal equilibrium with the Hawking temperature, without evaporating.
		\end{enumerate}
		
		The conclusion that at $T>T_c$ the solution is a thermodynamically stable black hole can be extended and interpreted as the entire spacetime being at equilibrium temperature $T$.
		This includes the boundary, where we expect the existence of a gauge theory at a finite temperature.
		
		Nevertheless, when taking the limit approaching the boundary ($r\rightarrow\infty$), the metric becomes
		\begin{flalign}
			ds^2 \simeq \frac{r^2}{L^2}dt^2 + \frac{L^2}{r^2} dr^2 + r^2 d\Omega_{d-1}^2,
		\end{flalign}
		which represents the topology of two spheres, one $S^1$ with radius $r/LT$ and another $S^{d-1}$ with radius $r$, that is, it is the metric of a $S^1\times S^{d-1}$ space. However, the construction so far anticipates that the gauge theory is hosted in a flat space, so we need to reformulate the metric to obtain a flat $d$-dimensional space $R^{d-1}$. Since the $S^1$ sphere is associated with the Euclidean time, it is responsible for maintaining the system at finite temperature. The objective, then, is to flatten $S^{d-1}$ and obtain a $R^{d-1}\times S^1$ space.
		
		A way to resolve this issue is by assuming the following limit for the ratio of the radii of both spheres:
		\begin{flalign}
			\frac{r}{\frac{r}{LT}} = LT \rightarrow \infty.
		\end{flalign}
		With $L$ fixed, this condition requires $T\rightarrow\infty$. According to (\ref{sads_temperature}), this can be achieved by taking $r_h\rightarrow\infty$ or $r_h\rightarrow0$. However, the latter is not a thermodynamically stable option, leaving us with the former condition.
		To circumvent these infinities, the coordinates are rescaled as
		\begin{flalign}
			r = \left(\frac{w_d	M}{L^{d-2}}\right)^{\frac{1}{d}}\rho, \qquad t = \left(\frac{w_d M}{L^{d-2}}\right)^{-\frac{1}{d}} \tau, \qquad
			d\Omega_i = \left(\frac{w_d M}{L^{d-2}}\right)^{-\frac{1}{d}}dx_i,
		\end{flalign}
		and thus we write the metric as
		\begin{flalign}
			ds^2 = \left[\left(\frac{w_d M}{L^{d-2}}\right)^{-2/d}+\frac{\rho^2}{L^2} - \frac{L^{d-2}}{\rho^{d-2}}\right] d\tau^2 + \left[\left(\frac{w_d	M}{L^{d-2}}\right)^{-2/d} + \frac{\rho^2}{L^2} - \frac{L^{d-2}}{\rho^{d-2}}\right]^{-1}d\rho^2 + \rho^2d\mathbf{x}^2.
			\label{general_sads}
		\end{flalign}
		Moreover, according to (\ref{outerhorizon}), taking $r\rightarrow\infty$ implies $M\rightarrow\infty$. In the way the metric is written above, after the coordinate change, this limit is trivial and thus we can rewrite it as	
		\begin{flalign}
			ds^2 = \left(\frac{\rho^2}{L^2} - \frac{L^{d-2}}{\rho^{d-2}}\right) d\tau^2 + \left(\frac{\rho^2}{L^2} - \frac{L^{d-2}}{\rho^{d-2}}\right)^{-1}d\rho^2 + \rho^2d\mathbf{x}^2,
		\end{flalign}
		where $d\mathbf{x}^2=dx_i dx^i$. With this configuration, we have $\Omega_i \sim x_i$, meaning that the sphere $S^{d-1}$ has been flattened as expected. This result continues to be a solution to the Einstein equations and represents a black hole in AdS spacetime. However, the planar horizon in this structure characterizes what is known as an \textit{AdS black brane}\index{black brane}.
		The Hawking temperature of the black brane can be obtained through the usual procedure of avoiding a conical singularity, which results in a period $\beta=4\pi L/d$ and thus
		\begin{flalign}
			T = \frac{d}{4\pi L}.
		\end{flalign}
		This temperature can also be obtained from (\ref{sads_temperature}), simply extend the rescaling of $\tau$ to $T$, i.e., $T=\left(w_d M/L^{d-2}\right)^{-1/d} T$, and then apply the limits $r_h\rightarrow\infty$ and $M\rightarrow\infty$.
		In any case, we finally have an expression for the temperature of the gauge theory living on the boundary.
		
		The physically most interesting case for us is $d=4$, with which we rewrite the metric (\ref{general_sads}) as
		\begin{flalign}
			ds^2 = \frac{\rho^2}{L^2}\left[\left(1-\frac{L^4}{\rho^4}\right)d\tau^2 + L^2d\mathbf{x}^2\right] + \frac{L^2}{\rho^2}\left(1-\frac{L^4}{\rho^4}\right)d\rho^2
		\end{flalign}
		and obtain a temperature for the gauge theory equal to
		\begin{flalign}
			T = \frac{1}{\pi L}.
		\end{flalign}
		
		There is one last step remaining, which is to rewrite the metric of the AdS black brane in Poincaré coordinates, with which we are more used to working with and which will be more interesting in future studies.
		Doing the following coordinate changes
		\begin{flalign}
			\rho=\frac{z_h L}{z}r, \qquad \tau=\frac{L}{z_h}t \qquad \mathbf{x}=\frac{1}{z_h}\tilde{\mathbf{x}},
		\end{flalign}
		where $z_h$ is the position of the event horizon, and Wick rotating back to the Lorentzian signature, we obtain
		\begin{flalign}
			ds^2 = \frac{L^2}{z^2}\left[-f(z)dt^2 + d\tilde{\mathbf{x}}^2 + \frac{1}{f(z)}dz^2\right],
			\label{adsblackholemetric}
		\end{flalign}
		with 
		\begin{flalign}
			f(z) = 1- \frac{z^4}{z_h^4}.
		\end{flalign}
		
		And the Hawking temperature is
		\begin{flalign}
			T=\frac{1}{\pi z_h}.
		\end{flalign}
		
		With this, we have a solution representing an AdS black hole in $d=4$ dimensions, at the same time a spacetime AdS version, written in Poincaré coordinates and with finite temperature.
		It is important to emphasize that the conjecture and the GKPW prescription remain valid and equal to the case of zero temperature. \autoref{table:dictionaryadscft} also summarizes how the AdS/CFT correspondence dictionary relates the duality between the Hawking temperature for black holes in the bulk and the finite temperature on the boundary.		
	
		This development can be further improved by including quantum corrections to the AdS-Schwarzschild black brane, following Ref. \cite{daRocha:2024lev}, or by implementing a self-consistent dynamical model for hadrons under conditions of finite temperature and density \cite{Toniato:2025gts}.
		
	\subsection{Phase-transition temperature}
	\label{sec:phasetransition}
	
	\index{phase-transition temperature}
		
	As seen in \Cref{section:qcd}, QCD at low energies (and now, more familiarly, we can also say at low temperatures) exhibits a property called confinement. At this scale, quarks and gluons do not appear "alone", only in bound states.
	The situation changes as the energy scale increases. Such high-energy conditions are created and observed in relativistic heavy-ion collisions at particle accelerators like the LHC and RHIC, and is also believed to have existed in the early universe, and may be present in the cores of some neutron stars.
	In this scenario of higher energy/temperature, quarks and gluons are deconfined and form a quark-gluon plasma.
	
	The QGP state is not observed directly. Being an extremely hot and short-lived medium, it quickly expands, cools, and undergoes a process known as \textit{hadronization}\index{quark-gluon plasma!hadronization}, the recombination of quarks and gluons into hadrons. Therefore, experiments can only measure the final-state particles -- the hadrons, leptons, and photons -- that emerge from the collision and carry imprints of the plasma they traversed. This experimental challenge underscores the importance of theoretically mapping the different states of nuclear matter. This map is conceptually organized in the QCD phase diagram, which plots temperature against baryon chemical potential ($\mu_B$) to delineate the confined hadronic phase from the deconfined QGP phase \cite{Fukushima:2010bq} . The nature of the transition between these phases remains a central topic of research.
	
	The concept of a limiting temperature for hadronic matter predates the consolidation of QCD.\index{phase-transition temperature!Hagedorn temperature} In 1965, Rolf Hagedorn, investigating the statistical thermodynamics of high-energy collisions, postulated that the density of hadronic states grows exponentially with mass, $\rho(m) \sim \exp(m/T_H)$ \cite{Hagedorn:1965st}. He concluded that such behavior causes the system's partition function to diverge at a critical temperature $T=T_H$. This value, later named the Hagedorn temperature ($T_H$), was interpreted as the point where hadronic matter undergoes a phase transition, "melting" into a new state of matter. Modern estimates for $T_H$ yield values in a range of approximately $140$ to $155$ MeV, which is equivalent to $\sim 10^{12}$ Kelvin \cite{Rafelski:2015cxa}.
	
	A remarkably similar result emerges in string theory, since the energy density of strings at high energies/temperature also exhibits an exponential behavior. This again leads to a divergence of the partition function near a limiting temperature $T_H$, obtained in this context as \cite{Zwiebach:2004tj}
	\begin{flalign}
		T_H = \frac{1}{4\pi \sqrt{\alpha'}}.
	\end{flalign}
	The main conclusion is that, as the system approaches this critical temperature, it reaches a turning point: instead of the added energy further exciting existing strings, it becomes energetically more favorable to create an exponentially growing number of new strings, particularly those of large length. This proliferation of states signals a phase transition \cite{Polchinski:1998rq}.
	
	Given that confining gauge theories such as QCD possess a critical temperature marking the confinement/deconfinement transition, and string theory features a limiting temperature that also induces a phase transition, it seems natural to attempt to establish an equivalence between them through the AdS/CFT correspondence. However, the relation between $T_c$ and $T_H$, if it exists, is far from straightforward and remains an open question.\\
	
	Setting aside the Hagedorn temperature for the moment, the AdS/CFT correspondence provides another important equivalence.
	In Ref. \cite{Witten:1998zw}, Witten pointed out that the Hawking-Page phase transition, present on the gravitational side of the AdS/CFT correspondence, would be equivalent to a transition between confinement and deconfinement on the gauge side. 
	It is relevant to bring this property under the scrutiny of holographic QCD because, even when deconfined, quarks and gluons still exhibit a strong interaction, rendering perturbative methods inefficient.
	This strongly-coupled nature of the QGP manifests itself in its collective behavior as a nearly perfect fluid, characterized by an extremely low shear viscosity to entropy density ratio, which makes it a subject of study in the fluid/gravity correspondence as well \cite{Son:2007vk,Rangamani:2009xk}, a relatively new area emerging from the gauge/gravity duality, but which relates hydrodynamics to solutions of the Einstein equation in AdS space. 
	
	Among other holographic models \cite{Critelli:2017oub,Rougemont:2017tlu,dePaula:2020bte}, the hard-wall and soft-wall models of AdS/QCD can be used to compute the transition temperature from their input parameters, as shown in \cite{Herzog:2006ra,Braga:2020opg,Braga:2021zyi,BallonBayona:2007vp}. The procedure consists of considering the two geometries involved in the transition: the thermal AdS spacetime in Euclidean signature, with metric
	\begin{flalign}
		ds^2 = \frac{L^2}{z^2}(dt^2 + d\mathbf{x}^2 + dz^2),
	\end{flalign}
	and the AdS black hole, whose metric is given in Eq. (\ref{adsblackholemetric}). The gravitational action considered in both HW and SW models has an integrand of the form
	\begin{flalign}
		I \sim R + \frac{12}{L^2}.
	\end{flalign}
	However, in both spacetimes the scalar curvature is given by $R=-20/L^2$, which allows us to write the actions as
	\begin{flalign}
		\mathcal{S}_\text{HW} = - \frac{1}{2\pi L^2} \int d^5x \sqrt{-g},
	\end{flalign}
	and
	\begin{flalign}
		\mathcal{S}_\text{SW} = - \frac{1}{2\pi L^2} \int d^5x \sqrt{-g} e^{-\Phi(z)}.
	\end{flalign}
	
	Let us focus on the soft-wall model, which we are most familiar with.
	
	For ThAdS, the integration limits include the periodicity of the temporal coordinate, $0 \le t < \beta'$, and $\epsilon \le z < \infty$ for the holographic coordinate, where $\epsilon \to 0$ is a regulator. For the black hole, the limits include the known periodicity of the temporal coordinate, $0 \le t < \pi z_h$, and the holographic coordinate bounded by the event horizon, $\epsilon \le z \le z_h$.
	To determine $\beta'$, one matches the two geometries where the periodicity in the time coordinate is locally the same, at $z=\epsilon$, which gives $\beta'=\pi z_h \sqrt{f(\epsilon)}$, recalling that $f(z)=1-(z/z_h)^4$.
	The chosen dilaton takes the usual quadratic form, $\Phi(z)=(\kappa z)^2$.

	With this, the action density is defined as $\mathcal{E}=\mathcal{S}/V$, where $V=\int d^3\mathbf{x}$ is the spatial volume. For ThAdS, this density is
	\begin{flalign}
		\mathcal{E}_\text{AdS} = \frac{L^3}{4\pi} \int_{0}^{\beta'}dt \int_{\epsilon}^{\infty} dz z^{-5} e^{-(\kappa z)^2}
	\end{flalign}
	while for the black hole it reads
	\begin{flalign}
		\mathcal{E}_\text{BH} = \frac{L^3}{4\pi} \int_{0}^{\pi z_h}dt \int_{\epsilon}^{z_h} dz z^{-5} e^{-(\kappa z)^2}.
	\end{flalign}
	
	Thus, the difference between the two densities is therefore
	\begin{flalign}
		\Delta\mathcal{E} &= \lim\limits_{\epsilon\to 0} (\mathcal{E}_\text{BH}-\mathcal{E}_\text{AdS}) \nonumber\\
		&= \frac{L^3}{16 z_h^3} \left[e^{-(\kappa z)^2} \big(-1+(\kappa z_h)^2\big) + \frac{1}{2} + (\kappa z_h)^4 \text{Ei}\big(-(\kappa z_h)^2\big) \right]
	\end{flalign}
	where $\text{Ei}(y)=-\int_{-y}^{\infty} e^{-t}/t dt$ is the exponential integral.
	The Hawking–Page transition temperature then emerges at the point where the two solutions coincide, i.e., when $\Delta\mathcal{E}=0$. Solving numerically yields
	\begin{flalign}
		T_c = 0.4917 \kappa.
	\end{flalign}
	
\begin{figure}
		\centering
		\includegraphics[width=11cm]{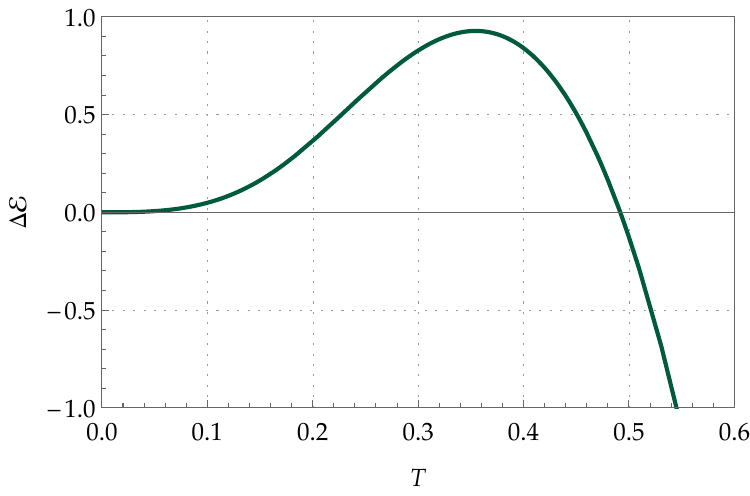}
		\caption[Action density difference $\Delta\mathcal{E}$ as a function of the temperature $T$.]{The difference between action densities of thermal AdS and AdS black hole, $\Delta\mathcal{E}$, as a function of the temperature $T$, assuming $L^3/16=\kappa^2=1$.}
		\label{fig:energyT}
	\end{figure}

	It is possible to notice that for $T<T_c$, one finds $\Delta\mathcal{E}>0$, which implies the dominance of the ThAdS geometry and, consequently, an instability of the black hole, as expected. However, for $T>T_c$, the black hole regains stability, in agreement with the conclusions regarding the transition.
	\autoref{fig:energyT} shows a plot of $\Delta\mathcal{E}$ as a function of $T$, where the transition becomes evident.
	
	In Ref. \cite{Herzog:2006ra}, the $\rho$-meson spectrum is used to fix the input parameters of the models. For the SW model, one finds $\kappa=338$ MeV, which yields $T_c=191$ MeV.
	For comparison, in the HW model, using a similar procedure, the result was $T_c=2^{1/4}/(\pi z_m) = 122$ MeV, with the IR cutoff $z_m$ determined from the mass of the lightest $\rho$ meson.
	
	We therefore observe that the estimated value of $T_c$ depends on the chosen model. This becomes even clearer when comparing with other results. In Ref. \cite{MartinContreras:2021bis}, using a model where the dilaton profile is adapted to more accurately account for the phenomenology of decay constants, the value obtained was $T_c=142$ MeV. In Ref. \cite{Afonin:2020crk}, where the transition temperature is derived as a function of the gluon condensate, the result was $T_c=156$ MeV, with the condensate value extracted from QCD sum rules.
	
	All these values can be compared with results from lattice QCD, such as $T_c=156.5\pm1.5$ MeV obtained by the HotQCD Collaboration \cite{HotQCD:2018pds} and $T_c=158.0\pm0.6$ MeV reported in Ref. \cite{Borsanyi:2020fev}, indicating a remarkable agreement not only among holographic QCD models but also with nonperturbative QCD methods.\\
	
	Configurational entropy can also be employed to study the phase transition, based on the ability of CIMs to serve as indicators of the stability of a given configuration.
		
	Before proceeding, it is necessary to address a crucial aspect regarding the treatment of the information paradox within the AdS/CFT correspondence. As previously discussed, large black holes in anti-de Sitter space are thermodynamically stable, meaning they do not evaporate via Hawking radiation. While this conclusion was justified earlier through thermodynamic quantities such as specific heat and free energy, there is an equally important geometric reasoning underlying this stability.
	
	From a geometric perspective, the stability arises from the intrinsic causal structure of anti-de Sitter space, which differs from that of flat Minkowski spacetime. In asymptotically flat space, radiation emitted by a black hole can propagate to infinity and escape the system forever, leading to mass loss and evaporation. In contrast, the negative curvature of AdS space generates a confining gravitational potential that acts, effectively, as a trapping mechanism, a "box" \cite{Susskind:2005js}.
	
	Analyses of the geodesic motion reveal that although the boundary of AdS is located at an infinite spatial distance, massless particles (such as photons comprising Hawking radiation) can traverse this distance and reach the boundary in a finite time \cite{Natsuume:2014sfa}. Since the spacetime acts as a closed system with a timelike boundary, energy cannot simply leak out. Hence, to ensure a well-defined evolution of the theory, one must impose conservative boundary conditions that force the radiation to reflect back into the bulk.
	This mechanism implies that the Hawking radiation emitted by the horizon does not dissipate. Instead, it bounces off the boundary and returns to the center, creating a thermal bath. Eventually, the rate of emission balances the rate of reabsorption, allowing the black hole to reach a stable thermal equilibrium with its own radiation atmosphere.	
	In this way, the reflective nature of the AdS boundary stabilizes the black hole and creates the thermal bath required to describe the boundary CFT at a finite temperature.
	
	A viable approach to circumvent this scenario and model systems that exhibit instabilities or phase transitions is through the breaking of conformal symmetry. This is precisely what is achieved in holographic QCD models by introducing a dimensionful scale (such as the IR cutoff in the HW model or the dilaton in the SW model), effectively changing the geometry of the box.

	In this context, the application of CIMs within AdS/QCD models becomes particularly relevant. 
	In Refs. \cite{Braga:2020opg,Braga:2021zyi}, it was found that the configurational entropy remains constant for the thermal AdS spacetime, consistent with the expected stability of this state, which represents the confined phase. In contrast, for the AdS black hole, which models the deconfined phase, configurational entropy was observed to increase with temperature, providing strong evidence of growing instability as the temperature rises.	
	This result supports models that attempt to incorporate AdS black hole evaporation into the AdS/CFT correspondence to address the information paradox \cite{Ong:2015fha}. In such cases, the increase in instability is expected as a consequence of the Hawking radiation evaporation process, which intensifies with temperature.
	
	However, the application of CE in combination with the HW model showed that the black hole entropy grows with temperature following a $\log(T)$ dependence. In contrast, for the SW model, the same relation is linear in $T$. It is also noteworthy that, for the SW model, the configurational entropy is not only constant but also vanishes for $T<T_c$, which further supports the idea that, since no black holes exist in the confined phase, there is no evaporation process capable of altering the stability of the system.
	
	Finally, a similar analysis can be performed using CIMs in conjunction with HW and SW models to investigate the confinement/deconfinement temperature and the phase diagram of the quark-gluon plasma, employing a charged and rotating black hole background to incorporate finite density and rotation, respectively \cite{Braga:2022yfe,Braga:2023qej,Braga:2023qee,Braga:2024nnj,Braga:2025eiz, Junqueira:2025xnx}.\\

	In \Cref{section:nucleons}, we will employ configurational entropy to estimate $T_c$ from the hadron mass spectrum.
	Before doing so, however, it is necessary to review the description of hadrons in thermal AdS/QCD, which will be carried out in the following section.

	\section{Hadrons in thermal AdS/QCD}
	\label{section:hadrons_termal}
	
		\subsection{Thermal soft-wall AdS/QCD model}
		
		To construct a thermal soft-wall model, we adopt as background the metric of the Schwarzschild-AdS$_5$ black hole with a warp factor suitable for the model:
		\begin{flalign}
			ds^2 = e^{2A(z)}\left[-f(z)dt^2 + d\mathbf{x}^2 + \frac{1}{f(z)}dz^2\right],
		\end{flalign}
		with $f(z)=1-z^4/z_h^4$, where the event horizon is related to the system temperature according to $T=(\pi z_h)^{-1}$. As before in the simplest SW model, $A(z)=-\ln(z)$ (with the AdS curvature radius $L=1$). From this point onward, all temperature dependence will be explicitly displayed.

		At first, the quadratic dilaton $\Phi(z)=\kappa^2z^2$ is sufficient to introduce the breaking of both chiral and conformal symmetries, as well as to reproduce the Regge trajectories.
		However, considering that the background of the model has been modified by functions of temperature, it is expected that the dilaton, which also defines the background, will also be $T$-dependent. The proposal was initially made in Ref. \cite{Vega:2018dgk}, which modified the two dilatons considered in \cite{Gherghetta:2009ac} to
		\begin{flalign}
			\Phi_1(z,T) &= \kappa^2(1+\alpha T)z^2, \\
			\Phi_2(z,T) &= \kappa_1^2(1+\alpha T)z^2 \tanh(\kappa_2^2(1+\alpha T)z^2),
		\end{flalign}
		where $\alpha$ is a parameter that can be fixed using the melting temperature of light mesons \cite{Vega:2018dgk}. For $\alpha=0$, the dilatons reduce to their usual forms at $T=0$.
		
		In order to make the model more consistent with QCD and based on the role of the parameter $\kappa$ in the spontaneous chiral symmetry breaking, Refs. \cite{Gutsche:2019blp,Gutsche:2019pls} proposes that the entire structure of the dilaton originates from the temperature behavior of the chiral condensate. This includes considering the scale parameter $\kappa$ as temperature-dependent as well.
			
		At zero temperature, the dependence of the parameter $\kappa$ on the chiral condensate is of the order of $\kappa \sim \Sigma$. Assuming that the relation for the $T$-dependent case remains the same for the case at zero temperature, we have
			\begin{flalign}
				\kappa^2(T) = \kappa^2 \frac{\Sigma(T)}{\Sigma}.
			\end{flalign}
		The expression for the $T$-dependent chiral condensate $\Sigma(T)=\bra{0}\bar{q}_L q_R\ket{0}_T$ can be obtained through chiral perturbation theory \cite{Gasser:1986vb}, which leads to
		\begin{flalign}
			\Sigma (T) = \Sigma \left(1+ \Delta_T +\mathcal{O}(T^6)\right),
		\end{flalign}	
		with
		\begin{flalign}
			\Delta_T = -\frac{(N_f^2 - 1)}{N_f}\frac{16 T^2}{9\kappa^2} - \frac{(N_f^2 - 1)}{2N_f^2} \left(\frac{16 	T^2}{9\kappa^2}\right)^2.
		\end{flalign}
		With this result in hand, the expression for the dilaton parameter up to $T^4$ is given by
		\begin{flalign}
			\kappa^2(T) = \kappa^2 \left(1+ \Delta_T\right).
		\end{flalign}
			
		In addition to the change in the parameter $\kappa$, the inclusion of a thermal prefactor is also proposed, which introduces corrections to the characteristic prefactor of the soft-wall model, i.e., $e^{\Phi(z)+\lambda_T(z)}$, with $\lambda_T$ given by
		\begin{flalign}
			\lambda_T(z) = \upalpha\frac{z^2}{z_h^2} + \upgamma\frac{z^4}{z_h^4} + \upxi \frac{\kappa^2z^6}{z_h^4}.
		\end{flalign}

		The dimensionless parameters $\upalpha$, $\upgamma$, and $\upxi$ regulate the contributions of the $z^2$, $z^4$, and $z^6$ terms, respectively, and will be determined in the course of the discussion.	
		The terms proportional to $z^2$ can be grouped together, making $\upalpha$ a small perturbative correction to the original quadratic dilaton. A new corrected thermal dilaton can thus be written as		
		\begin{flalign}
			\Phi_T(z,T) = \Phi(z,T) + \upalpha\frac{z^2}{z_h^2} = \kappa^2(T)z^2 + \upalpha\pi^2T^2z^2 = (1+\uprho_T)\kappa^2z^2 = \kappa_T^2 z^2,
		\end{flalign}
		where $\uprho_T$ is defined as
		\begin{flalign}
			\uprho_T = \left(\frac{9\upalpha\pi^2}{16} - \frac{N_f^2 - 1}{N_f}\right)\frac{16 T^2}{9\kappa^2} + \frac{(N_f^2 - 1)}{2N_f^2}\left(\frac{16 T^2}{9\kappa^2}\right)^2.
		\end{flalign}
		     	
	For convenience, we rewrite the holographic coordinate $z$ by introducing the Regge-Wheeler tortoise coordinate $r^*$, such that
	\begin{flalign}
		r^*= -\int \frac{dz}{f_T(z)} = \frac{z_h}{2}\left[-\arctan(\frac{z}{z_h}) + \frac{1}{2}\log(\frac{1-z/z_h}{1+z/z_h})\right],
	\end{flalign}
	and then writing $r\mapsto -r^*$ in order to simplify the notation.
	The four Minkowski coordinates $x=(t,\mathbf{x})$ remain unchanged.
	From this definition, we can expand $z$ in terms of $r$ as
	\begin{flalign}
		z = r\left[1-\frac{t_r^4}{5} + \mathcal{O}(t_r^8)\right],
	\end{flalign}
	with $t_r = r/z_h$, and thus rewrite the metric, 
	\begin{flalign}
		ds^2 = e^{2A(r)}f_T^{3/5}(r)\left[dt^2 - \frac{d\mathbf{x}^2}{f_T(r)}-dr^2\right],
		\label{metric_Tr}
	\end{flalign}
	where $f_T(r)=1-r^4/z_h^4$.
	
	The full thermal prefactor can also be rewritten:
	\begin{flalign}
		\Phi(r) + \lambda_T(r) = (1+\uprho_T)\kappa^2r^2 +\upgamma\frac{r^4}{z_h^4} + \frac{\kappa^2 r^6}{z_h^4}\left(\upxi-\frac{2}{5}\right)
	\end{flalign}
	As one can notice, the parameter $\upxi$ defines whether $r^6$ terms will be suppressed or not, which may happen depending on the hadron system considered. For our case, these terms are not necessary, so we will fix $\upxi=2/5$ from now on.	
	The remaining parameter $\upgamma$ will be fixed soon.
	
	With the preparations and the model concluded, the analysis from this point onward will be divided between mesons and baryons.
	
	\subsection{Mesons at finite temperature}
	
	To describe mesons in general with arbitrary spin $J$ at low temperature $T$, we consider a bosonic field $W_N = W_N(x,r,T)$ that satisfies an action analogous to that at zero temperature \cite{Branz:2010ub}, given by \cite{Gutsche:2019blp}
	\begin{flalign}
		\mathcal{S}_\text{W} = \frac{(-)^J}{2} \int d^4xdr \sqrt{-g} e^{\Phi_T -\upgamma\frac{r^4}{z_h^4}} \bigg[\left(\partial_N W_N\right)^2 - \left(m_J^2 + V_J\right)W^2 \bigg] 
	\end{flalign}
	with $\left(\partial_N W_N\right)^2 = \partial_N W_N \partial^N W^N$ and the potential $V_J = V_J (r,T)$ written as
	\begin{flalign}
		V_J (r,T) = \frac{e^{-2A(r)}}{f_T^{3/5}(r)} \left[\Phi''_T(r)+(3-2J)\Phi_T'(r) A'(r)\right],
	\end{flalign}
	where the prime indicates a derivative with relation to $r$, such that $h'=\partial_r h$.	
	The index $J$ indicates an explicit dependence on the total spin and should not be confused with a spacetime index.
	The metric is the same as in Eq. (\ref{metric_Tr}).
	
	The 5D mass of the field at finite temperature is related to its zero-temperature mass through the factor $f_T(r)$, in the form $m_J^2(r,T)=f_T^{-3/2}(r)m_J^2$. From the holographic dictionary that relates mass and conformal dimension, we can write
	\begin{flalign}
		m_J^2(r,T) = \frac{1}{f_T^{3/2}(r)}(\Delta-J)(\Delta+J-4).
	\end{flalign}
	
	The conformal dimension is given by $\Delta=N+L$, the sum of the number of partons $N$ and the maximum value of the z-component of the orbital angular momentum, $L=\max|L_z|$. 
	
	The procedure that follows is analogous to that performed for mesons in \Cref{subsection:softwall}, which includes performing a transformation of the type $W\mapsto e^{B_T/2}W$, such that
	\begin{flalign}
		B_T(r) = \Phi_T(r) + A(r)(2J-3) + \frac{r^4}{5z_h^4}(2J - 3 +5\upgamma).
	\end{flalign}	
	For scalar and pseudoscalar mesons ($J = 0$), the above expression reduces to $B_T=\Phi_T-3A$, while for vector and axial mesons ($J=1$) it becomes $B_T=\Phi_T-A$ (which is in accordance with what is observed in \Cref{subsection:softwall}). To ensure this remains valid, the parameter $\upgamma$ is fixed as $5\upgamma=J(J-3)+3$, in order to suppress the $r^4$ term in these cases. The expression then takes the form
	\begin{flalign}
		B_T(r) = \Phi_T(r) + A(r)(2J-3) + \frac{r^4}{5z_h^4}J(J-1).
	\end{flalign}
	Choosing an axial gauge $W_z(x,r,T)=0$, it is possible to expand the bosonic field through a Kaluza-Klein decomposition:
	\begin{flalign}
		W_J (x,r,T) = \sum_n W_{n,J} (x) w_{n,J}(r,T).
	\end{flalign}
	The functions $w_{n,J}(r,T)$ satisfy the same normalization and orthogonality conditions as in Eqs. (\ref{eq:uv_condition}) - (\ref{eq:ir_condition}). 
	After these steps, which also include taking the field to momentum space through a Fourier transform and identifying $k^2=-m_{w_{n,J}}^2(T)$, the equations of motion for $w_{n,J}(r,T)$, derived from the action, result in a Schrödinger-type equation given by
	\begin{flalign}
		\left[-\partial^2_r + U_J(r,T)\right]w_{n,J}(r,T) = m_{w_{n,J}}^2(T) w_{n,J}(r,T),
		\label{eq:gen_eom_mesonT}
	\end{flalign}
	with the effective potential defined as
	\begin{flalign}
		U_J(r,T) = \kappa^4r^2 + 2\kappa^2(J-1) + \frac{4\mu^2-1}{4r^2} + 2\rho_T \kappa^2(\kappa^2r^2 + J-1)
		+ \frac{4r^2}{5z_h^4}J(J-1)(\kappa^2r^2-J).
	\end{flalign}
	with the factor $\mu$ given by $\mu=N+\ell-2$ for mesons. The solution to the equation of motion (\ref{eq:gen_eom_mesonT}) is of the form
	\begin{flalign}
		w_{n,J}(r,T) = \sqrt{\frac{2\Gamma(n+1)}{\Gamma(n+\mu+1)}} e^{-\kappa_T^2r^2/2}r^{\mu+1/2}\kappa_\smallT^{\mu+1} L_n^\mu (\kappa_\smallT^2r^2),
	\end{flalign}
	where $\kappa_T^2=\kappa^2(1+\uprho_T)$, $\Gamma$ is the gamma function, and $L_n^\mu$ are the generalized Laguerre polynomials. The mass spectrum is then given by the following expression:
	\begin{flalign}
		m_{w_{n,J}}^2 (T) = & 4\kappa^2(1+\uprho_T)\left(n+\frac{L+J}{2}\right) \nonumber \\ 
		&+ \frac{4J(J-1)}{5}\bigg[(L+1)(L+2) + (6n-J)(n+\mu+1)-nJ\bigg]\frac{\pi^4 T^4}{\kappa^2}.
		\label{spectrum_mesonT}
	\end{flalign}
	It can be seen that, by setting $T = 0$, the solutions reduce to the case observed in the soft-wall model at zero temperature, as expected.

	\subsection{Baryons at finite temperature}
	
	Up to this point, we have somewhat deliberately avoided addressing baryons in detail. The reason is that baryons are, at first, more challenging to treat within the AdS/CFT correspondence, owing to the distinct behavior they exhibit compared to mesons when the large-$N_c$ limit is taken.	
	
	In \Cref{section:qcd}, baryons are studied within QCD under the SU$(3)$ gauge group, where they appear as bound states of $N_c=3$ quarks with different colors combined to form a color-neutral state. However, one may ask: what happens to this configuration when the number of colors is taken to infinity, i.e., $N_c \rightarrow \infty$?
	
	The first implication of this limit is quite straightforward. 
	The total baryon mass depends on the individual masses of its $N_c$ constituent quarks, as well as on the interaction energy among them. Consequently, the number of quarks (and, therefore, the number of colors) becomes the dominant factor in the final mass. Thus, one concludes that the baryon mass scales linearly with $N_c$, making baryons infinitely heavy in the large-$N_c$ limit \cite{Witten:1979kh}. Since the baryon size remains finite, constrained by the QCD confinement scale $\Lambda_{\text{QCD}}$, this corresponds to an object with extremely high energy density.
	
	Mesons, on the other hand, preserve their fundamental quark-antiquark structure, and their masses remain of order unity in the large-$N_c$ limit. Moreover, in the 't Hooft limit, their mutual interactions become exceedingly weak, scaling as $1/N_c$, rendering them effectively free and stable particles.
	
	In light of these results, Witten \cite{Witten:1979kh,Witten:1998xy} proposed a analogy between baryons and solitons. The mass of a soliton typically scales with the inverse of the coupling constant, thus diverging in the weak-coupling limit. So, considering that the coupling constant scales as $1/N_c$, then the mass of a soliton follows $M_\text{soliton}\sim 1/(1/N_c)\sim N_c$, a behavior that mirrors that of the baryon mass.
	One consequence of this insight was to revive the Skyrme model, originally proposed in the 1960s as a description of baryons and atom nuclei before the establishment of the quark model. In this framework, baryons are identified as skyrmions: non-trivial, topologically stable solutions in an effective field theory of mesons (specifically, pions) \cite{Brown:2010api}.	
	
	This description of baryons as solitonic objects finds a natural realization within top-down holographic approaches. In string theory, D-branes themselves are non-perturbative solitonic objects, and the holographic baryon is constructed from them. In the top-down Witten-Sakai-Sugimoto model \cite{Rebhan:2014rxa}, for instance, the baryon is realized as a "baryon vertex", a D-brane (specifically, a D4-brane) wrapped on a compact sphere within the extra-dimensional geometry. Then, $N_c$ strings emerges from this wrapped brane and ends on the stack of $N_c$ color D4-branes that generate the geometry, representing the valence quarks, and the wrapped brane binds them into a color-singlet state. From the four-dimensional perspective of the effective meson theory, this higher-dimensional soliton manifests precisely as a skyrmion, thus providing a compelling first-principles derivation of the soliton picture.
	
	While theoretically robust, this descriptions is unwieldy for phenomenological applications.
	In this work, we will continue with the bottom-up approach. Just like the meson, the physical baryon states in four dimensions, such as the nucleon and its resonances, are identified with the Kaluza-Klein tower of the dual field propagating in the five-dimensional bulk.
	According to this perspective, the procedure for baryons of spin $J$ at low temperature $T$ is not much different from what was done for mesons, only slightly more intricate due to the peculiarities of fermions. We consider a fermionic field $B_N=B_N(x,r,T)$ described by the action	\cite{Gutsche:2019pls}
	\begin{flalign}
		\mathcal{S}_\text{B} = \int d^4x dr \sqrt{-g}e^{-\Phi} \bar{B}_N (x,r,T)\mathcal{D}_{\pm} (r) B^N(x,r,T),
		\label{baryonaction}
	\end{flalign}
	where the derivative is given by
	\begin{flalign}
		\mathcal{D}_{\pm} = \frac{i}{2}\Gamma^M\left[\dvec\partial - \frac{1}{4}\omega_M^{ab}[\Gamma_a,\Gamma_b]\right] 
		\mp m_J(r,T) + V_J(r,T).
	\end{flalign}
	Here, the Dirac gamma matrices are arranged as $\Gamma^a=(\gamma^\mu,-i\gamma^5)$, with $\Gamma^M = \epsilon_a^M\Gamma^a$ and $\epsilon_a^M = z^{-1}\delta^M_a$. The potential is given by $V_J(r,T)=\Phi_T(r)/f_T^{3/10}(r)$, and $\omega_M^{ab} = (\delta^a_M\delta^b_r - \delta^b_M\delta^a_r)/(rf_T^{1/5}(r))$ denotes the spin connection term. Additionally, $\dvec\partial$ acts as
	\begin{flalign}
		\Gamma^M \dvec\partial_M = \Gamma^M \left(\cev{\partial}_M - \vec{\partial}_M\right) = g^{MN} \epsilon_N^a \Gamma_a\left(\cev{\partial}_M - \vec{\partial}_M\right).
	\end{flalign}

	After choosing the axial gauge $B_z(x,r,T)=0$, the fermion field can be expanded into its left- and right-chirality components, $B(x,r,T) = B^R(x,r,T)+ B^L(x,r,T)$, with
	\begin{flalign}
		B^{L,R}(x,r,T) = \frac{1\mp \gamma^5}{2} B(x,r,T),
	\end{flalign}
	and then expand each of them with a Kaluza-Klein decomposition,
	\begin{flalign}
		B^{L,R}(x,r,T) = \sum_{n} B^{L,R}_n(x) b_{n,J}^{L,R} (r,T),
	\end{flalign}
	the functions $b_{Jn}^{L,R}$ representing baryons with radial quantum number $n$ and total spin $J$ are obtained.
	
	After a simple transformation of the type $b_{n,J}^{L,R} \mapsto e^{A(r)(J-2)} b_{n,J}^{L,R}$, and by following the same procedure as previously applied to the mesons, the equation of motion for the fermionic field describing the baryons can be written as	
	\begin{flalign}
		\bigg[-\partial_r^2 + U^{L,R}(r,T)\bigg]b_{n,J}^{L,R}(r,T) = m_{b_{n,J}}^2(T)b_{n,J}^{L,R}(r,T),
		\label{schrodingerlikeT}
	\end{flalign}
	with the effective potential defined as
	\begin{flalign}
		U^{L,R}(r,T) &= \kappa^4r^2 + 2\kappa^2\left(\mu\mp\frac{1}{2}\right) + \frac{\mu(\mu\pm 1)}{r^2} +2\uprho_T\kappa^2\left(\kappa^2r^2+\mu\mp\frac{1}{2}\right).
	\end{flalign}
	For baryons, $\mu=N+L-3/2$. Defining $\mu^{L,R}=\mu\pm 1/2$, the solutions to the equation of motion at finite $T$ result in
	\begin{flalign}
		b_{n,J}^{L,R}(r,T) = \sqrt{\frac{2\Gamma(n+1)}{\Gamma(n+\mu^{L,R}+1)}} e^{-\kappa_\smallT^2r^2/2}r^{\mu^{L,R}+1/2}\kappa_\smallT^{\mu^{L,R}+1} L_n^{\mu^{L,R}} (\kappa_\smallT^2r^2),
	\end{flalign}
	with the mass spectrum of baryons resonances given by
	\begin{flalign}
		m_{b_{n,J}}^2(T) = 4\kappa^2(1+\uprho_\smallT)\left(n+\mu+\frac{1}{2}\right).
	\end{flalign}
	It is noteworthy that this expression is much simpler and more compact than for the meson case (\ref{spectrum_mesonT}), as it presents fewer temperature-induced corrections.
 	 	
 	\section{CIMs of light-flavor baryons at finite temperature}
 	\label{section:nucleons}
 		
 	In this section, configurational entropy is employed together with the finite-temperature AdS/QCD model to study baryons, specifically families of nucleons with different spin and parity values.
 	
 	In Ref. \cite{Ferreira:2020iry}, the nucleons $J^P=1/2^+$ and $J^P=3/2^+$ were analyzed using this same methodology, and the DCE was shown to successfully reproduce their Regge trajectories. Moreover, new resonances were estimated by extrapolating the trajectory equations, in the same manner as performed for mesons in the previous chapter.
 	
 	In this thesis, the analysis is extended to the nucleons with $J^P=3/2^-$, $J^P=5/2^-$, and $J^P=5/2^+$.
 	These are the last nucleon families that can be studied using the CIM-based AdS/QCD hybrid method, since obtaining the Regge-like trajectories requires interpolation of experimental data, and the remaining nucleon families have only a single known resonance each, namely: the $N(1990)$ with $J^P = 7/2^+$, the $N(2190)$ with $J^P = 7/2^-$, the $N(2220)$ with $J^P = 9/2^+$, the $N(2250)$ with $J^P = 9/2^-$, the $N(2600)$ with $J^P = 11/2^-$, and the $N(2700)$ with $J^P = 13/2^+$.
	 
	Since we are dealing with baryons, from this point onward we set the number of partons to three. We also fix $N_f = 2$, considering that nucleons are composed only of up and down quarks.
	 
	All other configurations remain the same as those established in the previous section, including the normalized solutions for light-flavor baryons, which can be written as
	\begin{subequations}
		\setlength{\abovedisplayskip}{2pt}
		\setlength{\belowdisplayskip}{2pt}
		\begin{alignat}{4}
			X^L_{n,\smallT}(r) = \sqrt{\frac{2n!}{\Gamma\left(n\!+\!\ell\!+\!\frac{5}{2}\right)}}{\color{black}e^{-\kappa_{\scalebox{.47}{\textsc{$T$}}}^2 r^2/2}}\kappa_\smallT^{\ell+\frac{5}{2}}r^{\ell+2}L_n^{\ell+\frac{3}{2}}(\kappa_\smallT^2 r^2),\\
			X^R_{n,\smallT}(r) = \sqrt{\frac{2n!}{\Gamma\left(n\!+\!\ell\!+\!\frac{3}{2}\right)}}{\color{black} e^{-\kappa_\smallT^2r^2/2}}\kappa_\smallT^{\ell+\frac{3}{2}}r^{\ell+1}L_n^{\ell+\frac{1}{2}}(\kappa_\smallT^2r^2).
		\end{alignat}
		\label{solutions_Tdependent}
	\end{subequations}
	 
	 The baryon mass spectrum is obtained by substituting these solutions into Eq. (\ref{schrodingerlikeT}), resulting in the following eigenvalue equation:
	 \begin{flalign}
		m_{n,\smallT}^2 = 4\kappa_\smallT^2\left(n+\ell+\frac{3}{2}\right).
		\label{spectra_Tdependent}
	\end{flalign}
	
 	In the zero-temperature limit, the $k_\smallT$ factor in the mass spectrum is simply reduced to $\kappa\equiv\lim_{\smallT\to0} \kappa_\smallT$. To analyze the mass spectrum in this section, we therefore adopt this limit.
 
 	The value of $\kappa$ is fixed in each case to best fit the experimental results, as done before in hadron spectroscopy analyses  \cite{FolcoCapossoli:2016uns}. Otherwise, using a single $\kappa$ value for all families would poorly reproduce the experimental baryon data. This approach causes the $\kappa$ value to carry its own uncertainty, which must be propagated in the mass spectrum.
  
\begin{figure}[H]
 		\centering
 		\includegraphics[width=11cm]{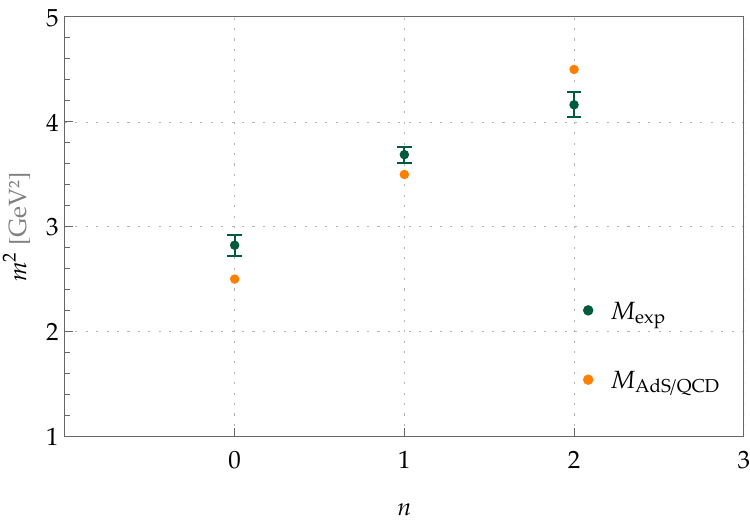}
 		\caption[Mass spectra for baryon resonances with $J^P=3/2^+$] {Mass spectra for baryon resonances with $J^P=3/2^+$ obtained from the AdS/QCD model and the experimental values in PDG.}
 		\label{fig:mass32p}
 	\end{figure}

	For the $J^P = 3/2^+$ baryon family, we fix $\kappa = 502 \pm 21$ MeV, which reduces the RMSD of the interpolation to $4.22$\%. \autoref{table:32p} presents the experimentally detected masses for the three resonances in this family, as well as the mass values obtained using Eq. (\ref{spectra_Tdependent}). \Cref{fig:mass32p} displays both sets of results. The uncertainties in the AdS/QCD-derived masses account for the statistical and measurement uncertainties of the experimental data, as well as the uncertainty in $\kappa$.
 
 	\begingroup
 	\renewcommand{\arraystretch}{1.2}
 	\begin{table}[H]
 		\centering
 		\setlength{\tabcolsep}{3mm}
 		\begin{tabular}{ccccc}
 			\rowcolor{ufabc!50}
 			\bf $n$ & \bf State & $M_\text{exp}$ (MeV) & $M_\text{AdS/QCD}$ (MeV) & RE (\%) \\
 			\rowcolor{ufabc!30} 
 			0 & $N(1720)$ & $1680^{+30}_{-20}$ & $1581 \pm 95$ & $5.88$ \\
 			\rowcolor{ufabc!10} 
 			1 & $N(1900)$ & $1920 \pm 20$ & $1870 \pm 99$ & $2.60$ \\
 			\rowcolor{ufabc!30} 
 			2 & $N(2040)$ & $2040^{+3}_{-4} \pm 25$ & $2121 \pm 119$ & $3.97$
 		\end{tabular}
 		\caption[Experimental and predicted mass spectra of the $J^P=3/2^+$ baryon family]{Experimental and predicted mass spectra of the $J^P=3/2^+$ baryon family. The experimental masses are taken from PDG \cite{ParticleDataGroup:2024cfk}, and the estimated values were obtained by solving Eq. (\ref{spectra_Tdependent}).}
 		\label{table:32p}
 	\end{table}
 	\endgroup
 	 	
\begin{figure}[H]
	 	\centering
	 	\includegraphics[width=11cm]{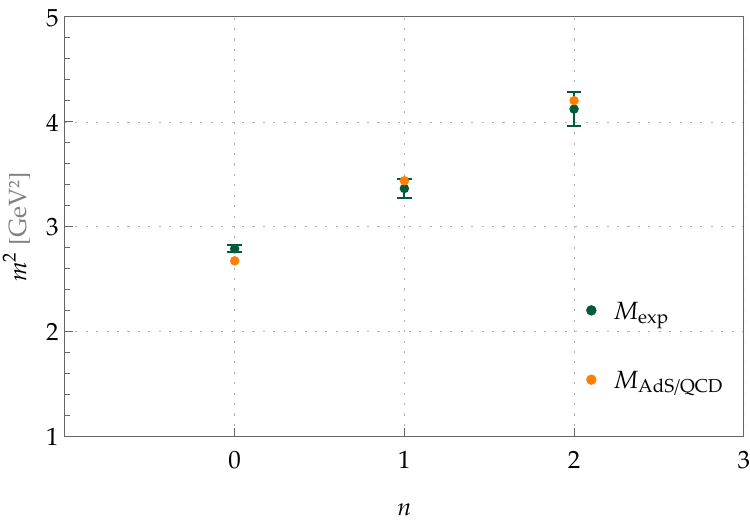}
	 	\caption[Mass spectra for baryon resonances with $J^P=5/2^+$] {Mass spectra for baryon resonances with $J^P=5/2^+$ obtained from the AdS/QCD model and the experimental values in PDG.}
	 	\label{fig:mass52p}
	 \end{figure}

	For the $J^P = 5/2^+$ baryon family, the fixed value is $\kappa = 437 \pm 6$ MeV, reducing the RMSD to $1.39$\%. The experimental and theoretical mass values are presented in \autoref{table:52p} and shown in \autoref{fig:mass52p}.
	 
	\begingroup
	 \renewcommand{\arraystretch}{1.2}
	 \begin{table}[H]
	 	\centering
	 	\setlength{\tabcolsep}{3mm}
	 	\begin{tabular}{ccccc}
	 		\rowcolor{ufabc!50}
	 		\bf $n$ & \bf State & $M_\text{exp}$ (MeV) & $M_\text{AdS/QCD}$ (MeV) & RE (\%) \\
	 		\rowcolor{ufabc!30} 
	 		0 & $N(1680)$ & $1670 \pm 10$ & $1635 \pm 33$ & $2.1$ \\
	 		\rowcolor{ufabc!10} 
	 		1 & $N(1860)$ & $1834 \pm 19.4$ & $1854 \pm 51$ & $1.1$ \\
	 		\rowcolor{ufabc!30} 
	 		2 & $N(2000)$ & $2030 \pm 40$ & $2049 \pm 68$ & $0.9$
	 	\end{tabular}
	 	\caption[Experimental and predicted mass spectra of the $J^P=5/2^+$ baryon family]{Experimental and predicted mass spectra of the $J^P=5/2^+$ baryon family \cite{ParticleDataGroup:2024cfk}. The estimated masses were obtained by solving Eq. (\ref{spectra_Tdependent}).}
	 	\label{table:52p}
	 \end{table}
	 \endgroup
	 
\begin{figure}[H]
		\centering
		\includegraphics[width=11cm]{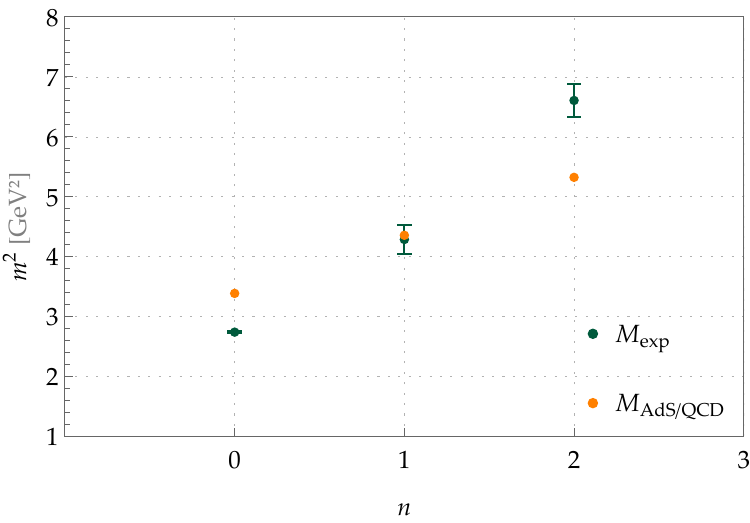}
		\caption[Mass spectra for baryon resonances with $J^P=5/2^-$] {Mass spectra for baryon resonances with $J^P=5/2^-$ obtained from the AdS/QCD model and the experimental values in PDG.}
		\label{fig:mass52n}
	\end{figure}

	 For the $J^P = 5/2^-$ baryon family, \autoref{table:52n} and \autoref{fig:mass52n} display the corresponding experimental and theoretical mass values. In this case, the parameter $\kappa = 492 \pm 44$ MeV reproduces the mass spectrum with an RMSD of $8.85$\%.
	
	\begingroup
	\renewcommand{\arraystretch}{1.2}
	\begin{table}[H]
		\centering
		\setlength{\tabcolsep}{3mm}
		\begin{tabular}{ccccc}
			\rowcolor{ufabc!50}
			\bf $n$ & \bf State & $M_\text{exp}$ (MeV) & $M_\text{AdS/QCD}$ (MeV) & RE (\%) \\
			\rowcolor{ufabc!30} 
			0 & $N(1675)$ & $1655 \pm 5$ & $1840 \pm 169$ & $11.2$ \\
			\rowcolor{ufabc!10} 
			1 & $N(2060)$ & $2070^{+60}_{-50}$ & $2087 \pm 245$ & $0.82$ \\
			\rowcolor{ufabc!30} 
			2 & $N(2570)$ & $2570^{+19}_{-10} \pm 34$ & $2307 \pm 251$ & $10.2$
		\end{tabular}
		\caption[Experimental and predicted mass spectra of the $J^P=5/2^-$ baryon family]{Experimental and predicted mass spectra of the $J^P=5/2^-$ baryon family. Experimental masses from PDG \cite{ParticleDataGroup:2024cfk}, and the estimated masses obtained by solving Eq. (\ref{spectra_Tdependent}).}
		\label{table:52n}
	\end{table}
	\endgroup
	
	 From the solutions (\ref{solutions_Tdependent}), we can also write the energy density
	  \begin{flalign}
	 	T_{00} = \frac{M_{n,\smallT}}{r} \left[\big(X_{n,\smallT}^L(r)\big)^2 + \big(X_{n,\smallT}^R(r)\big)^2\right],
	 	\label{energydensityT}
	 \end{flalign}
	from which the configurational entropy values are obtained. In the following section, we proceed with the analysis using the computed DCE results.
 	 	
 	\subsection{DCE of the $J^P=3/2^+$ baryon family}
 	
 	In the following, we analyze baryon resonances using the DCE, in the same manner as for mesons in the previous chapter. At first, we adopt the limit $T \to 0$.  
 	
 	For the baryon family with $J^P = 3/2$, the DCE is computed in the same way as before, and the results are presented in \autoref{table:dce32p}.  
 	
 	\begingroup
 	\renewcommand{\arraystretch}{1.2}
 	\begin{table}[H]
 		\centering
 		\setlength{\tabcolsep}{3mm}
 		\begin{tabular}{ccc}
 			\rowcolor{ufabc!50}
 			\bf $n$ & \bf State &  {\bf DCE} (nat)	  \\
 			\rowcolor{ufabc!30}	1 	& $N(1720)$ 	& $0.9960$ 	\\
 			\rowcolor{ufabc!10}	2	& $N(1900)$ 	& $1.1458$	\\
 			\rowcolor{ufabc!30}	3	& $N(2040)$ 	& $1.2290$
 		\end{tabular}
 		\caption{DCE of the radial resonances of the $J^P=3/2^+$ baryon family.}
 		\label{table:dce32p}	
 	\end{table}
 	\endgroup
 	 	
\begin{figure}[H]
 		\centering
 		\includegraphics[width=11cm]{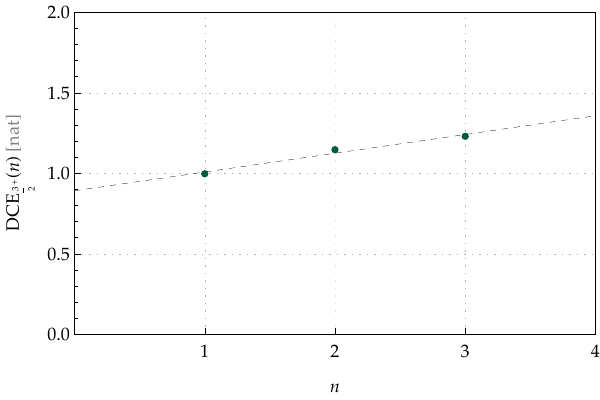}
 		\caption {DCE of the $J^P=3/2^+$ baryons as a function of the $n$ quantum number.}
 		\label{fig:dce32p_n}
 	\end{figure}

 	By interpolating the values obtained above, it is possible to construct the first type of DCE-Regge-like trajectory, which describes the DCE as a function of the radial quantum number. Using a linear interpolation to avoid overfitting, the expression for this trajectory is given by 
 	\begin{flalign}
 		\dce_{\frac{3}{2}^+} (n) = 0.1165  n + 0.8906,
 		\label{dcen32p}
 	\end{flalign}
 	with an RMSD of 1.5 \%.  \autoref{fig:dce32p_n} shows this interpolation.
 	
\begin{figure}[H]
 		\centering
 		\includegraphics[width=11cm]{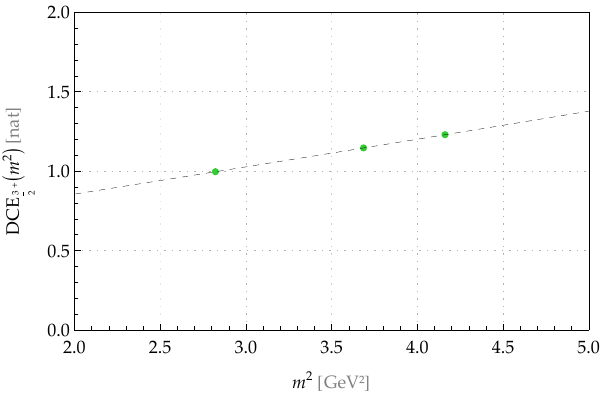}
 		\caption {DCE of the $J^P=3/2^+$ baryons as a function of their squared mass, for $n=1,2,3$.}
 		\label{fig:dce32p_m}
 	\end{figure}

 	From the calculated values, one can also derive an expression relating the DCE to the resonances masses, the second type of DCE-Regge-like trajectory, which results in  
 		\begin{flalign}
 			\dce_{\frac{3}{2}^+} (m^2) = 1.2728\cdot 10^{-3} m^4 + 0.1650m^2 + 0.5199,
 			\label{dcem32p}
 		\end{flalign}
 	with an RMSD of 1.5 \%. \autoref{fig:dce32p_m} shows this interpolation.
 	
 	Using both trajectories, the extrapolated masses for higher resonant states can be obtained. Inserting $n=3$ into Eq. (\ref{dcen32p}), we find a DCE equal to $1.3566$ nat. Substituting this value into the left-hand side of Eq. (\ref{dcem32p}) and solving the resulting algebraic equation, we obtain a mass for the $N_3^\star$ state equal to $2212\pm29$ MeV. Repeating the process for $n=4$, the $N_4^\star$ state yields a DCE value of $1.4731$ nat, corresponding to a mass of $2359\pm31$ MeV. Similarly, for $N_5^\star$, the DCE is $1.5896$ nat,  resulting in a mass $2497\pm33$ MeV. These results are summarized in \autoref{table:mass32p}.  

 	\begingroup
 	\renewcommand{\arraystretch}{1.2}
 	\begin{table}[H]
 		\centering
 		\setlength{\tabcolsep}{3mm}
 		\begin{tabular}{cccc}
 			\rowcolor{ufabc!50}
 			\bf $n$ & \bf State &  {\bf $M_{\text{exp}}$} (MeV) & {\bf $M_{\text{th}}$} (MeV) \\
 			\rowcolor{ufabc!30}	0 	& $N(1720)$ 	& $1680^{+30}_{-20}$ 	& $1581 \pm 28$ 	\\
 			\rowcolor{ufabc!10}	1	& $N(1900)$ 	& $1920 \pm 20$ 	& $1870 \pm 20$	\\
 			\rowcolor{ufabc!30}	2	& $N(2040)$ 	& $2040^{+3}_{-4}\pm25$ 	& $2121 \pm 4$	\\
 			\rowcolor{ufabc!10}	3	& $N^\star_3$ 	& --- 	& $2212 \pm 29$	\\
 			\rowcolor{ufabc!30}	4	& $N^\star_4$ 	& --- 	& $2359 \pm 31$	\\
 			\rowcolor{ufabc!10}	5	& $N^\star_5$ 	& --- 	& $2497 \pm 33$	\\
 		\end{tabular}
 		\caption[Masses of the radial resonances of the $J^P=3/2^+$ baryon family]{Masses of the radial resonances of the $J^P=3/2^+$ baryon family. The extrapolated masses for $n=3,4,5$, indicated with the "$\star$", were obtained through the combined use of Eqs. (\ref{dcen32p}, \ref{dcem32p})}
 		\label{table:mass32p}
 	\end{table}
 	\endgroup
 	
 	Unlike the meson case, few higher baryonic states have been experimentally observed. Therefore, no cataloged resonances with appropriate quark composition and main quantum numbers were found within the mass range corresponding to the predicted states in this analysis.
 	
 	\subsection{DCE of the $J^P=5/2^+$ baryon family}
 	
  	We now turn our attention to the baryon family with $J^P = 5/2^+$, following the same general methodology previously applied to the $J^P = 3/2^+$ states. All calculations are once again carried out in the limit $T \to 0$.  
 	
 	The configurational entropy values obtained for the $J^P = 5/2^+$ baryons are listed in \autoref{table:dce52p}. 
 	
 	\begingroup
 	\renewcommand{\arraystretch}{1.2}
 	\begin{table}[H]
 		\centering
 		\setlength{\tabcolsep}{3mm}
 		\begin{tabular}{ccc}
 			\rowcolor{ufabc!50} \bf $n$ & \bf State & {\bf DCE} (nat) \\
 			\rowcolor{ufabc!30} 0 & $N(1680)$ & $0.9257$ \\
 			\rowcolor{ufabc!10} 1 & $N(1860)$ & $1.0997$ \\
 			\rowcolor{ufabc!30} 2 & $N(2000)$ & $1.1822$
 		\end{tabular}
 		\caption{DCE of the radial resonances of the $J^P=5/2^+$ baryon family.}
 		\label{table:dce52p}
 	\end{table}
 	\endgroup
 	 	
\begin{figure}[H]
 		\centering
		\includegraphics[width=11cm]{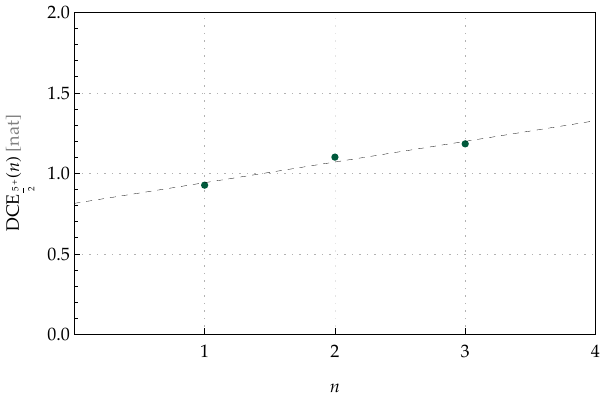}
 		\caption {DCE of the $J^P=5/2^+$ baryons as a function of the $n$ quantum number.}
 		\label{fig:dce52p_n}
 	\end{figure}

 	From these numerical values, we can construct the first type of DCE-Regge-like trajectory, obtaining
 	\begin{flalign}
 		\dce_{\frac{5}{2}^+} (n) = 0.1283 n + 0.8127,
 		\label{dcen52p}
 	\end{flalign}
 	within 1.5\% RMSD. This interpolation is shown in \autoref{fig:dce52p_n}.
 	
\begin{figure}[H]
 		\centering
 		\includegraphics[width=11cm]{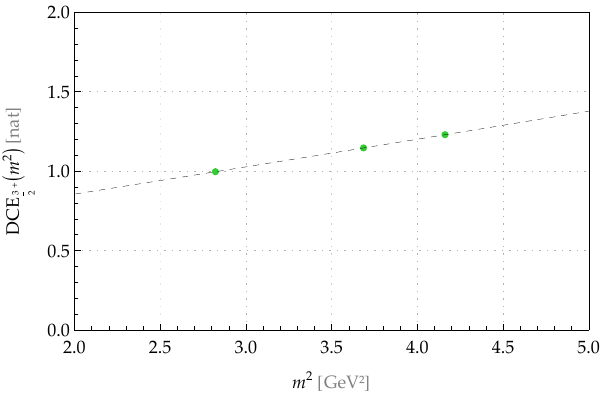}
 		\caption {DCE of the $J^P=5/2^+$ baryons as a function of their squared mass, for $n=1,2,3$.}
 		\label{fig:dce52p_m}
 	\end{figure}

 	In addition, a second type of trajectory can be derived by relating the DCE to the resonance masses. This trajectory can be expressed as
 		\begin{flalign}
	 			\dce_{\frac{5}{2}^+} (m^2) = - 0.1455 m^4 + 1.1982 m^2 - 1.2840,
 			\label{dcem52p}
 		\end{flalign}
  	within 1.5\% RMSD. \autoref{fig:dce52p_m} shows this interpolation.
 	 
 	By combining both relations, one can estimate the masses of higher excitations.  
 	For $n = 3,4,5$, corresponding to the extrapolated states $N_3^\star$, $N_4^\star$, and $N_5^\star$, respectively, we obtain the following DCE values from Eq. (\ref{dcen52p}): $1.3257$ nat, $1.4539$ nat, and $1.5822$ nat.
	By inserting these values into Eq. (\ref{dcem52p}), we derive the corresponding mass values summarized in \autoref{table:mass52p}.
 	
 	\begingroup
 	\renewcommand{\arraystretch}{1.2}
 	\begin{table}[H]
 		\centering
 		\setlength{\tabcolsep}{3mm}
 		\begin{tabular}{cccc}
 			\rowcolor{ufabc!50} \bf $n$ & \bf State & {\bf $M_{\text{exp}}$} (MeV) & {\bf $M_{\text{th}}$} (MeV) \\
 			\rowcolor{ufabc!30} 0 & $N(1680)$ & $1670 \pm 10$ & $1635 \pm 10$ \\
 			\rowcolor{ufabc!10} 1 & $N(1860)$ & $1834 \pm 19 \pm 6$ & $1854 \pm 25$ \\
 			\rowcolor{ufabc!30} 2 & $N(2000)$ & $2030 \pm 40$ & $2049 \pm 40$ \\
 			\rowcolor{ufabc!10} 3 & $N^\star_3$ & --- & $2187 \pm 29$ \\
 			\rowcolor{ufabc!30} 4 & $N^\star_4$ & --- & $2338 \pm 31$ \\
 			\rowcolor{ufabc!10} 5 & $N^\star_5$ & --- & $2479 \pm 32$
 		\end{tabular}
 		\caption[Masses of the radial resonances of the $J^P=5/2^+$ baryon family]{Masses of the radial resonances of the $J^P=5/2^+$ baryon family. The extrapolated masses for $n = 3, 4, 5$, indicated with the "$\star$", were obtained through the combined use of Eqs. (\ref{dcen52p}, \ref{dcem52p}).}
 		\label{table:mass52p}
 	\end{table}
 	\endgroup
 
 	As before, the number of experimentally detected states in this family remains limited. Consequently, no resonance has been identified so far within the mass intervals predicted by this extrapolation, though the results provide a useful reference for future investigations.
 
  	\subsection{DCE of the $J^P=5/2^-$ baryon family}
 	
 	Following the same procedure, we now analyze the baryonic resonances with $J^P = 5/2^-$. As before, the analysis is performed in the limit $T \to 0$.
 	
 	The DCE values obtained for each radial excitation are listed in \autoref{table:dce52n}.
 	
 	\begingroup
 	\renewcommand{\arraystretch}{1.2}
 	\begin{table}[H]
 		\centering
 		\setlength{\tabcolsep}{3mm}
 		\begin{tabular}{ccc}
 			\rowcolor{ufabc!50} \bf $n$ & \bf State & {\bf DCE} (nat) \\
 			\rowcolor{ufabc!30} 0 & $N(1675)$ & $1.0801$ \\
 			\rowcolor{ufabc!10} 1 & $N(2060)$ & $1.2542$ \\
 			\rowcolor{ufabc!30} 2 & $N(2570)$ & $1.3367$
 		\end{tabular}
 		\caption{DCE of the radial resonances of the $J^P=5/2^-$ baryon family.}
 		\label{table:dce52n}
 	\end{table}
 	\endgroup
 	
\begin{figure}[H]
 		\centering
 		\includegraphics[width=11cm]{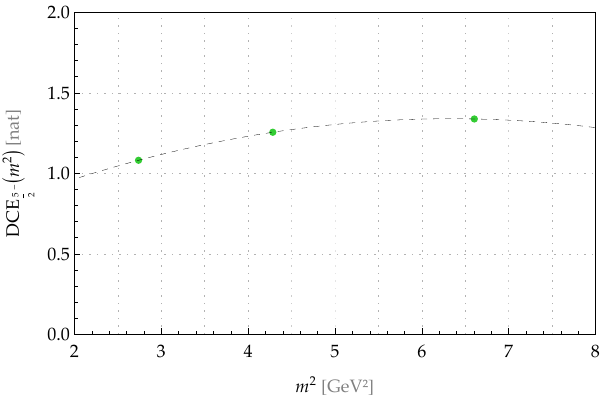}
 		\caption {DCE of the $J^P=5/2^-$ baryons as a function of their squared mass, for $n=1,2,3$.}
 		\label{fig:dce52n_m}
 	\end{figure}

 	Interpolating these results allows us to construct the first type of DCE-Regge-like trajectory, which describes the DCE as a function of the radial quantum number. Employing a linear interpolation to avoid overfitting, we find:
 	\begin{flalign}
 		\dce_{\frac{5}{2}^-} (n) = 0.1283 n + 0.9671
 		\label{dcen52n}
 	\end{flalign}
 	within 1.5\%  RMSD. \autoref{fig:dce52n_m} shows this interpolation.
 	
 	\begin{figure}[H]
 		\centering
 		\includegraphics[width=11cm]{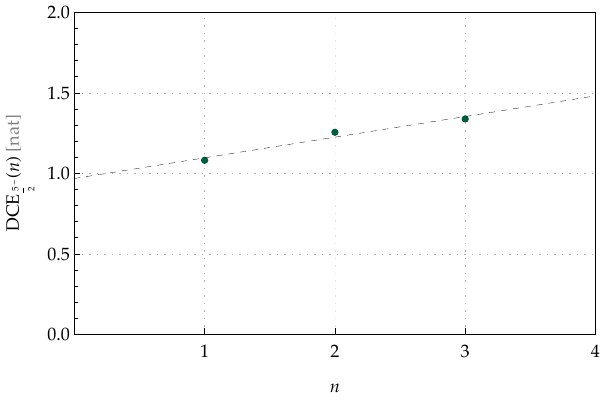}
 		\caption {DCE of the $J^P=5/2^-$ baryons as a function of the $n$ quantum number.}
 		\label{fig:dce52n_n}
 	\end{figure}
 	
 	From the DCE values it is also possible to obtain the second type of DCE-Regge-like trajectory, relating the DCE to the baryonic mass. The corresponding expression is given by
 	\begin{flalign}
 			\dce_{\frac{5}{2}^-} (m^2) = - 1.9933\cdot 10^{-2} m^4 + 0.2526 m^2 + 0.5377,
 			\label{dcem52n}
 	\end{flalign}
 	within 1.5\%  RMSD. This interpolation is shown in \autoref{fig:dce52n_n}.
 	
 	Using Eq. (\ref{dcen52n}), for $n = 3,4,5$, which correspond to the extrapolated states $N_3^\star$, $N_4^\star$, and $N_5^\star$, respectively, we find the following DCE values: $1.4803$ nat, $1.6086$ nat, and $1.7369$ nat.
 	Substituting these values into Eq. (\ref{dcem52n}), we obtain the corresponding masses summarized in \autoref{table:mass52n}.
 	
 	\begingroup
 	\renewcommand{\arraystretch}{1.2}
 	\begin{table}[H]
 		\centering
 		\setlength{\tabcolsep}{3mm}
 		\begin{tabular}{cccc}
 			\rowcolor{ufabc!50} \bf $n$ & \bf State & {\bf $M_{\text{exp}}$} (MeV) & {\bf $M_{\text{th}}$} (MeV) \\
 			\rowcolor{ufabc!30} 0 & $N(1675)$ & $1655 \pm 5$ & $1840 \pm 169$ \\
 			\rowcolor{ufabc!10} 1 & $N(2060)$ & $2070^{+60}_{-50}$ & $2087 \pm 245$ \\
 			\rowcolor{ufabc!30} 2 & $N(2570)$ & $2570^{+19}_{-10} \pm 34$ & $2307 \pm 251$ \\
 			\rowcolor{ufabc!10} 3 & $N^\star_3$ & --- & $2925 \pm 51$ \\
 			\rowcolor{ufabc!30} 4 & $N^\star_4$ & --- & $3250 \pm 57$ \\
 			\rowcolor{ufabc!10} 5 & $N^\star_5$ & --- & $3545 \pm 62$
 		\end{tabular}
 		\caption[Masses of the radial resonances of the $J^P=5/2^-$ baryon family]{Masses of the radial resonances of the $J^P=5/2^-$ baryon family. The extrapolated masses for $n = 3, 4, 5$, indicated with the "$\star$", were obtained through the combined use of Eqs. (\ref{dcen52n}, \ref{dcem52n}).}
 		\label{table:mass52n}
 	\end{table}
 	\endgroup
 	
 	Again, no cataloged resonances with suitable quark composition and quantum numbers were found within the mass range corresponding to the predicted states in this analysis.

 	\subsection{DCE of baryons at finite temperature}

\begin{figure}[H]
 		\centering
 		\includegraphics[width=11cm]{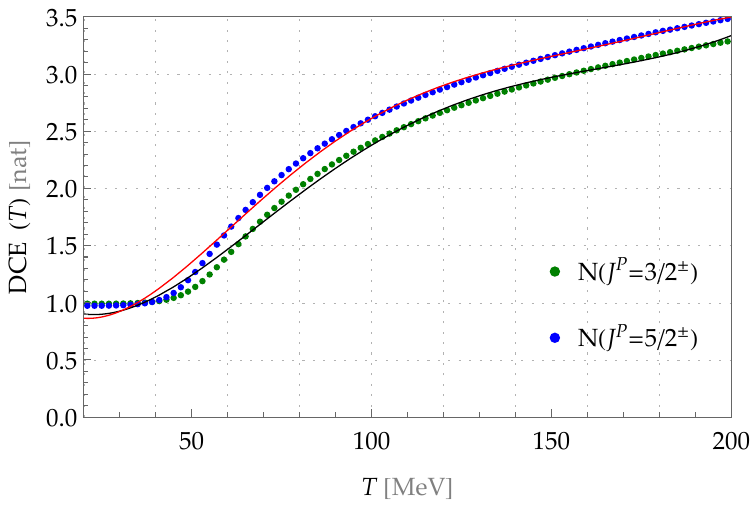}
 		\caption {DCE of the ground states of baryon resonances $J^P = 5/2^+$ and $J^P = 5/2^-$ as a function of the	temperature}
 		\label{fig:DCET}
 	\end{figure}

 	In the finite-temperature limit, the solutions (\ref{solutions_Tdependent}) are considered in their complete form, as well as the energy density (\ref{energydensityT}). Using these expressions, the DCE can be computed as a function of temperature.
 	In this case, the variables are minimized considering the ground state of the baryons in each family. Since parity is not involved in the analysis, there is no distinction between the $J^P = 5/2^+$ and $J^P = 5/2^-$ families.
 	The numerical results are presented in \autoref{fig:DCET}.
 	
 	A numerical interpolation of these results for the DCE as a function of temperature, up to fifth order, can be written as
 	\begin{flalign}
 		\text{DCE}_{\frac{3}{2}}(T) = &-3.4923\times10^{-11} T^5 + 3.1293\times10^{-8} T^4 \nonumber \\ &- 9.5495\times10^{-6} T^3 + 1.1930\times10^{-3} T^2 \nonumber \\  &- 4.0997\times10^{-2} T + 1.3159,
 	\end{flalign}
 	for the baryon resonances of $J^P=3/2^+$, and
 	\begin{flalign}
 		\text{DCE}_{\frac{5}{2}}(T) = &-1.0298\times10^{-10} T^5 + 6.8249\times10^{-8} T^4 \nonumber \\  &- 1.6699\times10^{-5} T^3 + 1.7596\times10^{-3} T^2 \nonumber \\  &- 5.4540\times10^{-2} T + 1.3742,
 	\end{flalign}
 	for the families with  $J^P=5/2^\pm$, within 2\% RMSD for both interpolations.
 	The corresponding interpolation curves are also displayed in \autoref{fig:DCET}.
 	
 	From the plot of DCE versus temperature, one can observe stable configurations for both families in the range from $0$ to $35$ MeV, with a DCE of approximately $0.99$ nat for the $J^P = 3/2$ baryons and $0.97$ nat for the $J^P = 5/2$ baryons.
 	Starting from around $40$ MeV, the DCE increases monotonically for both families, with a negative second derivative with respect to temperature, indicating a decrease in the system's stability. A sharper increase in instability occurs around $150$ MeV, coinciding with the Hagedorn temperature and the expected instability associated with hadron melting.
 	
\begin{figure}[H]
 		\centering
 		\includegraphics[width=11cm]{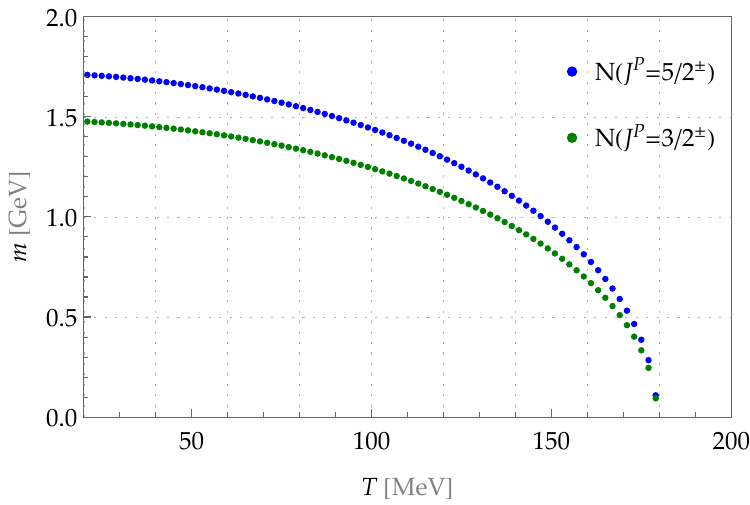}
 		\caption {Masses of the ground state of baryon resonances $J^P = 5/2^+$ and $J^P = 5/2^-$ as a function of the	temperature.}
 		\label{fig:mT}
 	\end{figure}

 	Using Eq. (\ref{spectra_Tdependent}) together with the explicit expression for $\kappa_T$, the baryon mass spectrum can be computed as a function of temperature. The results are shown in \autoref{fig:mT}.
 	 
 	In this plot, a decrease in the baryon mass is observed as the temperature increases, until it reaches zero. This point, located near $180$ MeV, may indicate a phase transition temperature corresponding to the transition from hadronic matter to the QGP.
 	The obtained value of $180$ MeV is consistent with the melting temperature of mesons found in Ref. \cite{Vega:2018dgk}, in an analysis using thermal dilatons.
 		
 	Furthermore, this result is in good agreement with the estimates for the transition temperature discussed in \Cref{sec:phasetransition}.
 	
 	\subsection{Concluding remarks}	
 	 	
  	In this section, the resonances of the $J = 3/2^+$, $5/2^+$, and $5/2^-$ baryons were analyzed using the soft-wall model at finite temperature, including the zero-temperature limit.
 	
 	The Regge-like trajectories were obtained as functions of the radial quantum number $n$ and the mass $m$, and were used to estimate the mass spectrum of higher excited states beyond those experimentally detected. Despite the consistent results, no possible correspondence was found with unidentified states in PDG \cite{ParticleDataGroup:2024cfk} that simultaneously matched the mass range, quantum numbers, and quark composition.
 	
 	On the other hand, the DCE proved to be a powerful tool for studying the stability of finite-temperature systems and to probe features of the hadronic phase transition. At low temperatures, the DCE for all analyzed baryon families remains nearly constant, indicating stability of hadronic states up to approximately $35$ MeV. Then, the DCE increases monotonically, implying growing instability as the temperature rises, with a sharper growth near the Hagedorn temperature, around $150$ MeV. This result reinforces previous calculations for the temperature at which hadrons melt \cite{Rafelski:2015cxa}.
 	
 	Similarly, the plot of baryon mass as a function of temperature shows a decrease in the spectrum until a point where the mass vanishes, at approximately $180$ MeV. Since this value is the same for all the light baryon families studied, it can be associated with the phase transition temperature. These results reinforce that the DCE is an effective tool to describe hadronic stability, even during the phase transition.
	 		
 	These results were published in \cite{daRocha:2025gcz}.

\chapter{Conclusion}
\label{chapter:conclusion}
	
	The work developed in this thesis consists of analyzing the mass spectrum of mesons and baryons using AdS/QCD models in conjunction with configurational information measures.  
	Three different scenarios were investigated, each arising from modifications of the soft-wall model in holographic QCD: light mesons in a dynamically consistent model with anomalous mass corrections; strange vector kaons taking into account the effects of constituent quark masses; and baryons within a finite-temperature model.  
	
	In all cases, the analysis was complemented by evaluating configurational information measures, such as entropy and complexity, of the hadronic systems. Within this approach, Regge-like trajectories were constructed, establishing spectra that relate configurational information to parameters such as the radial quantum number $n$ and the masses of hadronic resonances.  	
	In this way, the CIMs act as intermediaries, allowing for the construction of equations that directly connect the values of $n$ to the mass spectrum, enabling extrapolations to higher excited states. These estimated values were then compared with cataloged results for further states in the Particle Data Group, in search of possible correspondences.
	
	In the light meson sector, the application of the dynamical soft-wall model with a range of four different anomalous mass corrections allowed for the derivation of mass spectra for pseudoscalar, axial-vector, scalar, and vector families. The predicted masses for high radial excitations were compared against the further states listed by the PDG. This comparison revealed potential identifications for several unclassified resonances, notably, the $(a_1)_6^\star$ and the $\rho_8^\star$ and $\rho_9^\star$ states showed strong correspondence with the $X(2340)$ state, while the $\pi_6^\star$ was found to be compatible with the $X(2680)$ in all cases. In the scalar sector, the $(f_0)_{10}^\star$ and $(f_0)_{11}^\star$ states were found to exhibit possible correspondences with the $X(2600)$, $X(2632)$, and $X(2680)$ states.

	The investigation was subsequently extended to the strange vector kaon family. To account for the heavier strange quark, the standard SW model was adapted using a deformed dilaton profile whose main parameters depend on the constituent quark masses. This modification successfully reproduced the non-linear Regge trajectories expected for hadrons containing quarks beyond the light up and down flavors.  
	At this stage, the configurational complexity was also computed for the entire strange vector kaon family in order to corroborate the results obtained with the entropy, which was indeed confirmed. Among the identified correspondences, it is worth highlighting the possible match between the $n=6$ state and the $X(2210)$ state, which was further supported by both DCE and DCC.

	A consistent finding across all analyzed meson families is the monotonic increase of configurational entropy and complexity with the radial quantum number $n$. This behavior indicates that configurational instability grows with the excitation level. Such a result offers a theoretical interpretation for the scarcity of experimental detections regarding higher excited states, suggesting that their increased configurational instability makes them short-lived and difficult to isolate.

	Finally, the study addressed the thermal properties of baryons with spin $J = 3/2^+$, $5/2^+$, and $5/2^-$ at finite temperature. Although the spectroscopic analysis yielded no matches with current PDG data for these specific sectors, the DCE demonstrated its utility as a probe for phase transitions. The results indicate that hadronic states remain stable at low temperatures, up to approximately $35$ MeV, before exhibiting a sharp rise in entropy near the Hagedorn temperature ($\approx 150$ MeV). Coinciding with this instability, the baryon masses were observed to vanish at a critical temperature of approximately $180$ MeV, marking the deconfinement phase transition.
	
	In summary, the results presented in this thesis indicate that the combined application of AdS/QCD models with CIMs can provide consistent and predictive insights into hadronic properties across a wide range of scenarios. This approach facilitates the potential identification of new hadronic resonances and also proves capable of offering a detailed description of the stability and dynamical evolution of strongly coupled matter.

		\begin{appendices}

			\chapter{Full derivations of the soft-wall model}
			\label{appendix:fullsoftwall}
			
			\section{Vacuum expectation value of the chiral condensate}
			
				This section details the derivation of the equation of motion for the vacuum expectation value (VEV) of the scalar field $X$. This field is the holographic dual of the chiral condensate and is therefore responsible for the spontaneous breaking of chiral symmetry in the model.
			
			Assuming that the background field depends only on the holographic coordinate $z$, and setting all dynamical meson fluctuations to zero ($S=V=A=\pi=0$), the bulk scalar field reduces to its vacuum profile:
			\begin{flalign}
				X_0(z) = \frac{\chi(z)}{2}\mathds{1},
			\end{flalign}
			where $\mathds{1}$ is the $3\times3$ identity matrix in flavor space.		
			
			Substituting this parameterization into the general scalar action and computing the trace ($\Tr(\mathds{1})=3$) over the flavor indices, the vacuum action becomes:
			\begin{flalign}
				\mathcal{S}_0 &= \int d^5x \sqrt{-g} e^{-\Phi} \Tr\Big[\partial_M X_0 \partial^M X_0 + m_X^2 X_0^2\Big] \nonumber \\
				&= \int d^5x \sqrt{-g} e^{-\Phi} \frac{3}{4} \Big[g^{zz}(\partial_z \chi)^2 + m_X^2 \chi^2\Big].
			\end{flalign}
			
			Using the metric determinant $\sqrt{-g}=z^{-5}$ (setting the AdS radius $L=1$), the inverse metric component $g^{zz}=z^2$, and the standard bulk scalar mass squared $m_X^2=-3$, we can extract the effective Lagrangian density for $\chi(z)$:
			\begin{flalign}
				\mathcal{L}_\chi = \frac{3e^{-\Phi}}{4z^3}(\chi')^2 - \frac{9e^{-\Phi}}{4z^5}\chi^2,
			\end{flalign}
			where the prime denotes the derivative with respect to the holographic coordinate $z$.
			
			The equation of motion can be derived using the Euler-Lagrange equation,
			\begin{flalign}
				\partial_z \left( \frac{\partial \mathcal{L}_\chi}{\partial \chi'} \right) - \frac{\partial \mathcal{L}_\chi}{\partial \chi} = 0.
				\label{eulerlagrangeX}
			\end{flalign}
			Computing the terms
			\begin{flalign}
				\frac{\partial \mathcal{L}_\chi}{\partial \chi'} = \frac{e^{-\Phi}}{z^3}\chi', \quad \text{and} \quad
				\frac{\partial \mathcal{L}_\chi}{\partial \chi} = -\frac{3e^{-\Phi}}{z^5}\chi,
			\end{flalign}
			and substituting back into (\ref{eulerlagrangeX}) yields:
			\begin{flalign}
				\partial_z \left( \frac{e^{-\Phi}}{z^3}\chi' \right) + \frac{3e^{-\Phi}}{z^5}\chi = 0.
			\end{flalign}
			Expanding the derivative term considering $\Phi=\Phi(z)$,
			\begin{flalign}
				\frac{e^{-\Phi}}{z^3}\chi'' - \frac{3e^{-\Phi}}{z^4}\chi' - \frac{e^{-\Phi}\Phi'}{z^3}\chi' + \frac{3e^{-\Phi}}{z^5}\chi = 0,
			\end{flalign}
			and multiplying the entire equation by $z^3 e^{\Phi}$, we obtain the final equation of motion for the chiral condensate:
			\begin{flalign}
				\chi'' - \left(\Phi' + \frac{3}{z}\right)\chi' + \frac{3}{z^2}\chi = 0.
				\label{eom_chi}
			\end{flalign}	
			
			In the hard-wall model, the dilaton field is absent ($\Phi=0$), and the resulting equation of motion takes the form  
			\begin{flalign}
				\chi'' - \frac{3}{z}\chi' + \frac{3}{z^2}\chi =0,
			\end{flalign}  
			which corresponds to a Cauchy-Euler differential equation, whose general solution is a power law, $\chi\sim z^\alpha$. 
			Substituting this ansatz into the equation yields the characteristic polynomial
			\begin{flalign}
				\alpha(\alpha-1)z^{\alpha-2} - \frac{3}{z}\alpha z^{\alpha-1} + \frac{3}{z^2}z^\alpha &= 0 \nonumber\\
				\alpha^2 -4\alpha + 3 &= 0,
			\end{flalign}  
			whose roots are $\alpha=1$ and $\alpha=3$. Therefore, the general solution is a linear combination of the two independent solutions:  
			\begin{flalign}
				\chi(z) = c_1 z + c_2 z^3,
			\end{flalign}  
			where $c_1$ and $c_2$ are integration constants.
			
			To determine these constants, one must refer to the AdS/CFT dictionary. For a generic scalar field $\phi$ dual to an operator $\mathcal{O}$ with conformal dimension $\Delta$, as demonstrated in the process that led to Eq. (\ref{eq:phi_asymptotic_standard}) in the main text, the asymptotic solution near the AdS$_5$ boundary ($z \to 0$) is given by:
			\begin{flalign}
				\phi(z) \sim z^{4-\Delta} \phi_0 + z^\Delta \frac{\langle\mathcal{O}\rangle}{2\Delta-4},
				\label{eq:standard}
			\end{flalign}
			where $\phi_0$ is proportional to the source of $\mathcal{O}$, and $\langle\mathcal{O}\rangle$ is its vacuum expectation value \cite{Witten:1998qj}.
			
			In our case, $\chi$ is identified with the bulk scalar dual to the quark bilinear operator $\mathcal{O} = \bar{q}q$, which has a scaling dimension $\Delta = 3$. Applying the standard asymptotic form (\ref{eq:standard}) specifically for $\Delta=3$, we find:
			\begin{flalign}
				\chi(z) \sim \phi_0 z + \frac{\langle\bar{q}q\rangle}{2} z^3.
			\end{flalign}
			From this expression, we identify $c_1$ as the source term of the operator, which corresponds to the explicit quark mass $m_q$. As previously defined, the vacuum expectation value of $\bar{q}q$ is the chiral condensate, $\sigma \equiv \langle\bar{q}q\rangle$, which allows us to identify $c_2$. Therefore, the solution for the VEV in the hard-wall model is:  
			\begin{equation}
				\chi(z) = m_q z + \frac{\sigma}{2} z^3.
				\label{app:solution}
			\end{equation}
			
			In the standard soft-wall model, considering the commonly adopted quadratic dilaton profile $\Phi(z)=\kappa^2 z^2$, the equation of motion (\ref{eom_chi}) becomes:  
			\begin{flalign}
				\chi'' - \left(2\kappa^2 z + \frac{3}{z}\right)\chi' + \frac{3}{z^2}\chi = 0.
			\end{flalign}  
			Due to the introduction of the dilaton, the solution is no longer expressed as a simple power series in $z$, but rather as a combination of special functions. The VEV in this case is given by:  
			\begin{flalign}
				\chi(z) = C_1 \kappa z \, U\left(\frac{1}{2}; 0; \kappa^2z^2\right) + C_2 (\kappa z)^3 \, \prescript{}{1}{F}_1 \left(\frac{3}{2}; 2; \kappa^2z^2\right),
			\end{flalign} 
			where $C_1$ and $C_2$ are integration constants, $U$ is the Kummer's function of the second kind, and $\prescript{}{1}{F}_1$ is the Kummer's function of the first kind.
			
			The identification of these constants follows steps analogous to the hard-wall case. First, it is necessary to analyze the behavior of these special functions near the ultraviolet boundary to establish the proper connection with the AdS/CFT dictionary. Taking $z \to 0$, the Kummer functions can be expanded as:
			\begin{flalign}
				U\left(\frac{1}{2};0;x\right) &= \frac{1}{\Gamma(3/2)} - \frac{x}{\Gamma(1/2)} \left[ 1 - 2\gamma_E - \ln(x) - \Psi\left(\frac{3}{2}\right) \right]  + \mathcal{O}(x^2), \\
				\prescript{}{1}{F}_1\left(\frac{3}{2};2;x\right) &= 1 + \frac{3}{4}x + \mathcal{O}(x^2),
			\end{flalign}			
			where $\gamma_E$ is the Euler-Mascheroni constant and $\Psi(y)=\Gamma'(y)/\Gamma(y)$ is the digamma function.
			
			In a first-order approximation, the near-boundary expansion of the general solution can be directly compared to the expression of the VEV (\ref{app:solution}), which allows the straightforward identification of the integration constants as $C_1 = m_q \kappa^{-1} \Gamma(3/2)$ to properly normalize the source term, which subsequently implies that $C_2 = \sigma/2 \kappa^3$.
			
			However, this independence between the constants breaks down when analyzing the bulk interior. Upon imposing the IR boundary condition ($z \to \infty$), the Kummer function of the first kind, $\prescript{}{1}{F}_1$, diverges. 
			Therefore, for the general solution to be normalizable and retain its physical meaning,	it is necessary to assume $C_2 = 0$. This restriction leaves the exact physical solution in the following form:
			\begin{flalign}
				\chi(z) = m_q z \, \Gamma\left(\frac{3}{2}\right) U\left(\frac{1}{2};0;\kappa^2z^2\right).
			\end{flalign}
			
			By evaluating the asymptotic expansion of this function near the UV boundary, the vacuum profile becomes:
			\begin{tcolorbox}[title={},colback=white,colframe=ufabc!50,colback=ufabc!10,coltitle=black]
			\begin{equation}
				\lim_{z \to 0}\chi(z) \approx m_q z - \frac{m_q \kappa^2 z^3}{2} \left[1 - 2\gamma_E - 2\ln(\kappa z ) - \Psi\left(\frac{3}{2}\right)\right],
			\end{equation}
			\end{tcolorbox}
			This result corresponds exactly to Eq. (\ref{eq:vacuum_chi}) discussed in the main text.

			\section{Equations of motion of the dynamical soft-wall model}
			\label{ap:dsw}
			
			In this appendix, we present a detailed, step-by-step derivation of the background equations of motion for the dynamical soft-wall model. The five-dimensional gravitational action of the model, written in the Einstein frame, is given by
			\begin{align}
				\mathcal{S}_\text{G} =  \frac{1}{16\pi G_5} \int d^5x \sqrt{-g} \left(R - \frac{1}{2}\partial_M\Phi \partial^M \Phi - V_\Phi\right),
			\end{align}
			where $R$ is the Ricci scalar, $\Phi$ is the dilaton field and $V_\Phi$ is its associated potential. The background metric is expressed as
			\begin{equation}
				ds^2 = e^{2A(z)} (\eta_{\mu\nu}dx^\mu dx^\nu + dz^2),
			\end{equation}
			where $A(z)$ is the warp factor in the Einstein frame, which is related to the string frame warp factor $A^s(z)$ via the transformation $A(z) = A^s(z) - \frac{2}{3}\Phi(z)$.
			
			The equation of motion for the dilaton field can be obtained by applying the Euler-Lagrange equation to the scalar part of the Lagrangian density.
			
			First, we take the derivative of the Lagrangian with respect to the field $\Phi$:
			\begin{align}
				\frac{\partial \mathcal{L}_\Phi}{\partial\Phi} = -\sqrt{-g}\frac{\partial V_\Phi}{\partial \Phi} = -e^{5A}\frac{\partial V_\Phi}{\partial \Phi}.
			\end{align}
			Next, we evaluate the derivative with respect to the field gradient $\partial_R\Phi$:
			\begin{align}
				\frac{\partial\mathcal{L}_\Phi}{\partial(\partial_R\Phi)} &= -\frac{\sqrt{-g}}{2}  \left[g^{MN}\left(\frac{\partial(\partial_M\Phi)}{\partial(\partial_R\Phi)}\right)\partial_N\Phi + g^{MN}\partial_M\Phi\left(\frac{\partial(\partial_N\Phi)}{\partial(\partial_R\Phi)}\right)\right] \nonumber\\
				&=-\frac{\sqrt{-g}}{2} \left[g^{MN}\delta_{MR}\partial_N\Phi + g^{MN}\delta_{NR}\partial_M\Phi \right] \nonumber\\
				&=-\sqrt{-g}g^{RN}\partial_N\Phi.
			\end{align}
			Applying the total derivative $\partial_R$ to the above result, and assuming that $\Phi \equiv \Phi(z)$ depends solely on the holographic coordinate $z$, we obtain:
			\begin{align}
				\partial_R \left(\frac{\partial\mathcal{L}_\Phi}{\partial(\partial_R\Phi)}\right) &= - \partial_R (\sqrt{-g}g^{RN}\partial_N\Phi) \nonumber\\
				&= -\partial_z \left(e^{5A}e^{-2A}\partial_z\Phi\right) \nonumber\\
				&= - \left(\Phi'' + 3A'\Phi'\right)e^{3A},
			\end{align}
			where primes denote derivatives with respect to $z$. Equating $\partial_R \left(\frac{\partial\mathcal{L}}{\partial(\partial_R\Phi)}\right) = \frac{\partial \mathcal{L}}{\partial\Phi}$ and dividing both sides by $e^{3A}$, we arrive at the scalar equation of motion:
			\begin{tcolorbox}[title={},colback=white,colframe=ufabc!50,colback=ufabc!10,coltitle=black]
				\begin{equation}
					\Phi'' + 3A'\Phi' - e^{2A} \frac{\partial V_\Phi}{\partial \Phi} = 0.
				\end{equation}
			\end{tcolorbox}
			
			\vspace{5mm}
			
			To compute the backreaction of the scalar field on the geometry, we must derive its energy-momentum tensor. From standard field theory, the canonical energy-momentum tensor is constructed from the scalar Lagrangian function as:
			\begin{align}
				T_{MN} &= \frac{\partial L_\Phi}{\partial(\partial^M\Phi)}\partial_N\Phi - g_{MN}L_\Phi.
			\end{align}
			Using the relation $\frac{\partial L_\Phi}{\partial(\partial^R\Phi)} = - g_{MR} \partial^M \Phi$, the tensor takes the explicit form:
			\begin{align}
				T_{MN} = -\partial_M\Phi \partial_N\Phi + \frac{1}{2}g_{MN}g^{RS} \partial_R\Phi\partial_S\Phi + g_{MN}V_\Phi.
			\end{align}
			
			To evaluate the Einstein field equations, $G_{MN} = T_{MN}$, we must first compute the geometric properties of the AdS space. The Ricci tensor is constructed from the Christoffel symbols as follows:
			\begin{equation}
				R_{MN} = \partial_R\Gamma^R_{MN} - \partial_N\Gamma^R_{MR} + \Gamma^R_{MN}\Gamma^S_{RS} - \Gamma^R_{MS}\Gamma^S_{NR}.
			\end{equation}
			The Christoffel symbols are defined entirely by the metric:
			\begin{align}
				\Gamma_{MN}^L = \frac{1}{2}g^{LA} (\partial_M g_{AN} + \partial_N g_{AM} - \partial_A g_{MN} ).
			\end{align}
			For the conformally flat metric $g_{MN} = e^{2A}\eta_{MN}$, the non-vanishing Christoffel symbols are:
			\begin{align}
				\Gamma^z_{\mu\nu} &= \frac{1}{2}g^{zz}(- \partial_z g_{\mu\nu}) = \frac{1}{2}e^{-2A}(-2A'e^{2A}\eta_{\mu\nu}) = -A'\eta_{\mu\nu}, \\
				\Gamma^\mu_{z\nu} &= \frac{1}{2}g^{\mu\lambda}(\partial_z g_{\lambda\nu}) = \frac{1}{2}e^{-2A}\eta^{\mu\lambda}(2A'e^{2A}\eta_{\lambda\nu}) = A'\delta^\mu_\nu, \\
				\Gamma^z_{zz} &= \frac{1}{2}g^{zz}(\partial_z g_{zz}) = \frac{1}{2}e^{-2A}(2A'e^{2A}) = A'.
			\end{align}
			All other components, such as $\Gamma^\lambda_{\mu\nu}$, $\Gamma^\mu_{zz}$, and $\Gamma^z_{z\mu}$, are zero.
			
			With these symbols, the non-zero components of the Ricci tensor can be calculated. For the $zz$ component, taking into account the four-dimensional boundary indices, we have:
			\begin{align}
				R_{zz} &= \partial_M \Gamma^M_{zz} - \partial_z \Gamma^M_{zM} + \Gamma^M_{zz}\Gamma^N_{MN} -  \Gamma^M_{zN}\Gamma^N_{zM} \nonumber\\
				&= A'' - 5A'' + 5(A')^2 - 5(A')^2\nonumber\\
				&= - 4A'',
			\end{align}
			where we have used the properties of the Kronecker delta in four spacetime dimensions: $\delta^\mu_\mu = 4$ and $\delta^\mu_\nu\delta^\nu_\mu = 4$.
			
			Similarly, the components $R_{\mu\nu}$ are computed as:
			\begin{align}
				R_{\mu\nu} &= \partial_R\Gamma^R_{\mu\nu} - \partial_\nu\Gamma^R_{\mu R} + \Gamma^R_{\mu\nu}\Gamma^S_{RS} - \Gamma^R_{\mu S}\Gamma^S_{\nu R} \nonumber\\
				&=(-A'' - (A')^2 - 4(A')^2 + (A')^2 + (A')^2)\eta_{\mu\nu} \nonumber\\
				&=-\left[A'' + 3 (A')^2\right] \eta_{\mu\nu}.
			\end{align}
			
			Contracting the Ricci tensor yields the Ricci scalar $R$:
			\begin{align}
				R = g^{MN}R_{MN} &= g^{zz}R_{zz} + g^{\mu\nu}R_{\mu\nu} \nonumber\\
				&= -4A''e^{-2A} - 4e^{-2A}(A'' + 3(A')^2) \nonumber\\
				&= -4e^{-2A} \left[2A'' + 3(A')^2\right].
			\end{align}
			
			Finally, we construct the Einstein tensor $G_{MN} = R_{MN} - \frac{1}{2} g_{MN} R$ and match it with the energy-momentum tensor evaluated earlier. Starting with the $zz$ component:
			\begin{align}
				G_{zz} &= R_{zz} - \frac{1}{2}g_{zz}R \nonumber\\
				&=-4A'' + \left[4A'' + 6 (A')^2\right] \nonumber\\
				&=6(A')^2.
			\end{align}
			The corresponding component of the energy-momentum tensor is:
			\begin{equation}
				T_{zz} = \frac{(\Phi')^2}{2} - e^{2A}V_\Phi.
			\end{equation}
			Equating $G_{zz} = T_{zz}$ provides the first Einstein equation:
			\begin{tcolorbox}[title={},colback=white,colframe=ufabc!50,colback=ufabc!10,coltitle=black]
				\begin{equation}
					6(A')^2 - \frac{(\Phi')^2}{2} + e^{2A}V_\Phi = 0.
				\end{equation}
			\end{tcolorbox}
			
			Repeating the procedure for the $\mu\nu$ components of the Einstein tensor, we have:
			\begin{align}
				G_{\mu\nu} &= R_{\mu\nu} - \frac{1}{2}g_{\mu\nu}R \nonumber\\
				&=\left[-A'' - 3(A')^2 + 4A'' + 6(A')^2 \right]\eta_{\mu\nu} \nonumber\\
				&= \left[3A'' + 3(A')^2\right]\eta_{\mu\nu}.
			\end{align}
			Evaluating the $\mu\nu$ component of the energy-momentum tensor yields:
			\begin{equation}
				T_{\mu\nu} = -\frac{1}{2}\Big[(\Phi')^2 + 2e^{2A}V_\Phi \Big]\eta_{\mu\nu}.
			\end{equation}
			By equating $G_{\mu\nu} = T_{\mu\nu}$ and factoring out $\eta_{\mu\nu}$, we obtain the second Einstein equation:
			
			\begin{tcolorbox}[title={},colback=white,colframe=ufabc!50,colback=ufabc!10,coltitle=black]
				\begin{equation}
					3A'' + 3(A')^2 + \frac{(\Phi')^2}{2} + e^{2A}V_\Phi = 0.
				\end{equation}
			\end{tcolorbox}
			
			Together with the scalar equation of motion derived previously, these differential equations completely govern the gravitational background of the model.

		\end{appendices}

		{
			
			\pagebreak
			
		\small \singlespacing
		
		\addcontentsline{toc}{chapter}{Bibliography}	
		\bibliography{bibliography.bib}

\begin{thebibliography}{100}

\bibitem{Maldacena:1997re}
J.~M. Maldacena, ``{The Large N limit of superconformal field theories and
  supergravity},'' {\em Adv. Theor. Math. Phys.}, vol.~2, pp.~231--252, 1998.

\bibitem{Gubser:1998bc}
S.~S. Gubser, I.~R. Klebanov, and A.~M. Polyakov, ``{Gauge theory correlators
  from noncritical string theory},'' {\em Phys. Lett. B}, vol.~428,
  pp.~105--114, 1998.

\bibitem{Witten:1998qj}
E.~Witten, ``{Anti-de Sitter space and holography},'' {\em Adv. Theor. Math.
  Phys.}, vol.~2, pp.~253--291, 1998.

\bibitem{Aharony:1999ti}
O.~Aharony, S.~S. Gubser, J.~M. Maldacena, H.~Ooguri, and Y.~Oz, ``{Large N
  field theories, string theory and gravity},'' {\em Phys. Rept.}, vol.~323,
  pp.~183--386, 2000.

\bibitem{Gleiser:2011di}
M.~Gleiser and N.~Stamatopoulos, ``{Entropic Measure for Localized Energy
  Configurations: Kinks, Bounces, and Bubbles},'' {\em Phys. Lett. B},
  vol.~713, pp.~304--307, 2012.

\bibitem{Schwarzschild:1916ae}
K.~Schwarzschild, ``{On the gravitational field of a sphere of incompressible
  fluid according to Einstein's theory},'' {\em Sitzungsber. Preuss. Akad.
  Wiss. Berlin (Math. Phys. )}, vol.~1916, pp.~424--434, 1916.

\bibitem{Reissner:1916cle}
H.~Reissner, ``{\"Uber die Eigengravitation des elektrischen Feldes nach der
  Einsteinschen Theorie},'' {\em Annalen Phys.}, vol.~355, no.~9, pp.~106--120,
  1916.

\bibitem{Weyl:1917gp}
H.~Weyl, ``{The theory of gravitation},'' {\em Annalen Phys.}, vol.~54,
  pp.~117--145, 1917.

\bibitem{Nordstrom:1918}
G.~Nordström, ``On the energy of the gravitational field in einstein's
  theory,'' {\em Verhandlungen der Koninklijke Nederlandse Akademie van
  Wetenschappen, Afd. Natuurkunde}, vol.~26, pp.~1201--1208, 1918.

\bibitem{Jeffery:1921}
G.~B. Jeffery, ``The field of an electron on einstein's theory of
  gravitation,'' {\em Proc. R. Soc. Lond. A}, vol.~99, pp.~123--134, 1921.

\bibitem{Kerr:1963ud}
R.~P. Kerr, ``{Gravitational field of a spinning mass as an example of
  algebraically special metrics},'' {\em Phys. Rev. Lett.}, vol.~11,
  pp.~237--238, 1963.

\bibitem{Newman:1965my}
E.~T. Newman, R.~Couch, K.~Chinnapared, A.~Exton, A.~Prakash, and R.~Torrence,
  ``{Metric of a Rotating, Charged Mass},'' {\em J. Math. Phys.}, vol.~6,
  pp.~918--919, 1965.

\bibitem{Hawking:1971tu}
S.~W. Hawking, ``{Gravitational radiation from colliding black holes},'' {\em
  Phys. Rev. Lett.}, vol.~26, pp.~1344--1346, 1971.

\bibitem{Bekenstein:1972tm}
J.~D. Bekenstein, ``{Black holes and the second law},'' {\em Lett. Nuovo Cim.},
  vol.~4, pp.~737--740, 1972.

\bibitem{Hawking:1974rv}
S.~W. Hawking, ``{Black hole explosions},'' {\em Nature}, vol.~248, pp.~30--31,
  1974.

\bibitem{tHooft:1993dmi}
G.~'t~Hooft, ``{Dimensional reduction in quantum gravity},'' {\em Conf. Proc.
  C}, vol.~930308, pp.~284--296, 1993.

\bibitem{Susskind:1994vu}
L.~Susskind, ``{The World as a hologram},'' {\em J. Math. Phys.}, vol.~36,
  pp.~6377--6396, 1995.

\bibitem{Weinberg:1972kfs}
S.~Weinberg, {\em {Gravitation and Cosmology}: {Principles and Applications of
  the General Theory of Relativity}}.
\newblock New York: John Wiley and Sons, 1972.

\bibitem{Nastase:2015wjb}
H.~Nastase, {\em {Introduction to the ADS/CFT Correspondence}}.
\newblock Cambridge University Press, 9 2015.

\bibitem{Natsuume:2014sfa}
M.~Natsuume, {\em {AdS/CFT Duality User Guide}}, vol.~903.
\newblock 2015.

\bibitem{Bayona:2005nq}
C.~A. Bayona and N.~R.~F. Braga, ``{Anti-de Sitter boundary in Poincare
  coordinates},'' {\em Gen. Rel. Grav.}, vol.~39, pp.~1367--1379, 2007.

\bibitem{Polchinski:2001tt}
J.~Polchinski and M.~J. Strassler, ``{Hard scattering and gauge / string
  duality},'' {\em Phys. Rev. Lett.}, vol.~88, p.~031601, 2002.

\bibitem{Yang:1954ek}
C.-N. Yang and R.~L. Mills, ``{Conservation of Isotopic Spin and Isotopic Gauge
  Invariance},'' {\em Phys. Rev.}, vol.~96, pp.~191--195, 1954.

\bibitem{Weinberg:1996kr}
S.~Weinberg, {\em {The quantum theory of fields. Vol. 2: Modern applications}}.
\newblock Cambridge University Press, 8 2013.

\bibitem{Coleman:1967ad}
S.~R. Coleman and J.~Mandula, ``{All Possible Symmetries of the S Matrix},''
  {\em Phys. Rev.}, vol.~159, pp.~1251--1256, 1967.

\bibitem{Haag:1974qh}
R.~Haag, J.~T. Lopuszanski, and M.~Sohnius, ``{All Possible Generators of
  Supersymmetries of the s Matrix},'' {\em Nucl. Phys. B}, vol.~88, p.~257,
  1975.

\bibitem{Ammon:2015wua}
M.~Ammon and J.~Erdmenger, {\em {Gauge/gravity duality}: {Foundations and
  applications}}.
\newblock Cambridge: Cambridge University Press, 4 2015.

\bibitem{Zwiebach:2004tj}
B.~Zwiebach, {\em {A first course in string theory}}.
\newblock Cambridge University Press, 7 2006.

\bibitem{Green:1987sp}
M.~B. Green, J.~H. Schwarz, and E.~Witten, {\em {Superstring Theory. vol. 1:
  Introduction}}.
\newblock Cambridge Monographs on Mathematical Physics, 7 1988.

\bibitem{Becker:2006dvp}
K.~Becker, M.~Becker, and J.~H. Schwarz, {\em {String theory and M-theory: A
  modern introduction}}.
\newblock Cambridge University Press, 12 2006.

\bibitem{Polchinski:1998rq}
J.~Polchinski, {\em {String theory. Vol. 1: An introduction to the bosonic
  string}}.
\newblock Cambridge Monographs on Mathematical Physics, Cambridge University
  Press, 12 2007.

\bibitem{Polchinski:1998rr}
J.~Polchinski, {\em {String theory. Vol. 2: Superstring theory and beyond}}.
\newblock Cambridge Monographs on Mathematical Physics, Cambridge University
  Press, 12 2007.

\bibitem{Witten:1995ex}
E.~Witten, ``{String theory dynamics in various dimensions},'' {\em Nucl. Phys.
  B}, vol.~443, pp.~85--126, 1995.

\bibitem{Polchinski:1995mt}
J.~Polchinski, ``{Dirichlet Branes and Ramond-Ramond charges},'' {\em Phys.
  Rev. Lett.}, vol.~75, pp.~4724--4727, 1995.

\bibitem{tHooft:1973alw}
G.~'t~Hooft, ``{A Planar Diagram Theory for Strong Interactions},'' {\em Nucl.
  Phys. B}, vol.~72, p.~461, 1974.

\bibitem{Breitenlohner:1982bm}
P.~Breitenlohner and D.~Z. Freedman, ``{Positive Energy in anti-De Sitter
  Backgrounds and Gauged Extended Supergravity},'' {\em Phys. Lett. B},
  vol.~115, pp.~197--201, 1982.

\bibitem{Zaanen:2015oix}
J.~Zaanen, Y.-W. Sun, Y.~Liu, and K.~Schalm, {\em {Holographic Duality in
  Condensed Matter Physics}}.
\newblock Cambridge Univ. Press, 2015.

\bibitem{Klebanov:1999tb}
I.~R. Klebanov and E.~Witten, ``{AdS / CFT correspondence and symmetry
  breaking},'' {\em Nucl. Phys. B}, vol.~556, pp.~89--114, 1999.

\bibitem{Gross:1973id}
D.~J. Gross and F.~Wilczek, ``{Ultraviolet Behavior of Nonabelian Gauge
  Theories},'' {\em Phys. Rev. Lett.}, vol.~30, pp.~1343--1346, 1973.

\bibitem{Politzer:1973fx}
H.~D. Politzer, ``{Reliable Perturbative Results for Strong Interactions?},''
  {\em Phys. Rev. Lett.}, vol.~30, pp.~1346--1349, 1973.

\bibitem{Johnson:1999am}
G.~Johnson, {\em {Strange beauty: Murray Gell-Mann and the revolution in
  twentieth-century physics}}.
\newblock 1999.

\bibitem{Gell-Mann:1964ewy}
M.~Gell-Mann, ``{A Schematic Model of Baryons and Mesons},'' {\em Phys. Lett.},
  vol.~8, pp.~214--215, 1964.

\bibitem{Zweig:1964jf}
G.~Zweig, {\em {An SU(3) model for strong interaction symmetry and its
  breaking. Version 2}}, pp.~22--101.
\newblock 2 1964.

\bibitem{ParticleDataGroup:2020ssz}
P.~A. Zyla {\em et~al.}, ``{Review of Particle Physics},'' {\em PTEP},
  vol.~2020, no.~8, p.~083C01, 2020.

\bibitem{Yukawa:1935xg}
H.~Yukawa, ``{On the Interaction of Elementary Particles I},'' {\em Proc. Phys.
  Math. Soc. Jap.}, vol.~17, pp.~48--57, 1935.

\bibitem{Griffiths:2008zz}
D.~Griffiths, {\em {Introduction to elementary particles}}.
\newblock 2008.

\bibitem{Nakano:1953zz}
T.~Nakano and K.~Nishijima, ``{Charge Independence for V-particles},'' {\em
  Prog. Theor. Phys.}, vol.~10, pp.~581--582, 1953.

\bibitem{Fayyazuddin:1994wh}
Fayyazuddin and Riazuddin, {\em {A Modern Introduction To Particle Physics (3rd
  Edition)}}.
\newblock Oxford University Press, 2011.

\bibitem{Donoghue:1992dd}
J.~F. Donoghue, E.~Golowich, and B.~R. Holstein, {\em {Dynamics of the standard
  model}}, vol.~2.
\newblock CUP, 2014.

\bibitem{Chanowitz:1972da}
M.~S. Chanowitz and J.~R. Ellis, ``{Canonical Trace Anomalies},'' {\em Phys.
  Rev. D}, vol.~7, pp.~2490--2506, 1973.

\bibitem{Crewther:1972kn}
R.~J. Crewther, ``{Nonperturbative evaluation of the anomalies in low-energy
  theorems},'' {\em Phys. Rev. Lett.}, vol.~28, p.~1421, 1972.

\bibitem{Collins:1976yq}
J.~C. Collins, A.~Duncan, and S.~D. Joglekar, ``{Trace and Dilatation Anomalies
  in Gauge Theories},'' {\em Phys. Rev. D}, vol.~16, pp.~438--449, 1977.

\bibitem{Aguilar:2019teb}
A.~C. Aguilar {\em et~al.}, ``{Pion and Kaon Structure at the Electron-Ion
  Collider},'' {\em Eur. Phys. J. A}, vol.~55, no.~10, p.~190, 2019.

\bibitem{Shifman:1978bx}
M.~A. Shifman, A.~I. Vainshtein, and V.~I. Zakharov, ``{QCD and Resonance
  Physics. Theoretical Foundations},'' {\em Nucl. Phys. B}, vol.~147,
  pp.~385--447, 1979.

\bibitem{Ji:1995sv}
X.-D. Ji, ``{Breakup of hadron masses and energy - momentum tensor of QCD},''
  {\em Phys. Rev. D}, vol.~52, pp.~271--281, 1995.

\bibitem{Mathieu:2008me}
V.~Mathieu, N.~Kochelev, and V.~Vento, ``{The Physics of Glueballs},'' {\em
  Int. J. Mod. Phys. E}, vol.~18, pp.~1--49, 2009.

\bibitem{Djouadi:2005gi}
A.~Djouadi, ``{The Anatomy of electro-weak symmetry breaking. I: The Higgs
  boson in the standard model},'' {\em Phys. Rept.}, vol.~457, pp.~1--216,
  2008.

\bibitem{Gunion:1989we}
J.~F. Gunion, H.~E. Haber, G.~L. Kane, and S.~Dawson, {\em {The Higgs Hunter's
  Guide}}, vol.~80.
\newblock 2000.

\bibitem{Hoferichter:2025ubp}
M.~Hoferichter and J.~R. de~Elvira, {\em {Nucleon mass: trace anomaly and
  $\sigma$-terms}}.
\newblock 6 2025.

\bibitem{Gasser:1990ap}
J.~Gasser, H.~Leutwyler, and M.~E. Sainio, ``{Form-factor of the sigma term},''
  {\em Phys. Lett. B}, vol.~253, pp.~260--264, 1991.

\bibitem{Ji:1994av}
X.-D. Ji, ``{A QCD analysis of the mass structure of the nucleon},'' {\em Phys.
  Rev. Lett.}, vol.~74, pp.~1071--1074, 1995.

\bibitem{Regge:1959mz}
T.~Regge, ``{Introduction to complex orbital momenta},'' {\em Nuovo Cim.},
  vol.~14, p.~951, 1959.

\bibitem{Chew:1961ev}
G.~F. Chew and S.~C. Frautschi, ``{Principle of Equivalence for All Strongly
  Interacting Particles Within the S Matrix Framework},'' {\em Phys. Rev.
  Lett.}, vol.~7, pp.~394--397, 1961.

\bibitem{Collins:1977jy}
P.~D.~B. Collins, {\em {An Introduction to Regge Theory and High Energy
  Physics}}.
\newblock Cambridge University Press, 1977.

\bibitem{Schwarz:2012zc}
J.~H. Schwarz, ``{The Early History of String Theory and Supersymmetry},'' 1
  2012.

\bibitem{Tang:2000tb}
A.~Tang and J.~W. Norbury, ``{Properties of Regge trajectories},'' {\em Phys.
  Rev. D}, vol.~62, p.~016006, 2000.

\bibitem{Rickles:2014fha}
D.~Rickles, {\em {A Brief History of String Theory. From Dual Models to
  M-Theory}}.
\newblock The Frontiers Collection, Heidelberg, Germany: Springer, 2014.

\bibitem{ParticleDataGroup:2024cfk}
S.~Navas {\em et~al.}, ``{Review of particle physics},'' {\em Phys. Rev. D},
  vol.~110, no.~3, p.~030001, 2024.

\bibitem{DeRujula:1975qlm}
A.~De~Rujula, H.~Georgi, and S.~L. Glashow, ``{Hadron Masses in a Gauge
  Theory},'' {\em Phys. Rev. D}, vol.~12, pp.~147--162, 1975.

\bibitem{Godfrey:1985xj}
S.~Godfrey and N.~Isgur, ``{Mesons in a Relativized Quark Model with
  Chromodynamics},'' {\em Phys. Rev. D}, vol.~32, pp.~189--231, 1985.

\bibitem{Afonin:2014nya}
S.~S. Afonin and I.~V. Pusenkov, ``{Universal description of radially excited
  heavy and light vector mesons},'' {\em Phys. Rev. D}, vol.~90, no.~9,
  p.~094020, 2014.

\bibitem{Gleiser:2018jpd}
M.~Gleiser and D.~Sowinski, ``{Configurational information approach to
  instantons and false vacuum decay in $D$-dimensional spacetime},'' {\em Phys.
  Rev. D}, vol.~98, no.~5, p.~056026, 2018.

\bibitem{Gleiser:2018kbq}
M.~Gleiser, M.~Stephens, and D.~Sowinski, ``{Configurational entropy as a
  lifetime predictor and pattern discriminator for oscillons},'' {\em Phys.
  Rev. D}, vol.~97, no.~9, p.~096007, 2018.

\bibitem{Correa:2014boa}
R.~A.~C. Correa, A.~de~Souza~Dutra, and M.~Gleiser, ``{Information-Entropic
  Measure of Energy-Degenerate Kinks in Two-Field Models},'' {\em Phys. Lett.
  B}, vol.~737, pp.~388--394, 2014.

\bibitem{Gleiser:2017cro}
M.~Gleiser and N.~Jiang, ``{Predicting Atomic Decay Rates Using an
  Informational-Entropic Approach},'' {\em Int. J. Theor. Phys.}, vol.~57,
  no.~6, pp.~1691--1704, 2018.

\bibitem{Correa:2015lla}
R.~A.~C. Correa, R.~da~Rocha, and A.~de~Souza~Dutra, ``{Entropic information
  for travelling solitons in Lorentz and CPT breaking systems},'' {\em Annals
  Phys.}, vol.~359, pp.~198--212, 2015.

\bibitem{Correa:2015ako}
R.~A.~C. Correa, P.~H. R.~S. Moraes, A.~de~Souza~Dutra, and R.~da~Rocha,
  ``{Information content in F(R) brane models with nonconstant curvature},''
  {\em Phys. Rev. D}, vol.~92, no.~12, p.~126005, 2015.

\bibitem{Correa:2015qma}
R.~A.~C. Correa and P.~H. R.~S. Moraes, ``{Configurational entropy in
  $f\,(R,T\,)$ brane models},'' {\em Eur. Phys. J. C}, vol.~76, no.~2, p.~100,
  2016.

\bibitem{Correa:2015vka}
R.~A.~C. Correa and R.~da~Rocha, ``{Configurational entropy in brane-world
  models},'' {\em Eur. Phys. J. C}, vol.~75, no.~11, p.~522, 2015.

\bibitem{Barreto:2022mbx}
W.~Barreto, A.~Herrera-Aguilar, and R.~da~Rocha, ``{Configurational entropy of
  generalized sine-Gordon-type models},'' {\em Annals Phys.}, vol.~447,
  p.~169142, 2022.

\bibitem{Correa:2016noh}
R.~A.~C. Correa, D.~Monteiro~Dantas, P.~H.~R. da~Silva~Moraes,
  A.~de~Souza~Dutra, and C.~A.~S. de~Almeida, ``{Refinements of the Weyl Pure
  Geometrical Thick Branes From Information-Entropic Measure},'' {\em Annalen
  Phys.}, vol.~530, no.~7, p.~1700188, 2018.

\bibitem{Cruz:2017ems}
W.~T. Cruz, D.~M. Dantas, R.~A.~C. Correa, and C.~A.~S. Almeida, ``{Phase
  transitions in thick branes endorsed by entropic information},'' {\em Phys.
  Lett. B}, vol.~772, pp.~592--598, 2017.

\bibitem{Casadio:2016aum}
R.~Casadio and R.~da~Rocha, ``{Stability of the graviton
  Bose{\textendash}Einstein condensate in the brane-world},'' {\em Phys. Lett.
  B}, vol.~763, pp.~434--438, 2016.

\bibitem{Correa:2016pgr}
R.~A.~C. Correa, D.~M. Dantas, C.~A.~S. Almeida, and R.~da~Rocha, ``{Bounds on
  topological Abelian string-vortex and string-cigar from information-entropic
  measure},'' {\em Phys. Lett. B}, vol.~755, pp.~358--362, 2016.

\bibitem{Gleiser:2013mga}
M.~Gleiser and D.~Sowinski, ``{Information-Entropic Stability Bound for Compact
  Objects: Application to Q-Balls and the Chandrasekhar Limit of Polytropes},''
  {\em Phys. Lett. B}, vol.~727, pp.~272--275, 2013.

\bibitem{Gleiser:2015rwa}
M.~Gleiser and N.~Jiang, ``{Stability Bounds on Compact Astrophysical Objects
  from Information-Entropic Measure},'' {\em Phys. Rev. D}, vol.~92, no.~4,
  p.~044046, 2015.

\bibitem{Braga:2016wzx}
N.~R.~F. Braga and R.~da~Rocha, ``{Configurational entropy of anti-de Sitter
  black holes},'' {\em Phys. Lett. B}, vol.~767, pp.~386--391, 2017.

\bibitem{Gleiser:2015aav}
M.~Gleiser and D.~Sowinski, ``{Information-Entropic Signature of the Critical
  Point},'' {\em Phys. Lett. B}, vol.~747, pp.~125--128, 2015.

\bibitem{Sowinski:2016vxz}
D.~Sowinski and M.~Gleiser, ``{Information Dynamics at a Phase Transition},''
  {\em J. Stat. Phys.}, vol.~167, no.~5, pp.~1221--1232, 2017.

\bibitem{Bernardini:2016qit}
A.~E. Bernardini, N.~R.~F. Braga, and R.~da~Rocha, ``{Configurational entropy
  of glueball states},'' {\em Phys. Lett. B}, vol.~765, pp.~81--85, 2017.

\bibitem{Gleiser:2012tu}
M.~Gleiser and N.~Stamatopoulos, ``{Information Content of Spontaneous Symmetry
  Breaking},'' {\em Phys. Rev. D}, vol.~86, p.~045004, 2012.

\bibitem{Bernardini:2019stn}
A.~E. Bernardini and R.~da~Rocha, ``{Cosmological comoving behavior of the
  configurational entropy},'' {\em Phys. Lett. B}, vol.~796, pp.~107--111,
  2019.

\bibitem{Fernandes-Silva:2019fez}
A.~Fernandes-Silva, A.~J. Ferreira-Martins, and R.~da~Rocha, ``{Extended
  quantum portrait of MGD black holes and information entropy},'' {\em Phys.
  Lett. B}, vol.~791, pp.~323--330, 2019.

\bibitem{Barreto:2022len}
W.~Barreto and R.~da~Rocha, ``{Gravitational collapse in AdS: instabilities,
  turbulence, and information},'' {\em Eur. Phys. J. Plus}, vol.~137, no.~7,
  p.~845, 2022.

\bibitem{Casadio:2022pla}
R.~Casadio, R.~da~Rocha, P.~Meert, L.~Tabarroni, and W.~Barreto,
  ``{Configurational entropy of black hole quantum cores},'' {\em Class. Quant.
  Grav.}, vol.~40, no.~7, p.~075014, 2023.

\bibitem{daRocha:2021jzn}
R.~da~Rocha, ``{AdS graviton stars and differential configurational entropy},''
  {\em Phys. Lett. B}, vol.~823, p.~136729, 2021.

\bibitem{daRocha:2022bnk}
R.~da~Rocha, ``{Information entropy of nuclear electromagnetic transitions in
  AdS/QCD},'' {\em Nucl. Phys. B}, vol.~992, p.~116232, 2023.

\bibitem{Casadio:2023pmh}
R.~Casadio, R.~da~Rocha, A.~Giusti, and P.~Meert, ``{Thermodynamic and
  configurational entropy of quantum Schwarzschild geometries},'' {\em Phys.
  Lett. B}, vol.~849, p.~138466, 2024.

\bibitem{Erlich:2005qh}
J.~Erlich, E.~Katz, D.~T. Son, and M.~A. Stephanov, ``{QCD and a holographic
  model of hadrons},'' {\em Phys. Rev. Lett.}, vol.~95, p.~261602, 2005.

\bibitem{Abidin:2009aj}
Z.~Abidin and C.~E. Carlson, ``{Strange hadrons and kaon-to-pion transition
  form factors from holography},'' {\em Phys. Rev. D}, vol.~80, p.~115010,
  2009.

\bibitem{Ballon-Bayona:2017bwk}
A.~Ballon-Bayona, G.~Krein, and C.~Miller, ``{Strong couplings and form factors
  of charmed mesons in holographic QCD},'' {\em Phys. Rev. D}, vol.~96, no.~1,
  p.~014017, 2017.

\bibitem{Chen:2021wzj}
Y.~Chen and M.~Huang, ``{Holographic QCD model for Nf=4},'' {\em Phys. Rev. D},
  vol.~105, no.~2, p.~026021, 2022.

\bibitem{Karapetyan:2023sfo}
G.~Karapetyan, W.~de~Paula, and R.~da~Rocha, ``{Configurational information
  measure of mesonic states in 4-flavor AdS/QCD},'' {\em Phys. Lett. B},
  vol.~846, p.~138174, 2023.

\bibitem{Boschi-Filho:2002wdj}
H.~Boschi-Filho and N.~R.~F. Braga, ``{QCD / string holographic mapping and
  glueball mass spectrum},'' {\em Eur. Phys. J. C}, vol.~32, pp.~529--533,
  2004.

\bibitem{Boschi-Filho:2002xih}
H.~Boschi-Filho and N.~R.~F. Braga, ``{Gauge / string duality and scalar
  glueball mass ratios},'' {\em JHEP}, vol.~05, p.~009, 2003.

\bibitem{DaRold:2005mxj}
L.~Da~Rold and A.~Pomarol, ``{Chiral symmetry breaking from five dimensional
  spaces},'' {\em Nucl. Phys. B}, vol.~721, pp.~79--97, 2005.

\bibitem{DaRold:2005vr}
L.~Da~Rold and A.~Pomarol, ``{The Scalar and pseudoscalar sector in a
  five-dimensional approach to chiral symmetry breaking},'' {\em JHEP},
  vol.~01, p.~157, 2006.

\bibitem{Brodsky:2006uqa}
S.~J. Brodsky and G.~F. de~Teramond, ``{Hadronic spectra and light-front
  wavefunctions in holographic QCD},'' {\em Phys. Rev. Lett.}, vol.~96,
  p.~201601, 2006.

\bibitem{Evans:2005ip}
N.~Evans, J.~P. Shock, and T.~Waterson, ``{Towards a perfect QCD gravity
  dual},'' {\em Phys. Lett. B}, vol.~622, pp.~165--171, 2005.

\bibitem{Hirn:2005nr}
J.~Hirn and V.~Sanz, ``{Interpolating between low and high energy QCD via a 5-D
  Yang-Mills model},'' {\em JHEP}, vol.~12, p.~030, 2005.

\bibitem{Katz:2005ir}
E.~Katz, A.~Lewandowski, and M.~D. Schwartz, ``{Tensor mesons in AdS/QCD},''
  {\em Phys. Rev. D}, vol.~74, p.~086004, 2006.

\bibitem{Erlich:2006hq}
J.~Erlich, G.~D. Kribs, and I.~Low, ``{Emerging holography},'' {\em Phys. Rev.
  D}, vol.~73, p.~096001, 2006.

\bibitem{Hirn:2005vk}
J.~Hirn, N.~Rius, and V.~Sanz, ``{Geometric approach to condensates in
  holographic QCD},'' {\em Phys. Rev. D}, vol.~73, p.~085005, 2006.

\bibitem{deTeramond:2005su}
G.~F. de~Teramond and S.~J. Brodsky, ``{Hadronic spectrum of a holographic dual
  of QCD},'' {\em Phys. Rev. Lett.}, vol.~94, p.~201601, 2005.

\bibitem{Ballon-Bayona:2014oma}
A.~Ballon-Bayona, G.~Krein, and C.~Miller, ``{Decay constants of the pion and
  its excitations in holographic QCD},'' {\em Phys. Rev. D}, vol.~91,
  p.~065024, 2015.

\bibitem{Karch:2006pv}
A.~Karch, E.~Katz, D.~T. Son, and M.~A. Stephanov, ``{Linear confinement and
  AdS/QCD},'' {\em Phys. Rev. D}, vol.~74, p.~015005, 2006.

\bibitem{Csaki:2006ji}
C.~Csaki and M.~Reece, ``{Toward a systematic holographic QCD: A Braneless
  approach},'' {\em JHEP}, vol.~05, p.~062, 2007.

\bibitem{Cherman:2008eh}
A.~Cherman, T.~D. Cohen, and E.~S. Werbos, ``{The Chiral condensate in
  holographic models of QCD},'' {\em Phys. Rev. C}, vol.~79, p.~045203, 2009.

\bibitem{Colangelo:2008us}
P.~Colangelo, F.~De~Fazio, F.~Giannuzzi, F.~Jugeau, and S.~Nicotri, ``{Light
  scalar mesons in the soft-wall model of AdS/QCD},'' {\em Phys. Rev. D},
  vol.~78, p.~055009, 2008.

\bibitem{Gherghetta:2009ac}
T.~Gherghetta, J.~I. Kapusta, and T.~M. Kelley, ``{Chiral symmetry breaking in
  the soft-wall AdS/QCD model},'' {\em Phys. Rev. D}, vol.~79, p.~076003, 2009.

\bibitem{Li:2012ay}
D.~Li, M.~Huang, and Q.-S. Yan, ``{A dynamical soft-wall holographic QCD model
  for chiral symmetry breaking and linear confinement},'' {\em Eur. Phys. J.
  C}, vol.~73, p.~2615, 2013.

\bibitem{Li:2013oda}
D.~Li and M.~Huang, ``{Dynamical holographic QCD model for glueball and light
  meson spectra},'' {\em JHEP}, vol.~11, p.~088, 2013.

\bibitem{Kapusta:2009ab}
J.~I. Kapusta, T.~M. Kelley, and T.~Gherghetta, ``{Chiral Symmetry Breaking in
  a Soft-Wall Model of AdS/QCD},'' {\em Int. J. Mod. Phys. A}, vol.~25,
  pp.~453--463, 2010.

\bibitem{Batell:2008zm}
B.~Batell and T.~Gherghetta, ``{Dynamical Soft-Wall AdS/QCD},'' {\em Phys. Rev.
  D}, vol.~78, p.~026002, 2008.

\bibitem{dePaula:2008ph}
W.~de~Paula, T.~Frederico, H.~Forkel, and M.~Beyer, ``{Solution of the 5D
  Einstein equations in a dilaton background model},'' {\em PoS}, vol.~LC2008,
  p.~046, 2008.

\bibitem{Ballon-Bayona:2023zal}
A.~Ballon-Bayona, T.~Frederico, L.~A.~H. Mamani, and W.~de~Paula, ``{Dynamical
  holographic QCD model for spontaneous chiral symmetry breaking and
  confinement},'' {\em Phys. Rev. D}, vol.~108, no.~10, p.~106016, 2023.

\bibitem{dePaula:2008fp}
W.~de~Paula, T.~Frederico, H.~Forkel, and M.~Beyer, ``{Dynamical AdS/QCD with
  area-law confinement and linear Regge trajectories},'' {\em Phys. Rev. D},
  vol.~79, p.~075019, 2009.

\bibitem{Maldacena:1998im}
J.~M. Maldacena, ``{Wilson loops in large N field theories},'' {\em Phys. Rev.
  Lett.}, vol.~80, pp.~4859--4862, 1998.

\bibitem{Kinar:1998vq}
Y.~Kinar, E.~Schreiber, and J.~Sonnenschein, ``{Q anti-Q potential from strings
  in curved space-time: Classical results},'' {\em Nucl. Phys. B}, vol.~566,
  pp.~103--125, 2000.

\bibitem{He:2013qq}
S.~He, S.-Y. Wu, Y.~Yang, and P.-H. Yuan, ``{Phase Structure in a Dynamical
  Soft-Wall Holographic QCD Model},'' {\em JHEP}, vol.~04, p.~093, 2013.

\bibitem{Zhang:2021itx}
L.~Zhang, C.~Chen, Y.~Chen, and M.~Huang, ``{Spectra of glueballs and oddballs
  and the equation of state from holographic QCD},'' {\em Phys. Rev. D},
  vol.~105, no.~2, p.~026020, 2022.

\bibitem{Shannon:1948zz}
C.~E. Shannon, ``{A mathematical theory of communication},'' {\em Bell Syst.
  Tech. J.}, vol.~27, pp.~379--423, 1948.

\bibitem{Sowinski}
D.~Sowinski, {\em Complexity and Stability for Epistemic Agents: The
  Foundations and Phenomenology of Configurational Entropy}.
\newblock PhD thesis, Dartmouth College, 2016.

\bibitem{Kolmogorov:1963}
A.~N. Kolmogorov, ``On tables of random numbers,'' {\em Sankhyā: The Indian
  Journal of Statistics, Series A (1961-2002)}, vol.~25, no.~4, pp.~369--376,
  1963.

\bibitem{Kolmogorov:1968}
A.~N. Kolmogorov, ``Three approaches to the quantitative definition of
  information *,'' {\em International Journal of Computer Mathematics}, vol.~2,
  no.~1-4, pp.~157--168, 1968.

\bibitem{Kolmogorov:1968a}
A.~Kolmogorov, ``Logical basis for information theory and probability theory,''
  {\em IEEE Transactions on Information Theory}, vol.~14, no.~5, pp.~662--664,
  1968.

\bibitem{Cover:2012}
T.~Cover and J.~Thomas, {\em Elements of Information Theory}.
\newblock Wiley, 2012.

\bibitem{Braga:2018fyc}
N.~R.~F. Braga, L.~F. Ferreira, and R.~Da~Rocha, ``{Thermal dissociation of
  heavy mesons and configurational entropy},'' {\em Phys. Lett. B}, vol.~787,
  pp.~16--22, 2018.

\bibitem{daRocha:2025zth}
R.~da~Rocha and R.~D. Vilela, ``{Digit anomalies in the hadronic mass spectrum,
  classical and quantum information entropies, and the dynamical QCD scale},''
  {\em Eur. Phys. J. A}, vol.~62, no.~6, p.~117, 2026.

\bibitem{Barreto:2022ohl}
W.~Barreto and R.~da~Rocha, ``{Differential configurational entropy and the
  gravitational collapse of a kink},'' {\em Phys. Rev. D}, vol.~105, no.~6,
  p.~064049, 2022.

\bibitem{Braga:2019jqg}
N.~R.~F. Braga, ``{Information versus stability in an anti-de Sitter black
  hole},'' {\em Phys. Lett. B}, vol.~797, p.~134919, 2019.

\bibitem{Cruz:2018qby}
W.~T. Cruz, D.~M. Dantas, R.~V. Maluf, and C.~A.~S. Almeida, ``{Configurational
  Entropy and Newton's Law in Double Sine-Gordon Braneworlds},'' {\em Annalen
  Phys.}, vol.~531, no.~10, p.~1900178, 2019.

\bibitem{Bazeia:2018uyg}
D.~Bazeia, D.~C. Moreira, and E.~I.~B. Rodrigues, ``{Configurational Entropy
  for Skyrmion-like Magnetic Structures},'' {\em J. Magn. Magn. Mater.},
  vol.~475, pp.~734--740, 2019.

\bibitem{Bazeia:2021stz}
D.~Bazeia and E.~I.~B. Rodrigues, ``{Configurational entropy of skyrmions and
  half-skyrmions in planar magnetic elements},'' {\em Phys. Lett. A}, vol.~392,
  p.~127170, 2021.

\bibitem{Alves:2017ljt}
A.~Alves, A.~G. Dias, and R.~Silva, ``{Maximum Entropy Inferences on the Axion
  Mass in Models with Axion-Neutrino Interaction},'' {\em Braz. J. Phys.},
  vol.~47, no.~4, pp.~426--435, 2017.

\bibitem{Alves:2014ksa}
A.~Alves, A.~G. Dias, and R.~da~Silva, ``{Maximum Entropy Principle and the
  Higgs Boson Mass},'' {\em Physica A}, vol.~420, pp.~1--7, 2015.

\bibitem{Braga:2020opg}
N.~R.~F. Braga and O.~C. Junqueira, ``{Configuration entropy and
  confinement/deconfinement transiton in holographic QCD},'' {\em Phys. Lett.
  B}, vol.~814, p.~136082, 2021.

\bibitem{Lee:2018zmp}
C.~O. Lee, ``{Configurational entropy of tachyon kinks on unstable
  Dp-branes},'' {\em Phys. Lett. B}, vol.~790, pp.~197--204, 2019.

\bibitem{MartinContreras:2022lxl}
M.~A. Martin~Contreras, A.~Vega, and S.~Diles, ``{A holographic bottom-up
  description of light nuclide spectroscopy and stability},'' {\em Phys. Lett.
  B}, vol.~835, p.~137551, 2022.

\bibitem{Karapetyan:2016fai}
G.~Karapetyan, ``{Fine-tuning the Color-Glass Condensate with the nuclear
  configurational entropy},'' {\em EPL}, vol.~117, no.~1, p.~18001, 2017.

\bibitem{Karapetyan:2017edu}
G.~Karapetyan, ``{The nuclear configurational entropy impact parameter
  dependence in the Color-Glass Condensate},'' {\em EPL}, vol.~118, no.~3,
  p.~38001, 2017.

\bibitem{dePaula:2009za}
W.~de~Paula and T.~Frederico, ``{Scalar mesons within a dynamical holographic
  QCD model},'' {\em Phys. Lett. B}, vol.~693, pp.~287--291, 2010.

\bibitem{Frederico:2014bva}
T.~Frederico, K.~S. F.~F. Guimar\~aes, O.~Louren\c{c}o, W.~de~Paula,
  I.~Bediaga, and A.~C. dos Reis, ``{Heavy meson decay in three-mesons and
  FSI},'' {\em Few Body Syst.}, vol.~55, pp.~441--446, 2014.

\bibitem{Colangelo:2018mrt}
P.~Colangelo and F.~Loparco, ``{Configurational Entropy can disentangle
  conventional hadrons from exotica},'' {\em Phys. Lett. B}, vol.~788,
  pp.~500--504, 2019.

\bibitem{Braga:2021fey}
N.~R.~F. Braga, Y.~F. Ferreira, and L.~F. Ferreira, ``{Configuration entropy
  and stability of bottomonium radial excitations in a plasma with magnetic
  fields},'' {\em Phys. Rev. D}, vol.~105, no.~11, p.~114044, 2022.

\bibitem{Braga:2020hhs}
N.~R.~F. Braga and R.~da~Mata, ``{Configuration entropy description of
  charmonium dissociation under the influence of magnetic fields},'' {\em Phys.
  Lett. B}, vol.~811, p.~135918, 2020.

\bibitem{Braga:2021zyi}
N.~R.~F. Braga and O.~C. Junqueira, ``{Configuration entropy in the soft wall
  AdS/QCD model and the Wien law},'' {\em Phys. Lett. B}, vol.~820, p.~136485,
  2021.

\bibitem{GoncalvesdaSilva:2017bvk}
A.~Goncalves~da Silva and R.~da~Rocha, ``{Information-entropic analysis of
  Korteweg{\textendash}de Vries solitons in the quark{\textendash}gluon
  plasma},'' {\em Phys. Lett. B}, vol.~774, pp.~98--102, 2017.

\bibitem{Barbosa-Cendejas:2018mng}
N.~Barbosa-Cendejas, R.~Cartas-Fuentevilla, A.~Herrera-Aguilar, R.~R.
  Mora-Luna, and R.~da~Rocha, ``{Dynamical tachyonic AdS/QCD and information
  entropy},'' {\em Phys. Lett. B}, vol.~782, pp.~607--612, 2018.

\bibitem{Ma:2018wtw}
C.-W. Ma and Y.-G. Ma, ``{Shannon Information Entropy in Heavy-ion
  Collisions},'' {\em Prog. Part. Nucl. Phys.}, vol.~99, pp.~120--158, 2018.

\bibitem{LeThien:2019lxh}
Q.~Le~Thien and D.~E. Krause, ``{Spin-Independent Two-Neutrino Exchange
  Potential with Mixing and $CP$-Violation},'' {\em Phys. Rev. D}, vol.~99,
  no.~11, p.~116006, 2019.

\bibitem{Karapetyan:2020yhs}
G.~Karapetyan, ``{Baryon production probability via the nuclear configurational
  entropy},'' {\em Eur. Phys. J. Plus}, vol.~136, no.~1, p.~122, 2021.

\bibitem{daRocha:2021xwq}
R.~da~Rocha, ``{Holographic entanglement entropy, deformed black branes, and
  deconfinement in AdS/QCD},'' {\em Phys. Rev. D}, vol.~105, no.~2, p.~026014,
  2022.

\bibitem{Karapetyan:2021vyh}
G.~Karapetyan, ``{Total hadronic and photonic cross sections and the nuclear
  configurational entropy concept},'' {\em Eur. Phys. J. Plus}, vol.~136,
  no.~10, p.~1012, 2021.

\bibitem{Karapetyan:2021crv}
G.~Karapetyan, ``{Nuclear configurational entropy and high-energy hadron-hadron
  scattering reactions},'' {\em Eur. Phys. J. Plus}, vol.~137, no.~5, p.~590,
  2022.

\bibitem{Bernardini:2016hvx}
A.~E. Bernardini and R.~da~Rocha, ``{Entropic information of dynamical AdS/QCD
  holographic models},'' {\em Phys. Lett. B}, vol.~762, pp.~107--115, 2016.

\bibitem{Bernardini:2018uuy}
A.~E. Bernardini and R.~Da~Rocha, ``{Informational entropic Regge trajectories
  of meson families in AdS/QCD},'' {\em Phys. Rev. D}, vol.~98, no.~12,
  p.~126011, 2018.

\bibitem{Ferreira:2019inu}
L.~F. Ferreira and R.~Da~Rocha, ``{Pion family in AdS/QCD: the next generation
  from configurational entropy},'' {\em Phys. Rev. D}, vol.~99, no.~8,
  p.~086001, 2019.

\bibitem{daRocha:2021imz}
R.~da~Rocha, ``{Deploying heavier $\eta$ meson states: Configurational entropy
  hybridizing AdS/QCD},'' {\em Phys. Lett. B}, vol.~814, p.~136112, 2021.

\bibitem{Bernardini:2025wkb}
A.~E. Bernardini, W.~de~Paula, and R.~da~Rocha, ``{Spectroscopy of
  charmonium-like mesons, heavy-light mesons with charm, AdS/QCD, and
  configurational entropy},'' {\em Annals Phys.}, vol.~483, p.~170264, 2025.

\bibitem{Ferreira:2019nkz}
L.~F. Ferreira and R.~da~Rocha, ``{Tensor mesons, AdS/QCD and information},''
  {\em Eur. Phys. J. C}, vol.~80, no.~5, p.~375, 2020.

\bibitem{daRocha:2021ntm}
R.~da~Rocha, ``{Information entropy in AdS/QCD: Mass spectroscopy of isovector
  mesons},'' {\em Phys. Rev. D}, vol.~103, no.~10, p.~106027, 2021.

\bibitem{Braga:2017fsb}
N.~R.~F. Braga and R.~a. da~Rocha, ``{AdS/QCD duality and the quarkonia
  holographic information entropy},'' {\em Phys. Lett. B}, vol.~776,
  pp.~78--83, 2018.

\bibitem{MarinhoRodrigues:2020yzh}
D.~Marinho~Rodrigues and R.~da~Rocha, ``{Odd-spin glueballs, AdS/QCD and
  information entropy},'' {\em Phys. Lett. B}, vol.~811, p.~135943, 2020.

\bibitem{MarinhoRodrigues:2020ssq}
D.~Marinho~Rodrigues and R.~da~Rocha, ``{Configurational entropy and
  spectroscopy of even-spin glueball resonances in dynamical AdS/QCD},'' {\em
  Eur. Phys. J. Plus}, vol.~137, no.~4, p.~429, 2022.

\bibitem{Karapetyan:2021ufz}
G.~Karapetyan and R.~da~Rocha, ``{Configurational entropy of heavy-quark QCD
  exotica},'' {\em Eur. Phys. J. Plus}, vol.~136, no.~10, p.~993, 2021.

\bibitem{Ferreira:2020iry}
L.~F. Ferreira and R.~da~Rocha, ``{Nucleons and higher spin baryon resonances:
  An AdS/QCD configurational entropic incursion},'' {\em Phys. Rev. D},
  vol.~101, no.~10, p.~106002, 2020.

\bibitem{Guo:2024nrf}
X.~Guo, M.~A. Martin~Contreras, X.~Chen, and D.~Xiang, ``{Holographic bottom-up
  approach to {\ensuremath{\Sigma}} baryons*},'' {\em Chin. Phys. C}, vol.~49,
  no.~1, p.~013104, 2025.

\bibitem{Gubser:2008yx}
S.~S. Gubser, A.~Nellore, S.~S. Pufu, and F.~D. Rocha, ``{Thermodynamics and
  bulk viscosity of approximate black hole duals to finite temperature quantum
  chromodynamics},'' {\em Phys. Rev. Lett.}, vol.~101, p.~131601, 2008.

\bibitem{FolcoCapossoli:2016uns}
E.~Folco~Capossoli, D.~Li, and H.~Boschi-Filho, ``{Dynamical corrections to the
  anomalous holographic soft-wall model: the pomeron and the odderon},'' {\em
  Eur. Phys. J. C}, vol.~76, no.~6, p.~320, 2016.

\bibitem{Chen:2022pgo}
R.~Chen, D.~Li, K.~Bitaghsir~Fadafan, and M.~Huang, ``{Hadron spectra and pion
  form factor in dynamical holographic QCD model with anomalous 5D mass of
  scalar field*},'' {\em Chin. Phys. C}, vol.~47, no.~6, p.~063106, 2023.

\bibitem{daRocha:2024sjn}
R.~da~Rocha and P.~H.~O. Silva, ``{Deformed AdS/QCD, mesonic mass spectra, and
  the differential configurational entropy: Still a margin for heavier
  resonances},'' {\em Phys. Rev. D}, vol.~110, no.~12, p.~126019, 2024.

\bibitem{greiner1996gauge}
W.~Greiner and B.~M{\"u}ller, {\em Gauge Theory of Weak Interactions}.
\newblock Theoretical physics, Springer, 1996.

\bibitem{Ilner:2016xqr}
A.~Ilner, D.~Cabrera, C.~Markert, and E.~Bratkovskaya, ``{$K^{*}$ vector meson
  resonances dynamics in heavy-ion collisions},'' {\em Phys. Rev. C}, vol.~95,
  no.~1, p.~014903, 2017.

\bibitem{Rafelski:2001hp}
J.~Rafelski, J.~Letessier, and G.~Torrieri, ``{Strange hadrons and their
  resonances: A Diagnostic tool of QGP freezeout dynamics},'' {\em Phys. Rev.
  C}, vol.~64, p.~054907, 2001.
\newblock [Erratum: Phys.Rev.C 65, 069902 (2002)].

\bibitem{Markert:2002rw}
C.~Markert, G.~Torrieri, and J.~Rafelski, ``{Strange Hadron Resonances:
  Freeze-Out Probes in Heavy-Ion Collisions},'' {\em AIP Conf. Proc.},
  vol.~631, no.~1, p.~533, 2002.

\bibitem{ALICE:2014jbq}
B.~B. Abelev {\em et~al.}, ``{$K^*(892)^0$ and $\phi(1020)$ production in Pb-Pb
  collisions at $\sqrt{s{NN}}$ = 2.76 TeV},'' {\em Phys. Rev. C}, vol.~91,
  p.~024609, 2015.

\bibitem{ALICE:2021xyh}
S.~Acharya {\em et~al.}, ``{Measurement of K$^{*}$(892)$^{\mathrm{\pm}}$
  production in inelastic pp collisions at the LHC},'' {\em Phys. Lett. B},
  vol.~828, p.~137013, 2022.

\bibitem{Li:2019pzx}
Y.~Li, Z.~Rui, and Z.-J. Xiao, ``{$P$-wave contributions to $B_{(s)}\to\psi
  K\pi$ decays in perturbative QCD approach},'' {\em Chin. Phys. C}, vol.~44,
  no.~7, p.~073102, 2020.

\bibitem{Li:2018psm}
Y.~Li, W.-F. Wang, A.-J. Ma, and Z.-J. Xiao, ``{Quasi-two-body decays
  $B_{(s)}\to K^*(892)h\to K\pi h$ in perturbative QCD approach},'' {\em Eur.
  Phys. J. C}, vol.~79, no.~1, p.~37, 2019.

\bibitem{Chen:2018nnr}
J.-K. Chen, ``{Regge trajectories for the mesons consisting of different
  quarks},'' {\em Eur. Phys. J. C}, vol.~78, no.~8, p.~648, 2018.

\bibitem{Chen:2018bbr}
J.-K. Chen, ``{Concavity of the meson Regge trajectories},'' {\em Phys. Lett.
  B}, vol.~786, pp.~477--484, 2018.

\bibitem{Chen:2018hnx}
J.-K. Chen, ``{Regge trajectories for heavy quarkonia from the quadratic form
  of the spinless Salpeter-type equation},'' {\em Eur. Phys. J. C}, vol.~78,
  no.~3, p.~235, 2018.

\bibitem{Gershtein:2006ng}
S.~S. Gershtein, A.~K. Likhoded, and A.~V. Luchinsky, ``{Systematics of heavy
  quarkonia from Regge trajectories on (n,M**2) and (M**2,J) planes},'' {\em
  Phys. Rev. D}, vol.~74, p.~016002, 2006.

\bibitem{MartinContreras:2020cyg}
M.~A. Martin~Contreras and A.~Vega, ``{Nonlinear Regge trajectories with
  AdS/QCD},'' {\em Phys. Rev. D}, vol.~102, no.~4, p.~046007, 2020.

\bibitem{daRocha:2023nsb}
R.~da~Rocha and P.~H.~O. Silva, ``{Configurational entropy and shape complexity
  of strange vector kaons in AdS/QCD},'' {\em Eur. Phys. J. Plus}, vol.~138,
  no.~8, p.~729, 2023.

\bibitem{Susskind:2005js}
L.~Susskind and J.~Lindesay, {\em {An introduction to black holes, information
  and the string theory revolution: The holographic universe}}.
\newblock 2005.

\bibitem{Hawking:1982dh}
S.~W. Hawking and D.~N. Page, ``{Thermodynamics of Black Holes in anti-De
  Sitter Space},'' {\em Commun. Math. Phys.}, vol.~87, p.~577, 1983.

\bibitem{daRocha:2024lev}
R.~da~Rocha, ``{Deformations of the AdS{\textendash}Schwarzschild black brane
  and the shear viscosity of the quark{\textendash}gluon plasma},'' {\em Eur.
  Phys. J. Plus}, vol.~139, no.~11, p.~1006, 2024.

\bibitem{Toniato:2025gts}
B.~Toniato, D.~Dudal, S.~Mahapatra, R.~da~Rocha, and S.~S. Jena, ``{Holographic
  QCD model for heavy and exotic mesons at finite density: A self-consistent
  dynamical approach},'' {\em Phys. Rev. D}, vol.~111, no.~12, p.~126021, 2025.

\bibitem{Fukushima:2010bq}
K.~Fukushima and T.~Hatsuda, ``{The phase diagram of dense QCD},'' {\em Rept.
  Prog. Phys.}, vol.~74, p.~014001, 2011.

\bibitem{Hagedorn:1965st}
R.~Hagedorn, ``{Statistical thermodynamics of strong interactions at
  high-energies},'' {\em Nuovo Cim. Suppl.}, vol.~3, pp.~147--186, 1965.

\bibitem{Rafelski:2015cxa}
J.~Rafelski, ``{Melting Hadrons, Boiling Quarks},'' {\em Eur. Phys. J. A},
  vol.~51, no.~9, p.~114, 2015.

\bibitem{Witten:1998zw}
E.~Witten, ``{Anti-de Sitter space, thermal phase transition, and confinement
  in gauge theories},'' {\em Adv. Theor. Math. Phys.}, vol.~2, pp.~505--532,
  1998.

\bibitem{Son:2007vk}
D.~T. Son and A.~O. Starinets, ``{Viscosity, Black Holes, and Quantum Field
  Theory},'' {\em Ann. Rev. Nucl. Part. Sci.}, vol.~57, pp.~95--118, 2007.

\bibitem{Rangamani:2009xk}
M.~Rangamani, ``{Gravity and Hydrodynamics: Lectures on the fluid-gravity
  correspondence},'' {\em Class. Quant. Grav.}, vol.~26, p.~224003, 2009.

\bibitem{Critelli:2017oub}
R.~Critelli, J.~Noronha, J.~Noronha-Hostler, I.~Portillo, C.~Ratti, and
  R.~Rougemont, ``{Critical point in the phase diagram of primordial
  quark-gluon matter from black hole physics},'' {\em Phys. Rev. D}, vol.~96,
  no.~9, p.~096026, 2017.

\bibitem{Rougemont:2017tlu}
R.~Rougemont, R.~Critelli, J.~Noronha-Hostler, J.~Noronha, and C.~Ratti,
  ``{Dynamical versus equilibrium properties of the QCD phase transition: A
  holographic perspective},'' {\em Phys. Rev. D}, vol.~96, no.~1, p.~014032,
  2017.

\bibitem{dePaula:2020bte}
W.~de~Paula, C.-R. Ji, J.~P. B.~C. de~Melo, T.~Frederico, and
  O.~Louren{\c{c}}o, ``{The holographic paradigm of hadron dynamics for medium
  modified nuclear matters},'' {\em Phys. Lett. B}, vol.~803, p.~135339, 2020.

\bibitem{Herzog:2006ra}
C.~P. Herzog, ``{A Holographic Prediction of the Deconfinement Temperature},''
  {\em Phys. Rev. Lett.}, vol.~98, p.~091601, 2007.

\bibitem{BallonBayona:2007vp}
C.~A. Ballon~Bayona, H.~Boschi-Filho, N.~R.~F. Braga, and L.~A. Pando~Zayas,
  ``{On a Holographic Model for Confinement/Deconfinement},'' {\em Phys. Rev.
  D}, vol.~77, p.~046002, 2008.

\bibitem{MartinContreras:2021bis}
M.~A. Martin~Contreras, S.~Diles, and A.~Vega, ``{Heavy quarkonia spectroscopy
  at zero and finite temperature in bottom-up AdS/QCD},'' {\em Phys. Rev. D},
  vol.~103, no.~8, p.~086008, 2021.

\bibitem{Afonin:2020crk}
S.~S. Afonin, ``{A holographic relation between the deconfinement temperature
  and gluon condensate},'' {\em Phys. Lett. B}, vol.~809, p.~135780, 2020.

\bibitem{HotQCD:2018pds}
A.~Bazavov {\em et~al.}, ``{Chiral crossover in QCD at zero and non-zero
  chemical potentials},'' {\em Phys. Lett. B}, vol.~795, pp.~15--21, 2019.

\bibitem{Borsanyi:2020fev}
S.~Borsanyi, Z.~Fodor, J.~N. Guenther, R.~Kara, S.~D. Katz, P.~Parotto,
  A.~Pasztor, C.~Ratti, and K.~K. Szabo, ``{QCD Crossover at Finite Chemical
  Potential from Lattice Simulations},'' {\em Phys. Rev. Lett.}, vol.~125,
  no.~5, p.~052001, 2020.

\bibitem{Ong:2015fha}
Y.~C. Ong, ``{Hawking Evaporation Time Scale of Topological Black Holes in
  Anti-de Sitter Spacetime},'' {\em Nucl. Phys. B}, vol.~903, pp.~387--399,
  2016.

\bibitem{Braga:2022yfe}
N.~R.~F. Braga, L.~F. Faulhaber, and O.~C. Junqueira,
  ``{Confinement-deconfinement temperature for a rotating quark-gluon
  plasma},'' {\em Phys. Rev. D}, vol.~105, no.~10, p.~106003, 2022.

\bibitem{Braga:2023qej}
N.~R.~F. Braga and O.~C. Junqueira, ``{Inhomogeneity of a rotating quark-gluon
  plasma from holography},'' {\em Phys. Lett. B}, vol.~848, p.~138330, 2024.

\bibitem{Braga:2023qee}
N.~R.~F. Braga, L.~F. Ferreira, and O.~C. Junqueira, ``{Configuration entropy
  of a rotating quark-gluon plasma from holography},'' {\em Phys. Lett. B},
  vol.~847, p.~138265, 2023.

\bibitem{Braga:2024nnj}
N.~R.~F. Braga and O.~C. Junqueira, ``{Hawking-Page transition in holographic
  QCD at finite density},'' {\em Phys. Lett. B}, vol.~855, p.~138813, 2024.

\bibitem{Braga:2025eiz}
N.~R.~F. Braga and O.~C. Junqueira, ``{Holographic QCD phase diagram for a
  rotating plasma in the Hawking-Page approach},'' {\em Phys. Lett. B},
  vol.~868, p.~139669, 2025.

\bibitem{Junqueira:2025xnx}
O.~C. Junqueira and R.~da~Rocha, ``{Confinement/deconfinement at low
  temperatures and rotation in the exact soft wall model},'' 7 2025.

\bibitem{Vega:2018dgk}
A.~Vega and M.~A. Martin~Contreras, ``{Melting of scalar hadrons in an AdS/QCD
  model modified by a thermal dilaton},'' {\em Nucl. Phys. B}, vol.~942,
  pp.~410--418, 2019.

\bibitem{Gutsche:2019blp}
T.~Gutsche, V.~E. Lyubovitskij, I.~Schmidt, and A.~Y. Trifonov, ``{Mesons in a
  soft-wall AdS-Schwarzschild approach at low temperature},'' {\em Phys. Rev.
  D}, vol.~99, no.~5, p.~054030, 2019.

\bibitem{Gutsche:2019pls}
T.~Gutsche, V.~E. Lyubovitskij, I.~Schmidt, and A.~Y. Trifonov, ``{Baryons in a
  soft-wall AdS-Schwarzschild approach at low temperature},'' {\em Phys. Rev.
  D}, vol.~99, no.~11, p.~114023, 2019.

\bibitem{Gasser:1986vb}
J.~Gasser and H.~Leutwyler, ``{Light Quarks at Low Temperatures},'' {\em Phys.
  Lett. B}, vol.~184, pp.~83--88, 1987.

\bibitem{Branz:2010ub}
T.~Branz, T.~Gutsche, V.~E. Lyubovitskij, I.~Schmidt, and A.~Vega, ``{Light and
  heavy mesons in a soft-wall holographic approach},'' {\em Phys. Rev. D},
  vol.~82, p.~074022, 2010.

\bibitem{Witten:1979kh}
E.~Witten, ``{Baryons in the 1/N Expansion},'' {\em Nucl. Phys. B}, vol.~160,
  pp.~57--115, 1979.

\bibitem{Witten:1998xy}
E.~Witten, ``{Baryons and branes in anti-de Sitter space},'' {\em JHEP},
  vol.~07, p.~006, 1998.

\bibitem{Brown:2010api}
G.~E. Brown and M.~Rho, eds., {\em {The multifaceted skyrmion}}.
\newblock Singapore: World Scientific, 2010.

\bibitem{Rebhan:2014rxa}
A.~Rebhan, ``{The Witten-Sakai-Sugimoto model: A brief review and some recent
  results},'' {\em EPJ Web Conf.}, vol.~95, p.~02005, 2015.

\bibitem{daRocha:2025gcz}
R.~da~Rocha and P.~H.~O. Silva, ``{Higher-spin light-flavor baryonic
  spectroscopy in AdS/QCD at finite temperature},'' {\em Phys. Lett. B},
  vol.~870, p.~139917, 2025.

\end{thebibliography}

		}
		
		\printindex
	
\end{document}